# Development of Tools for the Classification of Peer Groups Geographies in the Analysis of Health Care Variation

by

Ludovico Pinzari

A DISSERTATION

submitted in partial fulfillment of the requirements for the degree

DOCTOR OF PHILOSOPHY

Translational Health Research Institute

University of Western Sydney

May 2019

# Copyright

# Acknowledgements

A PhD thesis is a highly personal achievement, but it would never be possible without interaction and support of other people. First of all, I would like to thank my supervisor, Dr. Federico Girosi, who has been an excellent supervisor throughout my studies. I would also like to thank my family, friends and other colleagues who have helped me through the process. I am especially grateful to my family in Italy, who supported me emotionally. Thank you to my parents for encouraging me in all of my pursuits and inspiring me to follow my dreams.

This research was co-founded by the Capital Markets CRC and the Australian Institute of Health and Welfare, and I would like to thank both organizations for their generous support.

**STATEMENT OF AUTHENTICATION**

The work presented in this thesis is, to the best of my knowledge and belief, original except as acknowledged in the text. I hereby declare that I have not submitted this material, either in full or in part, for a degree at this or any other institution.

Ludovico Pinzari

May 2019

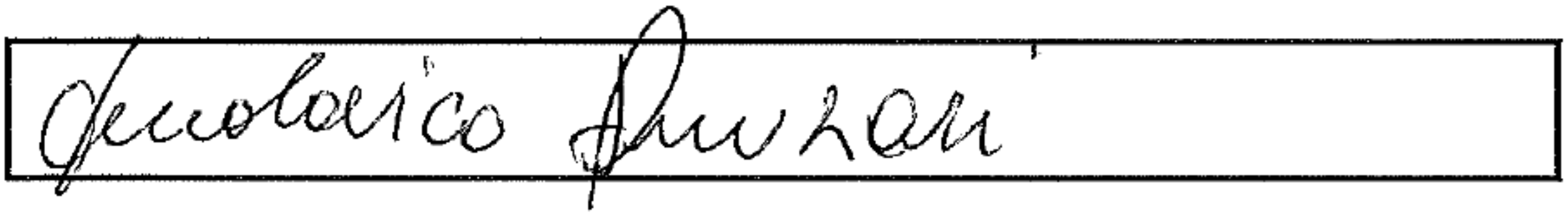

# Preface

This dissertation is based on a project co-founded by the Capital Markets CRC (CMCRC) and the Australian Institute of Health and Welfare (AIHW).

The scope of the project undertaken is the development of methods and geo-spatial software tools for use in reporting the variation of health-related indicators.

Studying geographic variation of health indicators can provide useful insight in many aspects of the health care system. Insights are derived by comparing indicators across similar regions that are socially and economically homogeneous, that is they represent areas with similar determinants of health needs. Normally government agencies use a variety of geographies to report performance indicators. However, the selection of these units is generally based on criteria that do not necessary reflect socioeconomic homogeneous areas and may not relate to health. As a response to these issues, I developed a method and a software tool to enable the presentation and reporting of comparable information of individual units with peers. The tool is the first and only application of its kind that retrieves population counts and features of a geographic area and allows users to compare and group geographies with similar characteristics.

The work has been presented at the Wennberg International Collaborative Conference: "Better Smarter Care Reducing Unwarranted Variation" hold in Melbourne on 3-4 May 2017. The most relevant conference on Health Care Variation. The presentation "Development of Tools for the Analysis of Geographic Variation and Classification of Geographic Peer Group" is available on the Wennberg web site (Wennberg International Collaborative, 2017), and a brief introduction to the use of the Tool is given at the end of this dissertation.

The research presented in this dissertation is partially based on the work delivered to AIHW and a publication in the International Journal of Health Geographics: "A framework for the identification and classification of homogeneous areas in the analysis of health care variation" (Pinzari, Mazumdar,

& Girosi, 2018). This work also informed the publication "A new generation of Primary Care Service Areas or general practice catchment areas", published in the Journal Transaction in GIS (Mazumdar, et al., 2017), and contributed to the development of a tool for designing customizable catchment areas. Part of this work was also presented at the 10th HSRAANZ Health Services & Policy Research Conference held in Gold Coast (Pinzari, 2017).

# Abstract

Detecting the variation of health indicators across similar areas or peer geographies is often useful if the spatial units are socially and economically meaningful, so that there is a degree of homogeneity in each unit. Indices are frequently constructed to generate summaries of socioeconomic status or other measures in geographic small areas. Small areas may be built to be homogenous using regionalization algorithms. However, there are no explicit guidelines in the literature for the grouping of peer geographies based on measures such as area level socioeconomic indices. Moreover, the use of an index score becomes less meaningful as the size of an area increases. This thesis introduces an easy to use statistical framework for the identification and classification of homogeneous areas. I propose the Homogeneity and Location indices to measure the concentration and central value respectively of an areas' socioeconomic distribution. I also provide a transparent set of criteria that a researcher can follow to establish whether a set of proposed geographies are acceptably homogeneous or need further refining.

I applied our framework to assess the socioeconomic homogeneity of the commonly used SA3 Australian census geography. These results showed that almost 60% of the SA3 census units are likely to be socioeconomically heterogeneous and hence inappropriate for presenting area level socioeconomic disadvantage. I also showed that the Location index is a more robust descriptive measure of the distribution compared to other measures of central tendency. Finally, the methodology proposed was used to analyse the age-standardized variation of GP attenders in a metro area. The results suggest that very high GP attenders (20+ visits) live in SA3s with the most socioeconomic disadvantage. The findings revealed that households with low income and families with children and jobless parents are the major drivers for discerning disadvantaged communities.

Reporting indicators rates for small area geographies grouped according to similarity may be useful for the analysis of geographic variation. The use of a framework for the identification of meaningful peer geographies would be beneficial to health planners and policy makers by providing realistic and valid peer group geographies.

# Table of Contents

# List of Figures

# List of Tables

# Chapter 1 - Introduction

## 1.1 – Background

The selection of an appropriate geographic unit of analysis is a key decision for analysing and interpreting the geographic variation of health-related indicators (Thygesen, et al., 2015), (M.Kim, Park, Kang, & Kim, 2016). It is commonly agreed that the unwarranted variation is not associated with the spatial effects of the geographic unit (Macintyre S. , 1997), and detecting the variation across geographic areas is often useful only if the units have similar characteristics. Therefore, it is important to define what it means for two geographic units to be comparable.

Take for example an indicator such as being a very high and frequent GP attender (12+ GP visits). In 2012-2013 there were 2.9 million (12.5 %) Australians in this category. These people were more likely to be older and live in rural areas with the most socioeconomic disadvantage (NHPA, Health Communities: Frequent GP attenders and their use of health services in 2012-13 (pp. 11,22), March 2015). In terms of socioeconomic status, for example, very high GP attenders (20+ GP visits) were almost twice as likely as low GP attenders (1-3 GP visits) to have lived in areas with the most socioeconomic disadvantage (29 % compared to 16 %). To adjust for these demographic and socioeconomic differences standardization for age and socioeconomic status is routinely undertaken to eliminate legitimate or warranted variations (Aylin, William, Jerom, & Bottle, 2005). However, these types of socioeconomic adjustments may not always be sufficient (Appleby, Raleigh, Frosini, Beven, & Geo, 2011) and the analysis can be hampered by the fact that geographic units of analysis are heterogeneous along many dimensions.

One approach to reduce the heterogeneity of the study area, while preserving the meaningfulness of the units of analysis, is to adopt a census geography which includes spatial units which are socially and economically meaningful so that there is a degree of homogeneity in each unit. The central idea behind this approach is that area effects on health have been observed to be stronger in more homogeneous areas (Haynes, Daras, Reading, & Jones, 2007) (Riva, Appericio, Gauvin, & Brodeun, 2008). A high level of homogeneity among people and households within each area on a given area level index, results in a strong relationship between that area level index and individual level indices (Wise & Matthews, 2011) (Baker & Adhikari, 2007). Accordingly, epidemiologists and geographers have argued that units with greater social homogeneity would be more appropriate for studying the association between unit characteristics and a given health indicator (Haynes, Lovett, Reading, Langford, & Gale, 1999). This property makes them suitable for interpreting the variation across similar units.

Following this approach, it is possible to remove the spatial effect of the factors which might influence health by presenting and reporting the variation of an indicator amongst units that have similar characteristics, better known as peer groups (NHPA, Health Communities: Australian's experiences with primary health care in 2010-11, p 9, 2013) (Canada, 2002). Here the emphasis is not necessarily on the explanation of the variation, but rather on producing a reliable picture of the variation in health needs across an area, allowing for the variation in other factors such as age and socioeconomic status. Age and socioeconomic status are the two standard confounders. Rates are usually adjusted for age. This leaves socioeconomic status, which may be accounted for by grouping similar status spatial units into peer groups. This enables a fairer comparison of individual units with peers and allows variation to be studied across peer groups that may be associated with socioeconomic status. Accordingly, reporting indicators rates by similar socioeconomically graded areas of residence provides a useful way to analyse health care variation.

In this context, one of the major challenges in choosing a suitable geographic unit relates to an adequate compromise between having a unit large enough to get stable indicator rates and not blurring meaningful local variation (Gregorio, Dechello, Samociuk, & Kulldorff, 2005), while preserving the homogeneous characteristics of the residential population (Osypuk & Galea, 2007) (Arsenault, Michel, Berke, Ravel, & Gosselin, 2013).

Unfortunately, the key drawback to the use of a census geography as a reporting unit is that the spatial scale of the indicator is not necessarily appropriate for analysing the characteristics of the population. This heterogeneity may affect the estimation of area effects and thereby the identification of the unexplained variation in geographic studies. As Boyle and Willms (Boyle & Willms, 1999) suggest place effects associated with large and heterogeneous administrative areas are relatively small and scientists should be cautious about using administrative boundaries as unit of analysis in their studies.

Therefore, the comparison of variation across a peer group should be accompanied by an evaluation of the degree to which those geographic units are internally homogeneous. For this reason, it is always recommended that the homogeneity of a given geography should be evaluated prior to any analysis (P.Boscoe & W.Pinckle, 2003) (Osypuk & Galea, 2007) (Riva, Appericio, Gauvin, & Brodeun, 2008).

Despite the relevance of this issue few researchers have attempted to assess the homogeneity of a geographic area (Duncan, Jones, & Moon, 1998) (Tranmer & Steel, 1998). Those methods helpfully introduced the concept of homogeneity but also have some weakness. They mostly rely on individual level data and lack a statistical framework which will assist the analyst with the subjective decision to take about a not homogeneous unit. However, not all jurisdictions have data available at the individual level. In this circumstance, one cannot determine how well the use of area-based indicators may appropriately measure social conditions and identify areas of social disadvantage which may be associated with poor health.

Another complexity in dealing with homogeneity measures found in the literature is their abstract quality. It is often not clear, for example, what meaning to give to a "homogeneous area" in terms of the socioeconomic characteristics of that area. For instance, the norm is to consider that if a particular area has, on average, a greater proportion of people with a relevant measure of disadvantage, then that area may be considered as disadvantaged. The choice of what proportion is an appropriate cut-off is difficult and often not well justified. Therefore, a critical decision is the selection of an appropriate socioeconomic classification and homogeneity measure when the data include categorical ordinal variables such as socioeconomic status.

Looking at how healthcare use varies across people living in different areas, across people with and without socioeconomic disadvantage, can show who in our community is missing out.

Variation in itself is not necessarily bad, and it can be good if it reflects health services responding to differences in patients' preferences or underlying needs.

In Australia and USA, for instance, the socioeconomic and demographic differences account for almost one-fourth of the variation in Medicare spending across region (NHPA, Health Communities: Frequent GP attenders and their use of health services in 2012-13 (pp. 11,22), March 2015) (Wennberg, Fisher, & Skinner, 2004) and are strongly correlated with the risk of hospitalization for ambulatory care sensitive conditions (NHPA, Health Communities: Potentially preventable hospitalisations in 2013-2014, 2015) (Pappas, Hadden, Kozak, & Fisher, 1997).

However, when a difference in the use of health services does not reflect these socioeconomic factors, it is unwanted variation, also known as unwarranted variation, and represents an opportunity for the health system to improve. For instance, it may reflect the fact that some people have less access to health care compared with others. It may also indicate that some people are having unnecessary and potentially harmful tests or treatments, while others are missing out on necessary interventions. Unwarranted variation may also mean that scarce health resources are not being put to best use. Thus, these insights are derived by comparing indicators across similar regions that are socially and economically homogeneous.

Currently there is no commercial or open-source software tool and algorithm that can be used to divide a set of geographies into groups with similar characteristics or “peer groups”.

As a result, AIHW entered in a collaborative agreement with CMCRC to develop methods and geo-spatial software tools for use in reporting on the performance of the health care system.

For all these reasons, it is necessary to define homogeneity precisely in order to determine how it should be operationalised and measured.

## 1.2 – Objectives

The overall objective of this work is to provide a framework and a software tool for the identification and classification of homogeneous geographic areas. As a result, this thesis addresses six key research questions:

1. How can we measure the homogeneity of a geographic area only by looking at the population distribution of area-based measures?
2. What criteria can be used to guide the selection of a homogeneity measure?
3. How can we formalize the notion of homogeneous area and how this relates to the selection of the geographic unit of analysis?
4. Which criteria and measure of central tendency should be used to classify a homogeneous area?
5. Can we use a combination of a homogeneity and central tendency measure to identify peer groups better?

6. How to design geo-spatial tools for reporting health-outcomes?

The main scope of this work is therefore to provide a suite of classification tools for the identification and classification of homogeneous geographic areas. This set of analyses use the Australian census geography, but the approach can be used to classify any geographic area and explore variation across any specified geographical boundaries.

Thus, to enable the presentation and reporting of comparable information of individual units with peers, I develop the Homogeneity and Location index, hereafter referred to as "HI" and "LI", to measure respectively the dispersion and central tendency of a probability distribution. The advantages of such indices include statistical efficiency and a simple presentation of results.

Finally, the combination of the HI and LI constitutes a clear and consistent framework for geographic variation studies. Such a framework can be useful to assist in the targeted delivery of health services by identifying areas where to direct efforts to deal with unwarranted variation. It can help investigators determine which variables should be included in their studies, such that sources of unwarranted variations may be identified.

## 1.3 – Thesis outline

This dissertation is organized as follows:

Chapter 2 address the first research question of this thesis.

In this chapter I briefly summaries the literature on measuring socioeconomic inequalities in health. I then discuss how the design of a homogeneity measure might be accomplished with a brief description of the issues that need to be considered when the data includes categorical ordinal variables.

Chapter 3 addresses the following research questions: second (section 3.1), third (section 3.2) and fourth (section 3.3).

In this chapter I introduce the HI as a new concentration measure of a probability distribution and the specifications used to evaluate the homogeneity degree of the spatial unit of analysis. Then, I propose the LI as a location parameter to classify the socioeconomic distribution of a geographic area.

Chapter 4 addresses the fifth research question (section 4.2).

In this chapter both LI and the HI will be implemented to illustrate how they can be used to better find homogeneous geographies and interpreting the variation of health indicators across peer groups.

Chapter 5 addresses the sixth research question.

In this chapter I present the main functionalities of the software Tool:SA3 Browser.

In chapter 6 and 7 I discuss the results presented in chapter 4 and summarize the most important conclusions.

# Chapter 2 - Literature review

In health geographic studies, the choice of the census unit is influenced by many factors that may act alone or in conjunction with others. Although some of the work on “place effects” on health has been undertaken by geographers and epidemiologist (Ellen, Mijanovich, & Dillman, 2001) (Pickett & Pearl, 2001), the literature does not give much guidance on what these might be and, if any, how they can be operationalised and measured (Macintyre, Ellaway, & Cummins, Place effects on health: how can we conceptualise, operationalise and measure them?, 2002). Complicating the selection of the census geography further is the fact that there are limited number of spatial scales to choose, and sometimes the analysis of several features may restrict their use.

To overcome all these limitations, the Australian National Health Performance Framework (NHPF) (AIHW, Meteor National Health system Performance, 2009) identifies at least three key dimensions to a geographic area that are quite distinct from each other and which are all important for policy-maker to take into account when they are allocating resources or designing programmes. These are: the demographic, socio-economic conditions and geographic remoteness. All three dimensions capture significantly different aspects of the geographic context which might influence health. These dimensions, however, need to be recognized and measured individually prior any more sophisticated analysis.

In recognizing the significance of contextual factors in influencing health related indicators, we need to become clear in the way that we specify context as an explanatory variable. We are particularly interested here in problems of measuring and operationalising the socioeconomic context.

## 2.1 – Measuring socioeconomic inequalities in health geographic studies

There is a growing literature documenting the effect of residential characteristics on health care variation (Curtis & jones, 1998). Many studies, particularly those who use census data, have shown that people living in areas with lower socio-economic status tend to have higher utilization rates of health care services and poorer health status (Reijneveld, Verheij, & Bakker, 2000). Implicit in these studies is that populations living in areas of higher social status are more likely to be at a lower risk of ill health. This phenomenon is observed in many industrialised and developing countries and is known as the socioeconomic-health status (SHS) gradient (Alder, et al., 1994) (Marmot, 2006).

In Australia and US, for instance, these differences accounts for almost one-fourth of the variation in Medicare spending across regions (NHPA, Health Communities: Frequent GP attenders and their use of health services in 2012-13 (pp. 11,22), March 2015) (Wennberg, Fisher, & Skinner, 2004), and are strongly correlated with the risk of hospitalization for ambulatory care sensitive conditions (NHPA, Health Communities: Potentially preventable hospitalisations in 2013-2014, 2015) (Pappas, Hadden, Kozak, & Fisher, 1997). Accordingly, measuring the size of the health gap between different social groups is important. This provides essential information for policies, programs and practices which seek to address social determinants in order to reduce health gaps (Harper & Lynch, 2016).

A common approach to measurement is to:

1. Rank the residential area by socioeconomic position.
2. Divide the residential areas into groups based on this ranking.
3. Compare each group on health indicators of interest.

To rank the residential area by socioeconomic position, factors such as population income, occupation or education are commonly used, although many other factors, such as housing, family structure or access to resources can also be used. These elements by themselves, however, do not provide an ideal measure, and their use often depends on type of health outcome in question (Geyer, Hemstrom, Peter, & Vagero, 2006). Therefore, these factors should not be used interchangeably (Macintyre, MacKay, Der, & Hiscock, 2003), because they may relate to different causal processes. This phenomenon is commonly referred to as status inconsistency in the sociological literature (Lenski, 1954). It has been shown that status inconsistency carries its own health risks (Faresjo, Svardsudd, & Tibblin, 1997). Accordingly, a social gradient measure that takes income, education, occupational class and other disadvantage dimensions into account simultaneously effectively ignores this hypothesis and is applicable to different health outcomes.

## 2.2 Deprivation indices

Deprivation indices are frequently constructed to generate summary of the socioeconomic status of residential areas. Most deprivation indices are calculated for geographic areas, rather than individuals, and they are based on census data. A number of indices have been devised for health research over the years (Fotso & Kuate-defo, 2005) (Krieger, et al., 2002) (Singh, 2003), including the notably Index of Multiple Deprivation, for identifying the most deprived areas in England (Smith, et al., 2015), and the Townsend's index designed to explain variation in health in terms of material deprivation (Morris & Carstairs, 1991)Another example is the Australian Bureau of Statistics (ABS) composite Index of Relative Socio-economic Disadvantage (IRSD), which is frequently used to stratify the population.

### 2.2.1 The Index of Relative Socioeconomic Disadvantage

The IRSD is one of four indices compiled by the ABS using information collected in the Census of Population and Housing (Pink, Socio-economic Indexes For Areas (SEIFA), 2011). The association between the IRSD and health has been widely research in the literature and by ABS (Cass, Cunningham, Wang, & Hoy, 2001) (Mather, et al., 2014) (ABS, Socio-Economic Indexes For Area Technical Paper, cat no 2039.0.55.001, 2008).

The IRSD scores each area by summarizing attributes of their populations, such as low income, low education attainment, high unemployment, and jobs in relatively unskilled occupations. A relatively low score on this index indicates the area is relatively more disadvantaged, and a relatively high score indicates the area is relatively less disadvantaged. For example, an area could have a low score if there are (among other things) many households with low income, many people with low qualifications or many people in low skill occupation. On the other hand, a high score on the IRSD occurs when the area has few households with low income and few people in unskilled occupation. The variables included in the index are listed in APPENDIX-F. For more information on how the IRSD's score is derived, the interested reader can refer to the ABS Technical Paper: Socio-economic Indexes for areas 2011, 2033.0.55.001.

For ease of interpretation, areas can be ranked by their IRSD score and are classified into groups (quantiles) based on their rank. Commonly used quantiles include decile and quintiles; which quantile to use in analysis depends to the specific geography. In this work, the IRSD score was classified into ten groups or decile.

In this case the IRSD commonly describes the low end of the scale – the first decile – as the population living in the 10% of areas with the greatest overall level of disadvantage or the "lowest socioeconomic group". The 10% at the other end of the scale – the top tenth – is described as the "living in the highest socioeconomic areas" or the "highest socioeconomic group".

The justification of this choice is that the first decile has a large spread of scores compared to the other deciles. This is because the index contains only disadvantaged indicators, so there is more scope to distinguish between disadvantaged areas than least disadvantaged areas. This means that using quintiles, if there is specific interest in identifying disadvantaged areas, the likelihood to mask the characteristics of larger geographies is high. The discriminating power of this index lies, therefore, in the lower and upper end of the distribution for identifying the relative disadvantage (lower decile), and the relative lack of disadvantage (upper decile) of people in an area. Then, if one's prime focus is on disadvantage and lack of disadvantage, one might choose to use the IRSD for targeting disadvantaged areas that require funding and services.  (ABS, Socio-Economic Indexes for Areas (SEIFA) Technical Paper, cat no 2033.0.55.001, 2011). Consequently, the IRSD is the primary socioeconomic proxy measure used to report the variation of health indicators by national departments and agencies (DOH, 2010) (AIHW, Meteor National Health system Performance, 2009).

To facilitate analysis, in 2011, ABS released the IRSD scores and rankings at the Statistical Area level 1 (SA1) with the corresponding decile and percentile distribution. The SA1 is the geography adopted in the dissemination of most census data and the smallest geographic level for which the Estimated Residential Population (ERP) count is available. In addition to these units the ABS also provide data at higher geographic level (i.e. SA2 and SA3) through clustering groups of SA1s. The structure of the Australian Geographical Classification (ASGC) is shown in Figure 1.The ASGC Structures are split into two broad groups, the ABS Structures (shown on the left-hand side of Figure 1) and the Non-ABS Structures (shown on the right-hand side of Figure 1).

The Non-ABS Structures are hierarchies of regions which are not defined or maintained by the ABS, but for which the ABS is committed in providing a range of statistics. They generally represent administrative units such as Postcode and Local Government Areas.

The ABS Structures are hierarchies of regions defined and maintained by the ABS. They comprise six interrelated hierarchies. They are: Main Structure; Indigenous Structure; Urban Centres and Localities

Section of State Structure; Remoteness Area Structures; Greater Capital City Statistical Area (GCSA) Structure; Significant Urban Area Structure.

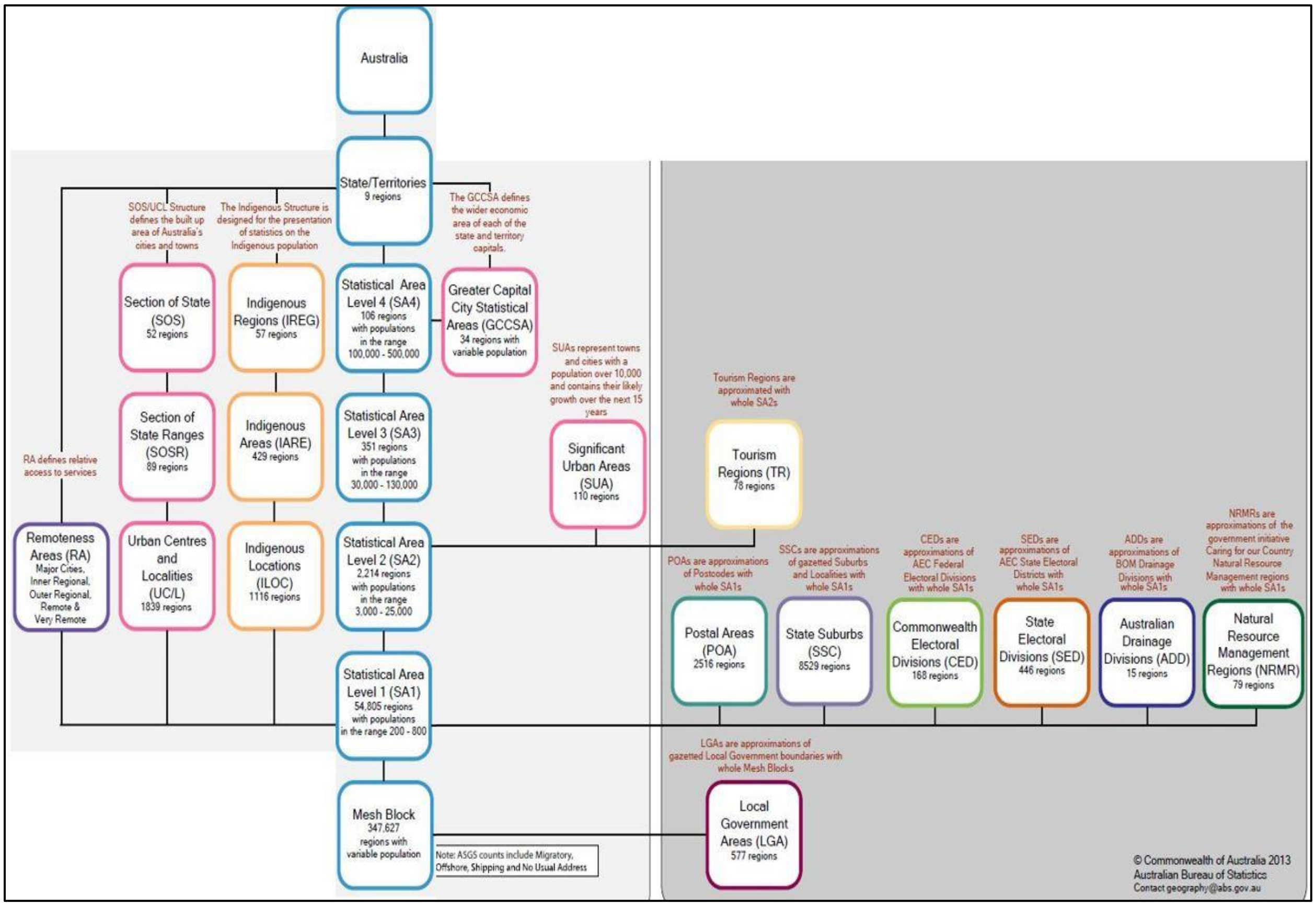

Figure 1 Australian Statistical Geography Standard (ASGS): structure and summary for census units.

The Main Structure of the ASGS is used to disseminate a broad range of ABS social, demographic and economic statistics. It is broadly based on the concept of a functional area. The functional area is the area from which people come to access services from a centre. The Structure has six hierarchical levels comprising in ascending order: Mesh blocks, SA1s, SA2s, SA3s, SA4s and S/Ts. Each level directly aggregates to the level above. Therefore, SA1s are aggregates of Mesh

Blocks and aggregate to SA2s. This principle continues up through the remaining levels of the hierarchy. At each hierarchical level, the component spatial units, for example SA1s, collectively cover all of Geographic Australia without gaps or overlaps.

However, the selection of these administrative boundaries as reporting units is generally based on criteria that do not necessarily reflect socioeconomic homogeneous areas.

In Australia, for instance, to report age-standardized results for Medicare Primary Care, the following conditions are usually required as a minimum: A denominator of at least 30 per age group; a total number of patients in the population of at least 20; and a total population count of at least 2,500 for the unit of analysis (NHPA, Healthy Communities: Frequent GP attenders and their use of health services in 2012-13, Technical supplement (p 10), 2015). In addition, the choice of geographical unit is often limited by disclosure issues, either because data on the geographic location of cases cannot be published for privacy protection issues (NHPA, Healthy Communities: Australians' experiences with access to health care in 2011-12, Technical Supplement, 2013) (NSW Health, 2015) or are not to be disclosed to the public for reasons related to Census Confidentiality Acts (Australian Government, 1905).

As a response to these issues, a common census geography used for reporting and map publication is the Statistical Areas Level 3 (SA3s) (AIHW, My Health Communities: Reporting local information, driving better health outcome, 2017), which are regions defined by ABS. They are designed to provide a regional breakdown of Australia. SA3s have generally a population of between 30,000 and 130,000 people. In major cities, they represent the area serviced by a major transport and commercial hub. In regional areas, they represent the area serviced by regional cities with populations of more than 20,000 people. In outer regional and remote areas, they represent areas which are widely recognized as having a distinct identity and have similar social and economic characteristics (ABS, ASGS: Volume I - Main Structure and Greater Capital City Statitstical Areas, 2011).

However, the majority of SA3s across Australia are in the major cities where the population composition is more likely to be heterogeneous within a geographic area. As a result, the spatial extent or scale at which the IRSD is calculated might have an effect on its value.

For larger geographies, like SA3s, an IRSD score is created from the population weighted average of the units scores within the larger areas (Pink, Socio-Economic Indexes for Areas (SEIFA): Technical Paper, 2011). However, a single score for an area does not take into account the socioeconomic diversity within that area. This index is an area-based level measure, means it will mask some diversity at finer level of disaggregation. Smaller spatial units located within the SA3 may thus be misclassified leading to their being grouped into incompatible peer-groups.

While ideally one would like to use regions in which deprivation levels are relatively uniform, the use of area aggregated measures usually entails a mixture of both more and less disadvantaged populated units. This raises the risk of the ecological fallacy, which is where errors are made when drawing conclusions about individuals based on the average characteristics of the area in which they live.

Consequently, this issue is most likely to be encountered in areas where the characteristics of individuals or other population subgroups are too diverse to be meaningfully represented by a single statistic. On the other hand, if there is a high level of homogeneity among people or households within each area, we will find a strong relationship between IRSD scores and individual level indices. This analysis has been fully conducted by the Methodology Division in the ABS (Wise & Matthews, 2011), (Baker & Adhikari, 2007).

### 2.2.2 Spatial statistical homogeneity

In the previous section we emphasized that the use of an index score may oversimplify reporting of an area's relative socioeconomic disadvantage. Despite the relevance of this issue few researchers have attempted to assess the homogeneity of a geographic area using a socioeconomic index. An attempt was made by Flowerdew (Flowerdew, Feng, & Manley, 2007) to build a new zone system to be used for the publication of Scottish Neighborhood statistics. For data zone construction, social homogeneity was assessed using the Townsend index of deprivation (Townsend, Philimore, & Beattie, 1988) and calculated by subtracting the number of the decile containing the lowest score from the decile containing the highest score. This value, however, is clearly affected by outliers and therefore inappropriate for the classification of highly skewed distribution. Moreover, the author does not provide an operational definition of homogeneous area.

A different approach to the definition of homogeneous socioeconomic areas consists in considering the social and economic variables separately. Following this method, Steel and Tranmer proposed a homogeneity measure for the distribution of a multi-category variable, using data from the UK census (Steel & Tranmer, 2012). This is a variance-based measure, weighted to account for differences in the population size of units across the geographic area. This definition, however, is not helpful to measure the homogeneity of an ordinal measure such as a socioeconomic index: the index scores do not represent an amount of disadvantage and as such there is no meaningful arithmetic relationship between values.

In this general setting, a range of summary measures are used to describe relative deprivation for higher-level geographies (Smith, et al., 2015). An example is the Extent measure of the Index of Multiple Deprivation (Smith, et al., 2015). This measure uses a weighted combination of the population that live in the three most deprived deciles of the distribution. This analytical approach is useful to compare disadvantaged areas, but it does not provide an operational definition of homogeneous area.

For all these reasons, it is necessary to define homogeneity precisely, in order to determined how it should be operationalized and measured.

### 2.2.3 A Framework for the classification of homogeneous areas

One approach to enhance the description of socioeconomic disadvantage or any other variable given by the index score, is to use the distributional information of the units' data within each area. Following this approach, this thesis aims at proposing a general framework to identify distributional properties of a set of data suitable for the presentation and reporting of comparable information of geographic regions with peers. More precisely, I propose the HI and LI to measure respectively the dispersion and central tendency of a discrete probability distribution.

Accordingly, our approach is related to the general theory of probability distributions, even if it is inspired by the analysis of spatial data. The reasoning behind the development of such a method, is to provide a natural benchmark for these measures in terms of defining what is a "high" (i.e. homogeneous) and "low" (i.e. heterogeneous) concentration of a probability distribution. Currently, there is no accepted benchmark that could be used to assess and compare the homogeneity of two geographic areas. To improve comparability and thereby identify areas of needs, I propose a collection of requirements specification meant to evaluate when the area-based variable does not truly represent all the people in the area (i.e. there is a significant variation within the area per that characteristics); In this case, there must be a follow up study for a more appropriate selection of the geographic area.

Furthermore, the are-level score computation at medium or large spatial scale is strongly affected by extreme values, leading to misclassification of the geographic area. Therefore, to distinguish the statistical properties of the data distribution, I propose the LI as a robust descriptive measure of the distribution location.

In this way, using the LI and HI, we can identify the location of where the distribution is mostly concentrated. This naturally leads to identify geographies where socioeconomic categories indicated by the LI are meaningful in terms of the HI concentration criteria. Therefore, using both indices and the specification criteria enable users to identify meaningful communities across geographic areas. Consequently, this framework is a valuable tool to show whether a geographic unit is acceptable homogeneous or needs further refining.

Thus, each index contains different information about the typical characteristics of a geographic area. Using either index in isolation will not give a complete picture.

Finally, this approach allows us to facilitate the visualization of multivariate data. Dealing with large numbers of small areas may be difficult for a user to see patterns in the data when confronted by maps and graph relating to larger zones. Individual features such as the number of people living in a specific geography and the characteristics of that area can be combined into a single thematic map that integrates the homogeneity and central location of the data. Moreover, it helps to present the characteristics of the geographic units in a way that makes easier to interpret the geographic variation of an indicator.

I therefore used the above-mentioned framework to answer the following questions.

- Are the SA3s an appropriate geography for reporting health care variation by socioeconomic status?
- How homogeneous are these units with regard to the IRSD?

In figure 2, I propose a conceptual framework that could be useful for the evaluation of homogeneous areas in health geographic studies. The first natural decision is the selection of the larger geographic area (e.g. SA3) and its subunits (e.g. SA1). Then, the contextual dimension along which one wishes to measure the homogeneity of the geographic area must be defined (e.g. SEIFA, socio-economic index for areas). Third, the selection of the variable used in the model must be specify since measuring the homogeneity among multiple unordered or multiple ordered categories of a variable needs a different set of measurement tools (e.g. IRSD decile). Finally, the selection of the statistical model used to represent the distributional characteristics of the area.

In this study particularly, we are looking at the population distribution in the SA3 IRSD decile categories. This is because the SA1s are specifically designed to include approximately equal population size and a small number of residents, on average 400. (ABS, ASGS: Volume I - Main Structure and Greater Capital City Statitstical Areas, 2011). This property allows users to aggregate these building blocks into their own specific purposes. In this way, the analysts can simply describe the complete set of SA1s' values (e.g. population counts, IRSD decile, etc) with a few numbers that represent the proportion of people (probability) in each category of the contextual variable.

Conceptually, the homogeneity measure of a spatial unit, which ranges between 0 to 1, can be therefore defined as the degree to which the residential population is concentrated among the set of categories for that area.

In this context, therefore, our major questions concern how to operationalize the homogeneity of a decile probability distribution and what measure of central tendency is more appropriate to classify the socioeconomic decile category of the geographic area.

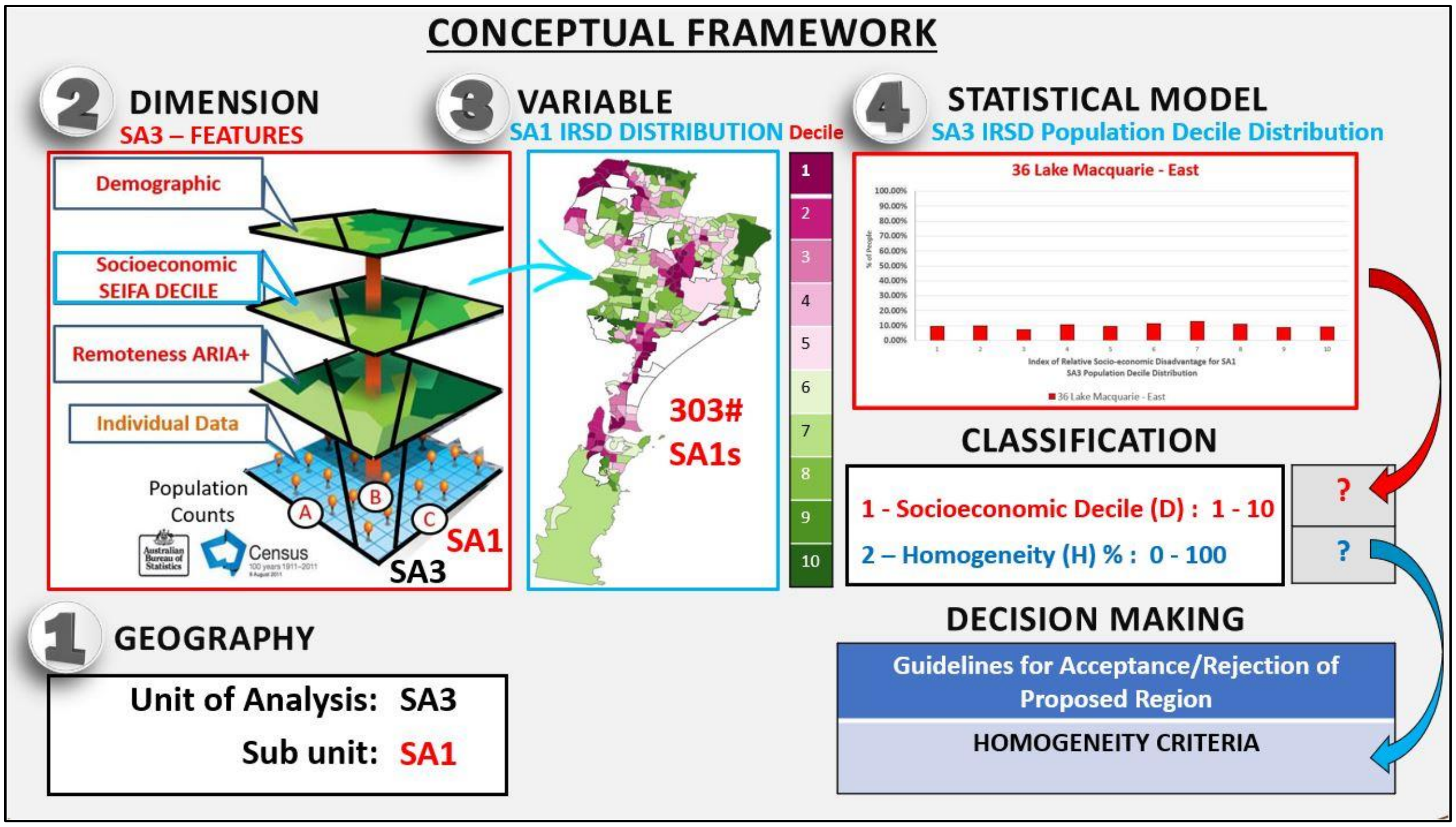

Figure 2 Conceptual Framework for the classification of homogeneous areas

## 2.2 – Homogeneity & central location measures

Different fields of science apply a wide range of indices to measure the statistical homogeneity or heterogeneity of a given data set. Numerous indices can turn all the data into one number between zero and one, between minimum and maximum homogeneity. Among these indices, measures of evenness are widely used (Eliazar & Sokolov, Measuring statistical evenness: A panoramic overview, 2012). In this regard, the homogeneity index is the complementary measure of evenness so that a coefficient of zero expresses minimal concentration and occurs when the population is equally distributed among the set of categories for the geographic area. Conversely, the maximal concentration (100 %) is attained if the whole mass of the distribution is concentrated in a single category (Kvalseth, A measure of homogeneity for nominal categorical data, 1993). In the latter case, there is no variation within the area per that characteristic and the geography is uniquely identified by the central value of the distribution.

However, the adoption of a particular evenness or homogeneity measure over others is based on its mathematical properties, and there is no general agreement on the index criteria (Tuomisto, 2012). Despite this disagreement, an intuitive and general agreed criterion is the Pigou-Dalton Transfer principle which specifies that a homogeneity measure should increase or decrease whenever a portion of the population is transferred from a less abundant category to a more abundant one or vice versa. This basic property is known in the mathematical literature as the Schur-concavity condition (Kvalseth, Evenness indicies once again: critical analysis of properties, 2015), that in the context of homogeneity measures should be replaced with convexity or strict convexity in the case of the strong version.

In general, not many indices satisfy the strict convexity condition but among those the most commonly used are the Shannon's entropy and the Gini's index or their different functional forms (W.Marshall, Olkin, & C.Arnold, 2011).

These indices focus on the measurement of dispersion among groups defined by nominal categorical variables, such as, for example, gender, race/ethnicity, religious etc. Many social and economic variables, however, are of ordinal nature. Examples include socioeconomic status, self-reported health status (Allison & Foster, 2004), level of access to primary care services (McGrail & Humphreys, 2015) and remoteness classification (ABS, Australia Standard Geographical Classification, 2005).

A principal difficulty in dealing with distributions of ordinal data is that the transfer principle is simply inappropriate for ordering concentration measures, since the position of data has no influence on its mathematical properties (Zheng, 2008). As a simple example, consider the socioeconomic quintile distribution $x = (0.3,0,0,0.5,0.2)$ with an amount of 0.1 being transferred from the last category to the first quintile to produce the distribution $y = (0.4,0,0,0.5,0.1)$. It is then clear from the exchange principle that the homogeneity measure of y should be increased. However, comparing the proportion of people below the median category of both distribution (i.e. the fourth quintile) reveals that y has 10% of the population with lower socioeconomic status, while the proportion of those above or equal the median is naturally less. Consequently, distribution y is on intuitive sense more "spread out" than x. This intuitive notion of concentration as "spread away from the median" has been proposed by Allison and Foster (Allison & Foster, 2004) in the context of health inequality indices to introduce a partial ordering based on a median-preserving spread of distributions. This approach, however, is suited only for the comparisons of distributions which share a common median and in some cases this reference point is not well defined (Cowell & Flachaire, 2017). These problems have been addressed recently by many researcher (Blair & Lacy, 2000) (Apouey, 2007) (Abul & T.Yalcin, 2008) (Lazar & Silber, 2013) (Cowell & Flachaire, 2017), and an axiomatic theory for measuring dispersion of ordinal data has emerged (LV & Xu, 2015).

This accumulating body of research, however, have been criticized for focusing more on polarization than homogeneity measure (Zheng, 2008) (Kobus, 2015). That raises an important distinction between homogeneity and polarization.

Polarization measures concentration around the extreme categories and corresponds to the distribution in which half of the population is concentrated in the lowest category and half in the top category, such a distribution will be consequently called a "two-point extreme distribution". For example, a geographic area where the residents are evenly distributed among the least and most disadvantaged socioeconomic quantile. The question then arises whether this divided population is also the most dispersed. Are we measuring the dispersion of a distribution as homogeneity or bimodality towards the ends?

The divergence between homogeneity and polarization is not merely a theoretical curiosity: it occurs in practice as well. In the inequality literature, this phenomenon is referred to as the "disappearing middle class" (Wolfson, 1994). A typical approach is to treat the two-point extreme distribution as the most unequal or dispersed distribution (Leik, 1966) (Berry & Mielke, 1992) (Blair & Lacy, 2000). This assumption, however, reopens the question of the axiomatic foundation of an evenness measure. Specifically, it raises question about the decision of whether the polarized distribution is more informative than the even distribution (Kobus, 2015).

Clearly, a geographic area where the population is equally distributed in all the socioeconomic categories is likely to be partitioned into a larger number of different zones than any distribution. Then, if the dispersion of the polarized distribution is less than the uniform distribution, how big is the gap between those values? Similarly, if the two-point extreme distribution is more dispersed than a bimodal distribution concentrated towards one end or the other of the socioeconomic scale, how big is the homogeneity gap?

Recently, researchers have acknowledged this problem and addressed the issue of measuring the dispersion of ordinal data based on frequency distribution (Kobus, 2015). In this general setting the major challenge is that the only information available is the ordering of the categories and, therefore, the concept of distance among categories is not defined. Accordingly, an ordinal measure should in principle be independent to the arbitrary scale applied to the categories of an ordinal variable (Allison & Foster, 2004). As a result, a natural approach is to work directly with the cumulative proportion of population in each category rather than values or weights assigned to categories.

Following this approach, I propose the HI as a measure of homogeneity for ordinal categorical data. This index is a generalization of the Gini's coefficient that incorporates the common properties of a polarised measure (LV & Xu, 2015) and meets the basic expectation of a homogeneity measure: it takes the value zero for the uniform distribution and the maximum for the distribution concentrated in one category.

Moreover, I provide additional criteria for validating and comparing measures of homogeneity when the involved variables assume ordinal nature. Specifically, I extend the Schur-convexity and value validity conditions of an evenness measure (Kvalseth, Evenness indicies once again: critical analysis of properties, 2015) to the homogeneity values of the lambda distribution (λ-distribution) (Kvalseth, The lambda distribution and its applications to categorical summary measures, 2011). The λ-distribution is a two-parameter distribution, recently introduced by Kvålseth, uniquely fit to establish necessary conditions for comparison of evenness measures. Additionally, considering the structural properties of the index, I establish a theoretical lower and upper bound for the two-point extreme distribution. Finally, the combination of this theoretical framework with the model criteria results in a family of generalized concentration indices for ordinal categorical data.

This axiomatic approach for the construction of a concentration measure has intuitive appeal but is impractical without knowing how to interpret a given homogeneity value. To further assist the analysis, I build a correspondence table between the concentration measures of the distribution range and the homogeneity values lower bounds. With this approach one is able to consider approximately not only the cumulative percentage of the population in the most concentrated categories but also around. This table serves as a basis for the definition of the homogeneity condition that must be met by a geographic unit in the analysis of the indicator.

To our knowledge, in the statistical literature no productive suggestions were made regarding how to provide a generalized version of the Gini's index for ordinal categorical data. Among those offered (Penaloza, 2016) (Van Doorslaer & Jones, 2003) (Giudici & Raffinetti, 2011), more information is needed to operationalize the statistical analysis of the selected variable.

Our aim in this paper is to fill out this gap by extending and operationalizing the Gini index to ordinal data.

In what follows, I show that the HI is a natural measure of statistical dispersion which is more useful and appropriate in the case of asymmetric, U-shaped and skewed probability laws than the entropy index. These probability laws (such as the Pareto, lognormal, exponential and others), are prevalent in various fields of science (Mitzenmacher, 2004) (Laherrere & Sornette, 1998) where the distribution is usually clumped around the extreme categories.

Moreover, when dealing with ordinal categorical data, the standard measures of center location, such as the mean or mode, are not appropriate (Allison & Foster, 2004).

The mode can be determined for ordinal and metric data, but it is especially valuable for nominal data. If the data are strictly nominal, then the only possible measure of center is assessing which category occurs most often. The mode measures a variable's center by pointing to the most typical category. An interpretation of the mode is that it provides the "best guess" as to the category a case has on the variable, if the goal is to be accurate as often as possible. That is, no other guess of a category for a random case would be correct as often as the mode is. The main advantage of the mode as a statistic is that it is easy to obtain and to interpret. Consequently, the mode is usually simple to communicate and explain to people. However, there are two main problems involved in dealing with the mode on non-numeric data. First, it may not be very descriptive of the data, because the most common category still may not occur very often. The second problem with the mode is that it may not be unique. For example, two categories may be equally likely and more common than any other category. A variable with such a distribution is termed bimodal. Indeed, several categories may be equally likely and may occur more often than any remaining category, in which case the variable is multimodal. In the most extreme case, if each category occurred with the same frequency, there would be no mode for the variable. Moreover, the mode's definition cannot be generalized to more variables as it is not an algebraic form.

On the other hand, the mean's definition is based on an algebraic form and can be extended to more variables. It's a measure of the central tendency for variables that are fully numeric. It summarizes the center of metric data by averaging the values on the variable. An important property for the mean is that the total sum of deviations around the mean is always zero. The mean is unique in that sense: the sum total of deviations around any other value would be higher. That the sum of deviations around the mean is zero implies also that the average signed deviations around the mean is zero. This property leads to an interpretation of the mean as a "best guess" statistic. Say we sought to guess the value of a particular score, such that the sum of the signed errors in guessing is minimized. Because the sum of the signed deviations from the mean is zero, the mean is the best guess of a score on the variable if the goal is to minimize the sum (or average) of signed errors. However, there are two problems with the mean. First, it is strongly affected by atypical outliers, it is considered nonresistant in contrast to more resistant measures of center such as the median. A second problem with the mean is that it can have fractional values, even when the variable itself can sensibly take on only integer values. Moreover, it would not be meaningful to take averages on ordinal data because category numbers are arbitrary.

An obvious choice is the median, but its computation takes much larger than computing the mean (Tibshirani, 2008). Additionally, its recursive formula is not easy to extend to more categorical variable and there is no absolute consensus about what the definition should be (G.Small, 1990).

Finally, it is possible that data may arise with more than a single median. Although, such distributions are rare and sometimes artefactual, it can be difficult and meaningless to manage multiple values for multimodal distributions.

Thus, I propose the LI as a measure of central tendency that is statistically efficient and reasonably straightforward to implement. In particular, I prove that it is an equivalent form of the median and comprehensible measure of the center value.

Lastly, I show that the combination of the LI and HI constitutes a clear and consistent framework for the classification and identification of homogeneous areas. It can help investigators to identify heterogeneous units and assist in determining the course of action needed to produce a homogeneous geography. Moreover, the combination of only two numbers and the classification criteria enables users to summarize easily the characteristics of a geography in a single table. This straightforward representation is suitable to show to health planners or policy makers the minimum or maximum number of possible peer groups.

# Chapter 3 - Methodology

## 3.1 – Homogeneity Index

Identifying a summary measure describing the distribution of an ordinal variable is not an easy task. Particularly problematic is the issue of the definition of dispersion, because there is no universally accepted measure of homogeneity for ordinal variables.

Using a concentration measure to describe an ordinal variable could bring counterintuitive results. In particular, the value of this function would be the same for all the distributions which have the same set of frequencies. Meanwhile, the order in which the frequencies are assigned to categories should be taken into account. On the other hand, the two-point extreme distribution exhibits the maximum value for a polarized function and violates, therefore, the basic definition of maximum dispersion for a heterogeneous geographic area.

Then, what is a proper way to specify the concept of dispersion in the case of ordinal variables and to define a measure that has suitable properties?

An intuitive idea is to combine a concentration measure and a polarized function into a single index to take the desired properties into consideration. However, there seems to be no obvious choice for such a function and there are no explicit guidelines in the statistical literature for the definition of such functional space. Thus, rather than looking for only a formula derived in an ad hoc manner, it is more desirable to derive a statistical model from independent theoretical criteria. The theoretical model should be reasonably simple to use for the construction of a concentration measure and applicable to any population independently of its distribution. This allows future comparability if different researchers selected different concentration and polarized functions in the model.

With this view, to coherently define the functional form of the HI, a definition of the functional space of these two functions is needed. In other words, given the general properties of such measures, what functional space they do define and more importantly how one can combine them in order to get an acceptable homogeneity measure which reflects the concentration and shape of the distribution.

In this general setting, the properties that are considered to be necessary for any acceptable concentration measure, hereafter referred to as Concentration Index (CI), are: Normalization, Continuity, Symmetry, Strict Schur-convexity and Value validity (Kvalseth, Evenness indicies once again: critical analysis of properties, 2015).

Most of the indices in the evenness literature satisfy the first three properties and only a few meet the requirements of Strict Schur-convexity and Value validity. As strongly argued by Kvålseth: "An index lacking this property may cause misleading results and incorrect conclusions" (Kvalseth, Evenness indicies once again: critical analysis of properties, 2015).

The concepts of Schur convex functions (and Schur concave functions as their duals) were introduced by Isac Schur as variants of convexity and concavity of real functions (Roventa, 2012). In fact, each symmetric convex function is Schur convex (and each symmetric concave function is Schur concave). The Schur convex functions are used in the study of majorization (Marshall & Olkin, 1974), a preorder on vectors of real numbers, and inequalities related to it. An example of a Schur convex function is the maximum. As indicated in chapter 2.2, not many indices satisfy the strict convexity condition but among those the most commonly used are the Shannon's entropy and the Gini's index or their different functional forms (W.Marshall, Olkin, & C.Arnold, 2011).

To determine the conditions for a CI to have Value validity, Kvålseth recently introduced the lambda distribution (λ-distribution). The term lambda distribution has been chosen for the distribution by Kvålseth because of its L-shape (Greek lambda: λ) form. For nominal categorical variable the λ-distribution is a single parameter distribution where the parameter λ reflects the uniformity or evenness of the distribution and $n$ the number of categories:

$$P_n^\lambda = \left(1 - \lambda + \frac{\lambda}{n}, \ \frac{\lambda}{n}, \cdots, \ \frac{\lambda}{n}\right) = \lambda \cdot P_n^1 + (1-\lambda) \cdot P_n^0 \ \ : \ \lambda \in [0 \quad 1] \qquad (1)$$

Basically, it is a weighted mean of the one-point distribution $P_n^0 = (1,0,\cdots,0)$, sometimes call singleton or degenerate distribution, and the uniform distribution $P_n^1 = (1/n\,, 1/n, \cdots, 1/n)$. Note that a concentration measure for nominal variables is a (permutation) symmetric function and therefore any choice of the one-point distribution could be made. However, we shall consider $1 - \lambda + \lambda/n$ to be the first element of $P_n^\lambda$ and 1 to be the first element of $P_n^0$. Then, λ is a rectangularity (uniformity) parameter with increasing λ indicating a decreasing approach of $CI(P_n^\lambda)$ from the singleton located in the first category to the rectangular (uniform) distribution $CI(P_n^1)$. Then, a concentration measure has Value validity if it regularly decreases from the one-point distribution to the uniform distribution. This means that adding or subtracting an amount δ to or from λ should cause the same absolute change in the value of CI. This is equivalent to define the CI as a function of a single variable λ, that is $CI\left(P_n^\lambda\right) = CI(\lambda) = 1 - \lambda$. It follows that all the CI values lies on the line segment between $P_n^0$ and $P_n^1$. This situation is better illustrated in fig 3A. The blue and yellow curves represent the Gini Index and the complementary value of the Shannon's entropy index, see Appendix E (Shannon, 1948) of $P_n^\lambda$ for $n = 10$. It is clearly indicated by these curves that the entropy function deviates substantially from the requirement of belonging to the segment $\overline{P^0P^1}$ and hence violates the value validity condition (Kvalseth, Entropy Evaluation Based on Value Validity, 2014). Obviously, the Gini index is not the only feasible concentration measure (Kvalseth, Evenness indicies once again: critical analysis of properties, 2015), but it is the most popular CI and it is easily related to the point-wise ordering of the Lorenz Curve (Gosselin, 2001). For this reason, this index has been selected as a concentration measure for this work. The CI is, therefore, defined as twice the area between the Lorenz Curve and the line of equality. This area is also referred as Lorenz Zonoid (Koshevoy & Mosler, 1996). For the definition, properties and computation of the CI see **Appendix A1**.

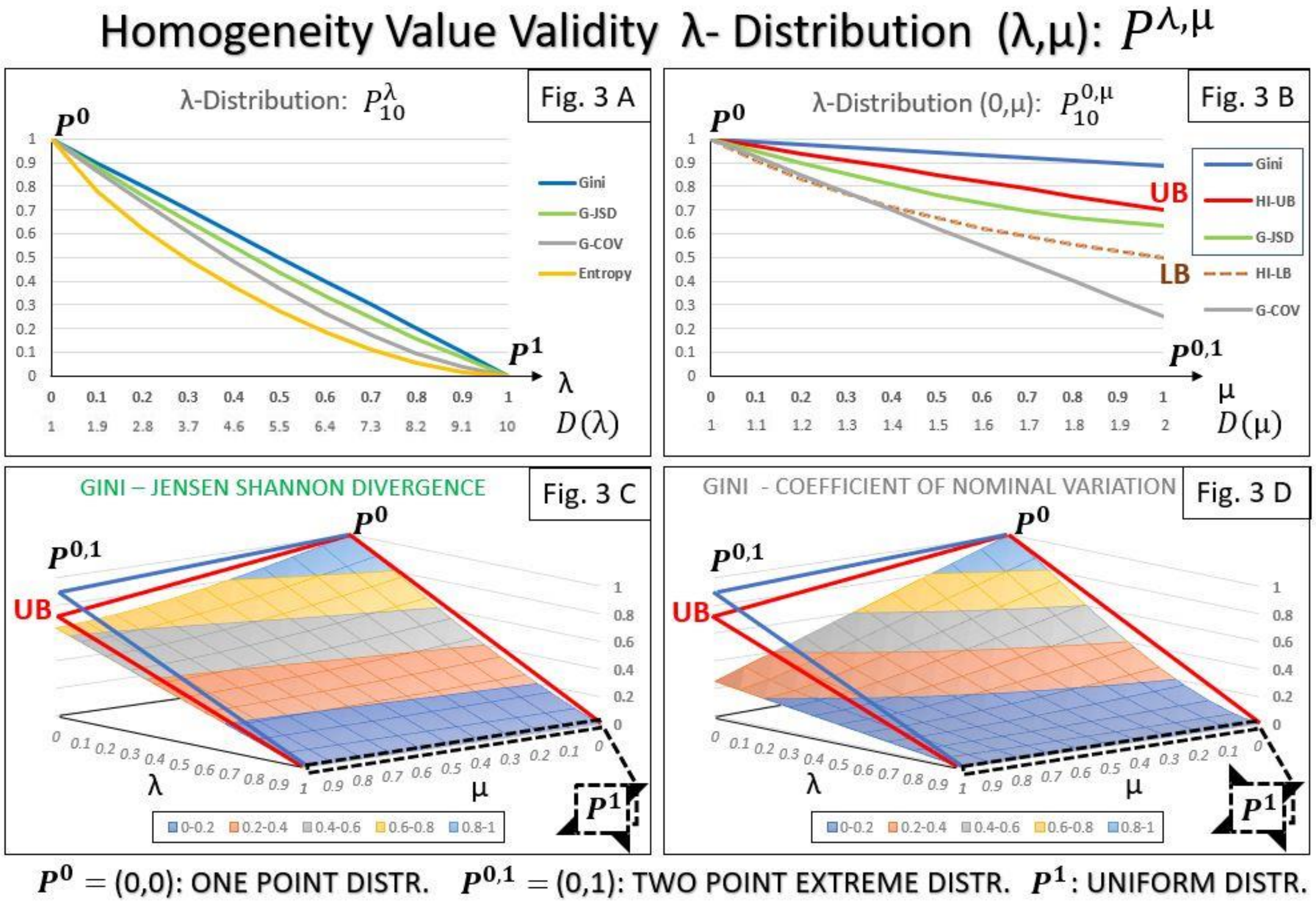

Figure 3 Homogeneity Index - Value Validity condition for lambda distribution

We note that the definition of the CI can be viewed as a study of the discrepancy between a probability distribution and the uniform distribution, but it does not completely reflect the amount of spread that the value of a random variable will take on. For example, different distributions with the same Lorenz Curve have different variance. As stated earlier, this means that the ordering of the distribution coefficients is not important. In view of this circumstance, I deemed it convenient to include a polarization measure in the computation of the HI.

The main properties of any polarization measure are: Normalization, invariance of parallel shifts and simple aversion to median-preserving spreads (LV & Xu, 2015).

Basically, the first two properties state that the minimum value (zero) should be taken for the one-point distribution and the "parallel" shift of the entire frequency distribution leaves the index's value unchanged. The last property states that the transfer of cases from one category into the next, which is closer to the extreme categories (i.e. the first or last category) should result in an increased value of the polarization measure as it will be closer to the two-point extreme distribution.

In the statistical literature, for ordinal types of data, are known lots of indicators to measure the degree of the polarization phenomenon. Typically, many of the widely used measures of distributional variability are defined as a function of a reference point, which in some "sense" could be considered representative for the entire population. This function indicates how much all the values differ from the point that is considered "typical". Of all measures of variability, the variance is a well-known example that use the mean as a reference point. However, mean-based measures depend to the scale applied to the categories (Allison & Foster, 2004) and are highly sensitive to outliers. An alternative approach is to compare the distribution of an ordinal variable with that of a maximum dispersion, that is the two-point extreme distribution. Using this procedure, three measures of variation for ordinal categorical data have been suggested, the Linear Order of Variation - LOV (Berry & Mielke, 1992), the Index Order of Variation - IOV (Leik, 1966) and the COV (Kvalseth, Coefficients of variations for nominal and ordinal categorical data., 1990). All these indices are based on the cumulative relative frequency distribution (CDF), since this contains all the distributional information of any ordinal variable (Blair & Lacy, 1996). Consequently, none of these measures rely on ordinal assumptions about distances between categories.

At this point, the reader might wonder if this approach to dispersion is adequate to define the functional form of the HI. "Why not measure the dispersion of the observed distribution as the distance from a point of minimal dispersion?". This question has been addressed by Blair and Lacy in the article *Statistics of ordinal variation* (Blair & Lacy, 2000). They argue that this approach is impractical since there are as many one-point distribution as the number of categories. Therefore, it would not be clear, from which one we should calculate the distance.

Then, is there another way to compare the dispersion of a distribution that does not depend on its location? Is the space of the cumulative frequency vector the only way to represent all possible distributions?

To address this challenge, I propose a new representation of probability measures, the Bilateral Cumulative Distribution Function (BCDF), which derives from a generalization of the CDF. Basically, it is an extended CDF that can be easily obtained by folding its upper part, commonly known as survival function or complementary CDF. Unlike the CDF, this functional has a finite constant area independently of the probability distribution (pdf) and, therefore, more convenient for any distribution comparison. For the definition, properties and computation of the BCDF see Appendix A2.1.

On this basis, to capture the amount of fluctuations about the mean and simultaneously the local variation around the median, we completely defined the shape of a probability distribution by its BCDF autocorrelation function (BCDFA). The BCDFA is a symmetric function that attains the MAD as its maximum and preserved the variance of the pdf. It follows that the maximum extent to which a BCDFA is stretched occurs when the mass probability is evenly concentrated at the end points of the distribution. In such a configuration, the variance of any bounded probability distribution is maximum (Bathia & Chander, 2000). On the other hand, the minimum support is attained when all the values fall in a single category. The main advantage of this representation is that it is invariant to the location of a distribution and therefore is only sensitive to the distribution shape. For example, any singleton distribution will have the same BCDFA curve. Similarly, distributions with same shape but different means and medians can be represented with a unique curve. For instance, Figure 4 shows the BCDFA for a family of two -point distributions over ten categories. The curve tau_0 represents the nine distributions uniformly distributed over two contiguous categories. On the other hand, the curve tau_8 represent the two-point extreme distribution. Lastly, the red and light blue curves represent the singletons and uniform distribution, respectively. In this way, to quantify the distance between a pdf and the singleton distribution, it is only a matter of choosing an appropriate metric, which would assign larger values to distributions with more dispersed BCDFA than the singleton. For the definition and computation of the BCDFA see Appendix A3.

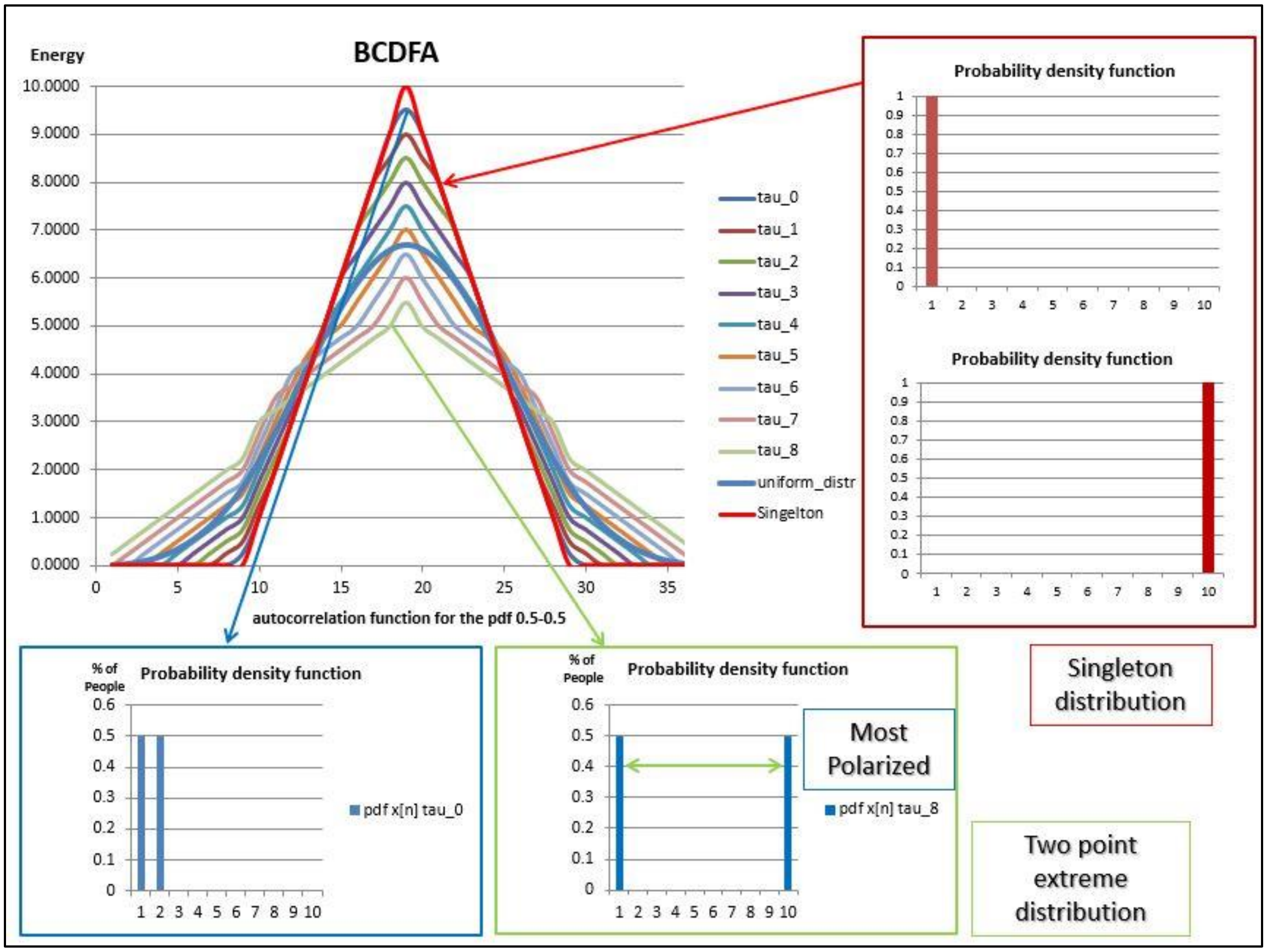

Figure 4 BCDFA of two-point distributions family

The selection of a measure to compare probability distribution is not a trivial matter and usually depends on the objectives. In this work, I propose the use of the Jensen-Shannon Divergence (Lin, 1991) (Topsoe, 2004). Since its definition is based on the BCDFA as opposed to density functions or CDF, is more regular than the variance (VAR), COV, IOV and LOV. In addition, unlike the other dispersion indices, this measure does not need to be normalized since is a bounded value in the unit interval. Another crucial characteristic of this metric is that it is infinite differentiable and its derivates are slowly decreasing than any power metric. A function with this characteristic is called tempered distribution or dual Schwartz functions (Stein & Shakarchi, 2003). Clearly, there are other functions that belong to this functional space (Taneja, 2001) (Jenssen, Principe, & Erologmus, 2006). However, the JSD is a well-known divergence and its square root is a metric (Endres & Schindelin, 2003). This last property will allow us to increase or reduce the magnitude of the divergence. Without loss of generality I call this class of functions, Divergence Index (DI).

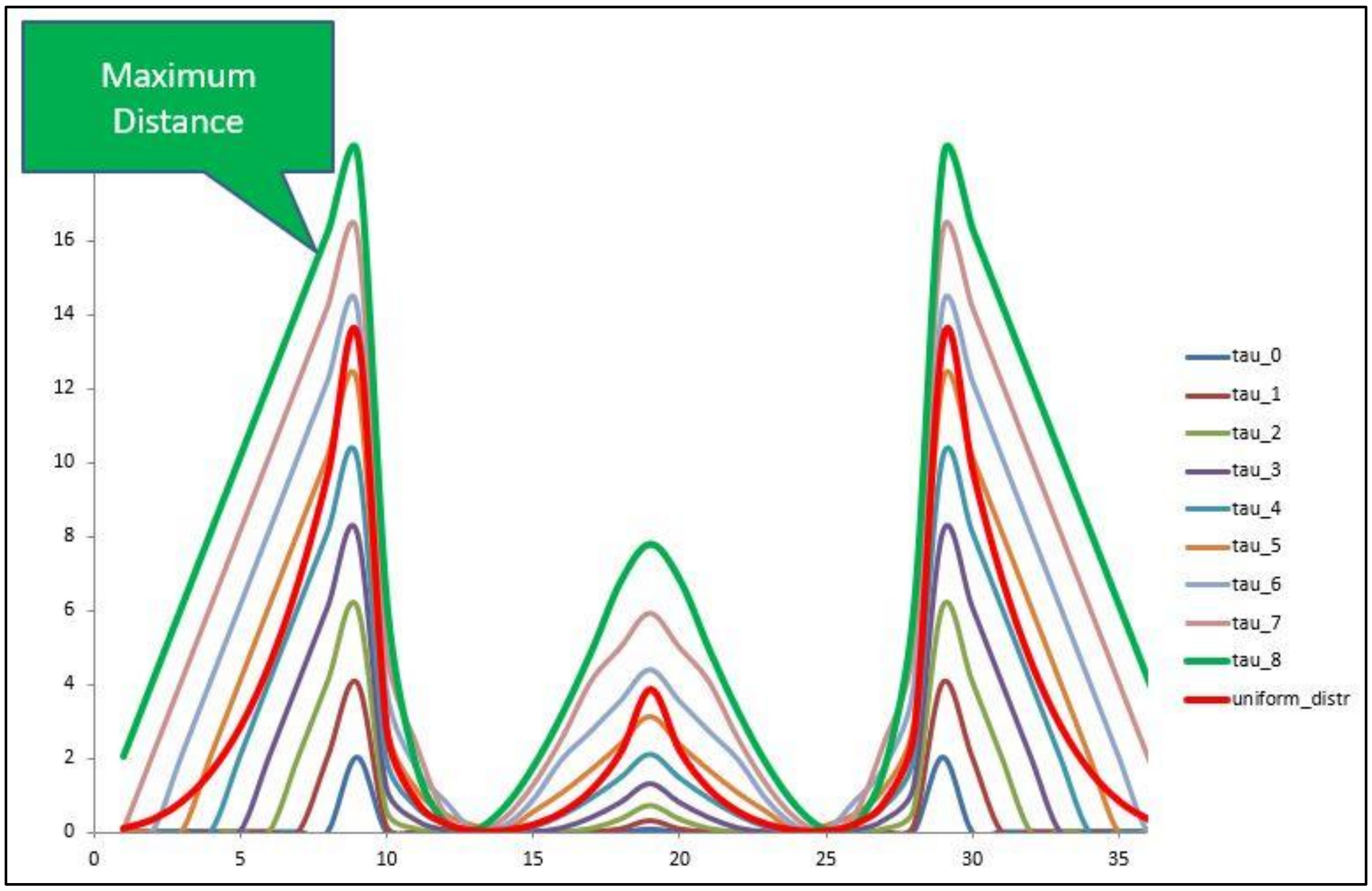

Figure 5 Jensen Shannon Divergence distance between the singleton and the two-point distributions

For instance, Figure 5 shows the JSD distance between the singleton and the family of two-point distributions represented in Figure 4. Since the DI is based on the JSD, I dub this new measure of polarization DI-JSD. For the definition, properties and computation of the DI-JSD see Appendix A4.

### 3.1.1 – Definition of the Homogeneity Index

In the following, we are in position to state the problem of the HI definition. As discussed previously, the idea is to combine a polarization measure, PI, and the CI to form the HI functional space. The first intuition is that both measures have a unique extreme value and opposing values for the singletons. The CI attains the minimum for the uniform distribution and maxima for the singletons.

On the other hand, a PI attains the maximum for the two-point extreme distribution and minima for the singletons. It is clear, then, one function is convex and the other is concave. These functions and the properties introduced previously identify a particular functional space called Fréchet space (Frechet Space, n.d.). In this context, the problem is to form a convex functional space that includes the properties of the PI and the extreme and maxima values of the CI. Since the difference of any two polarization function values in a Fréchet space does not affect its properties, I considered the degree of dispersion for a PI with respect to the uniform distribution. This function can be easily calculated by taking the following difference $PI(P_n^1) - PI(P_n) = \Delta(PI, P_n)$, where $P_n$ is a probability distribution vector of $n$ categories. Since the singleton distribution is the only distribution for which the polarization function is zero, it follows that the maximum value of this function is equal to the $PI$ calculated for the uniform distribution when $P_n = P_n^0$, that is $Max\{\Delta(PI, P_n)\} = PI(P_n^1) = \delta_M$. On the other hand, the minimum value is attained when the $PI$ function is applied to the two-point extreme distribution,$P_n^{0,1} = \{0.5,0,\cdots,0,0.5\}$, and it is equal to $min\{\Delta(PI, P_n)\} = PI(P_n^1) - PI(P_n^{0,1}) = \delta_m < 0$. In this way, we transformed a concave function into a convex function. Considering, therefore, the definitions of the CI and PI, the HI of $P_n$ is given by the combination of two convex functions:

$$HI(P_n) = \frac{CI(P_n) + \Delta(PI, P_n)}{1 + \delta_M} \; : \; \begin{cases} \Delta(PI, P_n) = PI(P_n^1) - PI(P_n) \quad \wedge \quad \delta_M = PI(P_n^1) \\ CI(P_n^{0,1}) > |\delta_m| \qquad \delta_m = PI(P_n^1) - PI(P_n^{0,1}) < 0 \end{cases} \tag{2}$$

At this point, it would be natural to ask if this convex function really represents a Fréchet space. Put in a different way, does this formula defines a positive function, between zero and one, that attains the minimum at the uniform distribution and the maximum when applied to a singleton?

One way to check this condition is to study the behavior of $HI(P_n)$ when $P_n = P_n^1$ and $P_n = P_n^0$. It is not difficult to verify that the numerator is equal to $1 + \delta_M$ for $P_n = P_n^0$ and zero for $P_n = P_n^1$. Thus, although there are many distributions with the same PI value of $P_n^1$, the CI is zero only if $P_n = P_n^1$. Then, the only sufficient condition for the HI to be a positive function is that the CI value for the two-point extreme distribution is larger than the difference between $PI(P_n^{0,1})$ and $PI(P_n^1)$. Most of the PI meet this requirement, such as the DI or the COV, LOV, IOV and any F-norm in the CDF space. Thus, the HI is always nonnegative and vanishes if and only if the distribution is uniform. This important result allows any researcher to try out different feasible concentration and polarized indices in the formula. However, the choice of the $PI$ function in the HI formula, cannot be completely arbitrary. As it will be shown later, the absolute value of $\delta_m$ must be reasonably small in order to get a meaningful value of $HI(P_n)$. This means that, in the case of an inappropriate $PI$, the negative values of the function $\Delta(PI, P_n)$ will compensate the $CI$ values. Then, there is no way, to discern between a polarized distribution and a concentration measure, as the numerator will be almost zero for most of the vectors $P_n$. From this observation, it follows that the main difficulty in the selection of $PI$, is how to trade off the "polarization effect" that comes from $\Delta(PI, P_n)$ with the $CI$ value. More precisely, the value of $\Delta(PI, P_n)$ should not be too large when the distributions have gaps in the value (i.e. areas where there are few or no values relative to surrounding areas) or variance greater than the uniform distribution. Conversely, when the samples are tightly packed around the central value or there is less variation in the data than the uniform distribution, this value should be closer to $\delta_M$. These properties are satisfied by the DI. It follows that the HI is given by the following formula:

$$HI(P_n) = \frac{CI(P_n) + DI(P_n^1) - DI(P_n)}{1 + DI(P_n^1)} \qquad (3)$$

Since the HI definition is based on the Gini Index and DI I named this functional space Gini-divergence space. In the case of the Jensen-Shannon Divergence (JSD) I dub this new measure of concentration: Gini-JSD. For the definition, computation and properties of the JSD refer to Appendix 4.1-4.7. In principle there could many reasonable choices of the PI. At first sight our choice of the Gini-JSD, might seem quite mysterious and since it is really the key to what follows, I shall now give a detailed explanation of our choice. The task is therefore to find a Value validity condition which incorporates the idea that a well behaved or "smoother" PI is more appropriate in the definition of the HI than a "rougher" PI.

To determine the condition for an HI to have Value validity, I shall use the two parameters $(\lambda, \mu)$ distribution, where the parameter $\mu$ reflects the polarization or dispersion of the distribution for a given value of $\lambda$. Thus, instead of $P_n^\lambda$ in equation $(1)$ I shall use:

$$P_n^{\lambda,\mu} = \left[\left(1 - \frac{\mu}{2}\right) \cdot (1 - \lambda) + \frac{\lambda}{n}, \frac{\lambda}{n}, \cdots, \frac{\mu}{2}(1 - \lambda) + \frac{\lambda}{n}\right] \;:\; \lambda, \mu \in [0 \quad 1] \quad n > 3 \qquad (4)$$

which reduces to equation $(1)$ for $\mu = 0$ and to:

$$P_n^{0,\mu} = \left(1 - \frac{\mu}{2}, 0, \cdots, 0, \frac{\mu}{2}\right) = \mu \cdot P_n^{0,1} + (1-\mu) \cdot P_n^0 \quad : \quad \mu \in [0 \quad 1] \qquad (5)$$

for $\lambda = 0$. For example, the two-point extreme distribution $P_n^{0,1}$ is the most dispersed $(\mu = 1)$ and the least heterogeneous $(\lambda = 0)$ configuration. On the other hand, the $P_n^0$ is the least dispersed and heterogeneous distribution $(i.e. HI(P_n^0) = 1)$. Lastly, the uniform distribution $P_n^1$ is the only distribution where $\lambda = 1$ and therefore the most heterogeneous $(i.e. HI(P_n^1) = 0 \ \forall \ \mu)$. It is assumed of course, that $HI(P_n^{\lambda,\mu})$ is strictly decreasing in μ for any given λ. This result is a natural consequence of the simple aversion to median-preserving spreads property. It is worth pointing out that, $P_n^{\lambda,\mu}$ represents only a "special" subset of the set of all probability distributions for a given n. However, "these distributions are uniquely suitable for establishing conditions that are necessary for valid comparisons involving the numerical values of summary measures for nominal and categorical variables" (Kvalseth, The lambda distribution and its applications to categorical summary measures, 2011). This is due to the fact that $P_n^0$ and $P_n^1$ are the extreme members of $P_n^{\lambda,\mu}$ for which the HI attains the maximum and minimum values. And $P_n^{\lambda,1}$ is the most polarised distribution for a given $\lambda$. Then, for any HI measure and any probability distribution $P_n \notin P_n^{\lambda,\mu}$, there exists one and only pair of non-polarized and polarized vectors $P_n^{\lambda,0}$ and $P_n^{\lambda,\mu}$ such that $HI(P_n) = HI\left(P_n^{\lambda,0}\right) = HI(P_n^{\lambda,\mu})$. This equivalence property between the set of distributions $P_n \notin P_n^{\lambda,\mu}$ and $(P_n^{\lambda,0}, P_n^{\lambda,\mu})$ ensures that the $(\lambda, \mu)$ distribution provides an accurate representation of the extent of $HI(P_n)$, for any polarised and not polarised distribution. Consequently, $P_n^{\lambda,\mu}$ can be used to determine whether or not the HI assumes reasonable values throughout its range. In this framework, then, what condition should be considered to define a HI that has adequate properties and outlined some index specifications that specify a criterion to compare any two HI measures?

There is no perfect answer to these questions, but we should expect that a valid HI will be closer to the CI curve for smaller value of the parameter $\mu$ and gives reasonably high values for the two-point extreme distribution. This insight leads to the problem of choosing a PI in the HI equation as a decision theory model. The problem can be represented as follows: we are looking for a HI which should vary as little as possible from a selected target function and be not too far away from the CI curve. In this problem, therefore, the HI is bounded above by the values assigned to the target function and below by a minimum curve. In this way, the upper and lower bound functions identify a feasible region for the HI. This situation can be easily illustrated by looking at the graph in Figure 2B, where the green and grey curves represent the Gini-JSD and Gini-COV (i.e. $PI(P_n) = COV(P_n)$), respectively. Both measures are below the target function (red curve), but only the G-JSD is above the lower bound of the HI (dotted brown curve). We also notice in Figure 3A that the G-JSD is closer to the CI curve, represented by the Gini function, compared to the G-COV. The behavior of the two curves throughout the (λ, μ)-distribution, for n=10, is further illustrated in Figure 3C and 3D. The G-JSD, intuitively, gives more reasonable intermediate values for the HI compared to the G-COV.

The use of the lower and upper bound functions represented in Figures 2, which seemed to work in a reasonable fashion in the example we considered, would seem to require some particular justification and explanation.

Then what are the criteria underlying the choice of these functions and most importantly how this choice is related to the DI functions?

As seen from the above discussion, a valid HI measure is a continuous function that slowly decreases from one to zero. This requirement implies that small changes in the distribution vector $P^{\lambda,\mu}$ should not lead to abrupt changes in the value of the index. Figure 2D clearly shows that the G-COV is a quadratic function and converges to zero faster than the G-JSD. This behaviour tends to overstate evenness and give poor discrimination over much of the range of λ and μ values (Molinari, 1989). As a consequence, the term $|\delta_m|$ in equation (2) assumes large values. It follows that the G-COV is not a slowing decreasing function and therefore it does not belong to the class of DI. This, therefore, leads us directly to the question of how to define a slowing decreasing function and to establish a necessary condition for a HI measure to have the Value validity property.

A simple intuition is that the lower and upper bound functions will have exactly the functional form of the CI for $\mu = 0$. This observation is justified by the fact that a valid evenness measure of a non-polarised vector $P_n^{\lambda,0}$ should regularly decrease with increasing values of the parameter λ and hence approaching the uniform distribution as $CI(P_n^{\lambda}) = 1 - \lambda$, as indicated by the blue curve in Fig 2A. This basic property gives to the target function $S(P_n^{\lambda,\mu})$, where the letter S stands for Superior, and the Inferior function, $I(P_n^{\lambda,\mu})$, the following forms:

$$S\left(P_n^{\lambda,\mu}\right) = (1-\lambda)\cdot f\left(P_n^{0,\mu}\right) \qquad I\left(P_n^{\lambda,\mu}\right) = (1-\lambda)\cdot g\left(P_n^{0,\mu}\right) \qquad (6)$$

where $f\left(P_n^{0,\mu}\right) = g\left(P_n^{0,\mu}\right) = 1$ if and only if $\mu = 0$. This condition is a logical consequence imposed by the maximum value of the HI at $P_n^0$. In this formulation, the problem is therefore to define $f\left(P_n^{0,\mu}\right)$ and $g\left(P_n^{0,\mu}\right)$, that are represented by the UB and LB curves respectively in the graph of Fig 2B. These functions set the boundaries of the feasible HI measures and therefore the class of the feasible slowing decreasing functions. Since $g(P_n^{0,\mu})$ identifies the lower bound of the DI functions, it is clear that the HI is any function that together with all its derivatives decreases slower than any power of $(1+\mu)^{-p}$, such that $p \geq 1$. From this definition it follows that a reasonable choice of the Inferior function is when $p = 1$:

$$I\left(P_n^{\lambda,\mu}\right) = (1-\lambda) \cdot g\left(P_n^{0,\mu}\right) \qquad g\left(P_n^{0,\mu}\right) = \frac{1}{1+\mu} \qquad (7)$$

One can easily note, therefore, that the lower function is basically a ratio, which is the evenness factor $(1-\lambda)$ divided by the polarization factor $(1+\mu)$. At first sight it would appear that these two factors are not related to each other. But its analysis brings out that the polarization parameter is closely related to the evenness of the distribution. More precisely, the denominator, $D(\mu) = 1 + \mu$, gives exactly one when applied to the degenerate distribution $P_n^0$ and two in the case of a population evenly distributed over two groups. It follows that this measure gives the number of equally populated groups in a distribution $P_n^{0,\mu}$. The number of equally abundant categories $(s)$ is also known as true diversity in ecology, where $s$ indicates the number of species (Jost, 2006). The true diversity reflects the "effective number of categories" of a distribution and simultaneously takes into account how evenly the population is distributed among these categories. It follows that $s$ lies between one and $n$. However, what happens when the groups aren't equally common?

This situation can be perhaps best illustrated by means of a simple numerical example. Consider, for instance, the distribution $P_{10}^{0,0.5} = (0.75, 0, \cdots, 0, 0.25)$ when $\mu = 0.5$. It seems completely natural to say that the number of equally populated groups should be a number between one and two. It is expected that the number of effective categories regularly increases from one to two, and since $\mu =$ $0.5$ is the mid-point of $P_{10}^{0,\mu}$ than $D(0.5) = 1.5$. This is so obvious that it seems odd to write it. But it is important to realize that this number is measured in units of number of categories and hence its scale does not depend on the choice of a specific index. In this way, it is possible to define a linear scale of evenness measure in terms of the number of groups along the μ direction, see figure 2B. Then for increasing values of the parameter μ the following condition is always met: $CI\left(P_n^{\lambda,0}\right) = D(\mu) \cdot I\left(P_n^{\lambda,0}\right)$.This observation means that the polarization effect of the function $\Delta(PI, P_n)$ in equation (2) does not affect much the value of $CI(P_n)$, and therefore it is a limiting case for an evenness measure. As a consequence, the area below the function $D(\mu) \cdot I(P_n^{\lambda,\mu})$ is equals to the area below the curve $CI\left(P_n^{\lambda,0}\right)$, indicated with the blue colour in Figure 2A. Since $CI\left(P_n^{\lambda,0}\right)$ is a linear function and depends only on the value of λ, the area below this curve is always equals to 0.5. This simple but important result gives a condition to define the set of slowing decreasing HI functions, or the DI space, as:

$$(C1): \qquad L(D, HI) = \iint D(\mu) \cdot HI\left(P_n^{\lambda,\mu}\right) \, d\lambda \, d\mu \, - 0.5 \, \geq \, 0 \qquad (8)$$

It follows that the area below the function $D(\mu) \cdot HI(P_n^{\lambda,\mu})$ must be greater or equal to 0.5. This condition is based on a loss function $L(D, HI)$, which indicates how far away is the HI from the lower function. The value of this function is clearly larger for HI functions closer to the CI and smaller for HI functions approaching the curve $I(P_n^{\lambda,\mu})$. Obviously, all the HI functions for which $L(D, HI)$ is negative do not belong to the class of the DI space. Therefore, the condition $(C1)$ is a necessary condition for any feasible HI. Then, what is a necessary and sufficient condition for a valid HI? In other words, how far the HI should be from $I(P_n^{\lambda,\mu})$ ?

From this analysis emerges that a $HI(P_n^{\lambda,\mu})$ satisfying the condition above could be any curve between $CI(P_n^{\lambda,\mu})$ and the lower function $I(P_n^{\lambda,\mu})$. However, a curve too close to the concentration curve would be insensitive to the variation of the distribution and therefore will be focusing more on evenness than polarization. Then, if $HI(P_n^{\lambda,\mu})$ is below $CI(P_n^{\lambda,\mu})$, how big is the gap between these functions? Or in decision theoretic terms, how can we choose the target function (S) of the various optimization problems, so that they could be compared?

Clearly the choice of the target function will depend on the requirements of the research topic. For example, if the two-point extreme distribution is not considered as homogeneous, then, how the upper bound $S(P^{0,1})$ can be interpreted as a concentration measure?

One solution is to compare this value to the CI value of a distribution that can be easily visualized and interpreted. The most easily visualized distribution is one whose groups are equally abundant. Anyone can imagine what a distribution of, say, four equally groups would be like. For instance, the CI of a distribution of 4 equally abundant categories over ten groups $P_{4,10} = \left(\frac{1}{4}, \frac{1}{4}, \frac{1}{4}, \frac{1}{4}, 0, \cdots, 0\right)$ is $CI\left(P_{4,10}\right) = 0.67$. Clearly, the value $0.67$ is a raw score of the distribution concentration. The conversion of this number to the number of equally abundant categories, or true diversity, is the key to an intuitive interpretation of the diversity in a community (Jost, 2006). In this way, finding a significant target function reduces to the problem of finding an equivalent distribution of $s$ equally populated groups with a frequency of ${}^{1}/_{s}$, where $s < n$.

Going back to the previous example, if the number four is chosen as a diversity index $(s = 4)$, then all communities that share a $CI(X) = 0.67$ are equivalent with respect to a distribution of four equally populated groups. In such a case the two-point extreme distribution $P^{0,1}$ will be equivalent to $P_{4,10}$. Consequently, the number of equally abundant categories creates equivalence classes among communities in which the representative member is a community whose demographic groups are equally common. Therefore, the number of equally abundant categories needed to give a particular value of an index, gives a simple way to specify the evenness of a distribution. The intuitive definition of true diversity is a simple means to transform any concentration index onto a linear and easy to visualize scale involving perfectly even communities (Jost, 2006).

In the above example we have easily computed the CI for a countable number of groups $(s)$, but how is possible to link the CI to any value of $s$ ? Since the underlying distribution is unknown, it is not immediately clear how one should apply this in practice. Furthermore, this transformation will depend on the choice of the concentration measure. However, the Value validity property of the CI can be quite useful for such problem. As discussed in the example of fig.2A, the CI values follows a straight line that basically is an arithmetic progression. In a similar way, the CI value can be mapped to a diversity index number in a linear manner. This transformation is clear if we look at the linear scale on the bottom of fig.2A. We clearly see that $CI(\lambda)$ and $D(\lambda)$ are regularly decreasing and increasing functions of the parameter λ respectively. This result further confirms the importance of the Value validity condition for the CI. Consequently, the direct and inverse formula are:

$$s = D(n, CI) = n - (n-1) \cdot CI \qquad CI(n,s) = \frac{n-s}{n-1} \quad CI \in [0 \quad 1] \ \wedge \ s \in [1 \quad n] \qquad (9)$$

It follows that calculating $s$ or $CI$ is a matter of simple algebra. To see how it work in practice I consider the example of the decile distribution. For instance, the maximum value of the diversity index is $s_{Max} = D(10{,}0.5) = 5.5$ that corresponds to the mid-point of the histogram, that is $\frac{n+1}{2}$. If $s_{Max}$ is chosen as a diversity index, then the two-point extreme distribution is equivalent to an ideal distribution with a frequency of ${}^{1}/_{s_{Max}} \approx 0.18$ over a support of $5.5$. Since we are considering histograms, a natural approach is to solve the inverse problem. Therefore, the rounded CI values for integer values of $s$ between two and five categories are $0.89, 0.77, 0.67$ and $0.56$ respectively. The practical importance of this transformation can be viewed as an attempt to calibrate the value of the target function for the two-point extreme distribution over the different optimization problems. Then, the ultimate tasks are the selection of $S(P_n^{\lambda,\mu})$ and a Value validity condition for the HI. Obviously, the target function should depend on the selection of the parameter $s$ and must be almost equal to the CI when $s = 2$. Since the CI is a linear function in the λ-distribution space, it follows that:

$$S\left(P_n^{\lambda,\mu}, s\right) = (1-\lambda)\cdot f\left(P_n^{0,\mu}\right) \qquad f\left(P_n^{0,\mu}\right) = 1 - \left(\frac{s-1}{n-1}\right)\cdot \mu \qquad s \in \left[\frac{n+5}{4} \quad \frac{n+1}{2}\right] \qquad n \geq 3 \quad (10)$$

The superior function always verifies the condition (C1) in equation (8) and is equal to the CI curve when s=2. Clearly, the choice of the parameter $s$ in the equation (10) depends on the number of categories in the distribution. For instance, the superior function for a distribution of three categories is simply the CI curve. Increasing the value of n, on the other hand, will increase the gap between the Lower and Upper Bound values of $s$.Now it remains to define a Value validity condition for the HI.

As indicated earlier, the superior function and any DI always verifies the condition (C1), this allows to establish a simple criterion to determine the set of feasible DI functions. Once the value of $s$ has been chosen, it is immediate to compute the value of $L(D, S)$. Then, the Value validity condition is given by the following equation:

$$V(s,n,HI) = L(D,S) \geq L(D,HI): \quad L(D,HI) \geq 0$$

$$L(D,S) = \iint D(\mu) \cdot S\left(P_n^{\lambda,\mu}, s\right) d\lambda d\mu - 0.5 \qquad (12)$$

It follows that a necessary condition is the non-negativity of the Loss function for the HI, and this value must be less than or equal to the Loss function for the target function. Then, for any valid HI measure, the one that minimizes the difference $\Delta L(D,S,,HI) = L(D,S) - L(D,HI) \geq 0$, is the optimal HI for the decision problem. On this basis, I computed the Loss function value for the target function $S(P_{10}^{\lambda,\mu}, 3.75)$, where $3.75$ is the minimum feasible value of $s$, to hundred equally spaced points of the $(\lambda,\mu)$ distribution, and the Loss function of the HI for four different polarization measures reported in the table 1 below.

| HOMOGENEITY VALUE VALIDITY λ-DISTRIBUTION $P_{10}^{\lambda,\mu}$ | | | | | |
|---|---|---|---|---|---|
| $CI(P^{0,1})$ | 0.88 | | % | **L (D, S)** | **12.01** |
| $PI$ | $HI(P^{0,1})$ | $\lvert\delta_m\rvert$ | $L(D,HI)$ | **C1** | $\boldsymbol{V(2,10,HI)}$ |
| G-JSD | **0.63** | **0.17** | **5.1** | **YES** | **YES** |
| G-LOV | 0.28 | 0.44 | -12.7 | NO | NO |
| G-COV | 0.25 | 0.52 | -13.8 | NO | NO |
| G-VAR | 0.21 | 0.59 | -21,1 | NO | NO |

Table 1 Homogeneity Index - Value Validity condition for decile distribution( n=10)

It can be seen that of these indices only the G-JSD loss function, $L(D, HI)$, is non-negative and therefore satisfies the necessary condition (C1) for Value validity. This value is also below the loss function of the target function, $L(D, S) = 12.01$, and hence can be used as valid polarization measure in the HI formula. Lastly, the value of the G-JSD for the two-point extreme distribution $HI(P^{0,1})$ is clearly above the minimum value of the lower function, that is 0.5, and hence the value of $|\delta_m|$ is smaller compared to the other polarization measures. This result further confirms the well-behavior of the G-JSD in fig 2B, that closely approaches the target function.

Evidently, the selection of the parameter $s$ in equation (10) determines the sensitivity of the HI to the polarization of the distribution. In this case, the analyst is interested in the use of a PI measure that can be easily adapted to different scenario. One possibility would be to assign a weight to the PI value which calibrate the polarization effect on the HI value. Following this approach, our proposal is to consider a parametric formula of the G-JSD:

$$HI(P_n) = \frac{CI(P_n) + [DI(P_n^1)]^\alpha - [DI(P_n)]^\alpha}{1 + [DI(P_n^1)]^\alpha} \qquad \alpha \in [\alpha_m \quad 1] \ : \ \alpha_m > 0 \quad (13)$$

Since the DI is a slowing decreasing function, it is possible to find an optimal value of the parameter α that minimizes the loss function of the optimization model. Therefore, the optimal solution of the optimization problem can be sought by simply tuning the value of α between $\alpha_m$ and $1$. In the case of $n = 10$ a feasible value of α, for which the Value validity condition holds, is $\alpha_m = 0.75$. Thus $\alpha$ may be treated as the degree of "polarization sensitivity" of the HI measure. The smaller is its value, the greater is the departure from the CI curve. It's worth noting, however, that setting values of $\alpha_m$ closer to zero is almost equivalent to divide the CI by two. This consideration further confirms the importance to check the necessary condition (C1) in equation (8).

The main contribution of our work, however, is not just the comparison of different indices, but to provide a general model to the theory of ordinal concentration measures. More precisely:

- It makes possible to understand and define more precisely what an index does;
- It provides simple means to assess analytically or numerically the validity of an index;
- It permits, to modify an index's behavior to increase or reduce the polarization effect on the concentration measure, as in equation (13);
- Lastly, these formulas have the advantage of being mathematically simple and easy to calculate.

In the next section I will show how this model can be adopted to the classification of a decile distribution.

## 3.2 – Requirement Specifications

A common problem with area-based analysis is to quantify whether a set of observed occurrences are clustered or not dispersed around some central value. It is, therefore, important for the analyst to provide a set of simple and transparent classification criteria to assess the homogeneity of a geographic unit. Accordingly, I propose a possible framework to map the homogeneity values into the specifications space.

A key issue in the classification process, however, is to identify the basic parameters describing the distribution concentration of the attribute for that geographic area. The intuitive definition of true diversity given in the previous section, as the number of equally abundant categories needed to give a particular value of an index, suggests a natural way to specify the evenness of a distribution. The importance of this number is that it is measured in units of number of categories ($s$) and hence its scale does not depend on the choice of a specific index. We have also emphasized the importance of linking this number to the CI to establish an equivalence relation between distributions with the same index value. Similarly, this operation can be extended to the HI. Since the true diversity is an ordinal scale, this transformation is monotonic decreasing for increasing values of $s$. Therefore, given the set of the HI's values which have been assigned to the number of effective categories $(s)$, we can carry out statistical operations on those values using the less than equal $(\leq)$ and greater than $(>)$ relationships. This allows us to partition the range of the index values into $n$ mutually exclusive and exhaustive equivalence classes in which the natural breaks among classes is determined by the number of categories in the distribution. In each of these classes the HI value is less than equal $(\leq)$ to the index value for $s$ equally abundant categories $HI\left(P_{s,n}\right)$, and greater than $(>)$ the index value for $s + 1$ categories $HI(P_{s+1,n})$. Thus, each distribution $P_n$ can be uniquely allocated to a single equivalence class $[s]$ according to the value of its concentration measure $HI(P_n)$. By these means we establish an isomorphism between the equivalence classes and the positive real numbers, this bijection is given by equation (9) for the CI, and a homomorphism between the distributions and the positive real numbers. This basic relation is, therefore, given by:

$$P_n \in [s] \quad \leftrightarrow \quad \begin{cases} HI\left(P_{s+1,n}\right) < HI(P_n) \leq HI\left(P_{s,n}\right) & : \; s \in \{1,2,\cdots,n-1\} \\ P_n = P_{n,n} & : \; s = n \end{cases} \tag{14}$$

For example, in the socio-economic decile distribution the HI range will be divided into ten classes. In this partition, the first and most homogeneous equivalence class $[1]$ identifies the set of distributions such that the HI is less than equal to one, corresponding to the singleton distribution $HI(P_{1,10})$, and greater than $HI(P_{2,10})$, where $P_{2,10} = \left(\frac{1}{2}, \frac{1}{2}, 0, \cdots, 0\right)$. On the other hand, the last equivalence class $[10]$ contains only the uniform distribution $P_{10,10}$ ,since it corresponds to the minimum value of the HI. This classification method creates, therefore, natural divisions among classes that represent the degree of diversity in the distribution of socio-spatial data. A remarkable feature of this procedure is that the homomorphism will be not unique. This means that there is more than a single mapping in which the relationship between the values of the concentration measure reflect the relationship between the distributions of $s$ equally abundant categories. Consequently, whatever HI or CI we choose in the equation (14), the order relation $HI(P_{s+1,n}) < HI(P_{s,n})$ is always verified. However, only a concentration measure that satisfy the Value validity property, like the CI, is capable to divide the range of values into $n-1$ equal width classes. This property comes directly from the linear transformation of this index to the true diversity, see equation (9). Consequently, the difference between any two CI values is constant, that is $\Delta(s, s+1) = CI(n, s) - CI(n, s+1) = \frac{1}{n-1}$. A simple example of an index that does not satisfy this property is the Entropy index. The distance between the maximum values of two adjacent classes is given by $\Delta(s, s+1) = \log_n\left(\frac{s+1}{s}\right)$. As we see, this distance is a decreasing function of $s$. For instance, when $n = 10$, the width of the first class is $\Delta(1,2) \approx 0.30$ and the shortest width is given by the last class, that is $\Delta(9,10) \approx 0.046$. Looking at these numbers, it clear that this index is highly non-linear and, therefore, does not follow the linearity of $s$. That is why the Value validity property is important in the formula of the HI.

Furthermore, the application of the Entropy index in the formula of the HI, or any other index of concentration for nominal categorical variable that does not satisfy the Value validity property, does not define a Fréchet space. This means that the HI is not a positive function and, therefore, it is not a concentration measure. Moreover, the combination of the CI and DI in the HI formula allows to almost preserve the linearity of the CI.

The reader may have noticed that the definition of true diversity represents concentration of the distributions referred to the number of effective categories but does not take into account any information about the order. This observation leads to a more specific concept of diversity in the case of ordinal variables as a distribution of $s$ equally abundant categories clustered on $s$ consecutive bins. In order to avoid any misunderstanding, we will refer to the diversity of an ordinal variable as the range of a distribution $(k_1)$. This parameter sets the smallest interval which contains all the data. For example, let us return to the decile distribution we discussed earlier. In the case of the first equivalence class, the value $k_1$ is 2 and a representative of this class is any distribution having all the observations in at most two consecutive socio-economic groups, except for the distribution $P_{2,10}$. In the same fashion, we could specify the distribution range using the diversity index. For instance, any distribution with a diversity index $(s)$ less than three is equivalent to a distribution concentrated in at most three adjacent groups (i.e. $k_1 = 3 \vee s < 3$). Therefore, the number of effective groups in a distribution is a powerful and intuitive parameter to specify the diversity of a geographic area. It is easy to visualize what this number means and can help analysists to make scientifically informed decisions.

This classification offers, however, only a partial picture of the distribution concentration. For example, if we want to compare the concentration of members of the same class is clear that we cannot make any statement in terms of real concentrations. Then, how is possible to combine the number of effective groups and the cumulative proportion of population within each distribution range?

Clearly, the true diversity is an unambiguous concept only when we are dealing with a countable number of groups. This number can be difficult to interpret when the effective number of groups is a fraction. For instance, what does it mean to have an ideal uniform distribution of $1.20$ groups? This value is not intuitive to health care policy makers, who prefer to think of spatial distribution in terms of proportion or percentage based on the histogram quantiles, groups or categories. To tackle this problem, we propose a simple operating procedure to express the diversity of a geographic area in terms of the cumulative percentage of the histogram quantiles.

### 3.2.1 – True Diversity & Lorenz Curves

A common framework to define evenness indices is the classical Lorenz curve (Gosselin, 2001). Our basic idea is to define a family of Lorenz curves without intersection points. In this way, we obtain a partial order where a Lorenz curve corresponds to a higher degree of evenness if it is situated above the other one (Rousseau, 2001). A Gini index (CI) is naturally associated to these curves and therefore gives rise to the same Lorenz curve ranking. This partial order has enabled us to introduce an additional parameter $(c)$ to further characterize the distribution range concentration. The combination of this parameter $(c)$ with the distribution range $(k_1)$ provides a simple and straightforward way to set a threshold value to the cumulative percentage of the most $(k_1 - 1)$ populated quantiles, groups or categories. For sake of simplicity we will refer to $(k_1 - 1)$ as $k$, or $k_1 = k + 1$, and the corresponding concentration index specification as $CIS = (n, k, c)$. This specification allows us to define a larger number of equivalence classes and hence more options in subsequent analysis.

However, the choice of the interval width and hence the value of $c$ cannot be arbitrary. The objective is to provide a reasonable number of classes for the evaluation and classification of the distributions. For smaller values of the parameter $k$ we should expect to have a larger number of classes and as we get closer to the uniform distribution, this number will get smaller. A natural choice is to form $(n-k)$ classes for each value of $k=\{1,2,\cdots,n-1\}$. Consequently, the total number of classes is $\frac{(n-1)n}{2}$ and the vector $c$ satisfying this condition is $c=\left(\frac{n-1}{n},\frac{n-2}{n-1},\cdots,\frac{n-i}{n-i+1},\cdots,\frac{2}{3},\frac{1}{2}\right)$. Implicitly, the concentration of the last category can be easily calculated by taking the complementary value of c, that is $\bar{c}=\left(\frac{1}{n},\frac{1}{n-1},\cdots,\frac{1}{n-i},\cdots,\frac{1}{3},\frac{1}{2}\right)$. This is an inevitable choice as these classes are subclasses of the distribution range $k_1$.

In this proposed schema, we developed a ready to use table based on the Lorenz partial order of the CI, in order to convert the range of $(s)$ to a quantitative measure of the population distribution. I shall illustrate the essential points easily without pretending to be analytically rigorous.

Table 2 shows, the mapping between the CI values and the true diversity. For sake of simplicity we named the true diversity as Diversity Index. In this table, the integer numbers on the top represent the parameter $k$, that indicates the first $k$ groups in the range. Accordingly, the numbers in the first column $(c)$, indicated with the blue and orange colour, represent the percentage of the population that is equally distributed in the first $k$ groups and the remaining proportion in the last group, respectively. For the purpose of illustrating the shape of the distribution, each cell of the table shows the corresponding histogram, where the blue bins represent the first $k$ categories and the bin filled with the orange colour represents the last category in the range. For instance, the first cell $[1,1]$ contains a histogram of two categories where the first category has $90\%$ of the population. On the other hand, the cell $[1,2]$ contains a histogram of three categories where the first and second category have $45\%$ of the population, or equivalently $90\%$ in the first two categories. It follows, that the last cell in each column, that is the class $[10-k,k]$, contains a histogram of $k+1$ equally abundant categories. This representation allows us to partition the Diversity index range into 45 equivalence classes. Each equivalence class specifies an interval of values for the true diversity. This classification method creates, therefore, natural divisions among classes that represent the degree of diversity in the distribution of socio-spatial data. For instance, the initial class $[1,1]$ specifies the set of distributions with a Diversity Index greater than or equal to one and less than $1.20$, $(1 \leq s < 1.20)$, or equivalently $CI \in (97.78 \;\; 100]$. This means that these set of distributions are equivalent to a distribution that has more than $90\%$ of the population in the first category over a range of two categories. They clearly represent the most homogeneous group in the table. On the other hand, the next class $[2,1]$ specifies the distributions with a true diversity $(s)$ that lies between $1.20$ and $1.21$, $(1.20 < s \leq 1.21)$, or equivalently $CI \in [97.54 \;\; 97.78)$. It follows that the last and most heterogeneous class $[1,9]$ includes the distributions in the following range $(9 < s \leq 10)$, or equivalently $CI \in [0 \;\; 11.11)$.

One way of thinking about the Lorenz partial order of the CI given by this classification system, is the idea of unevenness as smaller values of $k$ corresponds to more homogeneous distributions. This observation is clear if we examine the maximum value of the first equivalence class and the lower bound of the last class for a given value of $k$. Obviously, these values are exactly the same values given by equation (9) for a class $[s]=[k+1]$. It follows that the lower bound corresponds to a Lorenz curve, whose distribution vector $P_{s+1,n}$ is a community of $s+1$ equally abundant categories, that is closer to the diagonal (representing the uniform distribution for n categories) than the Lorenz curve of the maximum value attained for $P_{s,n}$. Then the partition of $[s]$ in $n-s+1$ subclasses represents a finite sequence of probability mass redistributions from the first $k$ categories to the last $k+1$ category, that "transform" the distribution vector $P_{s,n}$ to the vector $P_{s+1,n}$. For a formal definition and properties of the Requirement Specifications see Appendix A5.

| C \ K | 1 | 2 | 3 | 4 | 5 | 6 | 7 | 8 | 9 |
|---|---|---|---|---|---|---|---|---|---|
| 9/10<br>1/10 | 97.78 [1,1]<br>1.20 | 85.56 [1,2]<br>2.30 | 73.34 [1,3]<br>3.40 | 61.12 [1,4]<br>4.50 | 48.88 [1,5]<br>5.60 | 36.66 [1,6]<br>6.70 | 24.44 [1,7]<br>7.80 | 12.22 [1,8]<br>8.90 | 0.00 [1,9]<br>10 |
| 8/9<br>1/9 | 97.54 [2,1]<br>1.21 | 85.16 [2,2]<br>2.34 | 72.84 [2,3]<br>3.44 | 60.05 [2,4]<br>4.59 | 48.81 [2,5]<br>5.61 | 35.80 [2,6]<br>6.78 | 23.46 [2,7]<br>7.89 | 11.11 [2,8]<br>9 | |
| 7/8<br>1/8 | 97.22 [3,1]<br>1.25 | 84.72 [3,2]<br>2.37 | 72.22 [3,3]<br>3.50 | 59.72 [3,4]<br>4.63 | 47.22 [3,5]<br>5.75 | 34.72 [3,6]<br>6.88 | 22.22 [3,7]<br>8 | | |
| 6/7<br>1/7 | 96.82 [4,1]<br>1.28 | 84.12 [4,2]<br>2.43 | 71.42 [4,3]<br>3.57 | 58.74 [4,4]<br>4.71 | 46.02 [4,5]<br>5.86 | 33.33 [4,6]<br>7 | | | |
| 5/6<br>1/6 | 96.30 [5,1]<br>1.33 | 83.34 [5,2]<br>2.50 | 70.38 [5,3]<br>3.67 | 57.40 [5,4]<br>4.83 | 44.44 [5,5]<br>6 | | | | |
| 4/5<br>1/5 | 95.56 [6,1]<br>1.40 | 82.22 [6,2]<br>2.60 | 68.88 [6,3]<br>3.80 | 55.56 [6,4]<br>5 | | | | | |
| 3/4<br>1/4 | 94.44 [7,1]<br>1.50 | 80.56 [7,2]<br>2.75 | 66.67 [7,3]<br>4 | | | | | | |
| 2/3<br>1/3 | 92.60 [8,1]<br>1.67 | 77.78 [8,2]<br>3 | | | | | | | |
| 1/2<br>1/2 | 88.89 [9,1]<br>2 | | | | | | | | |

| | |
|---|---|
| $[i,k]$: $i \in \{1,..,10-k\}$ | Equivalence class |
| $CI(P_{10}(C_i,k))$: | Concentration Index |
| S= $D(10, P_{10}(C_i,k))$: | Diversity Index |

Table 2 True Diversity & Concentration Index for decile distributions ( n =10) – lambda distribution

However, the information provided by the Diversity and Concentration index in Table 2, does not take into account the "separation" between the groups of the distribution.

Whilst the Diversity index may be useful in telling us something about the number of different groups in a community, it does not help us understand how different these groups are from one other. For instance, a community of two effective categories with a population whose members live primarily in the first two socio-economic deciles of the IRSD clearly indicates a higher proportion of relatively disadvantage people in an area than a distribution equally concentrated on the lowest and highest socio-economic decile. If you are still hesitant, think of the IRSD as an index that ranks area on a continuum from the least disadvantaged to most disadvantaged, running from right (e.g. tenth decile) to left (e.g. first decile). Many would agree that under a two-spike distribution concentrated equally on the least disadvantaged and most disadvantaged deciles population is either "disadvantaged" or "not disadvantaged", and one may be inclined to perceive this situation as more "uncertain" than the initial one. This situation is likely to occur in geographic areas where diverse suburbs have high numbers of both the most and least disadvantaged SA1s. As a result, the averaging effect of the socio-economic index score chronically under-reports disadvantage and we cannot conclude that such an area has a relatively medium incidence of disadvantage. When users undertake analysis of the relationship between the level of disadvantage in one area and health, it is important to also look at these underlying characteristics as they can differ markedly between areas with similar IRSD score.

What measure of "separation", then, can be used to describe how spread out the individual groups of a distribution are from one other?

### 3.2.2 – True Diversity & Polarization

A governing approach to illustrate the phenomenon of polarization in societies was mapped out by Foster and Wolfson (Foster & Wolfson, 2010). In this framework, polarization is concerned as the process by which a distribution becomes bipolar. Consequently, the socio-economic bipolarism increases if the population groups are getting more homogeneous around the extreme groups and more separate one to the other. This approach is often called as "bipolarization". Bipolarization is, therefore, seen as a measure of the size of the middle class since it focuses on the concentration of the population around two poles and shows how the centre of the distribution of the specific variable is emptied.

Wolfson (Wolfson, 1994) and the authors who have followed his approach, (Chakravarty & D'Ambrosio, 2010) (Wang & Tsui, 2000), define the middle class using the national median income as a reference point. This definition, however, is not helpful to identify the middle class of a socio-economic index, such as the IRSD, since it summarizes a range of information about the economic and social conditions of people and households within an area. Also, the index scores are on an arbitrary numerical scale. The scores do not represent some quantity of disadvantage. For example, we cannot infer that an area with an IRSD value of 500 is twice as disadvantage as an area with an index value of 1000, it just had more markers of relative disadvantage. The IRSD, therefore, is an ordinal measure and as such there is no meaningful arithmetic relationship between the values. This means that the index score is used to order areas in terms of disadvantage and not for analyses which aim to somehow quantify socio-economic conditions. It can be linked to a soccer premiership table. Just as a team that finished the season with 100 points cannot claim to be twice as good as a team that finishes the season with 50 points. Similarly, it is incorrect to use the size of the gap between the scores to compare levels of disadvantage. For example, the difference in disadvantage between two areas with scores of 200 and 400, is not necessarily the same as the difference between two areas with scores of 1000 and 1200. Therefore, it is incorrect to state that an area with a low IRSD score is disadvantaged. It can only be determined that an area is disadvantage relative to other areas. Consequently, the index scores are used to rank areas in terms of disadvantage. For this reason, it is generally recommended using the index quantiles (e.g. decile) for analysis, rather than using the index scores. Then, how do we define the middle class of the IRSD variable by looking at the SA1's distribution?

As indicated in the conceptual framework section, the IRSD includes only variables related to relative disadvantage. This index is, therefore, appropriate for distinguishing between relatively disadvantaged areas. This is shown by the long-left tail in the IRSD distribution displayed in Figure 6. In this figure the horizontal axis shows the IRSD score, and the vertical axis shows how many SA1s have that score (52,575 SA1s in total). The values range from 121 to around 1200. The IRSD have been standardized to have a mean of 1000 and a standard deviation of 100. It is important to note that the distribution of the IRSD is not exactly a normal distribution even though it has been standardized. More precisely, the scores between 900 and 1100 represent almost 74 per cent of the SA1s.

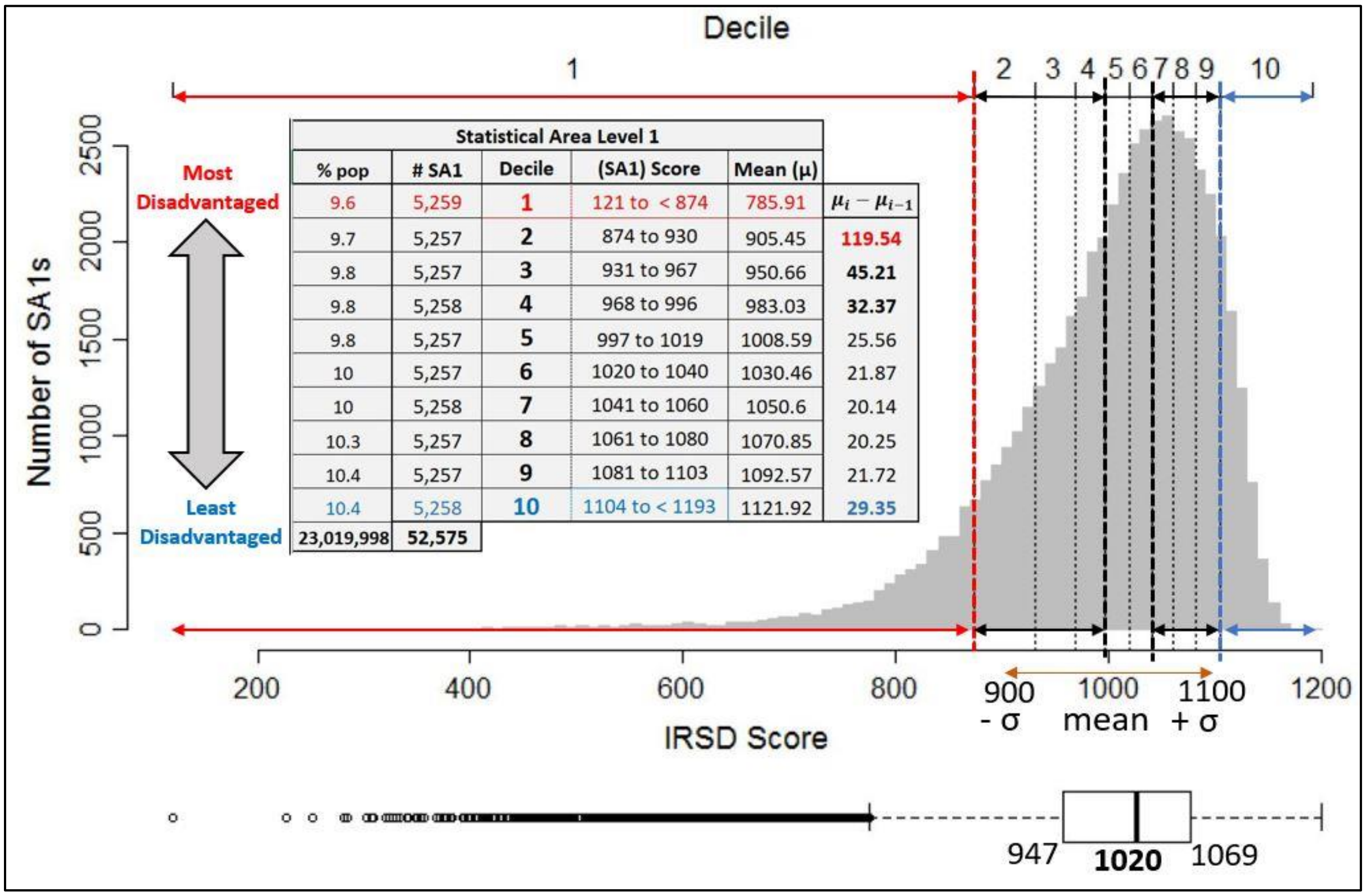

| Statistical Area Level 1 | | | | | |
|---|---|---|---|---|---|
| % pop | # SA1 | Decile | (SA1) Score | Mean (μ) | $\mu_i - \mu_{i-1}$ |
| 9.6 | 5,259 | 1 | 121 to < 874 | 785.91 | |
| 9.7 | 5,257 | 2 | 874 to 930 | 905.45 | 119.54 |
| 9.8 | 5,257 | 3 | 931 to 967 | 950.66 | 45.21 |
| 9.8 | 5,258 | 4 | 968 to 996 | 983.03 | 32.37 |
| 9.8 | 5,257 | 5 | 997 to 1019 | 1008.59 | 25.56 |
| 10 | 5,257 | 6 | 1020 to 1040 | 1030.46 | 21.87 |
| 10 | 5,258 | 7 | 1041 to 1060 | 1050.6 | 20.14 |
| 10.3 | 5,257 | 8 | 1061 to 1080 | 1070.85 | 20.25 |
| 10.4 | 5,257 | 9 | 1081 to 1103 | 1092.57 | 21.72 |
| 10.4 | 5,258 | 10 | 1104 to < 1193 | 1121.92 | 29.35 |
| 23,019,998 | 52,575 | | | | |

Figure 6 Index of Relative Socio-economic Disadvantage score distribution.

Clearly, the SA1s still remain in the same order but the scores are more condensed around the average compared to the normal distribution. As a consequence, the decile cut-offs (marked along the top axis) show that the middle deciles are close together and there is little difference in the scores of SA1s in the middle deciles. This indicates that SA1s with mid-range scores are neither particularly disadvantaged or lacking disadvantage relative to other areas. To further confirm this descriptive analysis, we computed the mean socio-economic score for each decile (indicated as $\mu$ in Figure 6), because, if the IRSD score is uniformly distributed, the difference in mean socio-economic score between adjoining deciles should be even. The mean difference, indicated in the last column in Figure 6, is higher between the most disadvantaged and second decile than any other adjoining decile. On the other hand, the difference is small among the middle deciles making differentiating between socio-economic groups difficult. Partly because of this, the interpretation of the index values is more straightforward for areas which have extreme values (i.e. very high or very low index values). For example, it is usually easy to see why a SA1 which is in the first or last decile has that status. That is, those SA1s which have large proportions of households with similar characteristics, will tend to have the lowest or highest index scores. For instance, all the IRSD population with stated annual household equivalized income between $1 and $20,799 live in SA1s with a score between the first and second decile (ABS, Socio-Economic Indexes for Areas (SEIFA) Technical Paper, cat no 2033.0.55.001, 2011). In contrast, areas with mid-range index values tend to contain a broader mix of people and households. This phenomenon is common in highly skewed socio-economic distributions and is generally referred in the literature as "clumping" (Vyas & Kumaranayake, 2006).

The discriminating power of this index lies, therefore, in the lower and upper end of the distribution for identifying the relative disadvantage (lower deciles), and the relative lack of disadvantage (upper deciles) of people in an area. Then, if one's prime focus is on disadvantage and lack of disadvantage say, one might choose to summarize the index by using the bottom and top decile of SA1 values in each area. Therefore, we define the middle class as the groups of individuals who live in areas where the scores fall within the second and ninth decile, i.e. $[874 \quad 1103]$. A justification of this choice is that the disadvantage of SA1s in the middle deciles is not likely to vary much compared to the first and last deciles. Due to IRSD being created at an area level, the population deciles are not exactly a tenth of the population. This difference, however, is not particular significant, as indicated in the first column of the table in Figure 6. This further confirms the SA1's property of equal population size. Thus, the population in the most disadvantaged decile, is the $9.6\%$ of the national population residing in the most disadvantaged areas, rather than the $10\%$ of the population.

Once the definition of the middle class is decided the final step is to analyse the concentration and polarization of a distribution by looking at the population percentage in the extreme and middle categories. The question, then arises as how the evenness and dispersion should be combined, and how to find a measure describing this combination. This is where the two-parameter $(\lambda, \mu) - distribution$ comes into play. The reason is that this distribution covers the concept of both evenness and polarization, and, hence, provides an accurate representation of the extent of the CI and HI functions.

We emphasized earlier the importance of linking the CI to the Diversity Index to establish an equivalence relation between distributions with the same CI value. Similarly, this operation can be extended to the single parameter $\lambda - distribution$ , equation (1). As illustrated in Figure 3A, the number of effective categories indicated as $D(\lambda)$ on the bottom, is monotonic increasing for increasing values of $\lambda$. Therefore, given the number of effective categories $(s)$, which have been assigned to the equivalence classes in Table 2, there exists one and only one value of the parameter $\lambda$ such that $D\left(n, CI\left(P_n^{\lambda}\right)\right) = s$. By these means we establish a correspondence between the equivalence classes in Table 2 and the non-polarized vector $P_n^{\lambda}$. This bijection is represented in Table 3.

In this table, the equivalence classes depicted in yellow, represent the set of “not polarised” distributions, i.e. $P_{10}^{\lambda} = \left(1 - \frac{9}{10}\lambda, \frac{\lambda}{10}, \ldots, \frac{\lambda}{10}\right)$.

**d1: *population%*** *HI*($P_{10}$): ***Homogeneity Index %*** *M*: *Middle Class population %* {*d*2, *d*3, .., *d*9}

| λ \ K | 1<br>$1 \le s \le 2$ | 2<br>$2 < s \le 3$ | 3<br>$3 < s \le 4$ | 4<br>$4 < s \le 5$ | 5<br>$5 < s \le 6$ | 6<br>$6 < s \le 7$ | 7<br>$7 < s \le 8$ | 8<br>$8 < s \le 9$ | 9<br>$9 < s \le 10$ |
|---|---|---|---|---|---|---|---|---|---|
| $\frac{11k-9}{90}$ | 98.00 [1,1]<br>97.37 | 87.00 [1,2]<br>82.97 | 76.00 [1,3]<br>68.92 | 65.00 [1,4]<br>55.41 | 54.00 [1,5]<br>42.59 | 43.00 [1,6]<br>30.60 | 32.00 [1,7]<br>19.50 | 21.00 [1,8]<br>9.3 | 10.00 [1,9]<br>0 |
| $\frac{10k-8}{81}$ | 97.78 [2,1]<br>97.08 | 86.67 [2,2]<br>82.53 | 75.56 [2,3]<br>68.53 | 64.44 [2,4]<br>54.74 | 53.33 [2,5]<br>41.84 | 42.22 [2,6]<br>29.79 | 31.11 [2,7]<br>18.65 | 20.00 [2,8]<br>8.4 | |
| $\frac{9k-7}{72}$ | 97.50 [3,1]<br>96.70 | 86.25 [3,2]<br>81.99 | 75.00 [3,3]<br>67.65 | 63.75 [3,4]<br>53.90 | 52.50 [3,5]<br>40.91 | 41.25 [3,6]<br>28.78 | 30.00 [3,7]<br>17.59 | $70 \le M < 80$<br>$s = 8 + \left(\frac{k+1}{10}\right)$ | 15.00 [3,9]<br>7.9 |
| $\frac{8k-6}{63}$ | 97.10 [4,1]<br>96.23 | 85.71 [4,2]<br>81.30 | 74.29 [4,3]<br>66.76 | 62.86 [4,4]<br>52.85 | 51.43 [4,5]<br>39.72 | 40.00 [4,6]<br>27.50 | $60 \le M < 70$<br>$s = 7 + \left(\frac{k+1}{10}\right)$ | 25.00 [4,8]<br>17.02 | 20.00 [4,9]<br>15.73 |
| $\frac{7k-5}{54}$ | 96.67 [5,1]<br>95.62 | 85.00 [5,2]<br>80.39 | 73.33 [5,3]<br>65.59 | 61.67 [5,4]<br>51.43 | 50.00 [5,5]<br>38.13 | $50 \le M < 60$<br>$s = 6 + \left(\frac{k+1}{10}\right)$ | 35.00 [5,7]<br>26.78 | 30.00 [5,8]<br>24.90 | 25.00 [5,9]<br>23.62 |
| $\frac{6k-4}{45}$ | 96.00 [6,1]<br>94.74 | 84.00 [6,2]<br>79.09 | 72.00 [6,3]<br>63.92 | 60.00 [6,4]<br>49.49 | $40 \le M < 50$<br>$s = 5 + \left(\frac{k+1}{10}\right)$ | 45.00 [6,6]<br>37.20 | 40.00 [6,7]<br>34.71 | 35.00 [6,8]<br>32.82 | 30.00 [6,9]<br>31.52 |
| $\frac{5k-3}{36}$ | 95.00 [7,1]<br>93.41 | 82.50 [7,2]<br>77.17 | 70.00 [7,3]<br>61.46 | $30 \le M < 40$<br>$s = 4 + \left(\frac{k+1}{10}\right)$ | 55.00 [7,5]<br>48.29 | 50.00 [7,6]<br>45.21 | 45.00 [7,7]<br>42.68 | 40.00 [7,8]<br>40.77 | 35.00 [7,9]<br>39.47 |
| $\frac{4k-2}{27}$ | 93.33 [8,1]<br>91.24 | 80.00 [8,2]<br>73.97 | $20 \le M < 30$<br>$s = 3 + \left(\frac{k+1}{10}\right)$ | 65.00 [8,4]<br>60.01 | 60.00 [8,5]<br>56.40 | 55.00 [8,6]<br>53.28 | 50.00 [8,7]<br>50.71 | 45.00 [8,8]<br>48.78 | 40.00 [8,9]<br>47.46 |
| $\frac{3k-1}{18}$ | 90.00 [9,1]<br>86.87 | $10 \le M < 20$<br>$s = 2 + \left(\frac{k+1}{10}\right)$ | 75.00 [9,3]<br>72.33 | 70.00 [9,4]<br>68.22 | 65.00 [9,5]<br>64.54 | 60.00 [9,6]<br>61.39 | 55.00 [9,7]<br>58.80 | 50.00 [9,8]<br>56.83 | 45.00 [9,9]<br>55.52 |
| | $0 \le M < 10$<br>$s = 1 + \left(\frac{k+1}{10}\right)$ | 85.00 [10,2]<br>85.16 | 80.00 [10,3]<br>80.63 | 75.00 [10,4]<br>76.46 | 70.00 [10,5]<br>72.75 | 65.00 [10,6]<br>69.55 | 60.00 [10,7]<br>66.94 | 55.00 [10,8]<br>64.95 | 50.00 [10,9]<br>63.62 |
| | $d10$: %<br>$[i, j]$: $\boldsymbol{i+j}$ | 15.00<br>12 | 20.00<br>13 | 25.00<br>14 | 30.00<br>15 | 35.00<br>16 | 40.00<br>17 | 45.00<br>18 | 50.00<br>19 |

Table 3 Homogeneity and equivalence classes classification - lambda distribution

Accordingly, for each row in the table there is a single expression of the parameter $\lambda$ as a function of $k$. The values of the parameter $\lambda$ are indicated with the blue colour in the first column. For instance, the $\lambda$ values of the equivalence classes $[1, k]$ in the first row are given by $\lambda = \frac{11k-9}{90}$. It follows that the $\lambda - distribution$ of the first equivalence class $[1,1]$ is $P_{10}^{2/90} = \left(0.98, \frac{2}{900}, \cdots, \frac{2}{900}\right)$ and the representative member of the last equivalence class $[1,9]$ is obviously the uniform distribution. The same reasoning can be easily applied to the other rows of the table. To facilitate the analysis of the distribution, the population percentage in the first decile and the HI value are indicated with the black and purple colour, respectively. A remarkable feature of this distribution is that the Lorenz partial order of the CI is preserved in the case of the HI. This means that there is more than a single mapping in which the relationship between the values of the HI measure reflect the relationship between the distributions of $s$ equally abundant categories. Furthermore, this distribution is useful to model skewed probability laws and hence appropriate to study the concentration around the extreme categories. This classification, however, offers only a partial picture of the distribution concentration.

As noted earlier, a configuration of two well separated poles at the right and at the left of the centre identifies a highly polarized distribution and, hence, it is more heterogenous than any community of two effective groups. However, the interesting question concerns the behavior of polarization through the entire range of the diversity index $(s)$. Bearing in mind that our aim is to measure the phenomenon of bipolarism, the basic idea is to look at the concentration of the least and most disadvantage decile for different sizes of the middle class. Clearly, a bimodal $(\lambda, \mu) - distribution$, i.e. $P_{10}^{\lambda,1} = \left(\frac{1-\lambda}{2} + \frac{\lambda}{10}, \frac{\lambda}{10}, \cdots, \frac{\lambda}{10}, \frac{1-\lambda}{2} - \frac{\lambda}{10}\right)$, has unambiguously more polarization exactly when it has a smaller middle class; since we move from a clustered distribution around two poles with an empty middle class to a uniform distribution $(s = 10)$ without any pole and a middle class size of $80\%$. For instance, the middle class size of a community formed by three groups is $10\%$ and the corresponding vector is $P_{10}^{1/8,1} = (0.45, 0.0125, \cdots, 0.0125, 0.45)$. Consequently, one can determine the size of the middle class of $P_{10}^{\lambda,1}$ for a given number of effective groups. Following this method, we computed the middle-class size of a countable number of groups. Then, for each vector $P_{10}^{\lambda,1}$, we transferred the population mass – a bit at a time – from the last category to the first category. The start of the process thereby moves us away from a bimodal or U-shape distribution to a unimodal distribution with most of the total population located in the first decile. Clearly, the final distribution is more homogeneous and less polarized than the initial one. This process is illustrated in the lower right part of Table 3. The light blue equivalence classes represent the set of "polarised" vector $P_{10}^{\lambda,\mu>0}$ and the stylized distributions in the last column identify the family of U-shape distributions. For example, the equivalence class $[10,9]$ is represented by the two-point extreme distribution $P_{10}^{0,1}$ and the representative members of the equivalence classes next to the left can be easily obtained by simply transferring $5\%$ of the population from the last decile to the first decile. It follows that the class $[10,8]$ has $55\%$ and $45\%$ in the least and most disadvantaged decile, respectively. Clearly, moving towards the first class of each row must increase the HI value and decrease the Diversity index. On the other hand, moving upwards along the columns is equivalent to decrease the population in the extreme deciles by $5\%$ and increase the size of the middle class by $10\%$. For example, the equivalence class $[9,8]$ has $50\%$ and $40\%$ in the outer deciles and a middle-class size of $10\%$. To make reading of the data in our table easier, the size of the middle class $(M)$ and the number of effective categories $(s)$ are indicated in the white cells below the last yellow equivalence class of each column. Then, the notation $0 \leq M < 10$ in the first column indicates that the size of $M$ is less than $10\%$ for the not polarized distributions $[i, 1]$ and is equal to zero for the set of polarized distributions $[10, k]$. As for the number of effective categories of $[10, k]$, indicated as $s = 1 + \left(\frac{k+1}{10}\right)$, it can be easily calculated by substituting the value of $k$. For example, the Diversity index of $[10,8]$ is $1.9$. Lastly, the row on the bottom shows the proportion of the population in the last category for a given sum of the row and column indices.

This classification framework gives extra information about the distribution of the IRSD, without doing analysis at the SA1 level. For example, given the number of groups and the HI value one can easily "guess" the concentration of a distribution. This can be done by partitioning the Diversity index range into nine sets, based on the values of $k$, indicated on the top of the table. For instance, when $k$ is equal to one the value of $s$ is any number between one and two. On the other hand, for $k > 1$ the value of $s$ is greater than $k$ and less than or equal to $k + 1$, i.e. $(k < s \leq k + 1)$. For each partition of $s$, we defined two typologies of equivalence classes. The classes in the same column, depicted in yellow, represent the set of "not polarised" distribution. On the other hand, the distributions on the bottom lines, next to the right of the white cells below the last equivalence class, represent "polarised distributions" with smaller HI values. As a consequence, a distribution with a Diversity index $s \leq 2$ belongs to the set of polarised distributions if and only if its HI value is less than the HI value of a distribution formed by two categories, $HI(P_{10}) < 86.87$, indicated by the class $[9,1]$. The same reasoning can be extended to the other intervals value of $s$. This allows us to partition the range of the HI into $19 - 2k$ mutually exclusive and exhaustive classes, for each value of $k \in \{1,2,\cdots,9\}$. In this way, one can easily assign a distribution $(P_{10})$ to an equivalence class in Table 3, just by looking at the Diversity index and HI values. A simplified explanation of this operating procedure runs as follows:

1) Given the CI value of a distribution $P_{10}$, $CI(P_{10})$, one can easily compute the corresponding Diversity index by the application of the equation (9): $s = 10 - 9 \cdot CI(P_{10})$.
2) Given the value of the Diversity index $(s)$, one can identify the corresponding column set $j$ of the distribution by checking the following condition:

$$j: \begin{cases} 1 & 1 \leq s \leq 2 \\ k & k < s \leq k + 1: \quad k \in \{2,3,\cdots,9\} \end{cases}$$

3) Given the value of the column set $j$, one can identify the distribution typology by comparing the HI value of $P_{10}$ to the HI value of a distribution of $j + 1$ groups $(P_{10,j+1})$.

$$Typology: \begin{cases} HI(P_{10}) \geq HI(P_{10,j+1}) & not\ polarised \\ HI(P_{10}) < HI(P_{10,j+1}) & polarised \end{cases}$$

4) Given the distribution typology and the column set $j$, one can assign the distribution $P_{10}$ to an equivalence class by simply comparing its HI value to the HI value of that class. Informally, for the first class of each column, one can check the following conditions:

$$P_{10} \in [1,j] \leftrightarrow \begin{cases} 97.37 \leq HI(P_{10}) \leq 1 & j = 1 \\ 82.97 \leq HI(P_{10}) < 86.87 & j = 2 \\ \quad .. & \\ 0 \leq HI(P_{10}) < 8.4 & j = 9 \end{cases}$$

So, for each equivalence class $[1,j]$ the Upper bound is given by the value of the HI of $j$ categories indicated in the class $[11-j, j-1]$ and the minimum value by the class $[1,j]$. For the remaining not polarized distribution $[i,j]$ the logic is the same with the only difference that the value of the Upper bound is given by the value of the HI in the class $[i-1,j]$. For instance, in the first column $P_{10} \in [2,1] \leftrightarrow 97.08 \leq HI(P_{10}) < 97.37$. As for the not polarised distribution, the Upper bound of the classes in the rows $11-j$ is given by the value of the HI in the previous column. For instance, the second polarised class in the last row is $[10,3]$, and $P_{10} \in [10,3] \leftrightarrow 80.63 \leq HI(P_{10}) < 85.16$.

To help get a feeling for how this classification method works, we show the application of this operating procedure to a polarised and not polarised SA3.

For example, the SA3 of East Arnhem situated in the far north-eastern of the Northern Territory has roughly a population of sixteen thousand people (15,959) and a socioeconomic decile distribution vector $P_{EA} = (\mathbf{0.659}, 0,0,0,0,0, \mathbf{0.032}, \mathbf{0.122}, \mathbf{0.167}, \mathbf{0.02})$. This is clearly a polarized distribution, in which $65\%$ of the population belong to the first and most disadvantaged decile and the rest to the last four and least disadvantaged deciles. The CI and HI are $86.89$ and $71.58$, respectively. Following the steps discussed earlier, it follows:

1. First, we convert the CI value to a Diversity index $s = 10 - 9 \cdot (0.869) \approx 2.18$.
2. **Identification of the column j**: $s = 2.18 \rightarrow 2 < s \leq 3 \rightarrow j = 2$.
3. **Identification of the typology:** First, we compare the HI value to the value given by the class $[8,2]$. $HI = 71.58 < 73.97 \rightarrow polarised\ distribution.$
4. **Assignment to an equivalence class:** We compare the HI value to the set of polarised classes in the second last row until we find the right interval. In this case, $68.22 \leq HI < 72.33 \rightarrow P_{EA} \in [9,4]$. This means that East Arnhem is equivalent to a polarized distribution of roughly two effective groups and almost $70\%$ of the population in the modal extreme category.

Let's us consider the case of a compact or not polarised distribution. For instance, the SA3 of Ku-ring-gai, situated in the Upper North share region of Sydney, has roughly a population of 120 thousand people (119,312) and is one of the least disadvantaged SA3 in Australia. The socioeconomic decile distribution vector is $P_{KG} = (0,0,0,0, \mathbf{0.004}, \mathbf{0.01}, \mathbf{0.021}, \mathbf{0.038}, \mathbf{0.170}, \mathbf{0.757})$. This is clearly a homogeneous SA3, in which $92.7\%$ of the population live in the last two least disadvantaged deciles. The CI and HI are $91.75$ and $91.77$, respectively. Following the same procedure as the previous example it is immediate to compute the number of effective categories, $s = 1.74$ and the equivalence class [8,1]. It follows that the SA3 of Ku-ring-gai is almost equivalent to a distribution of 1.74 groups and a population percentage of 93 % within a range of two groups. This result confirms that the residential population of this geographic unit is almost concentrated in the last two deciles.

This approach is suitable to study the concentration of skewed distributions located in the lower or higher categories of the IRSD and then it is insensitive to changes in the middle of the distribution. How can we further classify symmetric distributions in the categories of the middle class of the IRSD?

The main problem of the IRSD score distribution is that it is unable to distinguish the residential population among the middle deciles. As indicated previously, the scores are weighted combinations of disadvantaged indicators which have been standardized to a distribution with a mean of 1000 and standard deviation of 100. An area with all of its indicators equal to the national average will receive a score of 1000. The score for an area will decrease or increase if an area has an indicator of disadvantage that is greater or smaller than the national average. It follows that indicators which are further away from the national average have a larger impact on the score. It is important to remember, however, that the scores are an ordinal measure and therefore best interpreted as deciles rather than raw values. Then, the problem is to identify the most representative or "neutral" deciles of the IRSD score distribution.

One approach is to identify the decile where the mean value is located, that is the fifth decile in figure 3. A problem with this approach is that the scores of areas in the right deciles of a left-skewed distribution are less spread out than the scores of areas in the lower deciles, as evidenced by the steep right slope of the IRSD. A possible solution is to consider as "central deciles" the categories between the mean and the median scores.

Following this method, we propose a symmetric version of the $(\lambda,\mu) - distribution$, to study the concentration of symmetric distributions located in the middle categories. Thus, instead of $P_n^{\lambda}$ in equation (4) we shall use:

$$\hat{P}_{10}^{\lambda,\mu} = \left(d_1, \frac{\lambda}{10}, \frac{\lambda}{10}, \frac{\lambda}{10}, d_5, d_6, \frac{\lambda}{10}, \frac{\lambda}{10}, \frac{\lambda}{10}, d_{10}\right) \qquad \lambda,\mu \in [0 \quad 1] \qquad (15)$$

Where

$$d_1 = d_{10} = \frac{\mu}{2}(1-\lambda) + \frac{\lambda}{10} \qquad\qquad d_5 = d_6 = \frac{(1-\lambda)(1-\mu)}{2} + \frac{\lambda}{10}$$

An important characteristic of this probability distribution is that the central categories represent the median deciles of the distribution. This means that the first half and second half contain exactly 50% of the population. This property enables us to establish a correspondence between the equivalence classes in Table 2 and a family of symmetric distributions located in the fifth and sixth deciles. This bijection is represented in Table 4.

| %: $MD = \{d5, d6\}$; $HMD = \{d2, d3, d4\}$; $LMD = \{d7, d8, d9\}$ *Homogeneity Index* % | | | | | | |
|---|---|---|---|---|---|---|
| λ \ K | 2<br>$2 < s \leq 3$ | 3<br>$3 < s \leq 4$ | 4<br>$4 < s \leq 5$ | 5<br>$5 < s \leq 6$ | 6<br>$6 < s \leq 7$ | 7<br>$7 < s \leq 8$ |
| $\frac{11k-19}{80}$ | 97.00 86.34 [1,2] | 86.00 74.14 [1,3] | 75.00 61.87 [1,4] | 64.00 49.53 [1,5] | 53.00 37.17 [1,6] | 42.00 24.79 [1,7] |
| $\frac{10k-17}{72}$ | 96.67 85.97 [2,2] | 85.56 73.64 [2,3] | 74.44 61.25 [2,4] | 63.33 48.78 [2,5] | 52.22 36.30 [2,6] | 41.11 23.79 [2,7] |
| $\frac{9k-15}{64}$ | 96.25 85.51 [3,2] | 85.00 73.02 [3,3] | 73.75 60.46 [3,4] | 62.50 47.85 [3,5] | 51.25 35.21 [3,6] | 40.00 22.54 [3,7] |
| $\frac{8k-13}{56}$ | 95.71 84.91 [4,2] | 84.29 72.22 [4,3] | 72.86 59.48 [4,4] | 61.43 46.67 [4,5] | 50.00 33.81 [4,6] | |
| $\frac{7k-11}{48}$ | 95.00 84.13 [5,2] | 83.33 71.18 [5,3] | 71.67 58.13 [5,4] | 60.00 45.04 [5,5] | HL = 30<br>$s = 7$ | 35.00 28.41 |
| $\frac{6k-9}{40}$ | 94.00 83.02 [6,2] | 82.00 69.67 [6,3] | 70.00 56.27 [6,4] | HL = 22.5<br>$s = 6$ | 45.00 39.70 | 32.50 36.26 |
| $\frac{5k-7}{32}$ | 92.50 81.36 [7,2] | 80.00 67.45 [7,3] | HL = 15.5<br>$s = 5$ | 55.00 51.01 | 30.00 44.14 | |
| $\frac{4k-5}{24}$ | 90.00 78.59 [8,2] | HL = 7.5<br>$s = 4$ | 65.00 62.30 | 27.50 52.02 | *HMD*: *HIGH − MD* | |
| | HL = 0<br>$s = 3$ | 75.00 73.60 | 25.00 59.95 | | *LMD*: *LOW − MD*<br>*MD*: *Middle Disadvantage* | |
| HL = 0<br>$s = 2$ | 100 89.64 | 0 63.62 | %: $HL = \{d2, d3, d4, d7, d8, d9\}$ | | | |

Table 4 Homogeneity Index and equivalence classes classification - symmetric lambda distribution.

Following the same principles used for the construction of Table 3, we defined two typologies of distributions: the not polarised distributions $\hat{P}_{10}^{\lambda,0}$, represented by the yellow equivalence classes, and the polarised $\hat{P}_{10}^{\lambda,\mu}$, on the bottom rows of the table. Precisely, we start with most of the total population located in the inner deciles (i.e. 5th and 6th decile) and end with most of the population located in the outer deciles. The transfer from the middle of the distribution towards the extreme categories is often referred in the literature of polarization as "median preserving spread". To facilitate the reading of the table the percentage in the central categories is indicated with the black color and the cumulative percentage in the high and low middle decile of the polarized distribution (HL) with the green color. For instance, a symmetric distribution of three categories may have three different representative distributions based on its HI value. In such a case, the most homogeneous distribution is represented by the equivalence class $[8,2]$, and the least homogeneous is the outmost right cell in the last second row of the table. An example on how to use this table will be discussed in the Benchmark Geography section of chapter 5.

The essence of this approach is that it is a simple and effective method of representing the homogeneity of a community in “pictures”. It is easy to do this just by looking at the number of different groups $(s)$ and the “separation” among these groups, represented by the middle class of the distribution. The CI and HI classification represented in Tables 2,3 and 4 have the advantage of creating a link between the CI (that looks at the number of different groups), on one side, and the polarization or the phenomena of the disappearing middle class (that looks only at the separation between the extreme groups in a community), on the other side. Moreover, the classification represented in Tables 3 and 4 offer a simple mean to look at the concentration of skewed and symmetric distributions located in the extreme and middle categories. A key foundation for the quantitative analysis of these tables, however, is the definition of what is a “high” and “low” concentration of social disadvantage.

### 3.2.3 – Homogeneity degree classification

As discussed previously, the IRSD score distribution in the middle and upper deciles are less spread out then the scores of areas in the lower deciles. This insight leads to the problem of choosing cut-off points to differentiate the IRSD decile distribution into broad level of socio-economic disadvantaged group and determine the number of effective categories for a homogeneous group.

The most common cut-off points of the socio-economic index is the division into quintiles (Gwatkin, Rustein, Johnson, & et al., 2000), but it assumes a quite uniform distribution of the scores and may not be appropriate for the IRSD. Filmer and Pritchett (Filmer & Pritchett, 2001) used arbitrary cut-off points, such as the lowest $40\%$, the highest $20\%$ and the rest as the middle group. So, in our case, applying the $40-40-20$ split rule as in Filmer and Pritchett would be more realistic to reflect the "shape" of the underlying distribution for the IRSD. The Data and Research unit of Australian Capital Territory (ACT) proposed a classification of the IRSD into four groups to detect the level of disadvantage in the Australian Capital Territory (ACT gov, 2012) . This classification includes the lowest $40\%$ in the first two groups as the most disadvantaged deciles and the rest deciles as the least disadvantaged groups. It emerges from this analysis that a disadvantaged community or IRSD distribution has a greater proportion of people in the first four deciles. Therefore, researchers or policy analysts interested in the relationship between disadvantage on health status need to carefully consider the concentration in these deciles. As a result, a reasonable specification for a homogeneous geography would be an IRSD decile distribution formed by at most four effective categories.

To consider a very simple case, suppose to classify the homogeneity of a decile distribution by dividing the HI range into four groups of concentration, in which the natural breaks among groups is determined by the number of equally abundant categories in the distribution. In this partition, the first and most homogeneous group identifies the set of distributions such that the HI is less than or equal to $100$, corresponding to the singleton distribution, and greater than or equal to the HI value of a community formed by four groups, indicated as $HI(4) = HI(P_{4,10}) = 68.53$, where $P_{4,10} = \left(\frac{1}{4}, \frac{1}{4}, \frac{1}{4}, \frac{1}{4}, 0,0,0,0,0,0\right)$. The second and third group include communities of five and six groups, indicated as $HI(5) = 57.62$ and $HI(6) = 46.62$, respectively. Lastly, the set of heterogeneous units is represented by communities with a larger number of effective groups. This classification and the evaluation of the geographic unit are illustrated in table 5.

| Group | GUIDELINES FOR ACCEPTANCE/REJECTION OF PROPOSED REGION (N = 10 Categories) | | |
|---|---|---|---|
| j | Interval $I_j$ | $\alpha = 1$ | Decision Support System |
| A | $HI(4) \leq HI(P_{10}) \leq 100$ | $68.53 \leq HI(P_{10}) \leq 100$ | Proposed region is Acceptably homogeneous. |
| B | $HI(5) \leq HI(P_{10}) < HI(4)$ | $57.62 \leq HI(P_{10}) < 68.53$ | Marginal heterogeneity – reassignment of some units may be beneficial. |
| C | $HI(6) \leq HI(P_{10}) < HI(5)$ | $46.62 \leq HI(P_{10}) < 57.62$ | Judgment required whether to accept homogeneous region or to reassign units to other regions to improve homogeneity of current grouping units. |
| D | $HI(P_{10}) < HI(6)$ | $HI(P_{10}) < 46.62$ | Proposed region is likely heterogeneous – reassignment of some units is needed. |

Table 5 Homogeneity Index guidelines for acceptance/rejection of proposed region defined by socioeconomic decile distribution

The first two groups in the table include units that are reasonably homogeneous in terms of the categorical variable examined. The third group, on the other hand, requires further investigation. Lastly, the group D identifies the set of heterogeneous geographic units.

A straightforward way to look at the concentration of these groups is to identify the equivalence classes that fall within the interval of values for each group. For instance, the first group (A) includes all the equivalence classes in Tables 3 and 4 such that $HI(P_{10}) \geq 68.53$. On the other hand, the group of marginal heterogeneous distributions (B) includes all the equivalence classes that fall in the interval $57.69 \leq HI(P_{10}) < 68.53$, and so on. Clearly, other criteria can be chosen for the identification of homogeneous units. The criteria and guidelines used to classify the SA3 geography will be discussed in the Benchmark geography section.

The importance of this representational model lies essentially in its ability to serve as a guide for interpreting dimensionless concentration indices and provide a natural benchmark for these measures in terms of defining what is a "high" and "low" concentration of a probability distribution. The application of the operating procedure makes it easy to visualize what a concentration measure means and can help analysists to make scientifically informed decisions. This powerful method of illustrating and classifying a distribution is, therefore, a valuable tool that can act as an interface between the technical and policy disciplines as well as with the decision makers.

In the following section we shall discuss the criteria that can be used to guide the selection of a central tendency measure of a distribution and propose the LI as a location parameter to classify the socioeconomic decile of a geographic area. We also compare this index to other known location parameters.

## 3.3 – Location Index

An important task of spatial data analysis consists in identifying a part of data which represents the typical features of the population living in a geographic area.

Often the data points are assumed to vary around a centre that identifies a single value as representative of an entire distribution. The mean is the most commonly used descriptive statistic as it uses every value in the data. The irony in this is that most of the times this value never appears in the raw data, especially if the distribution is defined in terms of numerical/nominal ordinal variables. Moreover, this measure is not an appropriate measure of central tendency for skewed and high-variance distribution (e.g. U-shaped), and many times it does not give a meaningful value. An alternative measure is the mode, but the local concentration of a curve doesn't always correspond to a predominant peak. For example, skewed distribution with high concentrations of values near the middle are common in the SA3's IRSD decile distribution.

As a result, we propose a new measure of central tendency that is less sensitive to long tail skewed distribution and outliers, that we named it “Location Index” (LI). Furthermore, unlike the mean, this index has a clear interpretation and can be extended to more variables. This index identifies the position of the bin in the distribution where the values are mostly concentrated (i.e. the values in the surrounding areas of the bin are noticeably higher than the others). For example, in the case of the IRSD decile, a LI equals to one is representative of a very disadvantaged area. Conversely, a LI equals to ten indicates an area with the lowest level of disadvantage.

### 3.3.1 - Definition of the Location Index

The basic idea is to map a nested family of sets around the location of the bin into a single number. If the distribution is univariate the set is a symmetric interval formed by a finite number of categories and the concentration value is given by summing up the likelihood of these regions. In this way, each location has a concentration value and the one with the maximum value corresponds to the position of the LI. For a formal definition and properties of the LI the interested reader is referred to Appendix B.

Moreover, the definition of the function used in the computation of the bin concentration value, hereafter called Bin Concentration Function (BCF), serves as a basis in the evaluation and comparisons of other central tendency measures (e.g. mean and mode). Lastly, it provides simple rules to determine the location of a distribution. Consequently, we proved a set of necessary and sufficient conditions to determine the LI value. For example, considering once again the decile distribution, the set of necessary and sufficient conditions is given in the table 6.

| Location Index | $p[i,j] = \sum_{k=i}^{j} p_k \cdot 100$ | |
|---|---|---|
| Decile | $\geq 50\%$ | $\leq 50\%$ |
| 1 | [1,1] | [2,10] |
| 2 | [1,2] | [3,10] |
| 3 | [1,3] | [4,10] |
| 4 | [1,4] | [5,10] |
| 5 | [1,5] | [6,10] |
| 6 | [1,6] | [7,10] |
| 7 | [1,7] | [8,10] |
| 8 | [1,8] | [9,10] |
| 9 | [1,9] | [10,10] |
| 10 | [10,10] | [1,9] |

Table 6 Location Index conditions for decile distributions

In this table, the first and remaining columns indicate the LI and the deciles intervals that contain at least $50\%$ of the data and those with less or equal to $50\%$ of the samples. For example, the first and last row show that a distribution with a LI equal to one or ten have at least $50\%$ of the data in the first or last decile. Similarly, the value of this index for the other bins indicates the decile position separating the greater and lesser halves of a distribution.

Therefore, the LI turns out to be an equivalent form of the median, having a breakdown point of $50\%$: so long as no more than half the data are too much dispersed, the LI will not give an arbitrarily large or small result. The basic difference of the LI in describing data compared to the median is that the outcome is always a member of the distribution and it is easier to calculate. In addition, since more than one value can be at the median level, this location parameter identifies a unique interval of values for which the conditions hold. Lastly, the LI definition can be applied to a multivariate distribution. For the definition, properties and computation of the LI see Appendix B.

At this point, someone might ask: How "good" or "useful" are the LI and HI compared to other measures of central tendency and dispersion? What measure of variability is more suitable for a class of distributions? Lastly, given such measures, which measure of central tendency minimizes that variation from the centre among all choices of centre?

To demonstrate the effective use of these indices, we compared the LI to the mean, mode and SA1's population weighted average score (PWAVGS, see Appendix C) of three broad classes of SA3's IRSD distributions: Highly-Skewed (HS), Moderately-Skewed (MS) and Approximately-Symmetric (AS) distributions (Bulmer, 1979), for the definition see Appendix D. As representative members of the three classes of distribution we selected six SA3s, indicated in Table 7, from a dataset of 331 elements.

| | Distribution Centre | | | PWAVGS | | | Distribution |
|---|---|---|---|---|---|---|---|
| SA3 | Mean | Mode | LI | Score | Rank | Decile | Type |
| West Arnhem | 1.60 | 1 | **1** | 482.3 | 1 | 1 | HS |
| South Canberra | 8.55 | 10 | **10** | 1148.2 | 314 | 10 | |
| East Arnhem | 3.59 | 1 | **1** | 564.5 | 2 | 1 | MS |
| Weston Creek | 8.19 | 10 | **8** | 1130.0 | 304 | 10 | |
| Lake Macquarie East | 5.59 | 7 | **6** | 1009.8 | 173 | 5 | AS |
| West Torrens | 4.93 | 4 | **4** | 1008.458 | 172 | 5 | |

Table 7 SA3 distributions location parameter comparisons

The evaluation criteria used to measure the effectiveness of the central tendency measures depends mainly on two considerations: The level of measurement of the variable (i.e. variable type: nominal, ordinal and metric); and the minimization of a variable functional.

For this study, we considered an ordinal categorical variable (IRSD) and three variation functionals: standard deviation (Std), mean absolute deviation (MAD, Appendix) and the complementary value of the BCF used for the identification of the LI position. For sake of simplicity, we named the complementary value of the BCF as Compactness deviation ($Csd$,Appendix). The choice among these functional is based on the distribution of the values on the variable.

### 3.3.2 – Ordinal Data classification

Besides the shape of the distribution of values, the level of measurement must be considered first in choosing which of the measures of centre to use. There is considerably controversy over using numeric-based statistics, like means, on ordinal data. This problem is evident in Table 7 where the population mean is a fractional value, a value that cannot occur. For instance, finding that the IRSD decile mean of South Canberra is 8.5 would not be very meaningful, because the units between the ten socioeconomic categories are not equal. If the analyst instead rounded all fraction values down/up to the next lowest/largest integer there is a chance of making serious statistical fallacy. This situation is illustrated in Figure 7. In addition, this value is largely influenced by outliers when smaller values tend to cluster toward the tail of a skewed distribution. For example, the mean value of East Arnhem indicated in Table 7, it is clearly not representative for a population mostly concentrated in the first decile: $P_{EA} = (\mathbf{0.659}, 0,0,0,0,0, \mathbf{0.032}, \mathbf{0.122}, \mathbf{0.167}, \mathbf{0.02})$. As shown in Figure 8, the histogram of East Arnhem has more than 60 % of the population in the first decile and a smaller concentration around the third and fourth decile.

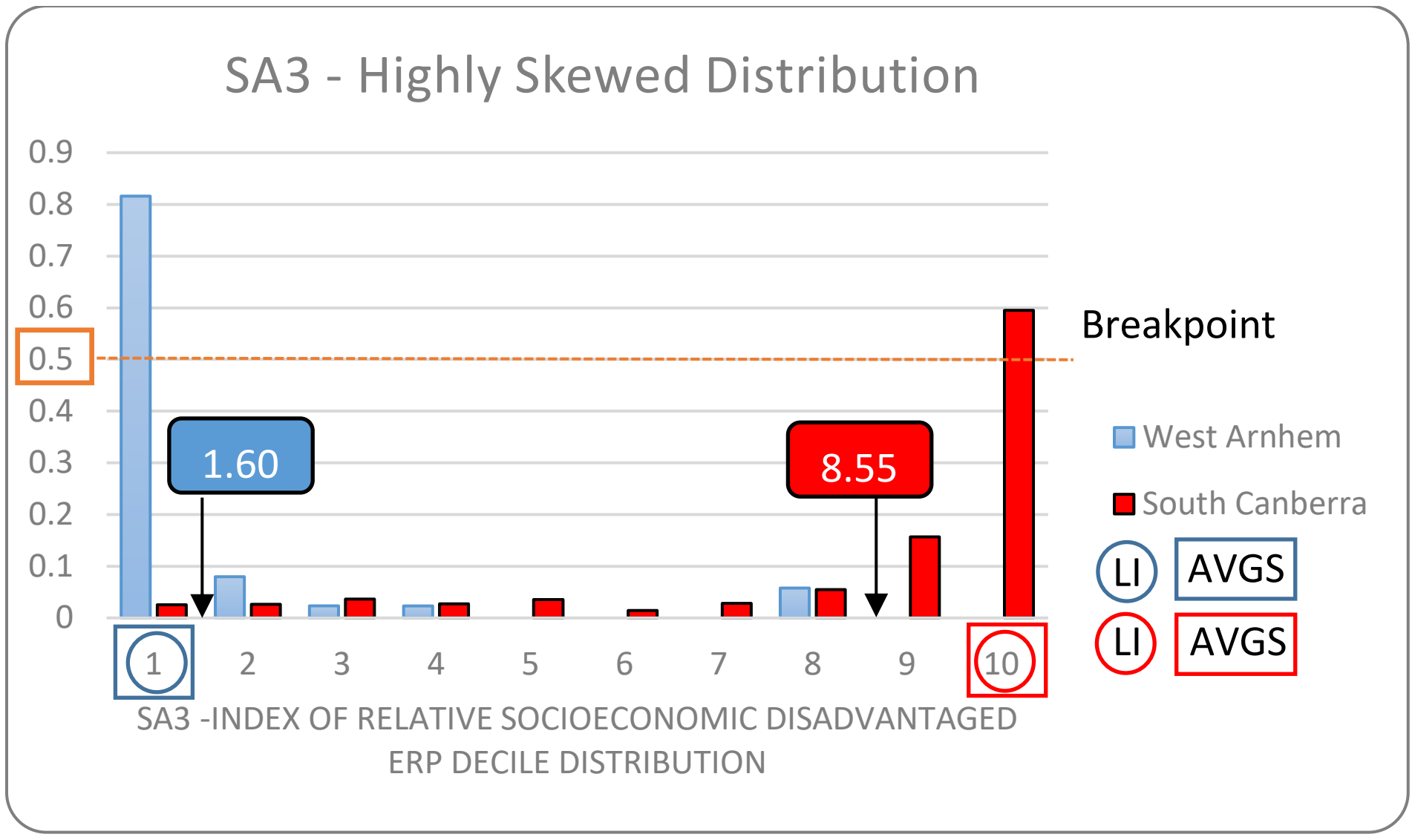

Figure 7 SA3 Highly-skewed Decile Distribution

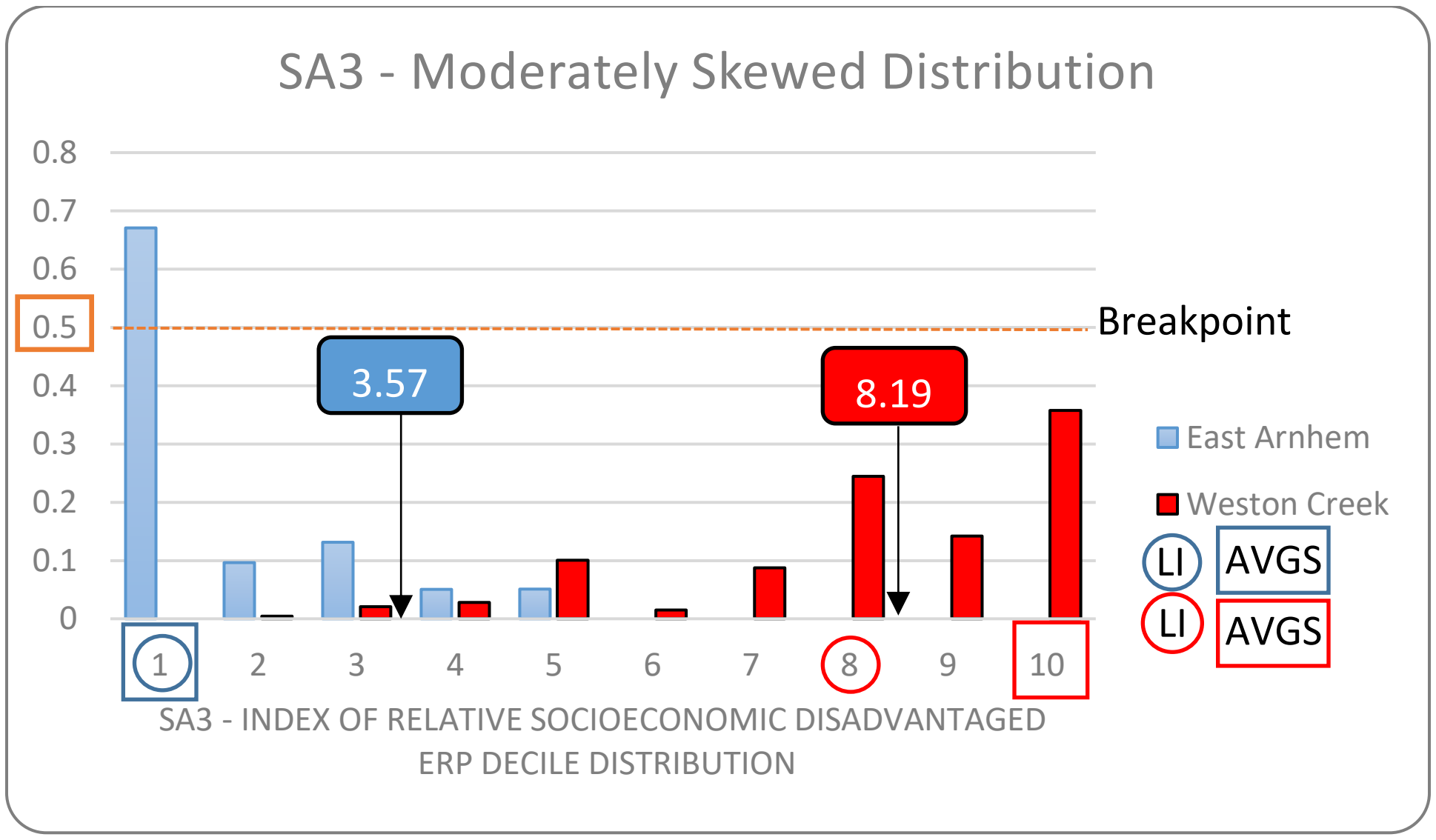

Figure 8 SA3 Moderately-skewed Decile Distribution

As a response to this issue, an alternative solution is to take the population weighted averages based on the true variable underlying scoring (PWAVGS). This has been achieved by multiplying each SA1 score by its Census population count for all SA1s that are contained within a SA3, adding the resulting figures together and dividing the total by the total population in the area. The SA3s are then ranked in order of their SA3 score and given a SA3 rank number (between 1 and 331). The final score is then standardized and classified according to the score table in Figure 6 (see Appendix). A problem associated with this methodology is that the averaging effect of this index can be misleading when there are "outlying scores" of SA1s with larger population, especially in the case of MS or AS distributions. Moreover, the score gives no indication of the diversity of socioeconomic conditions within the area.

This issue is best illustrated with an example. Consider the SA3 of Lake Macquarie-East, located in the Hunter Region of New South Wales. This Sa3 received a score of $1009.8\ (decile\ 5)$ on the IRSD, indicated in Table 7. The top chart in Figure 9 compares the population distribution of SA1 scores within this SA3 with that of Australia, depicted with the dotted black curve. This chart shows that almost all of the $25$ point ranges of SA1 scores are similar to the population distribution of Australia, suggesting that the SA3 of Lake Macquarie-East is not a homogeneous area. This qualitative analysis is further confirmed by looking at the population percentage in the extreme and middle deciles. In this SA3, approximately $9.45$ per cent of the residents live in SA1 with an IRSD score in decile 1 (between $501$ and $875$), $81.39$ per cent live in SA1 with a decile $2-9$ (between $876$ and $1100$) and $9.16$ per cent live in SA1s with a decile 10 ($1101$ and above). The score in the range $751-1075$ indicate $75$ per cent of the residents in the SA3 of Lake Macquarie-East live in a SA1 with an IRSD score between $751-1075$. The diversity of this geographic area is highlighted by the thematic map in Figure 9, which maps the decile distribution of the IRSD for the SA1s within Lake

Macquarie-East. Each SA1 is colored according to its IRSD decile; the legend shows the decile classification. Overall, the map contains many SA1s of varying socio-economic status and, therefore, it is not meaningful to assign an index score to this geographic area. This conclusion is further confirmed by the Diversity index ($s = 9.2$) and HI ($8.5\ \%$) of the IRSD decile distribution. It follows that the SA3 of Lake Macquarie-East belongs to the group D, represented in Table 5, (i.e. $8.5 < 46.62 = HI(6)$). This means that the unit of analysis is likely heterogeneous, and reassignment of some units is needed. Thus, whatever central tendency measure we use for the classification of this geographic area, the location value does not truly represent all the people in the area. As a consequence, it is more difficult to draw strong comparisons between this SA3 and a mid-ranked SA3. Therefore, the use of any location value should always be accompanied by the HI value to acknowledge the diversity within a geographic area.

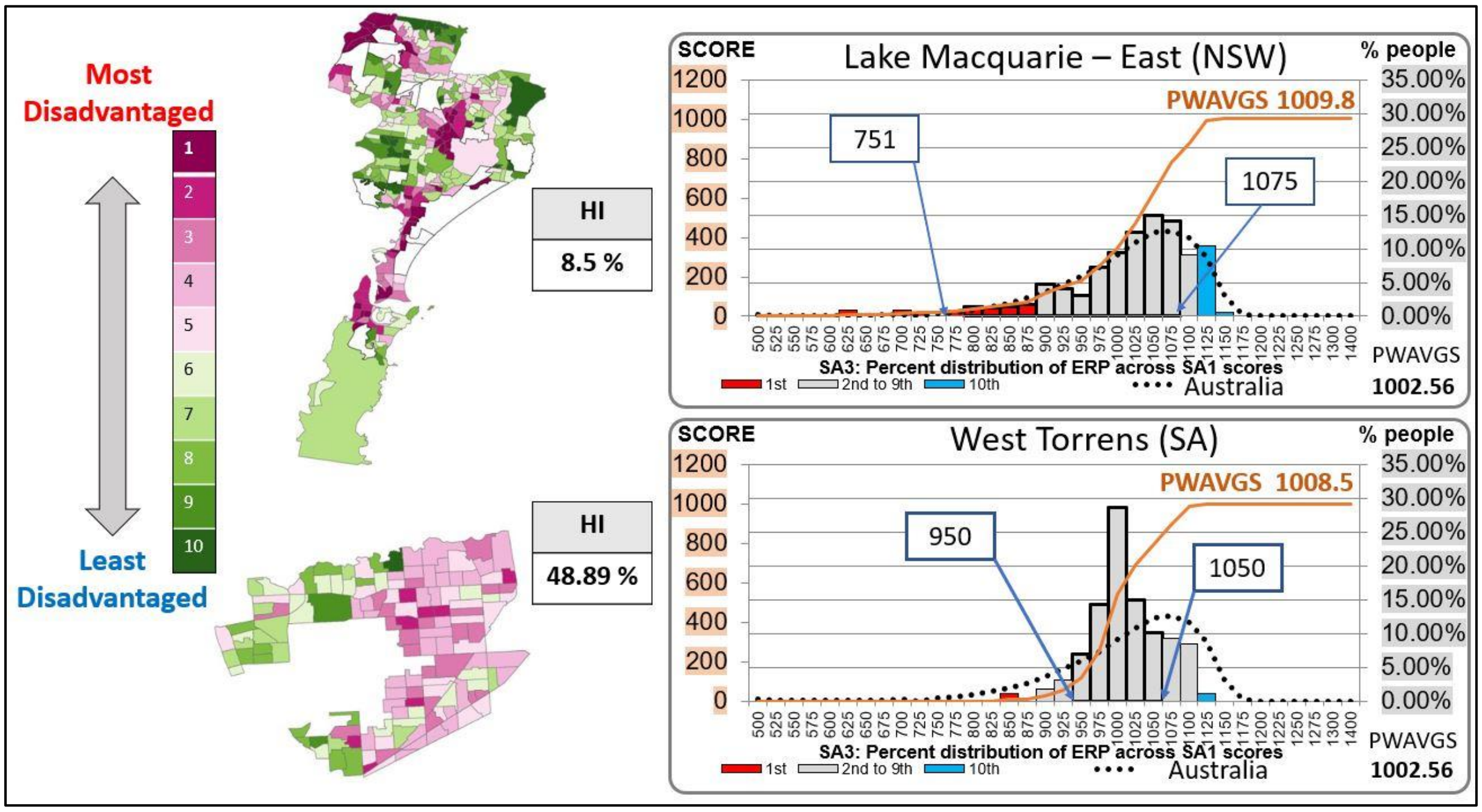

Figure 9 SA3 IRSD distribution comparison - Lake Macquarie-East & West Torrens.

The usefulness of the HI information becomes clear when comparing two geographic areas with similar score or location value. For instance, the SA3 of West Torrens located in the Western suburbs of Adelaide (SA), received a score of $1008{,}458\ (decile\ 5)$, which places it in the $172^{th}$ position of the IRSD ranking, which is one position lower than Lake Macquarie-East ($173^{th}$). Based on the IRSD score rank, one would conclude that West Torrens is not dissimilar to Lake Macquarie-East. However, the IRSD distribution in the SA3 of West Torrens, displayed in the bottom chart of Figure 9, shows that there is a greater proportion of its population living in the middle scores (i.e. $75$ per cent in the range $950 - 1050$) compared with Lake Macquarie-East. This simple analysis suggests that West Torrens has a higher HI value ($48.89\ \%$) compared with Lake Macquarie-East. This value places West Torrens in the third group of Table 5 (i.e. group C). In this circumstance, the SA3 can be included in the analysis or further refined to improve the homogeneity of the geographic area. This is shown in the map (on the bottom of Figure 9), showing the decile distribution of the IRSD for the SA1s within West Torrens. We can see that areas of high or low disadvantaged tend to be clustered together. Moreover, West Torrens has a relative high proportion of relatively disadvantaged SA1s, with a few SA1s in the upper deciles. Consequently, the location value is more meaningful when comparing two geographic areas. However, the IRSD score decile of West Torrens is not representative for a population mostly concentrated in the first four deciles ($53.29\%$), as indicated by the LI value in Table 7. Similarly, the SA3 of Weston Creek has at least $50$ per cent of the residential population concentrated in the first eight deciles (Table 7) and a HI value of $63.85$ per cent, whereas the IRSD decile is located in the least disadvantaged category. This information is further illustrated in Figure 10, which shows the decile distributions of Lake Macquarie-East and West Torrens. It is clear that the shapes of the histograms in Figure 10 are quite different. As a consequence, the IRSD score can be misleading even in the case of homogeneous areas.

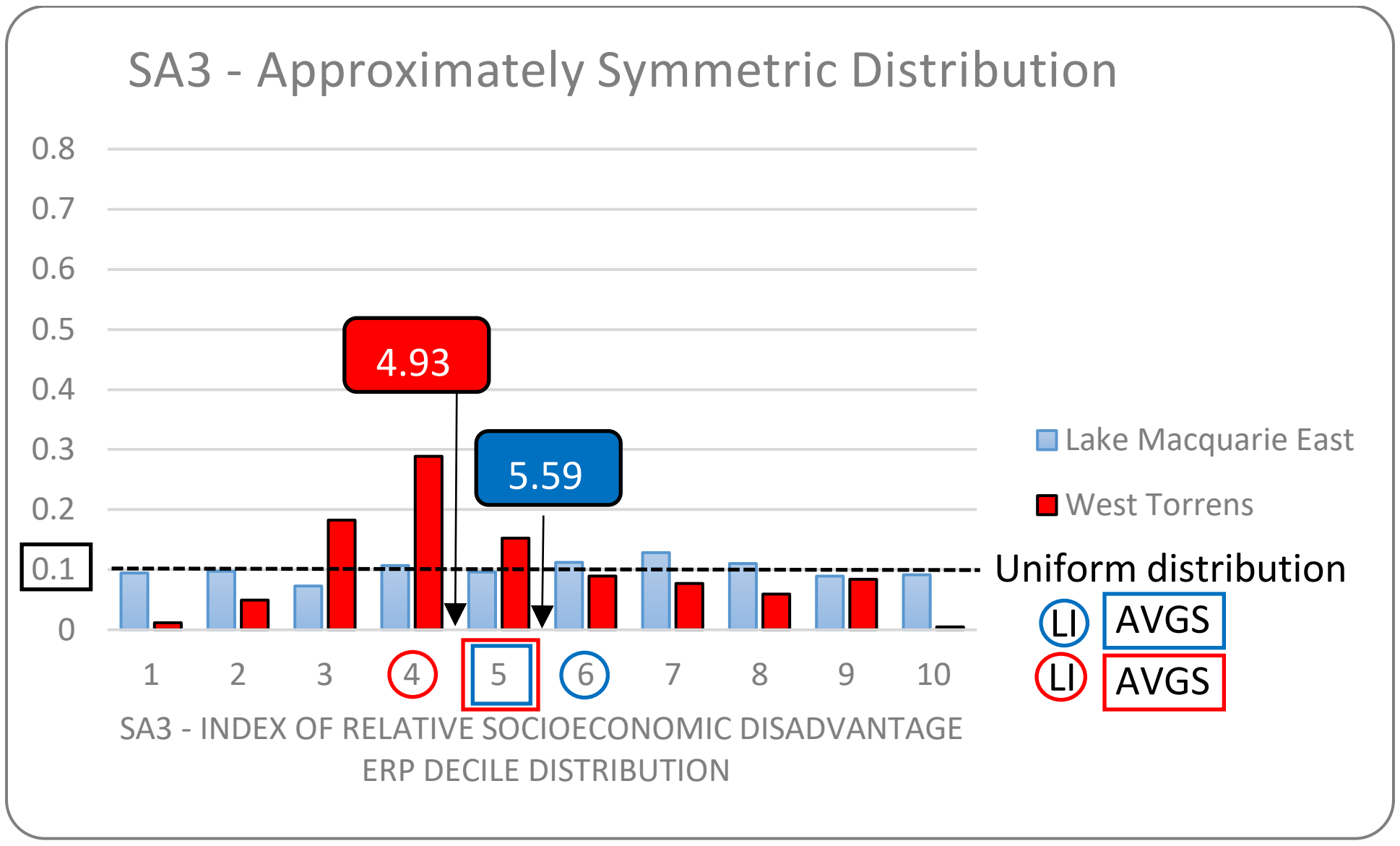

Figure 10 SA3 Approximately - symmetric Decile Distribution

From the analysis above, it is clear that the choice among the measure of centre can significantly affect the classification of a given geography and the consequent identification of Peer groups. That leads to the question of when a location value can be considered "typical" for the concentration of a distribution and most importantly how "typical" that value is. This concerns moves us along to consider what measure of variability is more appropriate for a given data set.

### 3.3.3 – Variation functionals

The most common measure of variability is the Std and the MAD. The first statistic is suitable to describe data that are normally distributed but not always these values follow a "bell shaped" curve. Moreover, it is a mean-based measure and therefore is not appropriate for categorical variables. On the other hand, the MAD shows how far off the values are, on average, from the location centre, when signs of deviation are ignored and thereby more appropriate to describe skewed and asymmetric distribution compared to the Std. Furthermore, the minimization of this functional is equivalent to finding the bin with the maximum concentration value or minimizing the $Csd$ (Appendix.). This special property gives the LI a "best guess interpretation": the LI is the value that is closest to all the other scores on a variable when the sign error in guessing does not matter but its magnitude does. Therefore, the LI is a better median that it is easy to obtain and to interpret.

For instance, table 8 and table 9 show the normalized values of the three variation functional for each SA3's location measure and the HI. By looking at the first and last column of the table 8, it is not difficult to see that the Mean minimizes the Std and the LI minimizes the Absd. But most importantly, the Csd value of the LI in table 7 is roughly the same as the Absd.

| N = 10 | **Std % (Max: 4.5)** | | | | **Absd % (Max: 0.45)** | | | |
|---|---|---|---|---|---|---|---|---|
| SA3 | Mean | Mode | AVGS | LI | Mean | Mode | AVGS | LI |
| **West Arnhem** | **37.53** | 38.57 | 38.57 | 38.57 | 27.39 | 13.33 | 13.33 | **13.33** |
| **South Canberra** | **55.12** | 63.75 | 63.75 | 63.75 | 47.44 | 32.00 | 32.00 | **32.00** |
| **East Arnhem** | **79.80** | 98.08 | 98.08 | 98.08 | 78.15 | 56.88 | 56.88 | **56.88** |
| **Weston Creek** | **42.88** | 58.74 | 58.74 | 43.08 | 33.75 | 40.13 | 40.13 | **33.75** |
| **Lake Macquarie East** | **61.97** | 69.41 | 62.62 | 62.62 | 52.75 | 56.33 | 52.75 | **52.75** |
| **West Torrens** | **44.66** | 49.22 | 50.57 | 49.22 | 36.24 | 34.77 | 44.44 | **34.77** |

Table 8 SA3 Standard Deviation and Mean Absolute Deviation comparisons

| N = 10 | Csd | | | | |
|---|---|---|---|---|---|
| SA3 | Mean | Mode | AVGS | LI | HI |
| **West Arnhem** | 27.40 | 13.44 | 13.44 | **13.44** | 89.65 |
| **South Canberra** | 47.45 | 32.06 | 32.06 | **32.06** | 65.99 |
| **East Arnhem** | 78.18 | 50.74 | 50.74 | **50.74** | 71.58 |
| **Weston Creek** | 33.77 | 40.14 | 40.14 | **33.77** | 63.84 |
| **Lake Macquarie East** | 52.78 | 56.36 | 52.78 | **52.78** | 8.49 |
| **West Torrens** | 36.25 | 34.79 | 44.49 | **34.79** | 48.88 |

Table 9 SA3 Compactness Deviation comparisons

Finally, Table 10 summarizes several of the properties of the central tendency measures that have been discussed in this section and section 2.2. Some of the conclusions in Table 10 are debatable, but it should be clear that the investigations based around the LI and HI provide extra information about a SA3 compared to the mean, mode and PWAVGS.

In the next section, we will show how to use the proposed framework to evaluate the socioeconomic disadvantage homogeneity of the SA3 geography.

| | Properties of Measures of Center | | | |
|---|---|---|---|---|
| | Mode | Median | Mean | LI |
| Level of measurement | Nominal or higher | Ordinal or higher | Metric usually | Ordinal or higher |
| Simple to understand | Best | Yes | yes | Yes |
| Easy to calculate | Best | Yes | yes | Yes |
| Algebraic | No | No | yes | Yes |
| Single valued | Not Always | Yes | best | Not Always |
| Resistant to outliers | Yes | Yes | no | Yes |
| Generalize to n-variables | No | No | yes | Yes |
| Easy interpretation for n variables | No | No | no | Yes |
| Equal to actual data values | Yes | Not always | Not Always | Yes |
| Interpretation | Most typical value | Middle value | Average value<br>Center of mass | Set of middle values |
| Bad guess interpretation | Highest % accuracy | Closest to all scores | Minimized sum of squared deviations or signed deviation | Closest to all scores. Highest concentration |

Table 10 Central Tendency Measure Properties - Adaption of 'Central tendency and Variability'

# Chapter 4 - Results

In the earlier chapters, we emphasized the importance of looking at both measures (LI, HI) in summarizing the "typical" values of a variable and how to use in practice, but there is still an open question about the effective use of these summary statistics on a given data set. In other words, does the use of the LI make a real difference in the classification of grouped data compared to the mean, mode and PWAVGS? Can we use a combination of HI and LI to better find homogeneous geographies and therefore identify peer-groups better?

## 4.1 – SA3 socioeconomic decile classification

In this section we address the following question: "Does the use of the LI make a real difference in the classification of grouped data compared to the mean, mode and PWAVGS ?"

To demonstrate the effectiveness of the LI, we compared the SA3's LI values to the rounded mean $\mu$ (i.e. rounded all fraction values to the largest integer when the fraction part is ≥ 0.5, as is usually done), mode $M$ and PWAVGS of two SA3 groups. More precisely, we partitioned the set of SA3s in two sets: $S_H$ (the set of SA3 such that $HI \geq 57.69$) and $S_{NH}$ (the set of SA3 such that $HI < 57.69$). For each group, we counted the number of SA3s incorrectly classified by the different summary statistics (μ, $M$, PWAVGS) in the three distribution classes (*HS, MS and AS*). The misclassification error ε is simply the absolute deviation of the location center from the LI value. For example, $\varepsilon_{\mu} = 2$ means that the rounded mean value differs two deciles from the LI value.

We start the analysis by looking at the Mean Decile Error Classification ($\varepsilon_{\mu}$) in the two SA3 groups ($i.e\ S_H\ and\ S_{NH}$), illustrated in Tables 11 and 12.

Looking at the table 11, we see that more than half of the SA3s $(51.43\ \%)$ in the group $S_H$ has not been properly classified by the mean value. Most of these distributions are *HS* $(27 \cong 38.58\%)$ and the maximum error classification is given by the case of East Arnhem $(\varepsilon_\mu = 3)$. On the other hand, the percentage of misclassified SA3 in the $S_{NH}$ group (table 12) is less $(30.65\ \%)$ and a substantial number resides in the *AS*$(35\ \cong 13.40\ \%)$ and *MS*$(29\ \cong 11.11\ \%)$.

| $S_H$ HI ≥ 57.69 | | DECILE ERROR $\varepsilon_\mu$ | | | | | | |
|---|---|---|---|---|---|---|---|---|
| | | 0 | 1 | 2 | 3 | Tot | % | $\varepsilon_\mu > 0$ |
| HS | Pos | 3 | 11 | 3 | 0 | 17 | 24.30 | 20.00 |
| | Neg | 21 | 13 | 0 | 0 | 34 | 48.50 | 18.58 |
| MS | Pos | 2 | 3 | 0 | 1 | 6 | 8.57 | 5.71 |
| | Neg | 1 | 4 | 0 | 0 | 5 | 7.14 | 5.71 |
| AS | | 7 | 0 | 1 | 0 | 8 | 11.42 | 1.43 |
| | Tot | 34 | 31 | 4 | 1 | 70 | 100 | 51.43 |
| | % | 48.57 | 44.29 | 5.71 | 1.43 | 100 | | |

Table 11 SA3 SH $(\boldsymbol{HI \geq 57.69})$ - Mean Decile Error Classification

| $S_{NH}$ HI < 57.69 | | DECILE ERROR $\varepsilon_\mu$ | | | | | |
|---|---|---|---|---|---|---|---|
| | | 0 | 1 | 2 | Tot | % | $\varepsilon_\mu > 0$ |
| HS | Pos | 1 | 8 | 0 | 9 | 3.45 | 3.07 |
| | Neg | 5 | 8 | 0 | 13 | 4.98 | 3.07 |
| MS | Pos | 22 | 20 | 1 | 43 | 16.48 | 8.04 |
| | Neg | 21 | 8 | 0 | 29 | 11.11 | 3.07 |
| AS | | 132 | 35 | 0 | 167 | 63.98 | 13.40 |
| | Tot | 181 | 79 | 1 | 261 | 100 | 30.65 |
| | % | 69.35 | 30.27 | 0.38 | 100 | | |

Table 12 SA3 SNH $(\boldsymbol{HI < 57.69})$ - Mean Decile Error Classification

The Mode Decile Error Classification ($\varepsilon_M$) for the two groups of SA3 ($i.e\ S_H\ and\ S_{NH}$), is illustrated in Tables 13 and 14.

Table 13 shows that there is still a high percentage of misclassified SA3s ($52.86\ \%$) in the $S_H$ group, mostly belonged to the *HS* class ($30\ \cong 42.86\ \%$) with limiting cases in the *MS*, such as Weston Creek ($\varepsilon_M = 2$). However, this measure is more sensitive to small changes in values for the case of flat distributions. As a result, roughly $76.63\ \%$ of the SA3s in the $S_{NH}$ group have been badly classified (table 14).

Lastly, the PWAVGS Decile Error Classification is illustrated in Tables 15 and 16.

Table 15 shows that there is a considerable number of misclassified SA3s in the $S_H$ group ($62.85\ \%$) compared to the mean and mode, mostly due to the presence of outliers in the *HS* distribution class ($45.71\ \%$). However, this measure is less affected by the fluctuation of values in the least homogeneous distributions ($67.05\ \%$: $table\ 16$) compared to the mode.

| $S_H$ HI ≥ 57.69 | | DECILE ERROR $\varepsilon_M$ | | | | | |
|---|---|---|---|---|---|---|---|
| | | 0 | 1 | 2 | Tot | % | $\varepsilon_\mu > 0$ |
| HS | Pos | 9 | 8 | 0 | 17 | 24.30 | 11.43 |
| | Neg | 12 | 21 | 1 | 34 | 48.57 | 31.43 |
| MS | Pos | 3 | 2 | 1 | 6 | 8.57 | 4.29 |
| | Neg | 1 | 3 | 1 | 5 | 7.14 | 5.71 |
| AS | | 8 | 0 | 0 | 8 | 11.42 | 0 |
| | Tot | 33 | 34 | 3 | 70 | 100 | 52.86 |
| | % | 47.14 | 48.57 | 4.29 | 100 | | |

Table 13 SA3 SH ($\boldsymbol{HI\ \geq 57.69}$) - Mode Decile Error Classification

| $S_{NH}$ HI < 57.69 | | DECILE ERROR $\varepsilon_M$ | | | | | | | | |
|---|---|---|---|---|---|---|---|---|---|---|
| | | 0 | 1 | 2 | 3 | 4 | 5 | Tot | % | $\varepsilon_\mu > 0$ |
| HS | Pos | 1 | 8 | 0 | 0 | 0 | 0 | 9 | 3.45 | 3.07 |
| | Neg | 4 | 5 | 4 | 0 | 0 | 0 | 13 | 4.98 | 3.45 |
| MS | Pos | 9 | 16 | 18 | 0 | 0 | 0 | 43 | 16.48 | 13.02 |
| | Neg | 9 | 11 | 8 | 1 | 0 | 0 | 29 | 11.11 | 7.66 |
| AS | | 38 | 59 | 39 | 26 | 4 | 1 | 167 | 63.98 | 49.43 |
| | Tot | 61 | 99 | 69 | 27 | 4 | 1 | 261 | 100 | 76.63 |
| | % | 23.37 | 37.93 | 26.44 | 10.34 | 1.53 | 0.39 | 100 | | |

Table 14 SA3 SNH ($\boldsymbol{HI < 57.69}$) - Mode Decile Error Classification

| $S_H$ HI ≥ 57.69 | | DECILE ERROR $\varepsilon_{PWAVGS}$ | | | | | |
|---|---|---|---|---|---|---|---|
| | | 0 | 1 | 2 | Tot | % | $\varepsilon_\mu > 0$ |
| HS | Pos | 10 | 7 | 0 | 17 | 24.30 | 10 |
| | Neg | 9 | 22 | 3 | 34 | 48.57 | 35.71 |
| MS | Pos | 4 | 2 | 0 | 6 | 8.57 | 2.86 |
| | Neg | 2 | 3 | 0 | 5 | 7.14 | 4.28 |
| AS | | 1 | 5 | 2 | 8 | 11.42 | 10 |
| | Tot | 26 | 39 | 5 | 70 | 100 | 62.85 |
| | % | 37.14 | 55.71 | 7.14 | 100 | | |

Table 15 SA3 SH ($\boldsymbol{HI \geq 57.69}$) - PWAVGS Decile Error Classification

| $S_{NH}$ HI < 57.69 | | DECILE ERROR $\varepsilon_{PWAVGS}$ | | | | | | |
|---|---|---|---|---|---|---|---|---|
| | | 0 | 1 | 2 | 3 | 4 | Tot | % | $\varepsilon_{\mu} > 0$ |
| HS | Pos | 4 | 5 | 0 | 0 | 0 | 9 | 3.45 | 1.92 |
| | Neg | 0 | 7 | 6 | 0 | 0 | 13 | 4.98 | 4.98 |
| MS | Pos | 10 | 29 | 4 | 0 | 0 | 43 | 16.48 | 12.64 |
| | Neg | 9 | 15 | 5 | 0 | 0 | 29 | 11.11 | 7.66 |
| AS | | 63 | 89 | 13 | 1 | 1 | 167 | 63.98 | 39.85 |
| | Tot | 86 | 145 | 28 | 1 | 1 | 261 | 100 | 67.05 |
| | % | 32.95 | 55.56 | 10.73 | 0.38 | 0.38 | 100 | | |

Table 16 SA3 SNH ( $\boldsymbol{HI} < \mathbf{57.69}$) - PWAVGS Decile Error Classification

The effectiveness of the LI becomes also evident by looking at the number of SA3s in each socioeconomic group decile for all the statistical measures, indicated in Table 17. It is clear that the mean tends to cluster the SA3s toward the middle and thereby there is a high risk in the misclassification of the lowest and highest socioeconomic categories. A contrasting case is the mode, that is largely concentrated at the end points of the socioeconomic scale. Similarly, the PWAVGS is biased by the skewed distributions of scores in the last and first categories. The LI, on the other hand, shows a more symmetric concentration of SA3s in the middle categories with a reasonable number of units in the first and last deciles. More insights about the SA3 distributions in these categories will be discussed in the next section. In conclusion, the LI provides a more realistic picture of the socioeconomic classification of the SA3 geography.

| | Socioeconomic Decile | | | | | | | | | | SA3 |
|---|---|---|---|---|---|---|---|---|---|---|---|
| | 1 | 2 | 3 | 4 | 5 | 6 | 7 | 8 | 9 | 10 | tot |
| LI | 7 | 21 | 51 | 55 | 57 | 37 | 39 | 26 | 29 | 9 | 331 |
| MEAN | 0 | 5 | 44 | 68 | 69 | 49 | 43 | 30 | 21 | 2 | |
| MODE | 55 | 41 | 37 | 25 | 24 | 27 | 14 | 33 | 26 | 49 | |
| PWAVGS | 22 | 49 | 44 | 40 | 32 | 24 | 26 | 30 | 13 | 51 | |

Table 17 SA3 Location Index (LI) and central measures classification of the IRSD

## 4.2 – Benchmark geography

In this section we address the following question: "Can we use a combination of HI and LI to better find homogeneous geographies and therefore peer-groups better?"

As been discussed throughout the previous section, the selection of the central tendency and concentration measure of a distribution is the starting point of any analysis regarding a set of attributes. In this process, the major challenge is related to the definition of the number of homogeneous regions to be designed. In this scenario, although the analysist does not know exactly how many regions need to be designed, she may know a condition that must be satisfied by every region in order to make them suitable for the analysis.

However, there is no common agreement as to which concentration measure is best. It is not intuitively clear, for example, why the HI should not be replaced by the Entropy Index (EI).

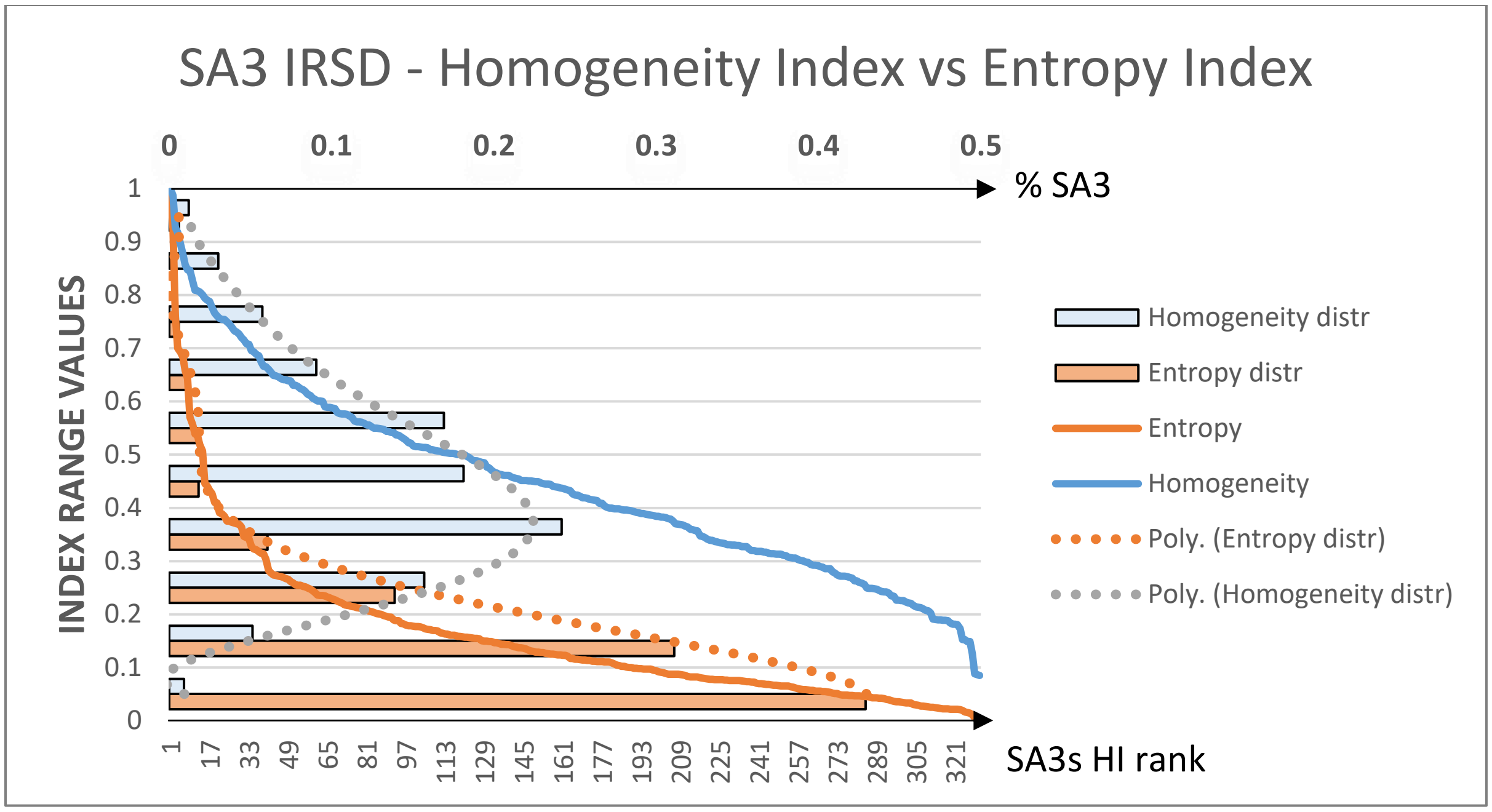

Figure 11 SA3 IRSD - Homogeneity vs Entropy Index

This situation is better illustrated in Figure 11. The thick blue and orange curves represent the HI and EI values sorted in descending order for the IRSD of the SA3 geography. Looking at the distributions of the SA3s within the index range, it is clear that, in the case of the EI (i.e. the orange bar chart), the large concentration of SA3s at the bottom of the interval can potentially map different classes of distributions in the same range. This case is better understood if we look at the number of SA3s for the HS, MS and AS distributions in Figures 12 and 13. In the case of the HI, the HS distributions are more concentrated toward the end of the interval followed by the MS in the middle and AS at the beginning. However, this "gradual" distribution of values is less evident in the case of the EI. As a result, the HI can be used to outline a larger class of probability distributions. Furthermore, the application of the EI in the formula of the HI, or any other index of concentration for nominal categorical variable that does not satisfy the Value validity property, does not define a Fréchet space.

This means that the resulting HI is not a positive function and, therefore, it is not a concentration measure.

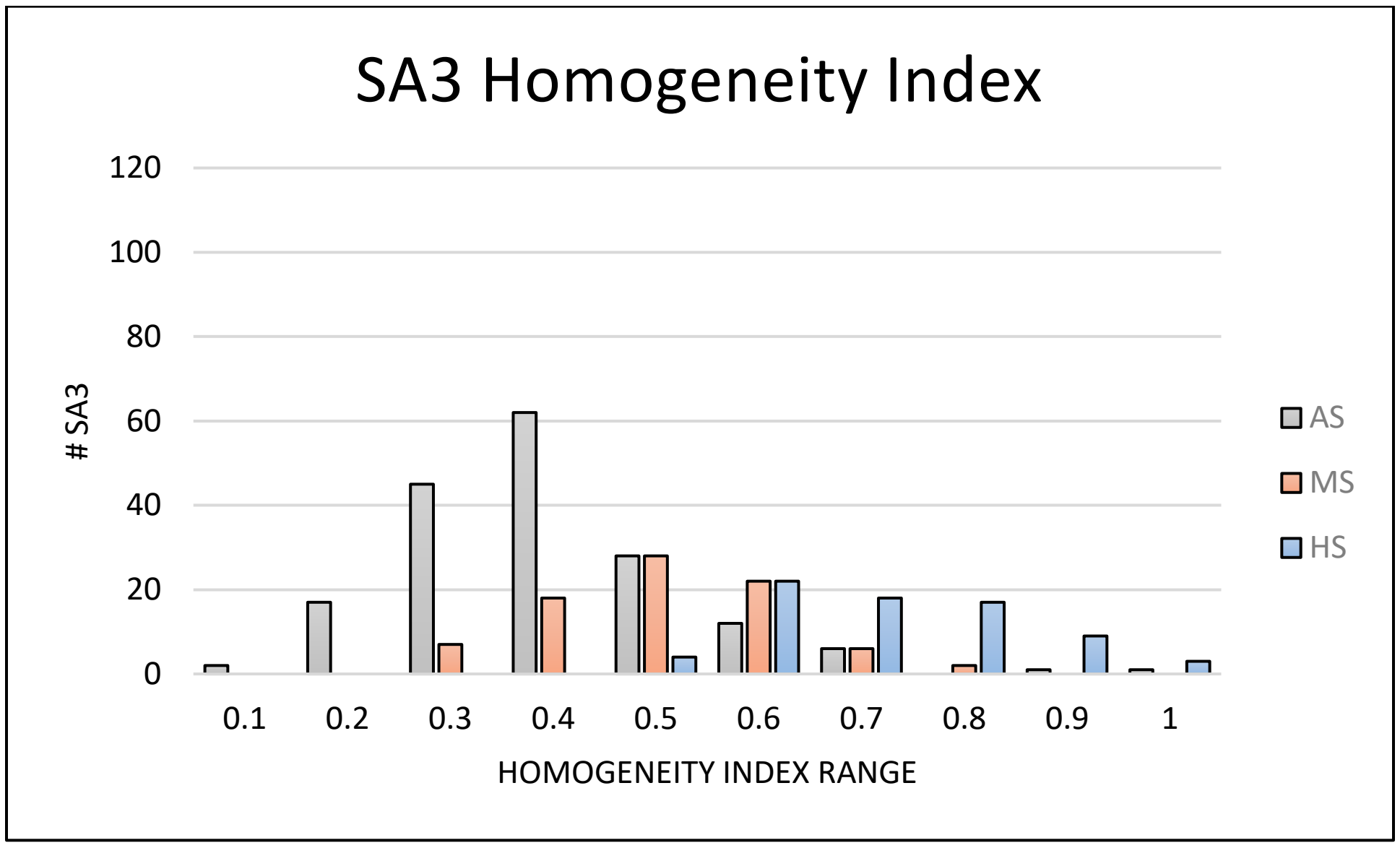

Figure 12 SA3 Homogeneity Index Distribution by skewness class

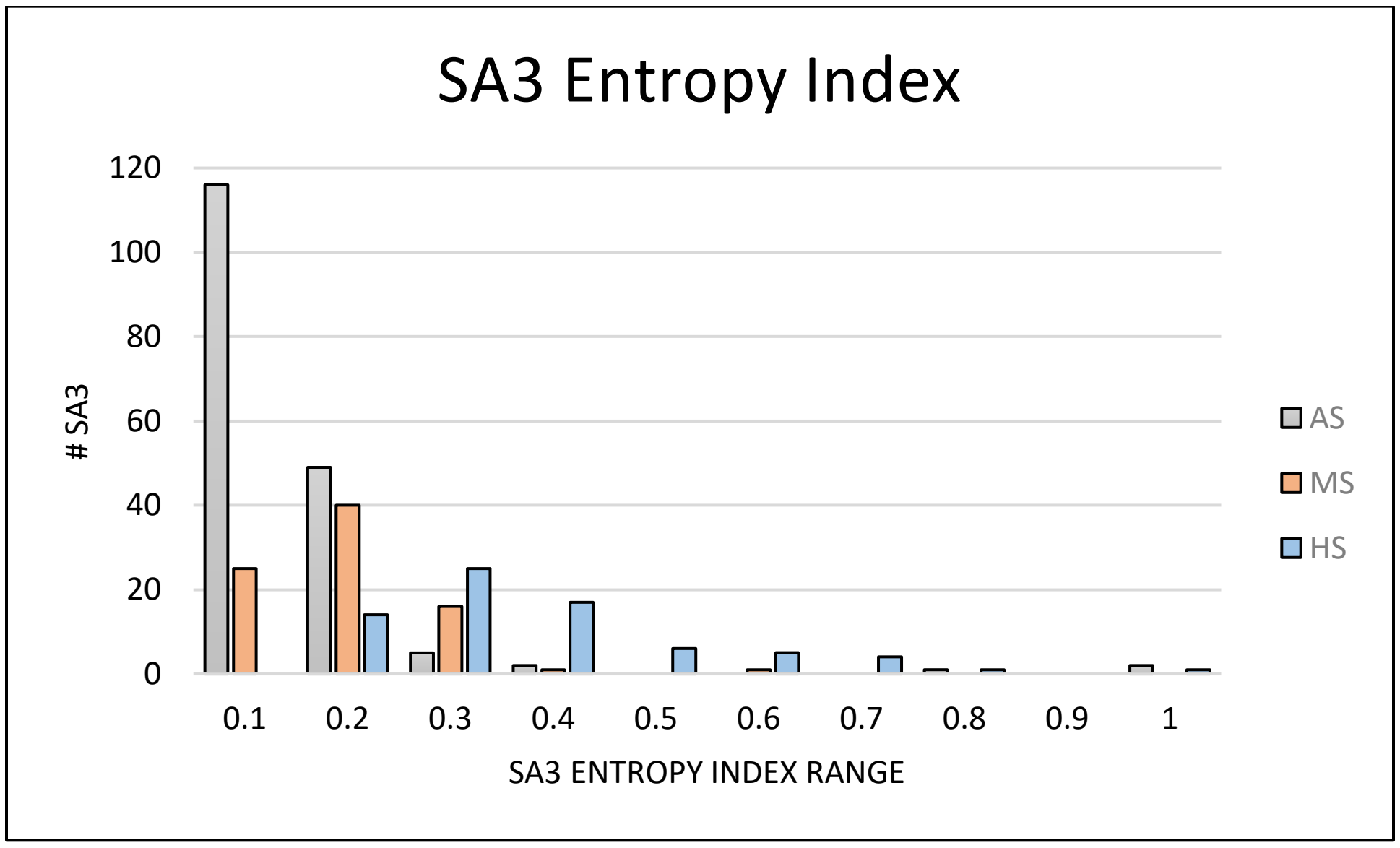

Figure 13 SA3 Entropy Index Distribution by skewness class

In light of the above considerations, the next problems to be solved are the selection of the homogeneity threshold value, that can be used by the analyst to easily identify the set of homogeneous regions, and how to assess the accuracy of the LI value classification.

For example, if there is a specific interest to identify all the areas with at least 50 per cent of the population in the two most disadvantaged deciles, we may simply count how many of them have a LI equal to one or two. However, this basic procedure cannot be applied if we increase the cut-off value and the number of categories. Taking, for example, the first three deciles and a cut-off value of 70 per cent, what meaning we may give to a HI value greater than 46.62 or 57.62, corresponding to the first three and two groups in Table 5?

The simplest and most intuitively way to numerically quantify the homogeneity threshold value into a percentage, is to choose a cut-off value for the distribution, and then calculate the proportion of the population which falls within this cut-off. However, trying to choose cut-off values can be difficult and sometimes meaningless.

To better understand why this might be the case, we partitioned the set of SA3s in three distinct ways: Low-Decile, Medium-Decile and High-Decile. The Low and High partitions analyse the concentration measure in the first and last three deciles, whilst the Medium one includes all deciles between the fourth and seventh. Each partition divides the full set of SA3s into ten groups according to various levels of concentrations, such as all the SA3s with less than 10 per cent of the population in the first three deciles (first group) or with at least 10 percent but less than 20 percent (second group) and so on. Then, the user specifies a cut-off value to identify the set of homogeneous regions in each partition.

A first natural choice is to develop a classification that is benchmarked, such as against the national incidence of the phenomenon. Thus, if we choose the whole Australia as our area, then the concentration will be equal to the percentage of the population distribution partition that is measured. So, if the Low-Decile or High-Decile partition is chosen to be the three most or least disadvantaged deciles, then by definition the population percentage for Australia will be 0.3 or 30 per cent. This proportion can then be compared to other areas. For instance, a distribution percentage for a particular SA3 that is less than 0.3 indicates an under-representation of the population in the most disadvantaged 30 per cent of the Australia IRSD population distribution. Similarly, values above 0.3 for specific SA3s indicate an over-representation of the population in the respective partitions. Thus, the extent to which the concentration exceeds 0.3 is indicative of the degree to which the population is spatially concentrated in a decile partition. Clearly, a distribution containing at least 70 per cent in the first three deciles and hence less or equal to 30 per cent in the last three deciles can be reasonably defined as a disadvantaged SA3.

In this scenario, we defined three sets of homogeneous categories which are not necessarily exhausted: High-Disadvantaged (HD), Medium-Disadvantaged (MD) and Low-Disadvantaged (LD). A HD or LD SA3 is characterized by an IRSD distribution with at least 70 per cent of the population in the first or last three deciles. On the other hand, a MD SA3 has at least 90 per cent of the population concentrated between the fourth and seventh decile. It follows that a MD SA3 is over-represented in the middle deciles and under-represented or not concentrated in the least and most disadvantaged deciles.

One approach to look at these groups is to build a Concentration Matrix whose elements are the number of SA3s with a given LI value (column) and concentration measure (row).

In this schema, the rows indicating the set of HD and LD SA3s with a concentration value greater or equal to the specified threshold $(i.e.\ 70\,\%)$ are the last three shading rows in Tables 18 and 19, respectively. As for the MD SA3s, the number of homogeneous units for each LI value is indicated in the last shading row in Table 20.

| | Decile | HIGH-DISADVANTAGED ( $TH_c \geq 70$ %) | | | | | | | | | | | |
|---|---|---|---|---|---|---|---|---|---|---|---|---|---|
| **LOW** | **1-3** | **LOCATION INDEX** | | | | | | | | | | **SA3** | |
| | **%pop** | **1** | **2** | **3** | **4** | **5** | **6** | **7** | **8** | **9** | **10** | **Tot** | **%** |
| | 0 to 10 | 0 | 0 | 0 | 0 | 0 | 8 | 21 | 22 | 29 | 9 | 89 | 26.89 |
| | 10 to 20 | 0 | 0 | 0 | 0 | 5 | 14 | 15 | 4 | 0 | 0 | 38 | 11.48 |
| | 20 to 30 | 0 | 0 | 0 | 1 | 23 | 14 | 3 | 0 | 0 | 0 | 41 | 12.39 |
| | 30 to 40 | 0 | 0 | 0 | 19 | 29 | 1 | 0 | 0 | 0 | 0 | 49 | 14.80 |
| | 40 to 50 | 0 | 0 | 0 | 35 | 0 | 0 | 0 | 0 | 0 | 0 | 35 | 10.58 |
| | 50 to 60 | 1 | 0 | 36 | 0 | 0 | 0 | 0 | 0 | 0 | 0 | 37 | 11.18 |
| | 60 to 70 | 2 | 6 | 13 | 0 | 0 | 0 | 0 | 0 | 0 | 0 | 21 | 6.34 |
| HD | **70 to 80** | **1** | **13** | **2** | 0 | 0 | 0 | 0 | 0 | 0 | 0 | **16** | **4.83** |
| | **80 to 90** | **2** | **2** | 0 | 0 | 0 | 0 | 0 | 0 | 0 | 0 | **4** | **1.21** |
| | **90 to 100** | **1** | 0 | 0 | 0 | 0 | 0 | 0 | 0 | 0 | 0 | **1** | **0.3** |
| | **Tot** | 7 | 21 | 51 | 55 | 57 | 37 | 39 | 26 | 29 | 9 | 331 | **6.34** |
| | ***% HD*** | 1.21 | 4.53 | 0.6 | 0 | 0 | 0 | 0 | 0 | 0 | 0 | 6.34 | **21** |

Table 18 HIGH - DISADVANTAGED SA3

| | Decile | LOW-DISADVANTAGED ( $TH_c \geq 70$ %) | | | | | | | | | | | |
|---|---|---|---|---|---|---|---|---|---|---|---|---|---|
| **HIGH** | **8-10** | **LOCATION INDEX** | | | | | | | | | | **SA3** | |
| | **%pop** | **1** | **2** | **3** | **4** | **5** | **6** | **7** | **8** | **9** | **10** | **Tot** | **%** |
| | 0 to 10 | 5 | 18 | 44 | 20 | 6 | 1 | 1 | 0 | 0 | 0 | 95 | 28.7 |
| | 10 to 20 | 0 | 2 | 6 | 28 | 24 | 5 | 0 | 0 | 0 | 0 | 65 | 19.64 |
| | 20 to 30 | 1 | 1 | 1 | 7 | 24 | 16 | 0 | 0 | 0 | 0 | 50 | 15.11 |
| | 30 to 40 | 1 | 0 | 0 | 0 | 3 | 14 | 13 | 0 | 0 | 0 | 31 | 9.36 |
| | 40 to 50 | 0 | 0 | 0 | 0 | 0 | 1 | 25 | 0 | 0 | 0 | 26 | 7.85 |
| | 50 to 60 | 0 | 0 | 0 | 0 | 0 | 0 | 0 | 15 | 0 | 0 | 15 | 4.53 |
| | 60 to 70 | 0 | 0 | 0 | 0 | 0 | 0 | 0 | 10 | 7 | 0 | 17 | 5.14 |
| L | **70 to 80** | 0 | 0 | 0 | 0 | 0 | 0 | 0 | **1** | **11** | 0 | **12** | **3.63** |
| D | **80 to 90** | 0 | 0 | 0 | 0 | 0 | 0 | 0 | 0 | **10** | **5** | **15** | **4.53** |
| | **90 to 100** | 0 | 0 | 0 | 0 | 0 | 0 | 0 | 0 | **1** | **4** | **5** | **1.51** |
| | **Tot** | **7** | **21** | **51** | **55** | **57** | **37** | **39** | **26** | **29** | **9** | **331** | **9.67** |
| | ***% HD*** | 0 | 0 | 0 | 0 | 0 | 0 | 0 | **0.3** | **6.65** | **2.72** | **9.67** | **32** |

Table 19 LOW - DISADVANTAGED SA3

| | Decile | MEDIUM-DISADVANTAGED ( $TH_c \geq 90$ %) | | | | | | | | | | | |
|---|---|---|---|---|---|---|---|---|---|---|---|---|---|
| **MED** | **4-7** | **LOCATION INDEX** | | | | | | | | | | **SA3** | |
| | **%pop** | **1** | **2** | **3** | **4** | **5** | **6** | **7** | **8** | **9** | **10** | **Tot** | **%** |
| | 0 to 10 | 3 | 1 | 0 | 0 | 0 | 0 | 0 | 0 | 3 | 4 | **11** | **3.23** |
| | 10 to 20 | 2 | 5 | 1 | 0 | 0 | 0 | 0 | 0 | 10 | 5 | **23** | **6.95** |
| | 20 to 30 | 1 | 14 | 5 | 0 | 1 | 0 | 0 | 8 | 14 | 0 | **43** | **13.00** |
| | 30 to 40 | 0 | 1 | 32 | 10 | 5 | 4 | 7 | 14 | 2 | 0 | **75** | **22.66** |
| | 40 to 50 | 1 | 0 | 13 | 28 | 17 | 13 | 18 | 4 | 0 | 0 | **94** | **28.40** |
| | 50 to 60 | 0 | 0 | 0 | 15 | 23 | 9 | 12 | 0 | 0 | 0 | **59** | **17.82** |
| | 60 to 70 | 0 | 0 | 0 | 2 | 9 | 5 | 1 | 0 | 0 | 0 | **17** | **5.14** |
| | 70 to 80 | 0 | 0 | 0 | 0 | 2 | 4 | 0 | 0 | 0 | 0 | **6** | **1.81** |
| | 80 to 90 | 0 | 0 | 0 | 0 | 0 | 1 | 0 | 0 | 0 | 0 | **1** | **0.3** |
| **MD** | 90 to 100 | 0 | 0 | 0 | 0 | 0 | **1** | **1** | 0 | **0** | **0** | **2** | **0.6** |
| | **Tot** | **7** | **21** | **51** | **55** | **57** | **37** | **39** | **26** | **29** | **9** | **331** | **0.6** |
| | ***% HD*** | 0 | 0 | 0 | 0 | 0 | **0.3** | **0.3** | **0** | **0** | **0** | **0.6** | **2** |

Table 20 MEDIUM - DISADVANTAGED SA3

To consider a very simple case, suppose that we want to classify the decile distribution of Ku-ring-gai $P_{KG} = (0,0,0,0,\mathbf{0.004},\mathbf{0.01},\mathbf{0.021},\mathbf{0.038},\mathbf{0.170},\mathbf{0.757})$, discussed in the Requirement Specification section. Then we need to consider two steps in the classification process: first we should identify the LI value, and finally compute the concentration value for each partition of the decile distribution. We see that more than half of the population of this SA3 is concentrated in the last decile ($i.e. 75.7 > 50$). This means that this unit belongs to the set of SA3s in the last column of each table $(i.e. LI = 10)$. Finally, for each partition we need to identify the concentration group of this SA3. This task can be easily accomplished by simply summing up the percentage of the population in the corresponding deciles partition and comparing this value to the concentration group range indicated in the first column of the tables.

In this case, the cumulative percentage in the first and last three deciles is $0$ and $96.5$ per cent, respectively. Lastly, $3.5$ per cent of the residents in the SA3 of Ku-ring-gai live in SA1 with a decile $4 - 7$. Consequently, this SA3 belongs to the sets of $9$ SA3s and $4$ SA3s in the first row and last column for the low-decile and medium-decile partition, in Tables 18 and 20 respectively. This means that Ku-ring-gai is under-represented in the LD and MD categories. On the other hand, the concentration group of the high-decile partition for this SA3 is clearly the set of $4$ SA3s in the last row of Table 19. It follows that Ku-ring-gai belongs to the most homogeneous concentration group of LD SA3s and, therefore, is over-represented in this category.

This simple example shows clearly that this classification method allows to partition the set of SA3s in three mutually exclusive homogeneous groups. That is an SA3 over-represented in a single partition must be under-represented in the other partitions. Intuitively, the numbers in these tables show that the least disadvantaged and most disadvantaged deciles are by far the most homogeneous with respect to the concentration value specification threshold. This is better illustrated by the last two rows and columns of each table. The rows on the bottom indicate the total number of SA3s and the percentage of homogeneous units for each LI value. The last two columns, on the other hand, show the total number and percentage of SA3s in the corresponding concentration group.

For example, the total number of SA3s in the first decile are $7$ (approximately $2.12$ per cent of SA3s), and $1.21$ per cent ($4$ SA3s) belong to the HD category. On the other hand, the SA3 of Daly-Tiwi-West Arnhem in Northern Territory is the only unit in the last homogeneous concentration group of HD SA3s and represent $0.3$ per cent of the SA3s. This SA3 is also the unit with the maximum $HI$ value ($89.65$) and minimum Diversity Index $s$ $(1.72)$ among the units with a LI equal to one. The distribution vector of this SA3 is $P = (\mathbf{81.59}, \mathbf{7.92}, \mathbf{2.34}, \mathbf{2.32}, 0,0,0, \mathbf{5.78}, 0,0)$ and the cumulative percentage in the first three deciles is $91.9$ per cent. This is clearly a highly-skewed distribution and is equivalent to a not polarised $\lambda - distribution$ (i.e. equivalence class [9,1]) with a concentration value of $90$ per cent. Lastly, the numbers indicated with the red and black colour in the last two cells on the bottom of each table, show the percentage and total number of homogenous SA3s in the respective Disadvantaged category.

It follows that the total number of homogeneous SA3s is 55, approximately 16.6 per cent of the SA3s. Almost all are in the HD$(21)$ and LD$(32)$, while only two units are in the MD category. This is mainly due to the high percentage of HS distribution concentrated in the first and last deciles. The majority have been classified in the second decile (4.53 %, LI=2; Table 18) and ninth decile (6.65 %, LI=9; Table 19).

This classification method is a valid measure for "relative disadvantage" and might be useful in answering different questions. However, the computation of this procedure is cumbersome and time consuming. Moreover, many other choices of cut-off value and partitions could be taken depending on the focus of the user. Nevertheless, these criteria are often arbitrary and not helpful for the identification of geographic areas that might be included in the analysis. For instance, the SA3 of West Torrens, analysed in the previous section, has the following decile distribution percentage: $P = (1.20,\ 4.93\ ,18.25\ ,\ \mathbf{28.91}\ ,15.22\ ,8.96\ ,\ 7.71\ ,\ 5.97\ ,\ 8.39\ ,\ 0.45)$. This distribution has 53.29 per cent of the population in the first four deciles and 60.8 per cent of people between the 4$^{th}$ and 7$^{th}$ decile, corresponding to the Medium-Decile partition. Clearly, these percentages are below the cut-off value of 70 per cent and therefore not feasible for a homogeneous SA3. However, the analysis of this SA3 indicate that the concentration of the IRSD distribution is not completely heterogeneous. This means that the proportion of people concentrated in the categories next to the LI is not negligible. In our case, the proportion of people between the third and fifth deciles is 62.38 per cent. This result can be inferred by looking only at the HI ($48.89$), Diversity index ($s = 5.63$) values and the distribution type (AS). Since this SA3 is almost a symmetric distribution, the user has to identify the equivalence class in Table 4. It follows that the SA3 of West Torrens belongs to the equivalence class [2,5] and it is equivalent to a symmetric distribution with almost 63 per cent of the samples.

From the preceding discussion, it should be evident that it is preferable to use a general approach that relies on a single parameter and suitable to any reasonable partition. Recalling the HD and LD categories, Tables 21 and 22 show the number of SA3s in the high and low disadvantage categories for a given combination of concentration level and HI values, indicated by the rows and columns. For instance, the SA3 of Daly-Tiwi-West Arnhem is the only unit in the last concentration group with a HI value greater or equal to $80$ and less than $90$, as indicated in the last row and column of Table 21. On the other hand, the number of disadvantaged SA3s with the minimum concentration and HI values are three. The minimum HI value is $57.01$, as indicated by the threshold value $TH_H$ in Table 21. By looking at the threshold values of the HI ($TH_H$) in Tables 21 and 22, it is clear that the “homogeneous” group is a subset of the SA3s with a $HI \in [57.01 \quad 100]$. Thus, a natural question to ask is how to select a representative $TH_H$ value of the “homogeneous” areas to be analysed?

Basically, the answer depends on the type of the analysis. Generally, the number of effective categories used in the computation of the homogeneity threshold should not exceed $\left\lfloor\frac{n}{2}\right\rfloor$ or $\left\lfloor\frac{n}{2}\right\rfloor + 1$, where n is the total number of categories in the distribution. In this study, we selected the least limiting conditions, by setting $s = 6$ and $\alpha = 1$, as indicated in Table 5. Other values of $s$ and α could be tried, but this choice will reduce the number of potential SA3 to be analysed.

The application of the criteria in Table 5 to the IRSD of the SA3 region, indicated in Table 23, shows that the percentage of heterogeneous SA3s is almost 59 per cent (59.82) and hence a reassignment of some subunits (i.e. SA1s or SA2s) is needed to improve the homogeneity in the study area. The proportion of SA3s in the first two homogeneous group is only 21 per cent. Lastly, the units to be evaluated are almost 19 per cent. To further illustrate this analysis, Table 24 shows the number of SA3s for a given combination of the LI and HI value as a function of $s$. In this table the grey shading rows indicate the set of SA3s in the first two groups $(HI(P) \geq HI(5))$, and the light shading row indicates all the SA3s in the third group $(HI(6) \leq HI(P) < HI(5))$. Lastly, the remaining light blue rows show the numbers of heterogeneous SA3s $(HI(P) < HI(6))$. As discussed previously, the most representative values of the LI are the first and last deciles. This is clear if we compare the number of SA3s in the first two groups with the total number of SA3s for each LI value. All the SA3s in the most and least disadvantage deciles belong to the homogeneous group, this information is further illustrated on the rows below the table that show the number and percentage of SA3s for each LI value as well as the number and percentage of SA3s in the homogeneous groups, indicated in the last two shading rows. Two representative members of these deciles are the SA3s of Daly-Tiwi-West Arnhem $(LI = 1; HI = 89.65)$ and Ku-ring-gai $(LI = 10; HI = 91.77)$. The SA3 of Daly-Tiwi-West Arnhem is the only unit with a HI value greater than a distribution of two consecutive categories $(HI = 89.64)$ among the most disadvantaged units, as indicated in the first row and column of Table 24. Similarly, Ku-ring-gai belongs to the set of 2 SA3s in the first row and last column of the Table. However, the proportion of SA3s in these categories is less than 5 per cent (i.e. 16 SA3s). A slightly bigger number of homogeneous SA3s is in the second and ninth decile, where they account for almost 11 per cent of the SA3s (i.e. 37 SA3s). The middle deciles, on the other hand, have a high proportion of heterogeneous units. Particularly, the least representative decile is the 5$^{th}$ decile and the most heterogeneous unit is the SA3 of Lake Macquarie-East ($LI = 5, HI = 8.5$), indicated in the last row of the 5$^{th}$ column. This result is also confirmed by the high percentage of heterogeneous AS distributions indicated in Figure 12.

To further illustrate this analysis, we defined three different classes of AS distributions: Highly Symmetric, Moderately Symmetric and Approximately Skewed (see Appendix D). For each class, we counted the number of SA3s in the corresponding HI range. The result of this analysis is represented in Figure 14. More than half ($57.41\%$) of the AS distributions have a HI value below 35.53 and the majority are in the Highly and Moderately Symmetric group. As a result, the most representative values of the LI are the first and last two deciles. These results are further illustrated in Figure 15 and 16. Figure 15 is simply a graphical representation of Table 24. Figure 16 shows the percentage of SA3s for each HI range and some representative members.

For instance, by looking at the HI of these SA3s, we see that the ones on the left are more homogeneous compared to the ones on the right. However, using the HI in isolation will not give a complete picture. Therefore, to better classify the geographic area we calculated the LI, indicated in the squared box. As we see this value is more meaningful for the Homogeneous SA3s.

In the next section we will show how to apply this classification framework to analyse the variation of an indicator.

| | Decile | HIGH- DISADVANTAGED ( $TH_C \geq 70$ %) | | | | |
|---|---|---|---|---|---|---|
| **LOW** | **1-3** | **HOMOGENEITY INDEX ($TH_H \geq 57.01$)** | | | | |
| | **%pop** | **50 to 60** | **60 to 70** | **70 to 80** | **80 to 90** | **Tot** |
| HD | **70 to 80** | **3** | **10** | **1** | **2** | **16** |
| | **80 to 90** | 0 | 0 | **3** | **1** | **4** |
| | **90 to 100** | 0 | 0 | 0 | **1** | **1** |
| | **Tot** | **3** | **10** | **4** | **4** | **21** |

Table 21 HIGH - DISADVANTAGED HOMOGENEITY MAPPING

| | Decile | LOW- DISADVANTAGED ( $TH_C \geq 70$ %) | | | | |
|---|---|---|---|---|---|---|
| **HIGH** | **7-10** | **HOMOGENEITY INDEX ($TH_H \geq 63.86$)** | | | | |
| | **%pop** | **60 to 70** | **70 to 80** | **80 to 90** | **90 to 100** | **Tot** |
| LD | **70 to 80** | 0 | 0 | 0 | 0 | **0** |
| | **80 to 90** | **12** | **1** | 0 | 0 | **13** |
| | **90 to 100** | 0 | **12** | **5** | **2** | **19** |
| | **Tot** | **12** | **13** | **5** | **2** | **32** |

Table 22 LOW - DISADVANTAGED HOMOGENEITY MAPPING

| IRSD | GROUP | | | |
|---|---|---|---|---|
| HI | **A** | **B** | **C** | **D** |
| Tot | **37** | **33** | **63** | **198** |
| % SA3 | **11.18** | **9.97** | **19.03** | **59.82** |

Table 23 SA3 IRSD Homogeneity Group percentage

| | | HOMOGENEOUS SA3 ($HI(P_{10}) \geq TH_{HI} = 57.69$) $s \leq 5$ | | | | | | | | | | | |
|---|---|---|---|---|---|---|---|---|---|---|---|---|---|
| | **RANGE** | **LOCATION INDEX** | | | | | | | | | | **SA3** | |
| **s** | ***HI (s)*** | **1** | **2** | **3** | **4** | **5** | **6** | **7** | **8** | **9** | **10** | **Tot** | **%** |
| **2** | **89.64** | **1** | 0 | 0 | 0 | 0 | **1** | **1** | 0 | 0 | **2** | **5** | **1.51** |
| **3** | **79.19** | **3** | **1** | 0 | 0 | 0 | 0 | 0 | 0 | **1** | **5** | **10** | **3.02** |
| **4** | **68.53** | **2** | **3** | 0 | **2** | 0 | 0 | 0 | 0 | **14** | **1** | **22** | **6.65** |
| **5** | **57.69** | **1** | **9** | **6** | 0 | 0 | **1** | **1** | **5** | **9** | **1** | **33** | **9.97** |
| **6** | **46.62** | 0 | 7 | 19 | 6 | 1 | 5 | 6 | 14 | 5 | 0 | **63** | **19.03** |
| **7** | **35.33** | 0 | 1 | 21 | 17 | 15 | 6 | 16 | 7 | 0 | 0 | 83 | 25.08 |
| **8** | **23.79** | 0 | 0 | 5 | 24 | 18 | 18 | 13 | 0 | 0 | 0 | 78 | 23.56 |
| **9** | **12.02** | 0 | 0 | 0 | 6 | 22 | 4 | 2 | 0 | 0 | 0 | 34 | 10.27 |
| **10** | **0** | 0 | 0 | 0 | 0 | 1 | 2 | 0 | 0 | 0 | 0 | 3 | 0.91 |
| | **Tot** | **7** | **21** | **51** | **55** | **57** | **37** | **39** | **26** | **29** | **9** | **331** | **21.15** |
| | **%** | **2.12** | **6.34** | **15.41** | **16.62** | **17.22** | **11.18** | **11.78** | **7.85** | **8.76** | **2.72** | **100** | **tot** |
| | **%** | **2.12** | **3.94** | **1.81** | **0.6** | **0** | **0.6** | **0.6** | **1.51** | **7.25** | **2.72** | **21.15** | **70** |
| | **Tot** | **7** | **13** | **6** | **2** | **0** | **2** | **2** | **5** | **24** | **9** | **70** | |

Table 24 SA3 Location Index and Homogeneity Index decile classification of the IRSD decile distribution

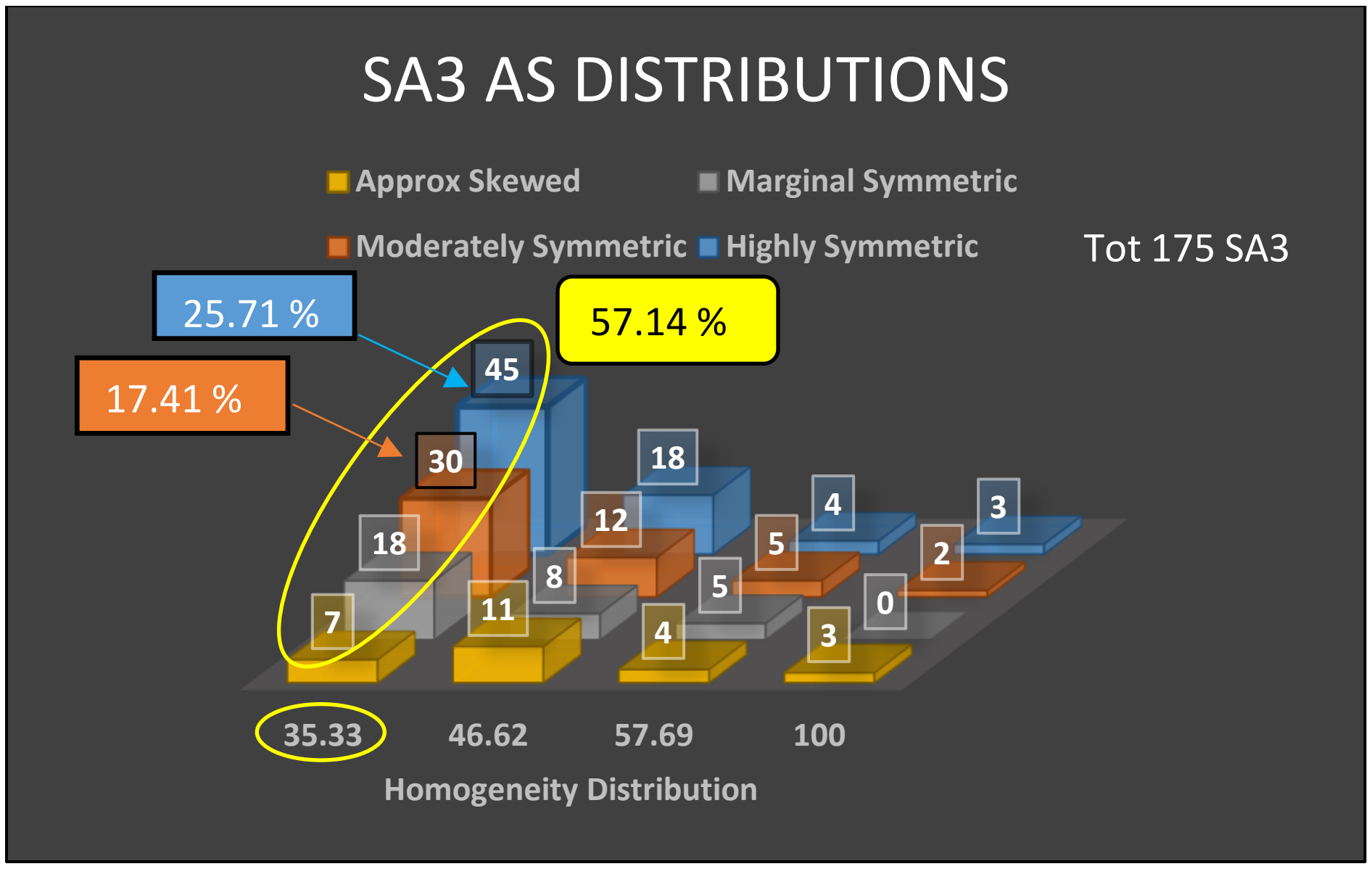

Figure 14 SA3 Homogeneity Index ofApproximately Symmetric Distribution

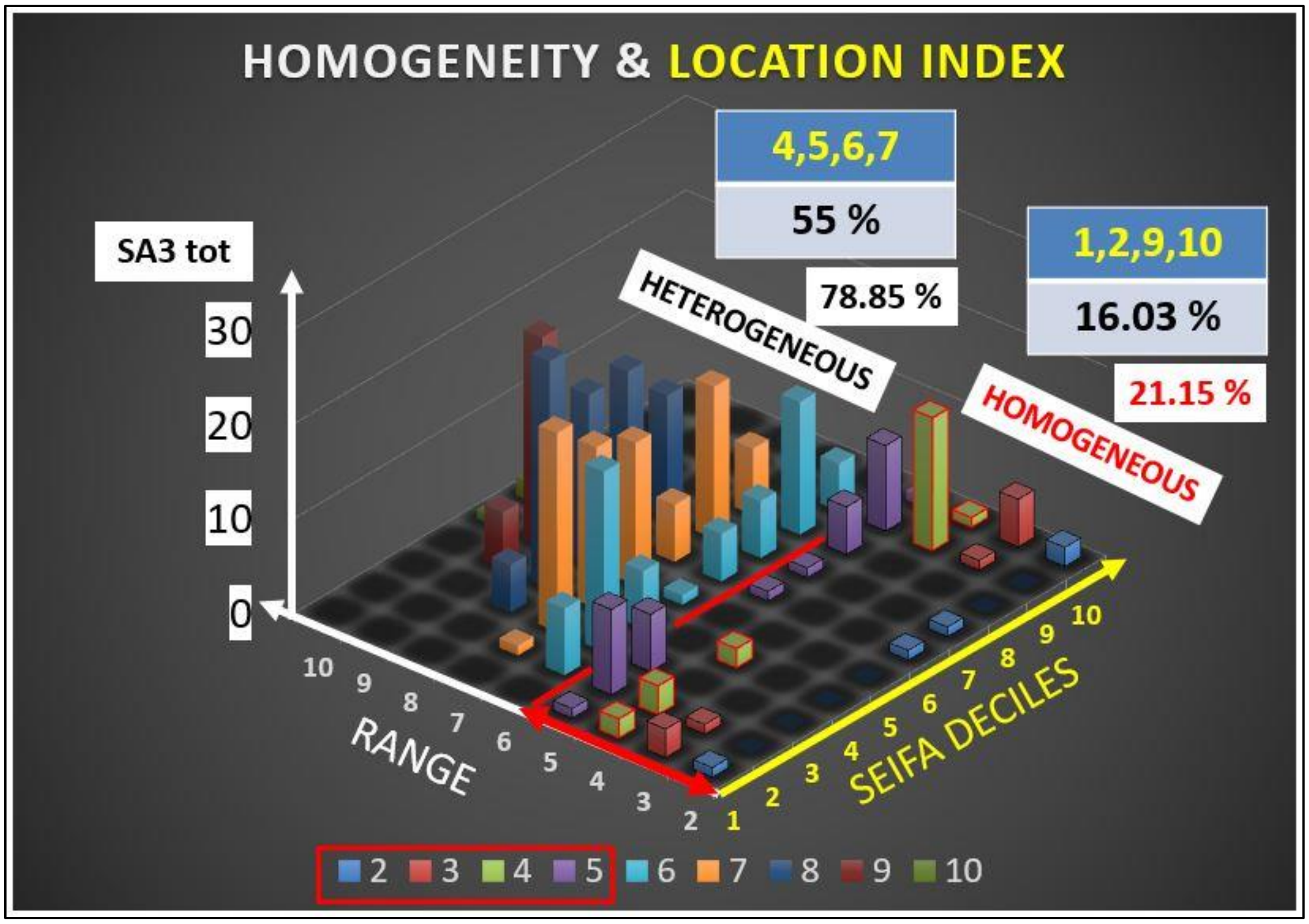

Figure 15 SA3 Homogeneity & Location Index Classification

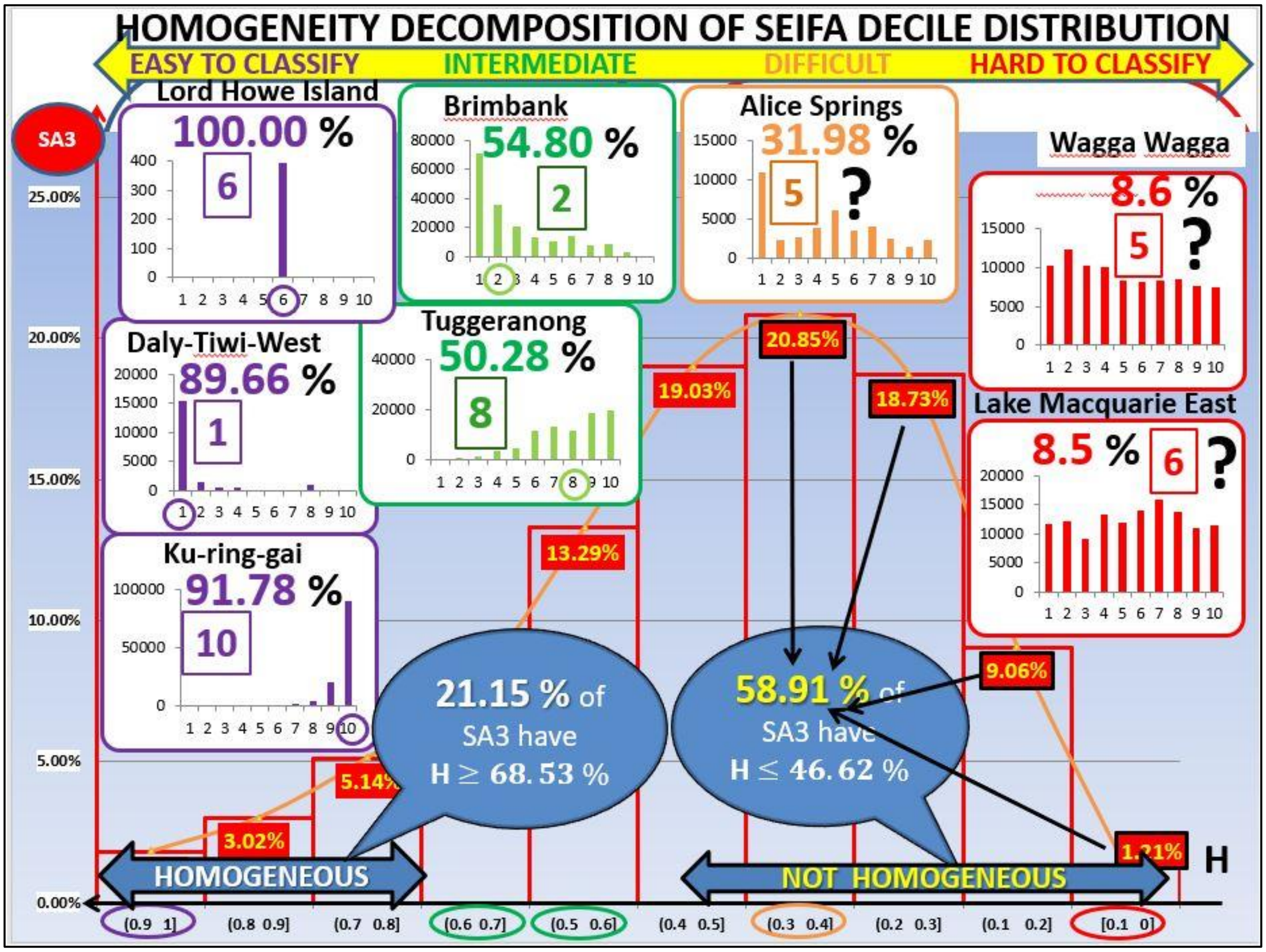

Figure 16 Homogeneity Decomposition of SEIFA decile distribution

## 4.3 – Case Study: Peer Groups

The purpose of this section is to demonstrate the usefulness of the HI and LI to identify peer groups. There are two valuables, yet different, analyses possible with the peer groups; health-related indicators can be compared between and within peer groups. Since peer groups are formed based on regions that have similar socio-economic characteristics, it is expected that differences between peer groups will arise. Peer groups with better socio-economic status indicators are likely to have better health status measures. The second analysis possible, are that may be of more relevant importance, is the comparison of regions within a peer group. Once the effects of the various social and economic characteristics known to influence health status have been removed, a more useful comparison of regions by measures of health status is possible. The example discussed in section 4.3.1 is a simple demonstration of how and when the HI and LI can be used to identify peer groups.

### 4.3.1 Example: Reporting health indicators across SA3s peer groups

To demonstrate the use of the HI and LI, we applied our methodology to the reporting of local variation among GP attenders across the SA3s in the inner and outer metropolitan area of Sydney. Data for this study were sourced from publicly available Medicare Benefits Schedule (MBS)n aggregated at the level of SA3. These data have two limitations: one is that they are mapped to the areas in which people live, rather than where services were provided, and the other is that they are collected at the level of postal area (POA), that do not have perfect correspondence to the SA3 and therefore require some adjustments.

Figure 17 shows the regional variation of age-standardized percentage of very high GP attenders across the SA3s in the NSW Metro area (NHPA, Health Communities: Frequent GP attenders and their use of health services in 2012-13 (pp. 11,22), March 2015), focusing on areas in the lowest (deciles 1-4) and highest (deciles 9-10) socioeconomic groups. The two maps on the top show that the most disadvantaged areas have the highest proportion of people who visited a GP more than 20 times in a year, compared to the least disadvantaged areas. In terms of regional variations, we also note that differences in the distribution of very high GP attenders can be seen even when restricting comparisons to local areas that share similar characteristics, such as the SA3 52 in the least disadvantaged group. In order to observe the characteristics of these socio-economic groups we will look at the residential population attributes for each of the 15 SA3s in the least heterogeneous groups (i.e. A, B, C). Specifically, we selected a subset of key census variables that contribute to the IRSD SA1 score: income, employment, occupation, education and some other miscellaneous indicators of disadvantage, indicated on the right-hand side of the graph in figure 17.

An area with most of the disadvantage indicators above the national average will have a low LI. Conversely, spatial units with values below the national average will have a high LI. It follows that indicators which are further away from the national average have a large impact on the LI value.

For example, the graph in figure 17 shows that the five most disadvantaged SA3s (i.e. 53, 60, 81, 83, 88) have on average a higher proportion of people with low income ($1\ INC\ LOW$: 51.36%) or families with children under 15 years of age and jobless parents ($2\ CHILDJOBLESS$: 20.34%) compared to the national average ($AUS, INC\ LOW$: 40.5% ; $CHILDJOBLESS$: 11.9%). The ten least disadvantaged SA3s (i.e. 52, 57, 66, 67, 68, 69, 70, 71, 72, 85), on the other hand, have on average a low percentage of $INC\ LOW$ (23.10%) and $CHILDJOBLESS$ (4.13%).

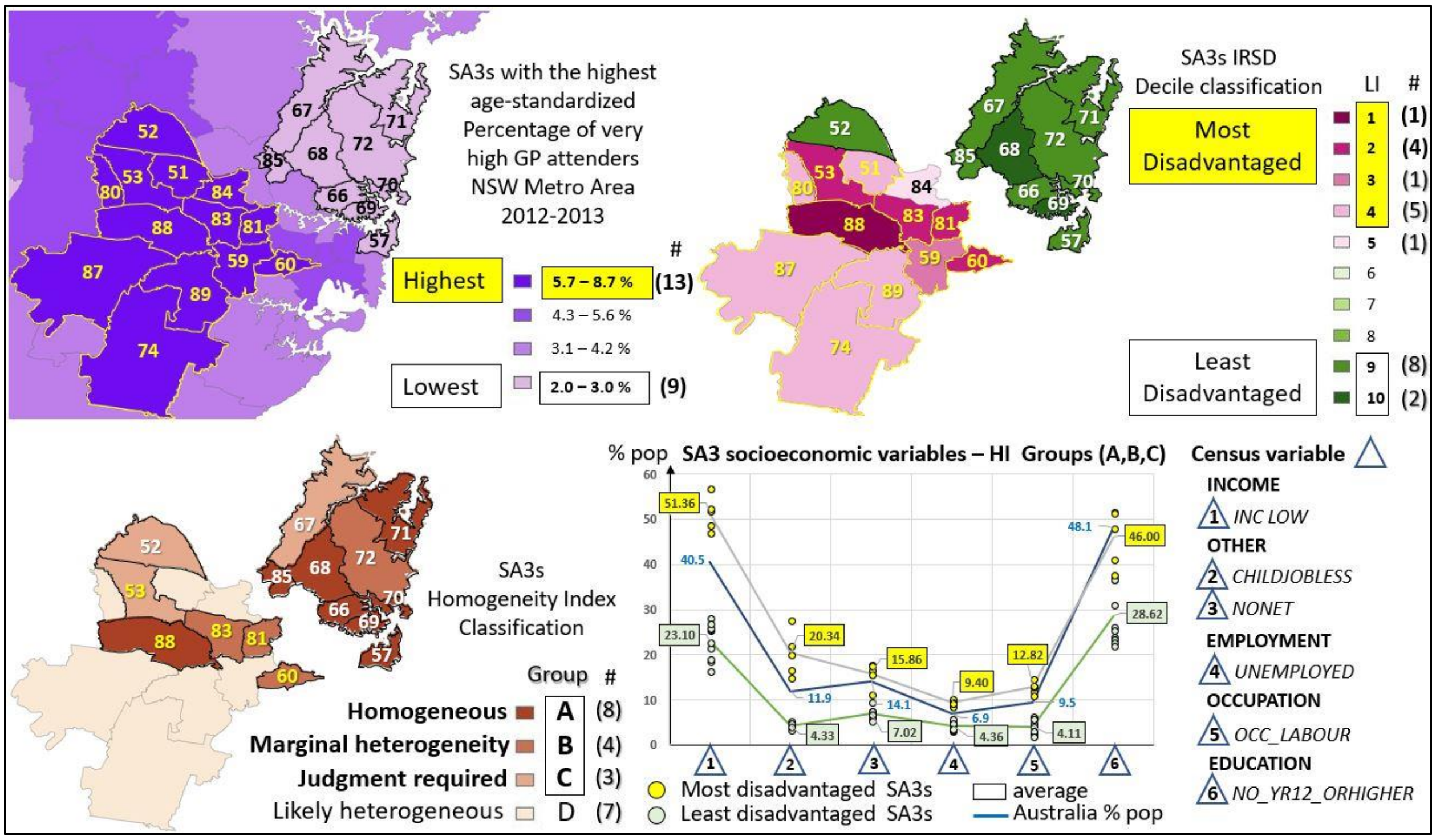

Figure 17 Homogeneous SA3s of Very high GP attenders NSW Metro Area 2012-2013

# Chapter 5 - SA3 BROWSER TOOL

The last release of the Australian Bureau of Statistics Census (2016), in conjunction with the establishment of the Primary Health Networks (PHN) (DOH, Primary Health Networks, 2015), has created the need to group the PHN geographies into peer groups with similar determinants of health needs.

However, the 31 PHN areas exhibit large variation in areal extent, population composition and population size. Therefore, government agencies reported the variation using smaller local areas (SA3). As a result, AIHW entered in a collaborative agreement with CMCRC to explore possible ways of identifying comparable SA3s. Consequently, I developed the **SA3 Browser Tool** (**SA3B**). **SA3B** is the first tool able to compare any two geographic areas and divide a set of geographies into groups with similar characteristics.

More precisely, each SA3 has been grouped into one of seven peer groups based on socioeconomic status and remoteness. The tool also includes different scenarios with larger and smaller number of peer groups.

This allows:

- PHN's SA3s to be compared to other PHN's SA3s within the same metropolitan, regional or rural socioeconomic peer group.
- SA3 to be compared with the average for their peer group.

This enables fairer comparisons of individual PHN catchments and provides a summary of the variation across Australia diverse metropolitan, regional and rural population by presenting aggregate results for each peer group.

In the next sections I give a brief description of the main functionalities available in the SA3B demo version.

## 5.1 - Software Architecture

The primary data source for the tool is the Census but additional data will be included to provide for more robust analysis.

The tool is composed of 4 modules: SA3 Compare, SA3 Cluster Atlas, SA3 Cluster Analysis and SA3 Distance DB. The modules are shown in Figure 18 below.

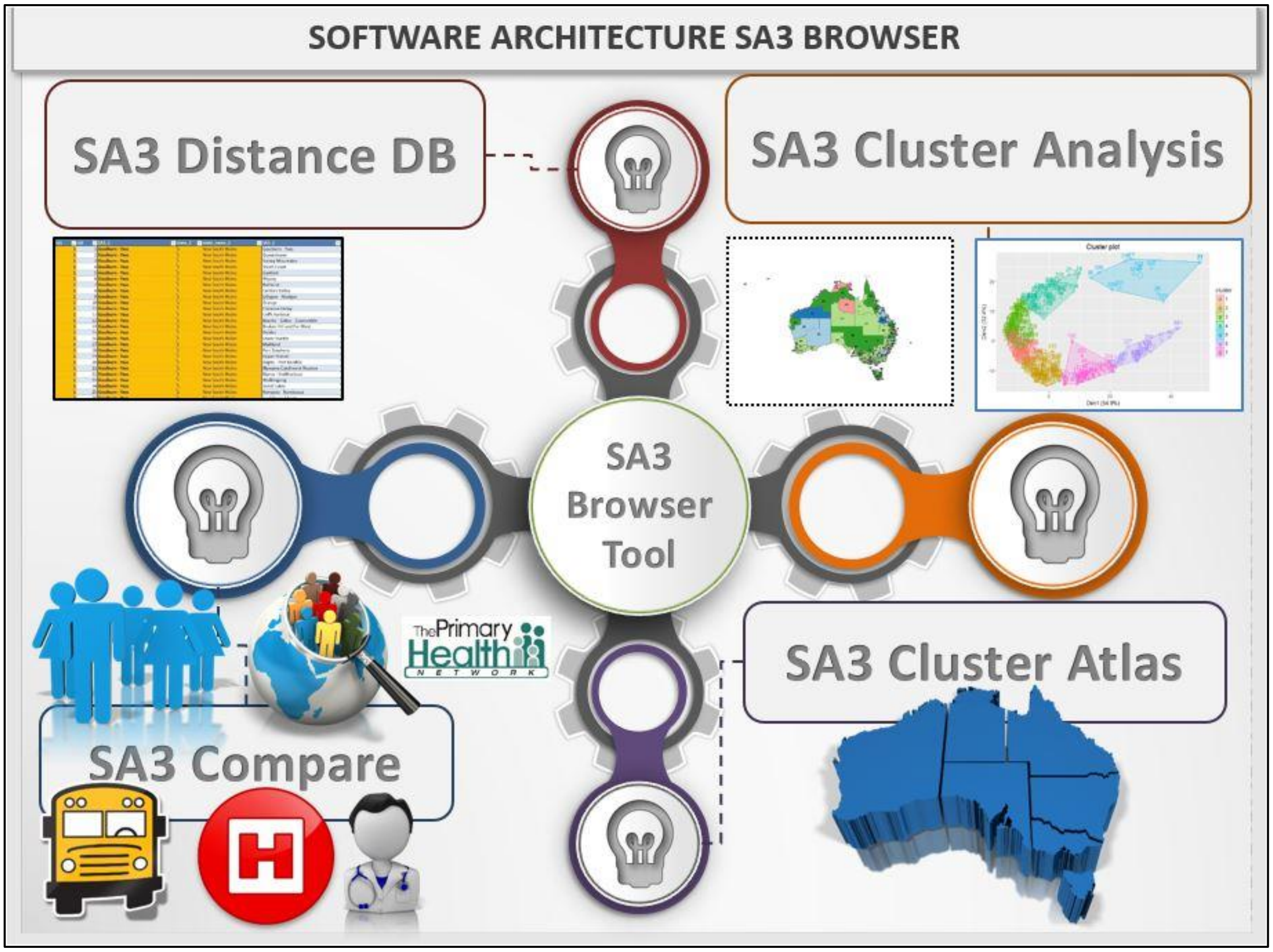

**Figure 18 SA3 Browser Software Architecture**

### 5.1.1 - SA3 Compare

This module provides a dashboard that allows users to compare any two SA3s and explore the IRSD decile distribution by looking at the summary statistics indices and related curves introduced in the previous chapters of this dissertation (e.g. CI, BCFA, LI, DI, HI). It also provides the SA3-PHN concordance and SA3s' peer group for the solution of 5 and 7 clusters.

When SA3 compare program starts (screen shot in Figure 19), a series of tabs are visible across the top of the screen. The interface is designed to guide the user in the socioeconomic analysis of the selected geographies. Let's just focus on the main functionalities of this module.

For instance, the two ribbon buttons allow the user to select the two SA3s to be compared. In this example, I selected the SA3s of Ku-ring-gai and Auburn. The state and SA3-PHN concordance grey windows show the SA3's state and PHN. In this case, the population living in Ku-ring-gai is located in the PHN of Northern Sydney, while 99.08 % of the population in Auburn is located in the PHN of Western Sydney and the remaining in South Western Sydney.

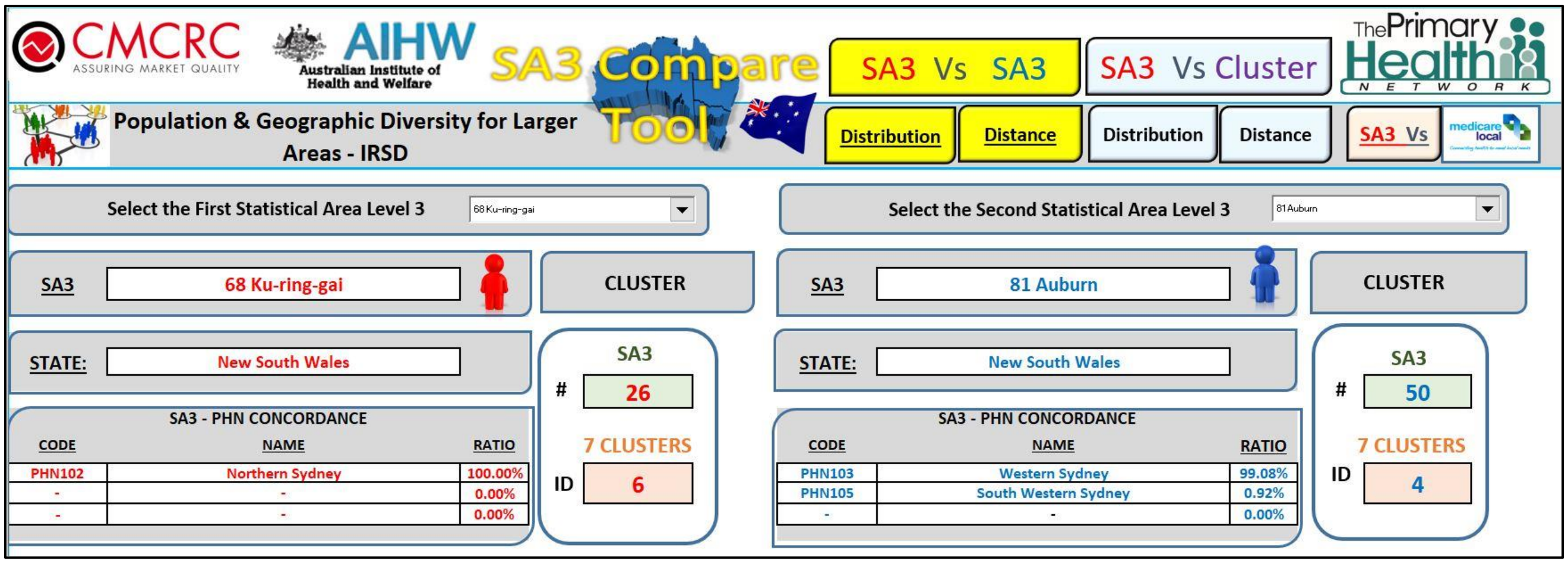

**Figure 19 SA3 Compare Main Menu**

Then, the numbers below the CLUSTRER grey area are the number of SA3s in the Peer group (green box) and the Peer group identification number (rose box), respectively. For example, there are 26 SA3s in the Peer Group number 6 and Ku-ring-gay is one of these SA3s. By clicking the distribution yellow button on the top of the Main Menu, the Compare module shows several information about the IRSD decile distributions, such as the IRDS score (PWAVGS, see Appendix), the socioeconomic disadvantage rank position, the decile and percentile. These information are indicated in the IRSD

SCORE area of the screen shot illustrated in Figure 20. This screen shot shows also the Concentration, Divergence, Location and Homogeneity indices of the two SA3s.

In this example, we see that Ku-ring-gai is one of the least disadvantaged SA3 (rank 331) among the 333 units analysed. This information is further confirmed by looking at the LI (10) and HI score (91.78%) shown on the right hand side of Figure 20. On the other hand, the population of Auburn is less homogeneous (57.94 %) and more disadvantaged (LI=2) than Ku-ring-gai.

To have an idea of what does the distribution of Ku-ring-gai or Auburn look like, the user can explore the Estimated Residential population (ERP) of the IRSD decile distribution by clicking on the IRSD DECILE ERP button on the left hand side of Figure 20.

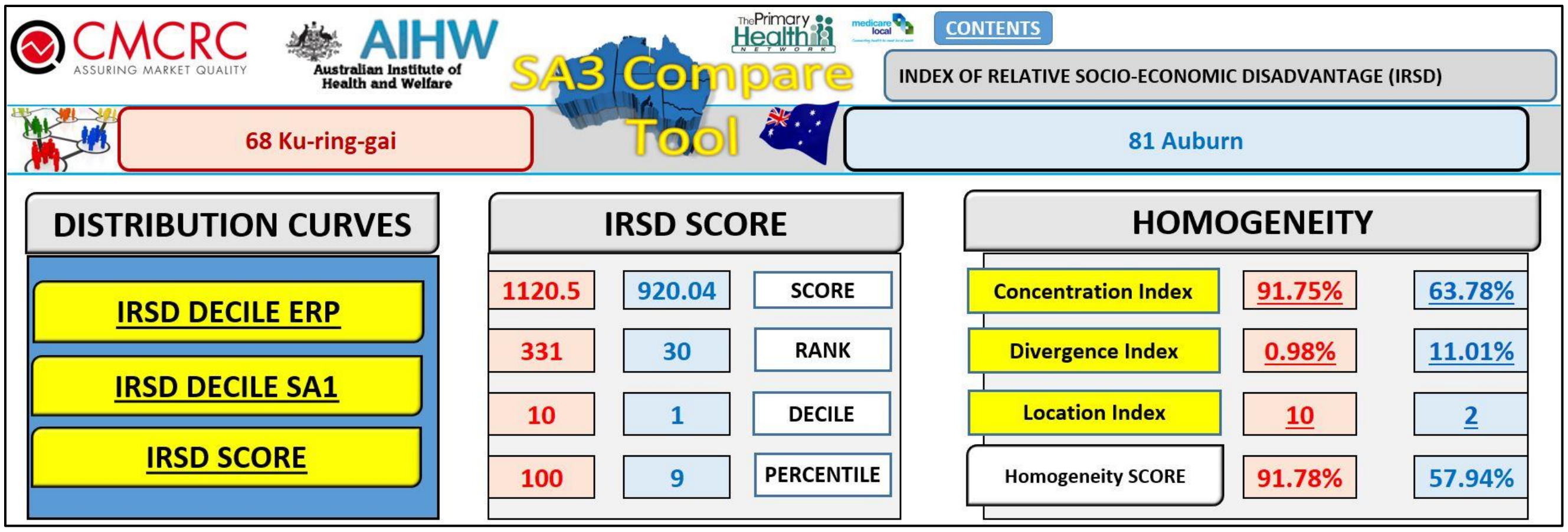

**Figure 20 SA3 Compare Distribution Menu**

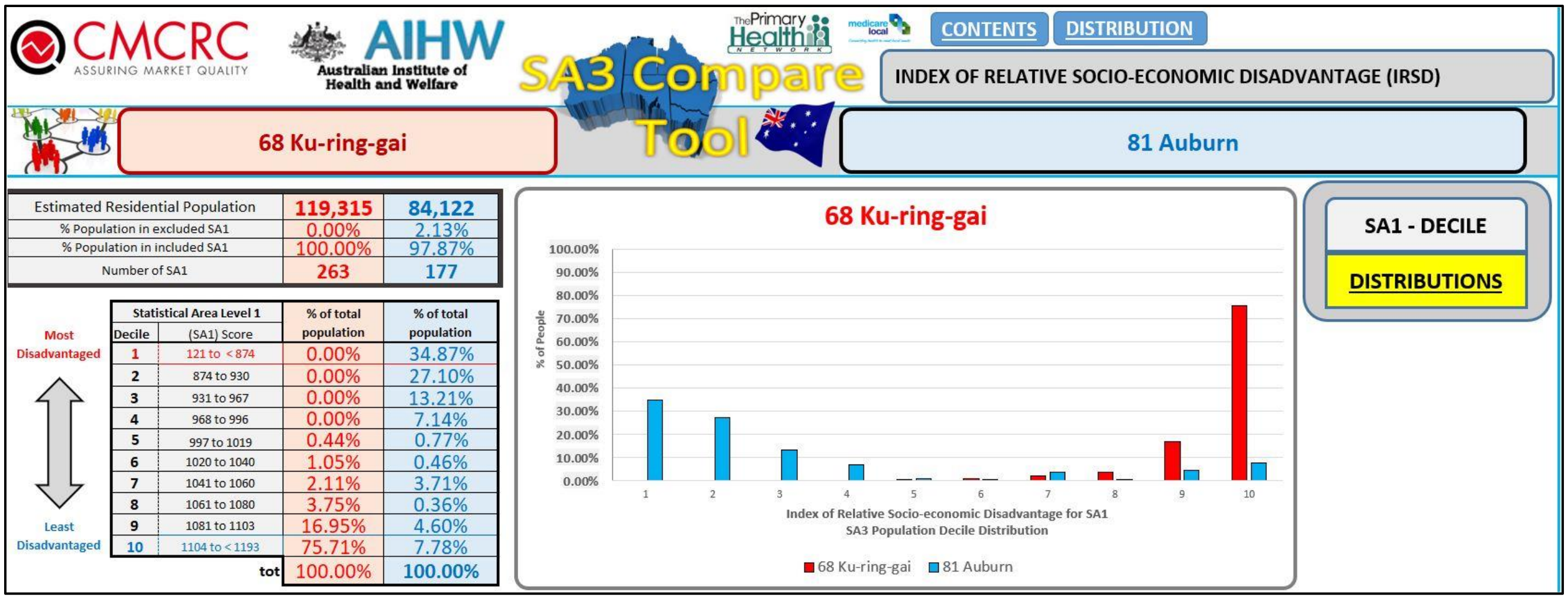

**Figure 21 SA3 Compare IRSD DECILE ERP**

The screen shot in Figure 21 shows the ERP IRSD decile distribution of the two SA3s. It also shows the population size and the number of SA1 within the SA3s. For instance, the SA3 of Auburn has a residential population of 84,122 people and 177 SA1s. We also notice that 2.31 % of the SA1s within Auburn have been excluded from the analysis for confidentiality reasons. Finally, to see the CI, DI and Bin Concentration function of the LI, the user can simply click on the numbers displayed in the Homogeneity grey area shown in Figure 20.

For instance, the screen shots in Figure 22 and 23 show the CI and DI curves of Auburn. Naturally, there are other information in these charts that I do not discuss in this document, but the objective of this presentation is to give an idea of the capabilities of this module.

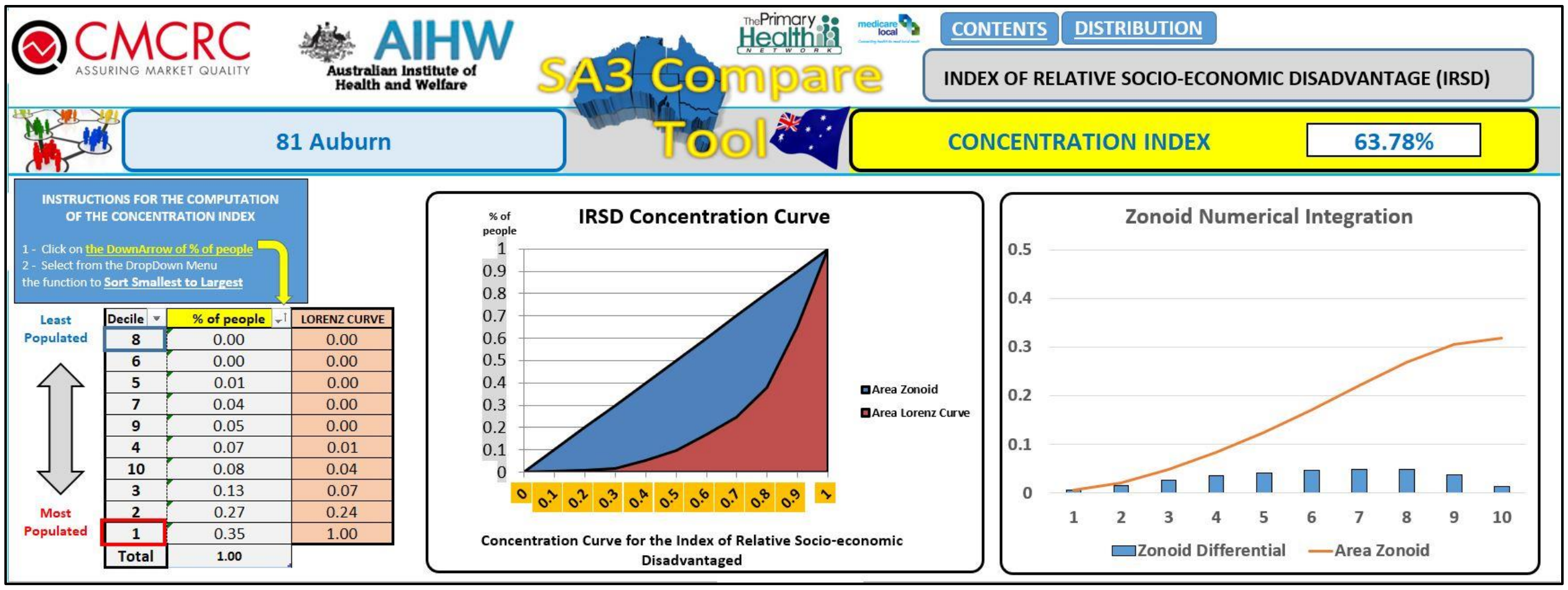

| Decile | % of people | LORENZ CURVE |
|---|---|---|
| 8 | 0.00 | 0.00 |
| 6 | 0.00 | 0.00 |
| 5 | 0.01 | 0.00 |
| 7 | 0.04 | 0.00 |
| 9 | 0.05 | 0.00 |
| 4 | 0.07 | 0.01 |
| 10 | 0.08 | 0.04 |
| 3 | 0.13 | 0.07 |
| 2 | 0.27 | 0.24 |
| 1 | 0.35 | 1.00 |
| Total | 1.00 | |

**Figure 22 SA3 Compare IRSD Concentration Index Curve**

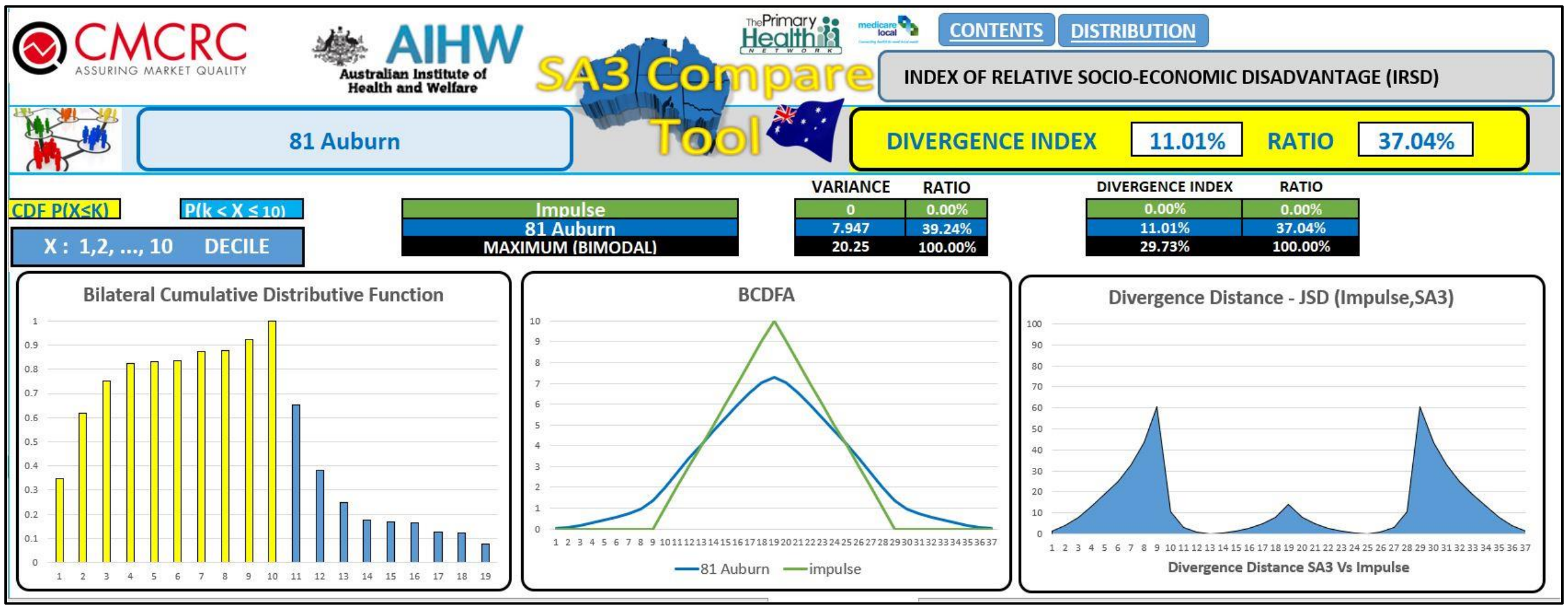

| | VARIANCE | RATIO | DIVERGENCE INDEX | RATIO |
|---|---|---|---|---|
| Impulse | 0 | 0.00% | 0.00% | 0.00% |
| 81 Auburn | 7.947 | 39.24% | 11.01% | 37.04% |
| MAXIMUM (BIMODAL) | 20.25 | 100.00% | 29.73% | 100.00% |

**Figure 23 SA3 Compare IRSD Divergence Index Curve**

Before I move on to the next module, it is worth showing two more functionalities of the SA3 compare module. The Primary Health Network (PHN) image and the distance tab in the main menu represented in Figure 20.

By clicking the PHN image on the top, the user can see the distribution of the selected SA3 and the corresponding PHN.

For example, the histograms in red and blue of Figure 24 show the IRSD decile distribution of Ku-ring-gai and the PHN of Northern Sydney, respectively.

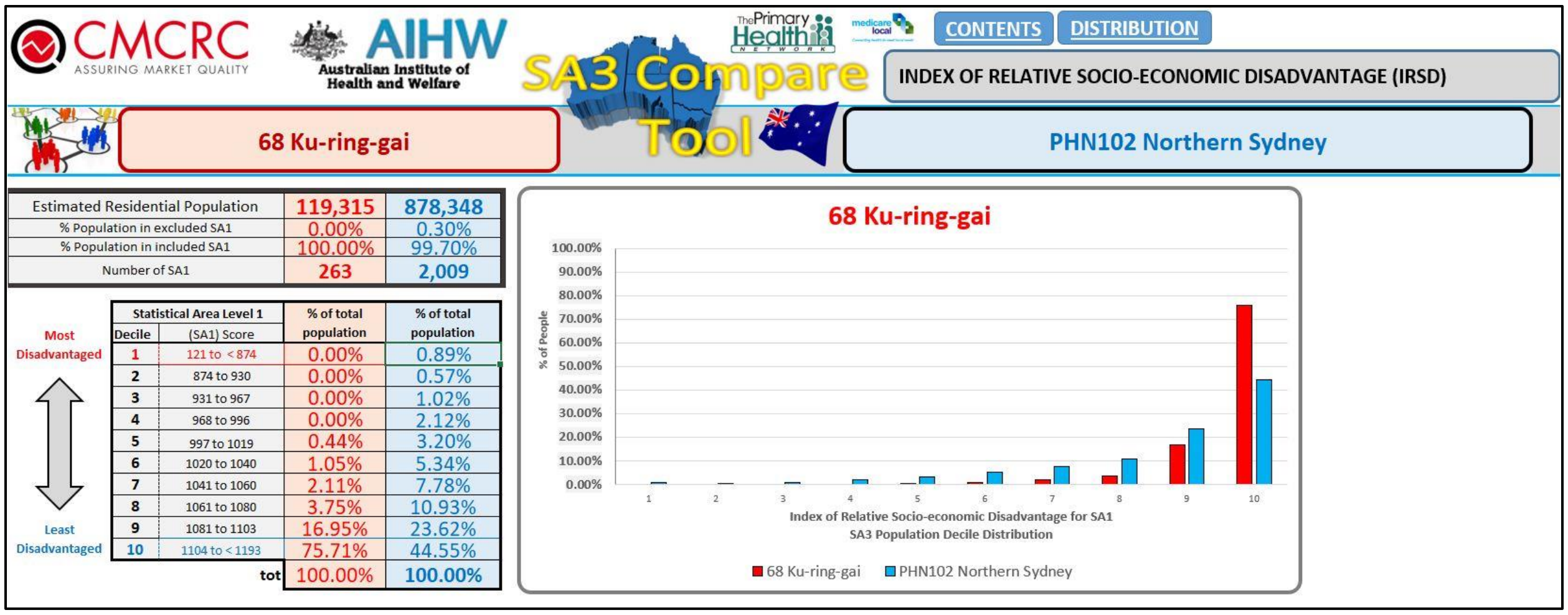

**Figure 24 SA3 Compare PHN distribution**

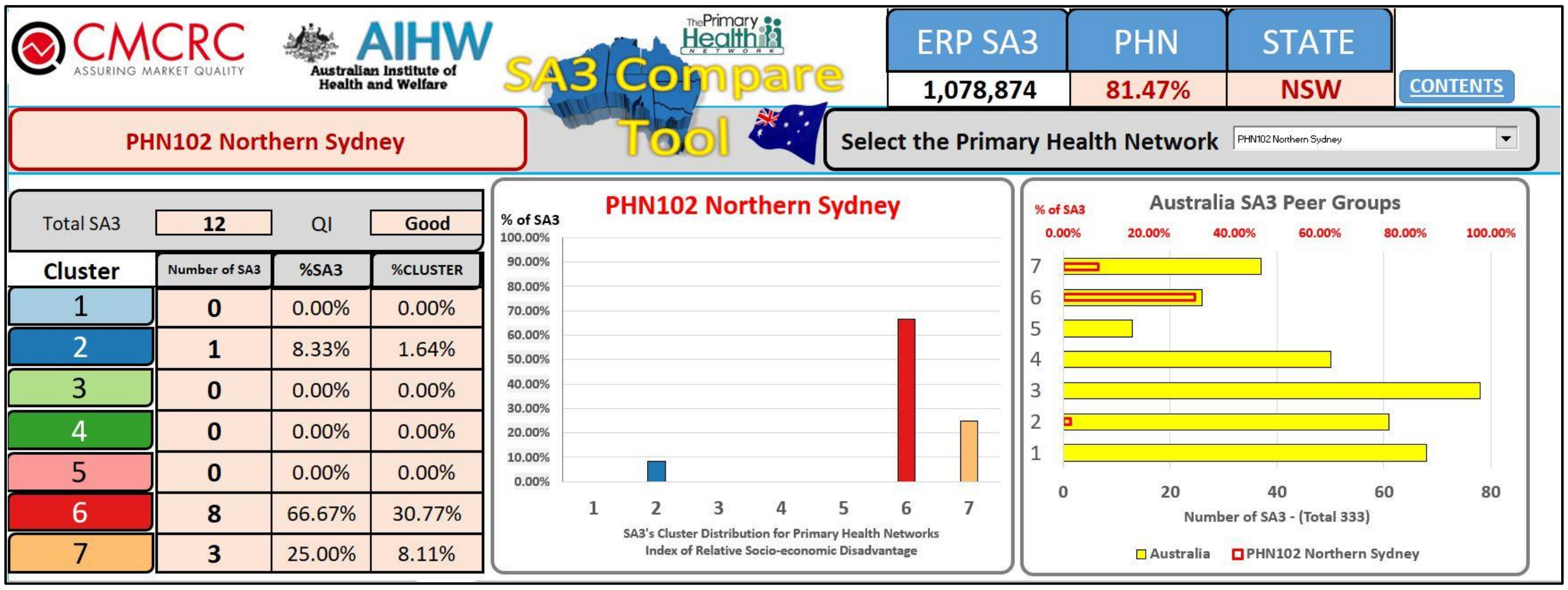

**Figure 25 SA3 Compare PHN characteristics**

Moreover, Figure 25 shows the number of SA3s within the PHN of Northern Sydney and how many of these SA3s are in the 7 peer groups. For example, this PHN has 12 SA3s and 8 of them are in the cluster number 5. This screen shot also show the total population of the SA3s within this PHN and the total correspondence. We see that the total population of the SA3s is 1,078,874 and 81.47 % live in the PHN of Northern Sydney. This information is further confirmed by the QI (Quality Index correspondence) shown on the left hand side of the screen shot (QI=good).

Lastly, the distance tab in the main menu allows the users to compare the two SA3s and tells how different they are. For instance, the dissimilarity pie chart in figure 26 shows that the population living in Ku-ring-gai is not comparable to the population living in Auburn, since the dissimilarity between the two SA3s is 77.51%.

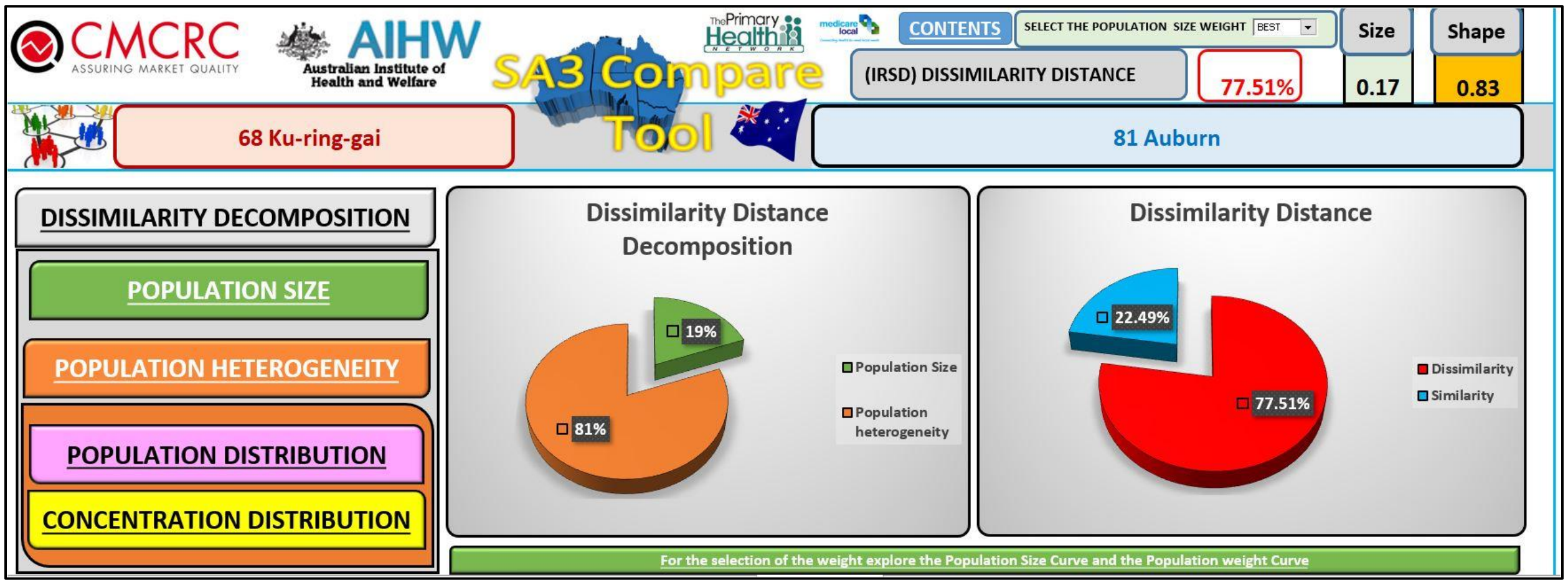

**Figure 26 SA3 Compare Distance**

## 5.1.2 - SA3 Cluster Atlas

In this module the user can explore the map of the SA3 peer grouping solution for 7 clusters of the IRSD variable across and within the 31 PHNs.

The module's dashboard shows the map of the PHNs within each state. It also shows the map of the SA3s' peer group within each PHN. For instance, Figure 27 shows the PHN within the state of New South Wales. Figure 28 shows the SA3s within the PHN of Northern Sydney and their corresponding peer group.

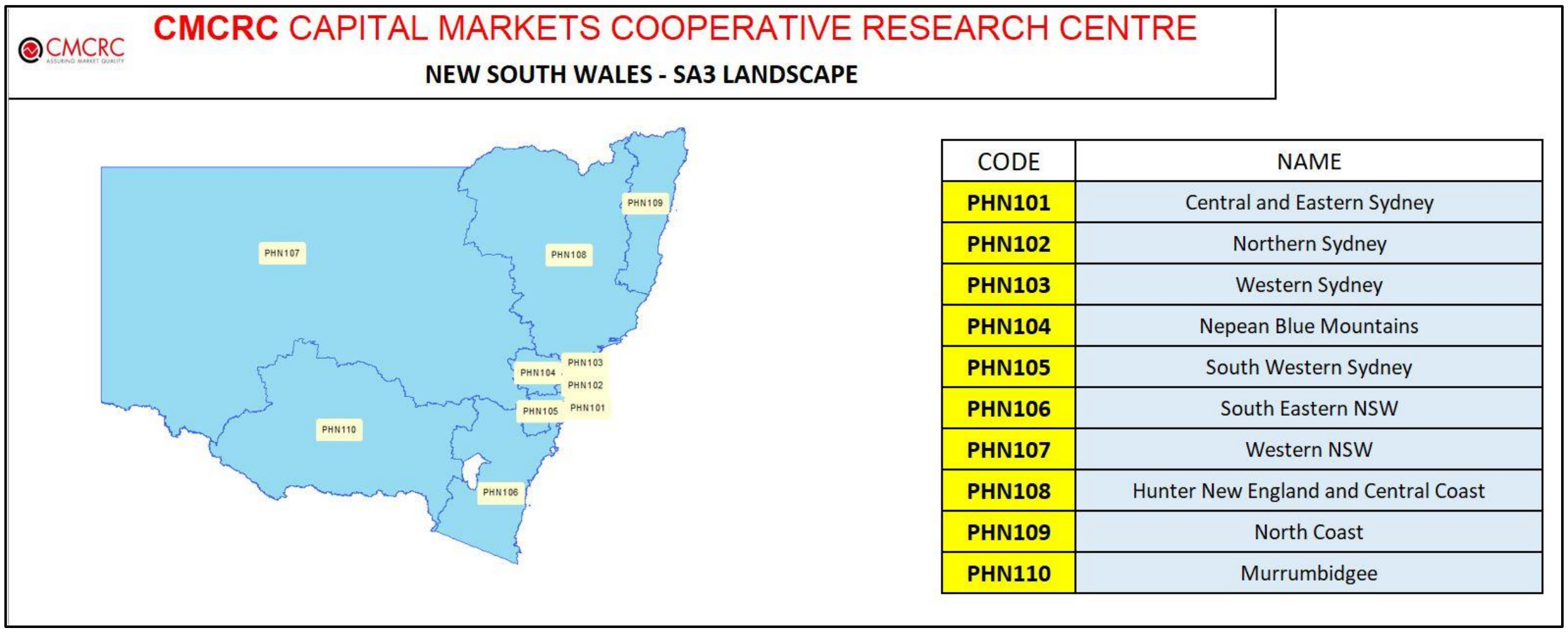

| CODE | NAME |
|---|---|
| PHN101 | Central and Eastern Sydney |
| PHN102 | Northern Sydney |
| PHN103 | Western Sydney |
| PHN104 | Nepean Blue Mountains |
| PHN105 | South Western Sydney |
| PHN106 | South Eastern NSW |
| PHN107 | Western NSW |
| PHN108 | Hunter New England and Central Coast |
| PHN109 | North Coast |
| PHN110 | Murrumbidgee |

**Figure 27 PHN Atlas mapping**

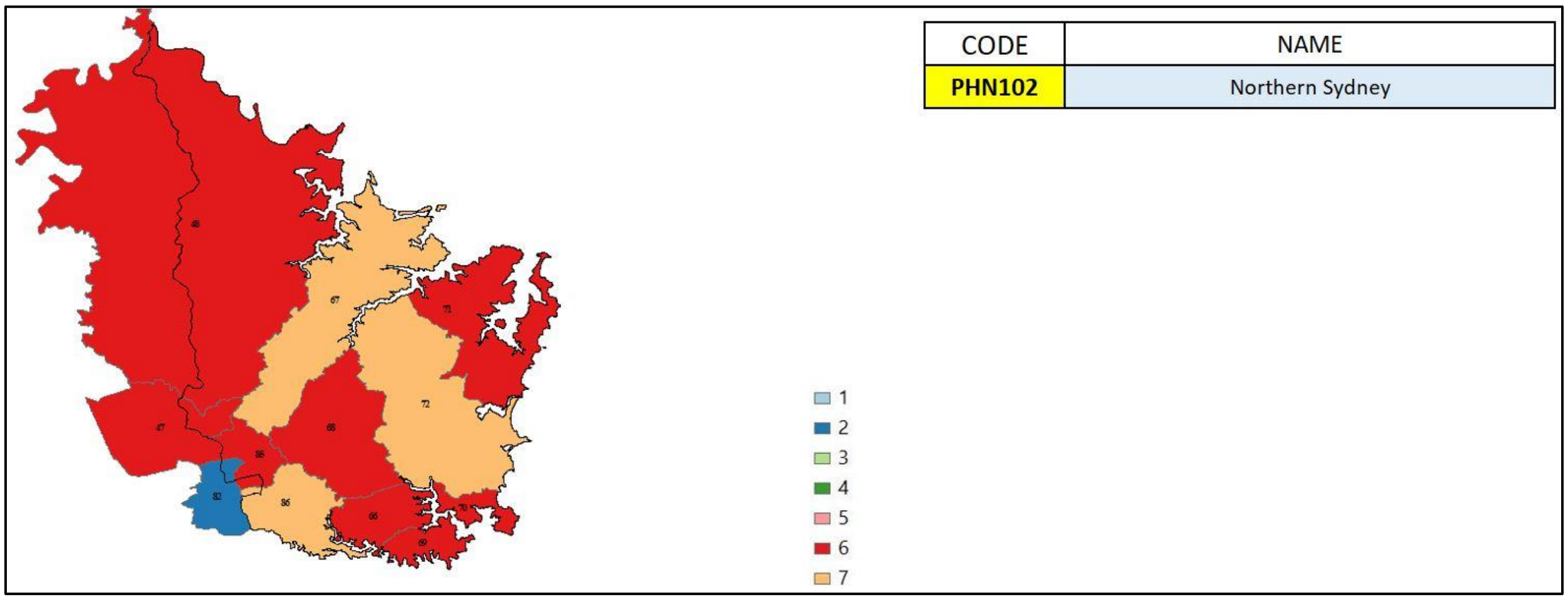

| CODE | NAME |
|---|---|
| PHN102 | Northern Sydney |

**Figure 28 SA3 Atlas - PHN mapping**

### 5.1.3 - SA3 Cluster Analysis

This module contains the cluster analysis of the IRSD SA3 peer groups for 5, 7, 10 and 15 clusters. It also contains an ATLAS of the SA3 classification in terms of the socioeconomic decile category (LI), HI and population size. The dashboard presents this information according to the state of the SA3.

The approach taken in this module was to perform cluster analysis based on a metric that embeds the notion of similarity in terms of population size, population age distribution, distribution of remoteness using the ARIA+ classification (ABS, Remoteness classification, 2006) and distribution of socio-economic status (IRSD).

In order to evaluate the degree of dissimilarity between any two SA3s, I designed a metric composed of three terms:

1. **Population size**: This term considers the size of the histogram population curve. The goal is to capture the property that the SA3s in the same cluster should have roughly the same number of residents.
2. **Population Heterogeneity**: The second term considers the socio-economic characteristics of the distribution curve.
3. **Concentration Location**: Finally, another important feature is where the concentration of the population is located in the distribution.

The clustering metric is defined as a linear combination of the three terms. The Population size term is based on the Sorensen dissimilarity (Sorensen, 1948). This dissimilarity measure computes the magnitudes of individual bins in the histograms. On the other hand, the Population Homogeneity, it is a L1 distance between probability distributions.

However, the Sorensen and the Population Homogeneity metrics are basically shuffling invariance (Sung-Hyuk Cha, 2002), meaning that the distances do not change when the values are permuted among themselves. Therefore, we deemed it convenient to include a metric that takes into account the order of the categories in the distribution. Consequently, we designed a non-shuffling invariance dissimilarity metric named Concentration location.

The clustering algorithm used was a k-medoids algorithm called the Partitioning Around Medoids (PAM) algorithm. The number of clusters, k, was determined by maximizing the average silhouette width, a measure of how well the data has been clustered. I split SA3s into 5,7,10 and 15 clusters.

As several clustering solutions have been provided, it is advised that the users consider each option and decide which, if any, sets of results conforms to their needs.

For instance, the snapshot in figure 29 shows the shape and the silhouette width of the clustering solutions for 5 and 7 peer groups.

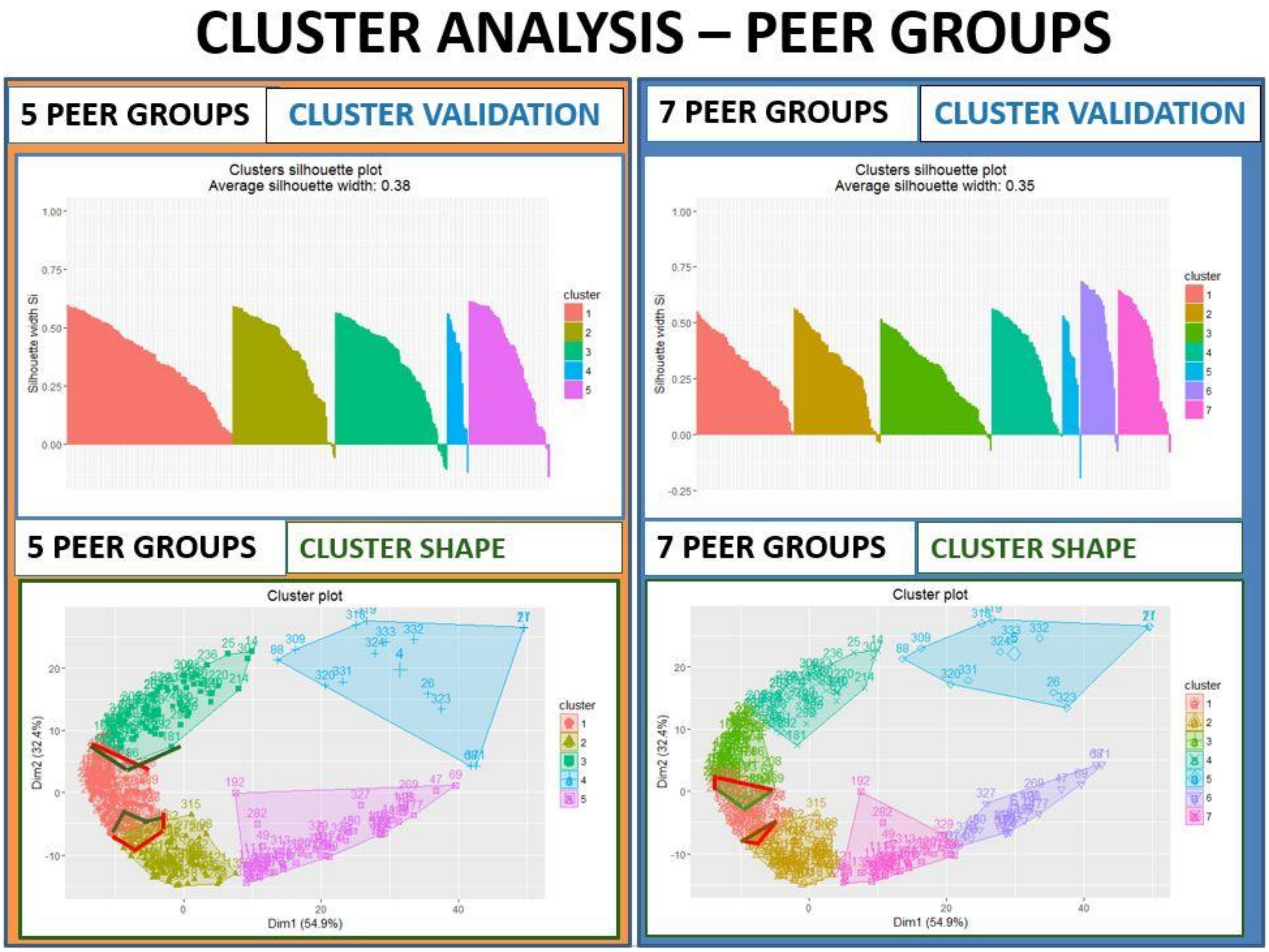

**Figure 29 SA3 Cluster Analysis**

### 5.1.4 - SA3 Distance DB

Lastly, this module contains all the pairwise distances (similarity measures) for each SA3. The Data Base allows the users to find the most similar or different geographic unit for each SA3.

This module contains an easy to use spreadsheet for the analysis of the dissimilarity between any two SA3s. The spreadsheet contains 13 columns illustrated in the snapshot of Figure 30. The orange and blue fields represent the attributes of the first and second SA3, while the last column contains the dissimilarity value. A dissimilarity value of zero clearly indicates that the two SA3s are identical. On the other hand, a value of 1 means maximum dissimilarity.

An easy way to start using this table is the comparison of a selected SA3 to the other SA3s in the table. The Data Filter option in Excel allows the user to select the SA3 to compare and the criteria for sorting the records in the table.

In the example of Figure 30, I filtered the first column of the table by selecting the SA3 of Ku-ring-gai and sorted the dissimilarity measure (Distance) in ascending order. Once the filtering and sorting criteria are selected, the user can easily identify the most similar SA3 for the filtered SA3 (Ku-ring-gai) and compare the population size, HI and LI.

For instance, the second row in the table shows that the most similar SA3 is Baulkham Hills located in the state of New South Wales. The population of this SA3 is 144,084 (pop_2) and the HI is 0.859 or 86% (Hom_2). Lastly, the socio-economic classification or LI is 10 (X0*_2). Clearly, the user can apply other criteria to filter and sort the table.

| SA3_1 | state_1 | state_name_1 | SA3_2 | state_2 | state_name_2 | pop_1 | pop_2 | Hom_1 | Hom_2 | x0*_1 | x0*_2 | Distance |
|---|---|---|---|---|---|---|---|---|---|---|---|---|
| Ku-ring-gai | 1 | New South Wales | Ku-ring-gai | 1 | New South Wales | 119315 | 119315 | 0.933430224 | 0.933430224 | 10 | 10 | 0 |
| Ku-ring-gai | 1 | New South Wales | Baulkham Hills | 1 | New South Wales | 119315 | 144084 | 0.933430224 | 0.859756209 | 10 | 10 | 0.1023462 |
| Ku-ring-gai | 1 | New South Wales | Kenmore - Brookfield - Moggill | 3 | Queensland | 119315 | 46829 | 0.933430224 | 0.944764295 | 10 | 10 | 0.112479615 |
| Ku-ring-gai | 1 | New South Wales | North Sydney - Mosman | 1 | New South Wales | 119315 | 97707 | 0.933430224 | 0.889542937 | 10 | 10 | 0.113077769 |
| Ku-ring-gai | 1 | New South Wales | Cottesloe - Claremont | 5 | Western Australia | 119315 | 71477 | 0.933430224 | 0.823055857 | 10 | 10 | 0.147447092 |
| Ku-ring-gai | 1 | New South Wales | Gungahlin | 8 | Australian Capital Territory | 119315 | 57051 | 0.933430224 | 0.795042241 | 10 | 10 | 0.221547552 |
| Ku-ring-gai | 1 | New South Wales | Brisbane Inner - West | 3 | Queensland | 119315 | 58695 | 0.933430224 | 0.808388637 | 10 | 9 | 0.228355648 |
| Ku-ring-gai | 1 | New South Wales | Boroondara | 2 | Victoria | 119315 | 170553 | 0.933430224 | 0.779874353 | 10 | 9 | 0.22836227 |
| Ku-ring-gai | 1 | New South Wales | Bayside | 2 | Victoria | 119315 | 98368 | 0.933430224 | 0.741303129 | 10 | 10 | 0.229570683 |
| Ku-ring-gai | 1 | New South Wales | Hills District | 3 | Queensland | 119315 | 85712 | 0.933430224 | 0.750638283 | 10 | 9 | 0.249928809 |
| Ku-ring-gai | 1 | New South Wales | Manly | 1 | New South Wales | 119315 | 44232 | 0.933430224 | 0.790541499 | 10 | 9 | 0.260889561 |

**Figure 30 SA3 Distance DB**

## 5.2 Future Developments

I am currently working on an open source and on-line version of the Tool. Furthermore, a second publication will be built on the work of this thesis and will include the definition and implementation of the dissimilarity metric and clustering algorithm used to produce the Peer grouping solutions. The source code and data set used in this dissertation are included in the submitted publication: A Framework for the identification and classification of homogenous areas in the analysis of health care variation.

There is considerable potential for wider application of this method in ecology, socioeconomic and community based clustering, such as customer segmentation.

# Chapter 6 - Discussion

Indexes are frequently constructed to generate summary of the socioeconomic status of residential areas. Although these measures are increasingly being used by national departments and agencies, little attention has been paid to the choice of the geography in presenting socioeconomic status measures. Most of the socioeconomic indices are designed primarily to be small-area measures (Noble, Wright, & Smith, 2006) (Cairstairs, 1995), they are also used to describe relative deprivation for higher-level geographies (Rezaeian, Dunn, & Appleby, 2005) (Strong, Mahesweron, Pearson, & Fryers, 2007).

A major problem facing researchers when constructing indices for larger areas is determining whether the index's score may be adequate to accurately describe the population living in that area. To facilitate this, a range of summary measures have been designed to help users understand deprivation patterns for higher-level geographies (Smith, et al., 2015). These methods are helpful to describe the overall intensity of deprivation across larger area and highlight different aspect of deprivation. However, these analyses focus only to the most deprived small areas and do not provide a full description of the entire distribution. In addition, they do not offer any guidance for the selection of the geographic area.

An additional problem that arises when working with socioeconomic indices is the classification of skewed distributions. Since the index score is a mean-based measure, the computed value is influenced by the extreme scores in the distribution. Consequently, skewed distribution with concentration of scores near the middle can be classified as disadvantaged areas.

As a result, the averaging effect of the socioeconomic index score chronically under-reports or over-reports disadvantage and can lead to incorrect socioeconomic groups. Similarly, the index score of skewed distributions with concentration around the extreme categories is biased towards the middle.

As a response to these issues, this research presented a framework and a software tool which is meant to assist in the analysis and reporting of health care variation by identifying homogeneous areas with similar socioeconomic characteristics better known as peer groups.

More precisely, this work addressed six research questions:

1. How can we measure the homogeneity of a geographic area only by looking at the population distribution of area-based measures?
2. What criteria can be used to guide the selection of a homogeneity measure?
3. How can we formalize the notion of homogeneous area and how this relates to the selection of the geographic unit of analysis?
4. Which criteria and measure of central tendency should be used to classify a homogeneous area?
5. Can we use a combination of a homogeneity and central tendency measure to identify peer groups better?
6. How to design geo-spatial tools for reporting health-outcomes?

The above research questions are briefly discussed in the following sections.

## 6.1 First research question

In chapter 2.1 we discussed how the design of a homogeneity measure might be accomplished with the introduction of a conceptual framework. The conceptual framework, illustrated in Figure 2, includes four main decisions:

1. The first decision is the selection of the larger geographic area and its subunits.
2. Then, the contextual dimension along which one wishes to measure the homogeneity of the geographic area.
3. Third, the selection of the variable used in the model.

4. Finally, the selection of the statistical model used to represent the distributional characteristics of the area.

This work uses The SA3 Australian census geography as geographic unit of analysis and the IRSD as area-based measure. However, the approach can be used to evaluate the homogeneity across any specified geographical boundaries. It is important to notice that the methodology does not require access to fine geographic scale data, and it is easily applied to any distribution of a categorical ordinal data.

Conceptually, the homogeneity measure of a spatial unit, which ranges between 0 to 1, can be therefore defined as the degree to which the residential population is concentrated among the set of categories for that area.

## 6.2 Second research question

Identifying a summary measure describing the distribution of an ordinal variable is not an easy task. Particularly problematic is the issue of the definition of dispersion, because there is no universally accepted measure of homogeneity for ordinal variables.

Then, what is a proper way to specify the concept of dispersion in the case of ordinal variables and to define a measure that has suitable properties?

In this work we propose a mathematical model that embodies a concentration measure and a polarized function into a single index to take the desired properties into consideration.

In this general setting, the properties that are considered to be necessary for any acceptable concentration measure are: Normalization, Continuity, Symmetry, Strict Schur-convexity and Value validity.

The main properties of any polarization measure are: Normalization, invariance of parallel shifts and simple aversion to median-preserving spreads.

These properties are discussed in chapter 3.1

## 6.3 Third research question

The use of a concentration measure is impractical without knowing how to interpret a given homogeneity value, and especially for ordinal variables there is not much guidance on this subject in the literature. Thus, to better classify and compare the diversity of a geographic area, we have proposed to specify and interpret the HI in terms of the number of equally populated groups in a distribution, also known as “true diversity” (Jost, 2006). The essence of this approach is that it is a useful and effective method of representing the homogeneity of a community in “picture”, and lets us compare the diversity of a community easily.

Clearly, the specification of the maximum number of equally populated groups that correspond to a homogeneous community depends on the index being analyzed. In this work we use the IRSD to represent the socioeconomic conditions of Australian geographic areas and capture aspects of disadvantage. Given the observed patterns in the distribution of the IRSD scores, the lowest 40% of scores was selected to identify the most disadvantaged deciles. As a result, the HI’s threshold for an “acceptably” homogeneous area was determined by a community of four equally populated contiguous deciles.

Following this approach, the HI’s range has been partitioned into four classes of concentration, as indicated in Table 5, and an application of these criteria to the Australian Sa3 has shown that almost 60% of the census units are likely heterogenous in terms of IRSD, making comparisons of indicators that correlate with socioeconomic disadvantage difficult to interpret. This result seems highly significant and clearly informs the discussion of which units are appropriate for reporting, suggesting that more work is needed to find ways that allow reporting at lower geographic level while preserving privacy.

## 6.4 Fourth research question

An important task of spatial data analysis consists in identifying a part of data which represent the typical features of the population living in a geographic area. Often the data points are assumed to vary around a center that identifies a single value as representative of an entire distribution. However, when dealing with ordinal categorical data (e.g. socioeconomic indices quantiles), the standard measures of central location, such as the mean or mode are not appropriate (Allison & Foster, 2004).

The mean is not an appropriate measure of central tendency for skewed and high-variance distribution, and many times it does not give a meaningful value. An alternative measure is the mode, but the local concentration of a curve doesn't always correspond to a predominant peak. For example, skewed distribution with high concentrations of values near the middle are common in the SA3's IRSD decile distribution. An obvious choice is the median, but its computation takes much larger than computing the mean. Additionally, its recursive formula is not easy to extend to more categorical variable and there is no absolute consensus about what the definition should be.

Finally, it is possible that data may arise with more than a single median. Although, such distributions are rare and sometimes artefactual, it can be difficult and meaningless to manage multiple values for multimodal distributions.

As a response to these issues, the central measure should be:

- Simple to understand
- Easy to calculate
- Resistant to outliers
- Generalize to n-variables
- Easy interpretation for n variables

— Equal to actual data values.

Thus, We propose the LI as a measure of central tendency that is statistically efficient and reasonably straightforward to implement. In particular, We prove that it is an equivalent form of the median and comprehensive measure of the center.

## 6.5 Fifth research question

As been discussed throughout the previous sections, the selection of the central tendency and concentration measure of a distribution is the starting point of any analysis regarding a set of attributes.

In Table 5 We illustrated the criteria for the classification of the IRSD decile distributions. Other criteria can be chosen for the identification of homogeneous units and there is no definitive or "optimal" HI's threshold value. However, we believe that greater clarity on this subject is obtained by bringing into the picture the LI, that can help users understand when a given area is "highly" homogeneous or heterogeneous. For instance, distributions of six or more equally populated deciles and middle LI values are likely to contain a broader mix of people and households. On the other hand, distributions which have extreme LI values (i.e. very high or very low) and three or less populated deciles are likely to have large proportions of households with similar characteristics.

Therefore, the importance of this representational model lies essentially in its ability to serve as a guide for interpreting dimensionless concentration indices and provide a natural benchmark for these measures in terms of defining what is a "high" and "low" concentration of a probability distribution. This naturally leads to identify geographies where socioeconomic deciles indicated by the LI are meaningful in terms of the HI concentration criteria.

In table 24, we extend the analysis to include the true diversity index (s), and show the number of SA3 for a given combination of LI and HI value as a function of the true diversity index, which is easier to interpret than the value of homogeneity,

To summarize, we believe that the combination of the LI and the HI's classification criteria enables users to summarize easily the characteristics of a geography in a single table. This straightforward representation is helpful to evaluate the number of homogeneous areas for each socioeconomic disadvantage group and select a suitable region to identify peer groups geographies.

## 6.6 Sixth research question

In chapter 5, We illustrated how to design a geo-spatial software tool for reporting health-outcomes in geographic variation studies. This guide can be used to structure the Software Architecture of a general-purpose tool aimed to group and compare geographic regions. The Architecture should include four basic modules. Each module provides a specific functionality that is illustrated in Figure 18 and thoroughly discussed in section 5.1.

The tool enables fairer comparisons of individual region and provides a summary of the variation across Australia diverse metropolitan, regional and rural population by presenting aggregate results for each peer group.

### 6.6.1 Potential application of SA3 Browser

The use of a software tool for mapping variation across Peer groups is an invaluable tool for understanding how our health care system is providing care. The SA3 Browser Tool could be integrated into projects undertaken by geospatial groups at government agencies. An example is the Australian Atlas of Healthcare published by the Australian Commission on Safety and Quality in Healthcare.

The Australian Atlas of Healthcare variation project illuminates variation by using data to map the use of health care according to where people live.

The Atlas provides information on clinical items grouped into four clinical themes, covering procedures, tests, investigations, treatments or hospitalizations. The geographic local areas used are ABS SA3. The data specifications for each item can be accessed on the AIHW Metadata Online Registry (MeteOR) at www.meteor.aihw.gov.au.

## 6.7 Limitations and Future recommendation for research

As discussed in the previous sections, the HI is a concentration measure applied to univariate distributions of ordinal variables (i.e. variables for which there are no numbers but only the ordering). However, in many applications it is natural to consider several variables instead of a single one and to measure their dispersion in a multivariate setting.

Furthermore, bivariate distributions of ordinal indicators are often considered while measuring socioeconomic inequalities in health. In empirical applications the SES is often proxied by income classes, educational attainment, social or occupational class and health is proxied by self-reported health status; thus bivariate distribution of ordinal indicators are analyzed (Conwell, Kobus, & Kurek, 2017).

The definition of the Homogeneity Index is based on the CI and DI. The extension of the DI in a multivariate setting can be obtained by the application of the convolution operator for a multidimensional distribution. However, the CI's definition is based on the Gini's Index and Lorenz Curve. Generally, the Lorenz Curve measures sort of scatter or dispersion with respect to a single variable.

A few attempts have been made in the literature to generalize the Lorenz Curve and Lorenz dominance for multidimensional data (Koshevoy & Mosler, 1996). But besides a few rather special cases (Eichorn & Krtscha, 2007), none homogeneity measure has been proposed to extend the univariate Gini coefficient to a multivariate distribution. Furthermore, these definitions are difficult to interpret in a multivariate situation. Consequently, the major problem we face is the extension of the Lorenz Curve to a multivariate distribution and introduce a new definition of the Gini coefficient.

# Chapter 7 - Conclusion

Reporting on variation across similar areas aims to assist health care planners in the targeted delivery of health services by identifying areas where to direct efforts to deal with unwarranted variation. Importantly, a key step in studying health outcomes or services utilization is the investigation of spatial patterns of community characteristics by mapping them and by assessing the degree of homogeneity using statistical methods based on individual level information.

Most data collections, however, do not include information regarding individual-level socioeconomic position, leading to a reliance on area-based information. While area-based measures are relatively easy to collect and utilize, there is no general definition of homogeneous areas based on area level socioeconomic indices. Moreover, the use of the index score becomes less and less meaningful as the size of an area increases. Hence, the misclassification resulting from the use of area-based measures to direct support for areas of needs creates a risk for resource misallocations. As a result, this paper introduced an easy to use statistical framework for the identification and classification of homogenous areas.

I applied this framework to assess the socioeconomic homogeneity of SA3, a census geography commonly used in Australia for reporting health indicators. Results from our investigation prompt us to discourage the use of this geography as unit of analysis in the socioeconomic context. The findings also suggest that the proposed framework is a useful tool for strategic planning purpose through its ability to identify areas of disadvantage within broader regions and capture key characteristics of the regions.

Therefore, this tool can be useful in raising discussion on the selection and use of geographic regions (grouping of geographic units) which may be indicative of socioeconomic status. If the region is not acceptably homogeneous, a different definition of the region should be considered. The region could be divided in more subregions or a different assignment of the geographic units to regions could be tried. Another option is the selection of a completely different geographic aggregation unit.

Finally, the combination of the Homogeneity and Location index constitutes a clear and consistent framework for geographic variation studies. The advantages of such indices include statistical efficiency and a simple presentation of results. They facilitate the visualization of socioeconomic characteristics of geographic areas in a way that can be combined into a dashboard, integrating the homogeneity and central location of the data. This powerful method of illustrating and classifying a geographic area is, therefore, a valuable tool that can act as an interface between the technical and policy disciplines as well as with the decision makers, so they can make scientifically informed decisions.

# Appendix A - Homogeneity Index (HI)

The *Homogeneity index* (**HI**) is a generalization of the Gini index for ordinal categorical variables. This function captures the diversity of a geographic area in a single number, 0 to 1 (or 100%). A Homogeneity coefficient of zero expresses minimal concentration and occurs when the population is equally distributed among the set of attributes for that area. For example, a heterogeneous area is where the residents are evenly distributed across different age groups or socioeconomic status quantiles (e.g. quintiles or deciles). Conversely, the maximal concentration (100%) is attained if the distribution concentrates at one point (a.k.a. singleton). In the latter case, there is no variation within the area per that characteristic and the geography is uniquely identified by its central value.

The two key components underlying the HI are the *Concentration index* (**CI**) and the *Divergence index* (**DI**).

The **CI** is a concentration measure for nominal categorical variables. The properties to be considered necessary for any acceptable CI are: Normalization, Continuity, Symmetry, Strict Schur-Convexity and Value validity (Kvalseth, Evenness indicies once again: critical analysis of properties, 2015).

Most of the indices in the evenness literature satisfy the first three properties but only a few meet the requirements of Strict Schur-Convexity and Value validity. Among these indices, the Gini index is the most popular and is easily related to the pointwise ordering of the Lorenz curve (Gosselin, 2001).

Therefore, the CI is defined with reference to the Lorenz curve (LC). The LC plots the cumulative percentage of the population (y-axis) against quantiles or groups of the attribute variable. If the population is uniformly distributed, the LC will be a 45-degree line, running from the bottom left-hand corner to the top right-hand corner. This is known as line of equality. Otherwise, the curve is convex and lies below this straight line. Thus, the CI is defined as twice the area between the LC and the line of equality. This area is also referred as Lorenz Zonoid (Koshevoy & Mosler, 1996). For the computation and definition of CI see A1.

We note that the definition of the CI can be viewed as a study of the discrepancy between a probability distribution (pdf) and the uniform distribution, but it does not completely reflect the amount of spread that the values of a random variable will take on. For example, different distributions with same LC have different variance. This means that the ordering of the pdf coefficients is not important. In view of this circumstance, we deemed it convenient to include the DI in the computation of the HI.

The **DI** is a polarization measure for ordinal categorical variables. The main properties of the DI are: Normalization, invariance of parallel shifts and simple aversion to median-preserving spreads (LV & Xu, 2015).

Basically, the first two properties state that the minimum value (zero) should be taken for the one-point distribution and the "parallel" shift of the entire frequency distribution leaves the index's value unchanged. The last property states that the transfer of cases from one category into the next, which is closer to the extreme categories (i.e. the first or last category) should result in an increased value of the polarization measure as it will be closer to the distribution in which half of the population is concentrated in the lowest category and half in the top category (a.k.a. two-point extreme distribution). It follows that the DI is maximized when the probability is concentrated at the end points of the distribution. In such a configuration, the variance of any bounded probability distribution is maximum (Bathia & Chander, 2000). For example, a geographic area where the residents are evenly distributed among the least and most disadvantage socioeconomic quantile or the youngest and oldest age group. On the other hand, the minimum value is attained when there is no variation within the area (i.e. singleton). This bounded value, therefore, captures the amount of fluctuations about the central location of a distribution and simultaneously the local variation in the data. For example, distributions with same mean but different variance should have different DI. On the other hand, distributions with different mean or median but same variance must have the same DI value. For the definition and properties of the DI see A2.

Considering, finally, the definition of the CI and DI, the HI of a probability distribution vector of n categories ($P_n$) is given by the following formula:

$$HI(P_n) = \frac{CI(P_n) + DI(P_n^1) - DI(P_n)}{1 + DI(P_n^1)} \qquad n \geq 3 \quad P_n^1 = \left(\frac{1}{n}, \frac{1}{n}, \cdots, \frac{1}{n}\right)$$

At this point, it would be natural to ask if this formula defines a positive function, between zero and one, that attains the minimum at the uniform distribution ($P_n^1$) and the maximum when applied to a singleton ($P_n^0$)?

One way to check this condition is to study the behavior of $HI(P_n)$ when $P_n = P_n^1$ and $P_n = P_n^0$. It is not difficult to verify that the numerator is equal to $1 + DI(P_n^1)$ for $P_n = P_n^0$ and zero for $P_n = P_n^1$. Then, the only sufficient condition for the HI to be a positive function is that the CI value for the two-point extreme distribution is larger than the difference between $DI(P_n^1)$ and $DI(P_n)$. This property is satisfied by the DI. Thus, the HI is always nonnegative and vanishes if and only if the distribution is uniform.

## A1: Concentration Index (CI)

Consider a resident population of P individuals and let X= $\{X_1, X_2, \dots, X_i, \dots, X_n\}$ be a specific characteristic of the residential population grouped in n categories (e.g. age groups, IRSD quantiles, etc.). Let P(X)= $\{p_{X1}, p_{X2}, \dots, p_{Xi}, \dots, p_{Xn}\}$ be the probability mass function of X (i.e. $p_{Xi} = \frac{f_i}{P}, f_i\ absolute\ frequency\ of\ X_i$ ). The increasing order of the observations is denoted:

$$\text{Ш}(P(X))^1 = \{\boldsymbol{p}_{(1)}, \boldsymbol{p}_{(2)}, \dots, \boldsymbol{p}_{(i)}, \dots, \boldsymbol{p}_{(n)}: \boldsymbol{p}_{(i)} \leq \boldsymbol{p}_{(j)} \wedge i < j\}$$

Then the Lorenz Curve is the piecewise linear connection of points:

$$(x, y) = \left(\frac{k}{n}, \sum_{i=1}^{k} p_{(i)}\right), \quad k = 0, \dots, n \wedge p_{(0)} = 0$$

in the unit square. Accordingly, the *Lorenz Function* is:

$$\boldsymbol{LF}(\boldsymbol{x}) = \left\{\frac{y_{k+1} - y_k}{x_{k+1} - x_k} \cdot (x - x_k) + y_k : y_k = \sum_{i=1}^{k} p_{(i)} \wedge y_{k+1} = y_k + p_{(k+1)}, \quad x \in [0\ 1]\right\}$$

It follows that the area under the Lorenz Curve is:

$$A = \int_0^1 LF(x)dx$$

(See Theorem A1.2)

Therefore, the *Lorenz Zonoid* is

$$\boldsymbol{LZ} \stackrel{\text{def}}{=} 2\left[\int_0^1 x\,dx - A\right] = 1 - 2A$$

Finally, the CI is given by the following formula:

$$CI = \frac{LZ}{\widehat{LZ}} \qquad \widehat{LZ} = \frac{n-1}{n}$$

The Upper Bound of the Concentration Index is $(n-1)/n$ (it is attained when all the observation except one are equal to zero). As a result, the Lorenz Zonoid is divided by $\widehat{LZ}$ (See **Theorem A1.1**). It follows that the CI is a function bounded between zero and one.

---

[1] Π(P(X)) ∈ $\text{Ш}(X) \stackrel{\text{def}}{=}$ all the permutation over $X = \{X_1, \dots, X_N\}$. It is clear $|\text{Ш}(X)| = N!$

**Theorem A1.1**: Let P(X) be a probability density function of the random variable X= $\{X_1, X_2, \dots, X_i, \dots, X_n\}$. Then $\widehat{LZ} = \max\{CI\} = \frac{n-1}{n}$.

*Proof*:

$\max\{CI\} = \max\{LZ\}$

$$\max\{LZ\} = 1 - 2 \cdot \min\{A\}$$

$\min\{A\} = \int_0^1 LF(x, \underset{y_k}{\arg\min}\{A\})\, dx \rightarrow \underset{y_k}{\arg\min}\{A\} = \{y_k = 0\ \forall\ k \neq n\ \wedge\ y_n = 1\}(i.e. singleton)$

$$= \int_{1-1/n}^{1} (nx - n + 1)\, dx = n \cdot \left[\frac{x^2}{2}\right]_{1-\frac{1}{n}}^{1} - n \cdot [x]_{1-\frac{1}{n}}^{1} + [x]_{1-\frac{1}{n}}^{1} = \frac{1}{2n}$$

It follows that:

$\max\{CI\} = 1 - 2 \cdot \frac{1}{2n} = \frac{n-1}{n}$ ***Q.E.D***

**Theorem A1.2**: Let LF(X) be the Lorenz Function of a random variable X= $\{X_1, X_2, \dots, X_i, \dots, X_n\}$. Then, the area under the Lorenz Curve is $A = \sum_{k=0}^{n-1} \frac{1}{n} \cdot \left(\frac{y_k + y_{k+1}}{2}\right)$.

*Proof*:

Reminding the definition of LF(X):

$$LF(x) = \left\{\frac{y_{k+1} - y_k}{x_{k+1} - x_k} \cdot (x - x_k) + y_k :\ y_k = \sum_{i=1}^{k} p_{(i)} \wedge y_{k+1} = y_k + p_{(k+1)}\ , \qquad x \in [0\ \ 1]\right\}$$

Considering the following identities:

$$\begin{cases} x_k = \dfrac{k}{n} \quad k = 0, \dots, n \quad \rightarrow \quad x_{k+1} = x_k + \dfrac{1}{n} \\ y_{k+1} = y_k + p_{k+1} \end{cases}$$

We can compute the Lorenz Curves' Area:

$$A = \int_0^1 LF(x)dx = \sum_{k=0}^{n-1} \int_{x_k}^{x_k+\frac{1}{n}} n \cdot (p_{k+1}) \cdot (x - x_k) + \ y_k \ dx$$

$$= \sum_{k=0}^{n-1} \left\{ n \cdot (p_{k+1}) \cdot \left[\frac{x^2}{2}\right]_{x_k}^{x_k+\frac{1}{n}} - n \cdot (p_{k+1}) \cdot x_k [x]_{x_k}^{x_k+\frac{1}{n}} + \ y_k \cdot [x]_{x_k}^{x_k+\frac{1}{n}} \right\}$$

$$= \sum_{k=0}^{n-1} \frac{1}{n} \left( \frac{y_k + y_{k+1}}{2} \right)$$

***Q.E.D***

**Corollary A1.1:** Let LZ(X) be the Lorenz Zonoid of a random variable X= $\{X_1, X_2, \dots, X_i, \dots, X_N\}$. Then LZ(X) = LZ(Y) = $1 - \frac{1}{N} \cdot [1 + 2 \cdot \sum_{i=1}^{N-1} y_i]$.

*Proof*:

Reminding the definition of CI(X) and LZ(X):

$$LZ(X) = \ 1 - 2 \int_0^1 LF(x)dx = \ 1 - 2 \ \sum_{j=0}^{N-1} \left( \frac{1}{N} \cdot \frac{y_i + y_{i+1}}{2} \right)$$

$$= \ 1 - \frac{1}{N} \sum_{i=0}^{N-1} (y_i + y_{i+1})$$

Solving the summation:

$$\sum_{i=0}^{N-1} (y_i + y_{i+1}) = y_0 + \overbrace{y_1 + y_1}^{2y_1} + \cdots + \overbrace{y_i + y_i}^{2y_i} + \overbrace{y_{i+1} + y_{i+1}}^{2y_{i+1}} + \cdots + \overbrace{y_{N-1} + y_{N-1}}^{2y_{N-1}} + y_N$$

$$= 1 + 2 \sum_{i=1}^{N-1} y_i : \quad y_0 = 0 \ \wedge \ y_N = 1$$

It follows that:

$$LZ(X) = LZ(Y) = 1 - \frac{1}{N} \cdot \left[1 + 2 \cdot \sum_{i=1}^{N-1} y_i\right]$$

***Q.E.D***

**Lemma A1.1:** Let $\delta(X)$ and $U(X)$ be the singleton and uniform distribution of the random variable $X = \{X_1, \dots, X_N\}$, such that $CI(X) \in [0 \ \ 1]$. Then

$$arg_{m,M}\, CI(X) = \begin{cases} \arg\min_X \{CI(X)\} = CI\{U(X)\}: & U(X_i) = {}^{1}/_{N} \quad \forall\, i \\ \arg\mathrm{Max}_X\{CI(X)\} = CI\{\delta(X)\}: & \delta(X_i) = 1 \ \wedge \ \delta(X_j) = 0 \ \wedge \ i \neq j \end{cases}$$

*Proof*:

Reminding the definition of Lorenz Zonoid:

$$LZ(X) = LZ(Y) = 1 - \frac{1}{N} \cdot \left[1 + 2 \cdot \sum_{i=1}^{N-1} y_i\right]$$

Suppose $P(X) = U(X)$ then $y_i = y_{i-1} + p_i$. It follows that:

$$\sum_{i=1}^{N-1} y_i = y_1 + y_2 + \cdots + y_i + \cdots + y_{N-1}$$

$$= \frac{1}{N} + \overbrace{\left(\frac{1}{N} + \frac{1}{N}\right)}^{\mathbf{2}/\boldsymbol{N}} + \cdots + \overbrace{\left(\frac{1}{N} + \cdots + \frac{1}{N}\right)}^{\boldsymbol{i}/\boldsymbol{N}} + \cdots + \overbrace{\left(\frac{1}{N} + \cdots + \frac{1}{N}\right)}^{(\boldsymbol{N}-\mathbf{1})/\boldsymbol{N}}$$

$$= \frac{1}{N} \sum_{i=1}^{N-1} i = \frac{1}{N}\left(\sum_{i=1}^{N} i - N\right) = \frac{1}{N}\left[\frac{N(N+1)}{2} - N\right] = \frac{N+1}{2} - 1$$

It follows that:

$$\mathbf{2}\sum_{\boldsymbol{i=1}}^{\boldsymbol{N-1}}\boldsymbol{y_i} = \boldsymbol{N} - 1 \xrightarrow{yields} LZ\{U(X)\} = 1 - \frac{1}{N}[1 + N - 1] = 0 \xrightarrow{yields} CI(X) = {}^{LZ}/_{\widehat{LZ}} = 0$$

Suppose $P(X) = \delta(X)$ then $y_k = 0 \;\forall\; k \neq n \wedge y_N = 1$. It follows that:

$$\sum_{i=1}^{N-1} y_i = 0 \xrightarrow{yields} LZ\big(\delta(X)\big) = 1 - {}^{1}/_{N} = \frac{N-1}{N} \xrightarrow{yields} CI(X) = 1$$

Since $LZ(X)$ is a linear function, $CI(X)$ is strictly convex:

$$\frac{\partial LZ(Y)}{\partial y_i} = -\frac{2}{N} < 0\,.\;\forall\, y_i$$

It follows that CI(X) is bounded:

$$0 = CI\{U(X)\} < \; arg_X\{CI(X)\} < \; CI\{\delta(X)\} = 1 \qquad \boldsymbol{Q.E.D}$$

**Lemma A1.2:**

Let $LF(X) = \{y_1, \dots, y_N\}$ be the Lorenz curve of the random variable $X = \{x_1, \dots, x_N\}$. Then the set of feasible Lorenz curves is defined by the convex hull:

$$\boldsymbol{P} = \{2y_i \leq \; y_{i+1} + \; y_{i-1} \;\wedge\; y_i \;\in\; [0\;\;1)\colon\; i = \{\,1, \dots, N-1\} \;\wedge\; y_0 = 0 \;\wedge\; y_N = 1\}$$

*Proof*:

Since $LF(X)$ is a convex function, it follows that:

$$\frac{y_{i+1} - y_{i-1}}{x_{i+1} - x_{i-1}} \cdot (x_i - x_{i+1}) + \; y_{i+1} \geq \; y_i \qquad\qquad \forall\, x_{i-1} < \; x_i < \; x_{i+1} \;\; i = 1, \dots, N-1$$

In other words, the secant line of a convex function lies above the graph. Then substituting the following identities in the above equation:

$$\begin{cases} x_{i+1} - x_{i-1} = {}^{2}/_{N} \\ x_i - x_{i+1} = {}^{-1}/_{N} \end{cases} \rightarrow \; 2y_i \le y_{i+1} + y_{i-1} \quad y_i \qquad i = 1, \dots, N-1$$

Based on the definition of $LF(X)$:

$$y_i = y_{i-1} + p_i : \quad i = 1, \dots, N-1 \;\; \wedge \;\; y_0 = 0 \;\wedge\; p_i \in \; \mathbb{III}(P(X))$$

Therefore, given that $p_i \le p_j \;\; \forall\, i < j$ :

$$\begin{cases} \sum_{i=1}^{N} p_i = 1 \\ p_i \in [0 \;\; 1) \end{cases} \rightarrow \;\; \begin{cases} y_i \in [0\; 1) \;\; i \in \{1, \dots, N-1\} \\ y_N = 1 \end{cases}$$

***Q.E.D***

**Lemma A1.3:** Let LZ(X) be the Lorenz zonoid of the random variable $X = \{x_1, \ldots, x_N\}$. Then min/max $\{CI(X)\}$ = min/max $\sum_{i=1}^{N-1} y_i$.

*Proof*:

$$\min\{CI(X)\} = LZ(Y^*): \quad Y^* = \left\{ \begin{matrix} \arg min \\ Y \in P \end{matrix} \; LZ(Y) \right\}$$

$$\max\{CI(X)\} = LZ(Y^*): \quad Y^* = \left\{ \begin{matrix} \arg Max \\ Y \in P \end{matrix} \; LZ(Y) \right\}$$

It follows that:

$$\underset{Y \in P}{\arg Max} \; LZ(Y) = \underset{y \in P}{max}\{1 - \frac{1}{N}[1 + 2\sum_{i=1}^{N-1} y_i]\} = \underset{y \in P}{max}\{-\sum_{i=1}^{N-1} y_i\}$$

Since $\max[-f(x)] = \min[f(x)]$:

$$\begin{cases} min/\max CI(P(X)) \\ \sum_{i=1}^{N} p_i \\ p_i \in [0 \;\; 1] \end{cases} \Leftrightarrow \begin{cases} max/\min \sum_{i=1}^{N-1} y_i \\ 2y_i \le y_{i-1} + y_{i+1} \quad i = 1, \ldots, N-1 \\ y_i \in [0 \;\; 1) \;\wedge\; y_0 = 0 \;\wedge\; y_N = 1 \end{cases}$$

$\boldsymbol{Q.E.D}$

# A2: Divergence Index (DI)

In the statistical literature, for ordinal types of data, are known lots of indicators to measure the degree of the polarization phenomenon. Typically, many of the widely used measures of distributional variability are defined as a function of a reference point, which in some "sense" could be considered representative for the entire population. This function indicates how much all the values differ from the point that is considered "typical". Of all measures of variability, the variance is a well-known example that use the mean as a reference point. However, mean-based measures depend to the scale applied to the categories and are highly sensitive to outliers.

Then, is there another way to compare the dispersion of a distribution that does not depend on its location?

To address this challenge, we propose a new measure of polarization, the Divergence Index (DI). The definition of the DI is based on a new representation of probability measures, the Bilateral Cumulative Distribution function (**BCDF**), which derives from a generalization of the Cumulative Distribution Function (CDF). Basically, it is an extended CDF that can be easily obtained by folding its upper part, commonly known as Survival function or complementary CDF. Unlike the CDF, this functional has a finite constant area independently of the probability distribution and, therefore, more convenient for any distribution comparison. For the definition of the **BCDF** see A2.1. For the area invariance property see theorem A3.1.2.

Accordingly, the DI is a non-parametric measure that compares the BCDF of two data sets. More specifically, the idea is to compare the distribution shape with that of minimum dispersion, that is the singleton distribution. On this basis, we completely defined the shape of a probability distribution by its BCDF autocorrelation function (**BCDFA**). The main advantage of this representation is that it is invariant to the location of a distribution and therefore is only sensitive to the distribution shape. For example, any singleton distribution will have the same BCDFA curve. Similarly, distribution with same shape but different means and medians can be represented with a unique curve. In this way, to quantify the distance between a pdf and the singleton distribution, it is only a matter of choosing an appropriate metric, which would assign large values to distribution with more dispersed BCDFA than the singleton. For the definition of the **BCDFA** see A3.

The selection of a measure to compare probability distribution is not a trivial matter and usually depends on the objectives. In this work we propose the use of the Jensen Shannon Divergence (JSD) (Lin, 1991). Since the DI is based on the JSD we dub this new measure of polarization **DI-JSD**. This measure does not need to be normalized since is a bounded value in the unit interval.

For the definition, properties and computation see A4.

## A2.1 – The Bilateral Cumulative Distribution Function (BCDF)

Let $f(X) = P(X = X_i)$ be the probability mass function of the discrete random variable X = $\{X_1, \dots,\ X_i, \dots, X_n\}$ such that $X_i$ is a categorical ordinal variable of x (e.g. Age, IRSD score, etc.). It follows that:

$$X_i = \left\{x_{m_i} \le x \ \le \ x_{M_i}\colon \ \ f(X_i) = P\left(x_{m_i} \le x \ \le \ x_{M_i}\right) = P(X_i) = \ p_i \right\} \quad i \ \in \{1, \dots, n\}$$

Let Y = $\{Y_1, \dots\ , Y_n\ , Y_{n+1}\ , \dots\ , Y_{2n-1}\}$ be a discrete random variable such that:

$$Y_i = \begin{cases} X_i & i \ \in \{1, \dots, n\} \\ X \ \setminus \bigcup_{j=1}^{k} X_j = \bigcup_{j=1}^{n} X_j \setminus \bigcup_{j=1}^{k} X_j & i = n + k \ \ \wedge \ \ k \in \{1, \dots, n-1\} \\ \emptyset & otherwise \end{cases}$$

Then the *Bilateral Cumulative Distribution Function* **BCDF** of X is given by the following formula:

$$F(Y_i) \begin{cases} P\left(x \ \le \ x_{M_i}\right) = P\left(\bigcup_{j=1}^{i} X_i\right) = \sum_{X=X_1}^{Y_i} f(X) & i \ \in \{1, \dots, n\} \\ P\left(x_{m_{k+1}} \le x \ \le \ x_{M_n}\right) = P(X_n) - P\left(\bigcup_{j=1}^{k} X_j\right) = 1 - \sum_{X=X_1}^{Y_k} f(X) & i \ \in \{n+1, \dots, 2n-1\} \\ P\left(x < \ x_{m_1} \ \vee \ \ x > \ x_{M_n}\right) = P(\emptyset) = 0 & otherwise \end{cases}$$

It follows that the recursive definition is:

$$\boldsymbol{F(Y_i)} \stackrel{\text{def}}{=} \begin{cases} \boldsymbol{F(Y_{i-1}) + P(X_i)} & \boldsymbol{i} \ \in \{\boldsymbol{1}, \dots, \boldsymbol{n}\} \\ \boldsymbol{1 - F(Y_k)} \quad \boldsymbol{i} \in \{\boldsymbol{n+1}, \dots, \boldsymbol{2n-1}\} & \rightarrow \boldsymbol{k} \ \in \{\boldsymbol{1}, \dots, \boldsymbol{n-1}\} \\ \boldsymbol{0} & \boldsymbol{otherwise} \end{cases}$$

It is worth noting that: $F(Y_i) = CDF(X)$ for $i \ \in \{1, \dots, n\} \rightarrow \int f(X)dx = CDF(X)$ and $\int 1 - CDF(X)dx = E(X)$ . Basically, the BCDF= $CDF \ \cup CCDF$ Complementary CDF. Considering, therefore, the recursive definition and the integral relation we can compute the BCDF in terms of the convolution operator (see **Theorem A2.1**).
**Theorem A2.1**:

Let $F(Y)$ be the BCDF of the discrete random variable X [**see A2**] and P(X) = $\{p_1, \ldots, p_i, \ldots, p_N\}$ the corresponding probability mass function vector. Let $h(X) = \{1 : X = X_i\}$ be a n-dimensional unitary vector, then $F(Y_i) = p[X] * h[X] = \sum_{k=-\infty}^{+\infty} p[X_k] \cdot h[i-k]$.

*Proof*:

Since $h(i-k)$ is simply the folded sequence $h(-k)$ shifted to the right by $k$ units (if k is positive) or to the left (if k is negative), it follows:

$$F(Y_i) = Y[i] = \sum_{k=1}^{N} p[X_k] \cdot h[i+1-k] \qquad i \in \{1, \ldots, 2n-1\}$$

$$Y[1] = p[X_1] \cdot h[1] = p_1 \qquad i = 1$$

$$Y[2] = p[X_1] \cdot h[2] + p[X_2] \cdot h[1] = Y[1] + p_1 \qquad i = 2$$

$$\ldots$$

$$Y[N] = p[X_{n-1}] + p_N = Y[N-1] + p_N = 1 \qquad i = N$$

$$Y[N+1] = \sum_{k=1}^{N+1} x[k] \cdot h[N+2-k] \qquad i = N+1$$

$$= x[1] \cdot h[N+1] + x[2] \cdot h[N] + \cdots + x[10] \cdot h[2]$$

$$= \sum\nolimits_{i=2}^{N} p_i = 1 - Y[i]$$

$$\ldots$$

$$Y[2N-1] = 1 - Y[1] \qquad i = 2N+1$$

***Q.E.D***

**Theorem A2.2**:

Let $F(Y) = p[X] * h[X]$ be the BCDF of the discrete random variable $X = \{X_1, \dots, X_N\}$, and $h(X) = \{1: X = X_i\}$ a N-dimensional unitary vector. Then $\int_{-\infty}^{\infty} F(Y)dy = N$.

*Proof*:

Let A$[F(Y)]$ denote the area of BCDF:

$$A[F(Y)] = \sum_{i=-\infty}^{\infty} F(Y_i) = \sum_{i=1}^{2N-1} F(Y_i)$$

Then

$$A[F(Y)] = A\big[p[X] * h[X]\big] = A\big[p[X]\big] \cdot A\big[h[X]\big] = \sum_{k=1}^{N} p[X] \cdot \sum_{k=1}^{N} h[X] = N$$

This can be proved by:

$$A[F(Y)] = \sum_{i=1}^{2N-1} F(Y_i) = \sum_{i=1}^{2N-1} p[X_i] * h[X_i]$$

$$= \sum_{i=1}^{2N-1} \sum_{k=1}^{N} p[X_k] \cdot h[i+1-k]$$

$$= \sum_{k=1}^{N} p[X_k] \sum_{i=1}^{2N-1} h[i+1-k]$$

$$= \sum_{k=1}^{N} p[X_k] \sum_{i=1}^{2N-1} h[i+\tau] \qquad \tau = 1-k$$

Since the area $A[h[X]] = A[h[i+\tau]]$ (do not change with a shift in τ). It follows that:

$$A[F(Y)] = \sum_{k=1}^{N} p[X_k] \sum_{k=1}^{N} h[X_k] = N$$

***Q.E.D***

## A3: The Bilateral Cumulative Distributive Function Autocorrelation (BCDFA)

Let denote $\widehat{F}\,(Y)$ the folding sequence of $F(Y)$:

$$\hat{F}(Y_i) = \begin{cases} F(Y_N) & i = 1 \\ F(Y_{N-i+1}) & i = \{2, .., N\} \end{cases}$$

Then the autocorrelation function of $F(Y)$ is given by the following formula:

$$\boldsymbol{R_f}(\boldsymbol{W}) = r_f(w_i) = F(Y) * \hat{F}(Y) = \sum_{k=1}^{N} F[Y_k] \cdot \hat{F}[i + 1 - k] \qquad i \in \{1, \dots, \ 4N - 3\}$$

Since the maximum value of the autocorrelation function occurs when the sequence correlates perfectly with itself with zero delay and the symmetry property, we can use the following relations:

$$r_f(w_i) = \begin{cases} \sum_{k=1}^{N} F[Y_k] \cdot \hat{F}[i + 1 - k] & i \in \{1, \dots, \ 2N - 1\} \\ r_f(w_{4N-2-i}) & i \in \{2N, \dots, 4N - 3\} \end{cases}$$

It follows that the maximum occurs for $i = 2N - 1$.
Let denote $\delta(X)$ the singleton distribution of the random variable $X = \{X_1, \dots, X_N\}$ :

$$\delta(X) = \begin{cases} 1 \quad X_i & i \in \{1, \dots, N\} \\ 0 \quad X_j & j \neq i \end{cases}$$

Then the autocorrelation function of $\delta(X)$, is given by the following formula:

$$\boldsymbol{R_\delta}(\boldsymbol{W}) \stackrel{\text{def}}{=} r_\delta(w_i) = \begin{cases} 0 & i \in \{1, \dots, N - 1\} \\ i - N + 1 & i \in \{N, \dots, 2N - 1\} \\ r_f(w_{4N-2-i}) & i \in \{2N, \dots, 4N - 3\} \end{cases}$$

## A4: The Divergence Index - Jensen Shannon Divergence (DI-JSD)

Let $R_f(W)$ and $R_\delta(W)$ be the autocorrelation functions of the random variable $X = \{X_1, \dots, X_N\}$ and the impulse distribution $\delta(X)$ [see A3].

Based on the Theorem A2.2:

$$\int_{-\infty}^{\infty} R_f(w)\,\mathrm{dw} = \int_{-\infty}^{\infty} R_\delta(W)dw = A[F(Y)] \cdot A[F(Y)] = N^2$$

It follows that the normalized sequences are:

$\widehat{R_f}(W) = \frac{1}{N^2} \cdot R_f(W)$ and $\widehat{R_\delta}(W) = \frac{1}{N^2} \cdot R_\delta(W)$

For sake of simplicity let denote $P(W) = \widehat{R_f}(W)$ and $Q(W) = \widehat{R_\delta}(W)$. Then, based on the Jensen-Shannon Divergence (JSD), we propose the following divergence measure:

$$\boldsymbol{DI}(\boldsymbol{X}) \stackrel{\text{def}}{=} 2 \cdot JSD(P\,||Q) = D(P||M) + D(Q||M)$$

Where:

$M = \frac{1}{2}(P + Q) \to M(w_i) = \frac{1}{2}(P(w_i) + Q(w_i))$ $\qquad 0 \leq DI(X) < 1$

$$D(P||M) = \sum_{w=1}^{4N-3} P(w) \cdot \log_2 \frac{P(w)}{M(w)} = \sum_{w=1}^{4N-3} P(w) \cdot [\log_2 P(w) - \log_2 M(w)]$$

$$D(Q||M) = \sum_{w=1}^{4N-3} Q(w) \cdot \log_2 \frac{Q(w)}{M(w)} = \sum_{w=1}^{4N-3} Q(w) \cdot [\log_2 Q(w) - \log_2 M(w)]$$

Since $\lim_{w \to 0^+} w \cdot \log w = 0$ , we assume $\log 0$ = 0.

**Theorem A4.1:**

Let $R_f(W)$ be the autocorrelation function of the random variable $X = \{X_1, \dots, X_N\}$ and $h(X) = \{1: X = X_i\}$ a sequence of unit impulses spaced at unit interval. Let's denote $\sigma_{Rf}^2, \sigma_h^2$ $and$ $\sigma_x^2$ the variance of the autocorrelation function, the impulses sequence and the probability distribution respectively. Then:

$$\sigma_{Rf}^2 = k + 2\sigma_x^2 \quad : \quad k = \frac{N^2 - 1}{6}$$

*Proof*:

$$h(X_i) = 1 \qquad i = 1, \dots, N$$

$$\sigma_h^2 = E[X^2] - (E[X])^2 \qquad\qquad E[X] = \frac{1}{N}\sum_{i=1}^{N} X_i = \frac{1}{N}\sum_{i=1}^{N} i = \frac{N+1}{2}$$

$$\sigma_h^2 = \frac{(N+1)(2N+1)}{6} - \left(\frac{N+1}{2}\right)^2 \qquad\qquad E[X^2] = \frac{1}{N}\sum_{i=1}^{N} X_i = \frac{1}{N}\sum_{i=1}^{N} i^2 = \frac{(N+1)(2N+1)}{6}$$

$$\sigma_h^2 = \frac{N^2 - 1}{12}$$

It is known that the variance $\sigma_X^2$, which is a measure of spread of the probability distribution $p(x)$, is increased under convolution $h(x)$ according to the result: $\sigma_{p[X]*h[X]}^2 = \sigma_h^2 + \sigma_X^2$.

It follows that the BCDF of x (i.e. F(Y)) has variance: $\sigma_{F(Y)}^2 = \sigma_h^2 + \sigma_X^2$.

Similarly, the variance of the autocorrelation function is given by the formula:

$$\sigma_{Rf}^2 = \sigma_{F(Y)}^2 + \sigma_{F(Y)}^2 = 2\sigma_{F(Y)}^2 \rightarrow \sigma_{F(Y)}^2 = \frac{\sigma_{Rf}^2}{2}$$

It follows that:

$$\frac{\sigma_{Rf}^2}{2} = \sigma_h^2 + \sigma_X^2 \quad \rightarrow \quad \sigma_{Rf}^2 = 2\left[\left(\frac{N^2 - 1}{12}\right) + \sigma_X^2\right] \rightarrow \sigma_{Rf}^2 = k + 2\sigma_x^2 \quad k = \frac{N^2 - 1}{6}$$

***Q.E.D***

**Theorem A4.2:**

Let $R_f(W)$ be the autocorrelation function of the random variable $X = \{X_1, \ldots, X_N\}$. Then:

$$\max\{R_f(2N-1)\} = N \qquad : \delta(X) = \{p_i = 1 \ \wedge \ p_j = 0 \ \forall \ \ j \neq i\} \ i.e.singleton$$
$$\min\{R_f(2N-1)\} = \frac{N-1}{2} + 1 \ : \quad \rho(X) = \{p_1 = p_N = {}^1/_2 \ \wedge \ p_j = 0 \ \forall \ \ j \neq 1, N\}$$

*Proof*:

Based on theorem A2.2 and the definition of the BCDF (A2):

$$\int_{-\infty}^{\infty} R_f(w)dw = [F(Y)]^2 = N^2$$

For sake of simplicity, let's denote $F(Y_i) = x_i$. It follows that:

$$N^2 = \sum_{i=1}^{N} x_i^2 + 2x_1\left(\sum_{i=2}^{2N-1} x_i\right) + \cdots + 2x_j\left(\sum_{i=j+1}^{2N-1} x_i\right) + \cdots + 2x_N\left(\sum_{i=N+1}^{2N-1} x_i\right) + \cdots + 2x_{N-1}x_N$$

Substituting the following relations:

$$\sum_{i=1}^{2N-1} x_i = N \ \rightarrow \ \sum_{i=j+1}^{2N-1} x_i$$
$$= N - \sum_{i=1}^{j}(1 - x_j) \quad \wedge \ \ x_N = \sum_{i=1}^{N} p_i = 1 \quad \wedge \ \ x_i = 1 - x_k \begin{cases} i \in \{1, .., 2N-1\} \\ k \ \in \{1, \ldots, N-1\} \end{cases}$$

$$R_f(2N-1) = \sum_{i=1}^{N} x_i^2$$

It follows that:

$$R_f(2N-1) = 2\sum_{i=1}^{N-1} x_i^2 - 2\sum_{i=1}^{N-1} x_i + N$$

The autocorrelation function of the BCDF is therefore a conic. In order to find the maximum and minimum value of the autocorrelation function, it is convenient to reduce the conic in a canonical form:

$$\begin{cases} x_i = \gamma_i + c_i \\ x_i \leq x_j \quad j > i \end{cases}$$

$$R_f(2N-1) = 2\sum_{i=1}^{N-1}\gamma_i^2 + \sum_{i=1}^{N-1}\gamma_i(-2+4c_i) + \lambda \quad : \quad \lambda = 2\sum_{i=1}^{N-1}c_i^2 - 2\sum_{i=1}^{N-1}c_i + N$$

In order to eliminate the linear terms and operate the translation to the origin, we consider the following conditions:

$$-2+4c_i = 0 \rightarrow c_i = {}^{1}\!/_{2} \quad Therefore \;\; \lambda = \frac{N-1}{2} + 1$$

$$\begin{cases} max/\min f(\gamma) = 2\sum_{i=1}^{N-1}\gamma_i^2 + \frac{N-1}{2} + 1 \\ \gamma_i \leq \gamma_j \quad j > i \\ \gamma_i \in [0 \quad {}^{1}\!/_{2}] \end{cases} \Rightarrow \begin{cases} \max \; f(\gamma^*) = N \quad \wedge \; \gamma_i^* = {}^{1}\!/_{2} \;\; \forall\, i \\ \min f(\gamma^*) = \frac{N-1}{2} + 1 \;\; \wedge \;\; \gamma_i^* = 0 \;\; \forall i \end{cases}$$

Since $Y_i = \gamma_i + c_i$:

$$\max\{R_f(w)\} = N \qquad Y_i^* = \begin{cases} 1 & i = 1, \dots, N \\ 0 & otherwise \end{cases} \rightarrow \;\; \delta(X) = \begin{cases} 1 & i = 1 \\ 0 & otherwise \end{cases}$$

$$\min\{R_f(w)\} = \frac{N-1}{2} + 1 \quad Y_i^* = \begin{cases} {}^{1}\!/_{2} & i \neq N \\ 1 & i = N \end{cases} \rightarrow \;\; \rho(X) = \begin{cases} {}^{1}\!/_{2} & i = 1, N \\ 0 & otherwise \end{cases}$$

These two distributions correspond to the minimum and maximum variance of a bounded probability distribution:

$$\sigma^2(\delta(X)) = \;\; 0 \; < \;\; \sigma^2(P(X)) < \left(N - \frac{N+1}{2}\right)\left(\frac{N+1}{2} - 1\right) \; = \; \sigma^2(\rho(X))$$

It's worth noting that in the case N=2 $\rho(X) = U(X)$.

***Q.E.D***

**Lemma A4.2:**

Let DI(X) be the divergence index of the random variable $X = \{X_1, \dots, X_N\}$. Then the DI(X) is a bounded metric such that:

$\arg\underset{P(x)}{Max}$ {DI(X)} $=\rho(x) \quad \rho(x) = \{p_1 = p_N = \frac{1}{2} \;\; \wedge \;\; j \neq i, N\}$

*Proof:*

Let $R_{fp}$ and $\tau_0$ be the autocorrelation function of the BCDF and its central coordinate (i.e. $\tau_0 = 2N - 1$). Then we introduce the autocorrelation width as:

$$W_{Rfp} = \frac{\int_{-\infty}^{\infty} R_{fp}(w)}{R_f(\tau_0)} = \frac{N^2}{R_f(\tau_0)}$$

That is the area of the autocorrelation function divided by its central ordinate. Based on Theorem A 4.2, it follows that:

$\arg\underset{P(x)}{Max}\{W_{Rfp}\} = \delta(x) \qquad \arg\underset{P(x)}{min}\{W_{Rfp}\} = \rho(x)$

Then:

$W_{Rf}\big(\delta(x)\big) = N$

Since $R_\delta(w_i) = 0 \;\; i \in \{1, \dots, N-1\}$ then:

$$\int_{-\infty}^{\infty} R_\delta(w)dw = \int_{\tau_0 - W_{Rf}+1}^{\tau_0 + W_{Rf}-1} R_\delta(w) = \sum_{i=N}^{3N-2} r_\delta(w_i) = N^2$$

We therefore define as a measure of compactness for certain shapes of distribution:

$$S(N) = \frac{\sum_{i=N}^{3N-2} r_{fp}(w)}{N^2} = \begin{cases} \max\{S\} = 1 & \leftrightarrow P(X) = \delta(X) : \;\; N > 1 \\ \min\{S\} = \dfrac{3N+1}{4N} & \leftrightarrow P(X) = \rho(X): \end{cases}$$

The minimum value can be easily computed, considering the autocorrelation of the BCDF $\rho(x)$.

$$R_f\big(\rho(X)\big) = \begin{cases} \dfrac{i}{4} \;\; i \in \{1, \dots, N-1\} \\ \dfrac{1}{2} + \dfrac{i}{4} \quad i \in \{N, \dots, 2N-2\} \\ \dfrac{N+1}{2} \qquad i = 2N-1 \end{cases}$$

$$\sum_{i=N}^{3N-2} R_f(w_i) = 2\sum_{i=N}^{2N-2} r_f(w_i) + \frac{N+1}{2}$$

$$2\sum_{i=N}^{2N-2} r_f(w_i) = 2\sum_{i=N}^{2N-2}\left(\frac{1}{2}+\frac{i}{4}\right)$$

$$= \sum_{i=N}^{2N-2} 1 + \sum_{i=N}^{2N-2}\frac{i}{2} = (N-1) + \frac{(2N-2)(2N-1)}{4} - \frac{(N-1)N}{4}$$

$$\sum_{i=N}^{3N-2} R_f(w_i) = \frac{N(3N+1)}{4} \quad \Rightarrow \quad S_\rho = \frac{3N+1}{4N}$$

It follows that:

$$\lim_{N\to\infty} S_\rho(N) = \lim_{N\to\infty}\frac{3N+1}{4N} \cong \frac{3}{4} = 0.75 \quad \wedge \quad \lim_{N\to 1} S_\rho(N) = 1 \quad \to DI\big(\rho(x)\big) = 0$$

The Divergence index is therefore a bounded metric.

***Q.E.D***

**Lemma A4.3**

Let DI(X) be the divergence index of the random variable $X = \{X_1, \dots, X_N\}$. Then

$$UB\big(DI(X)\big) = \lim_{N\to\infty} DI_{\rho(x)}(N) \cong 0.34$$

*Proof:*

Let $\hat{R}_{\delta(x)}$ and $\hat{R}_{\rho(x)}$ be the autocorrelation function of the impulse and $\rho(x)$ distribution. For sake of simplicity let's denote:

$$\hat{R}_{\delta(x)} = P(w_i) = \begin{cases} 0 & i \in \{1, \dots, N-1\} \\ \dfrac{i-N+1}{N^2} & i \in \{N, \dots, 2N-2\} \\ \dfrac{1}{N} & i = 2N-1 \end{cases}$$

$$\hat{R}_{\rho(x)} = Q(w_i) = \begin{cases} \dfrac{i}{4N^2} & i \in \{1, \dots, N-1\} \\ \dfrac{1}{N^2}\left(\dfrac{1}{2} + \dfrac{i}{4}\right) & i \in \{N, \dots, 2N-2\} \\ \dfrac{1}{N}\left(\dfrac{N+1}{2}\right) & i = 2N-1 \end{cases}$$

Then $M(w_i) = \frac{1}{2}[P(w_i) + Q(w_i)]$

$$M(w_i) = \begin{cases} \dfrac{i}{8N^2} & i \in \{1, \dots, N-1\} \\ \dfrac{5i-4N+6}{8N^2} & i \in \{N, \dots, 2N-2\} \\ \dfrac{3N+1}{4N^2} & i = 2N-1 \end{cases}$$

It follows:

$$\sum_{i=1}^{N-1} DI(X_i) = \sum_{i=1}^{N-1} Q(X_i) \log_2\left[\frac{Q(X_i)}{M(X_i)}\right] = \sum_{i=1}^{N-1} \left(\frac{i}{4N^2}\right) \log_2 2 = \frac{N-1}{8N}$$

Since the autocorrelation is an even function, it follows:

$$\sum_{i=1}^{N-1} DI(X_i) + \sum_{i=3N-1}^{4N-3} DI(X_i) = \frac{N-1}{4N}$$

Then for $i = 2N - 1$

$$DI(X_{2N-1}) = D(P||M) + D(Q||M)$$

$$D(P||M) = P(w_{2N-1}) \cdot \log_2 \left[\frac{P(w_{2N-1})}{M(w_{2N-1})}\right] = \frac{1}{N}\log_2 \left[\frac{4N}{3N+1}\right]$$

$$D(Q||M) = Q(w_{2N-1}) \cdot \log_2 \left[\frac{Q(w_{2N-1})}{M(w_{2N-1})}\right] = \frac{N+1}{2N^2}\log_2 \left[\frac{2(N+1)}{3N+1}\right]$$

Let’s solve for $i = N, \dots, 2N - 2$

$$\sum_{i=N}^{2N-2} DI(X_i) = D(P||M) + D(Q||M)$$

$$D(P||M) = \sum_{i=N}^{2N-2} \left[\frac{i-N+1}{N^2}\right] \cdot \log_2 \left[\frac{8(i-N+1)}{5i-4N+6}\right]$$

$$= \sum_{i=N}^{2N-2} \left[\frac{i-N+1}{N^2}\right] \log_2 8 + \sum_{i=N}^{2N-2} \left[\frac{i-N+1}{N^2}\right] \log_2 \left(\frac{i-N+1}{5i-4N-6}\right)$$

$$\sum_{i=N}^{2N-2} \left[\frac{i-N+1}{N^2}\right] \log_2 8 = \frac{3(N-1)}{2}$$

$$\sum_{i=N}^{2N-2} \left[\frac{i-N+1}{N^2}\right] \log_2 \left(\frac{i-N+1}{5i-4N-6}\right)$$

$i = N$ $\quad \frac{1}{N^2}\log_2 \left(\frac{1}{N+6}\right)$

$i = N + 1$ $\quad \frac{2}{N^2}\log_2 \left(\frac{1}{N+11}\right)$

$i = N + 2$ $\quad \frac{3}{N^2}\log_2 \left(\frac{1}{N+16}\right)$

$$\sum_{i=N}^{2N-2}\left[\frac{i-N+1}{N^2}\right]\log_2\left(\frac{i-N+1}{5i-4N-6}\right)=\sum_{i=1}^{N-1}\left(\frac{i}{N^2}\right)\cdot\log_2\left(\frac{i}{5i+N+1}\right)$$

Since the logarithm function is convex on $I=\{1,\ldots,N-1\}$, then the following inequality (Jensen's discrete inequality) holds:

$$\sum_{i=1}^{N}p_i\cdot f(x_i)\geq f\left(\sum_{i=1}^{N-1}p_i\cdot x_i\right)\geq 0$$

An upper bound (depending on f and I only) of Jensen's difference:

$$\sum_{i=1}^{N-1}p_i\cdot f(x_i)\leq f\left(\sum_{i=1}^{N-1}p_i\cdot x_i\right)+f(a)+f(b)-2f\left(\frac{a+b}{2}\right)\ :\qquad a=1\,,\ \ b=N-1$$

Is given by S. Simic (Simic.S, 2009).

Indeed, the relation holds if

$$p\cdot f(a)+(1-p)\cdot f(b)-f(p\cdot a+(1-p)\cdot b)\leq f(a)+f(b)-2f\left(\frac{a+b}{2}\right)\quad\forall\, p\in[0\ \ 1]$$

In this case, can be easily shown that for $p=1/2$ the relation doesn't hold. Therefore, we propose a tight bound for the computation of the Divergence index.

$$\sum_{i=1}^{N-1}p_i\cdot f(x_i)\leq f\left(\sum_{i=1}^{N-1}p_i\cdot x_i\right)+p_1 f(a)+p_{N-1}f(b)-2f\left(\frac{a+b}{2}\right)\ :$$

It follows that:

$$f\left(\sum_{i=1}^{N-1}p_i\cdot x_i\right)=\left(\sum_{i=1}^{N-1}i\right)\cdot\log_2\left[\frac{\sum_{i=1}^{N-1}i}{\sum_{i=1}^{N-1}(5i+n+1)}\right]$$

$$\sum_{i=1}^{N-1}(N+5i+1)=(N-1)\left(\frac{7N+2}{2}\right)$$

$$f\left(\sum_{i=1}^{N-1}p_i\cdot x_i\right)=\frac{(N-1)N}{2}\cdot\log_2\left[\frac{N}{7N+2}\right]$$

$$f_p(a)=f_p(1)=\log_2\left(\frac{1}{N+6}\right)$$

$$f_p(b)=f_p(N-1)=\log_2\left(\frac{N-1}{6N-4}\right)$$

$$f_p\left(\frac{a+b}{2}\right)=f_p\left(\frac{N}{2}\right)=\frac{N}{2}\log_2\left(\frac{N}{7N+2}\right)$$

$$D(P||M) = \frac{1}{N^2}\left[f\left(\sum_{i=1}^{N-1} p_i \cdot x_i\right) + N \cdot \left(f_p(a) + f_p(b) - 2 \cdot f\left(\frac{N}{2}\right)\right)\right]$$

$$= \frac{(N-1)}{2N}\left[3 + \log_2\left(\frac{N}{7N+2}\right)\right] + \frac{1}{N} \cdot \left[f_p(a) + f_p(b) - 2 \cdot f\left(\frac{N}{2}\right)\right]$$

Similarly, we proceed in the computation of the second term:

$$D(Q||M) = \sum_{i=N}^{2N-2} Q(w_i) \cdot \log_2\left[\frac{Q(w_i)}{M(w_i)}\right]$$

$$= \sum_{i=N}^{2n-2} \frac{1}{N^2}\left(\frac{1}{2} + \frac{i}{4}\right) \cdot \log_2\left[\frac{1}{N^2}\left(\frac{1}{2} + \frac{i}{4}\right)\left[\frac{8N^2}{5i - 4N + 6}\right]\right]$$

$$= \sum_{i=N}^{2N-2} \frac{1}{N^2}\left(\frac{1}{2} + \frac{i}{4}\right)\log_2\left[\frac{2(2+i)}{5i - 4N + 6}\right]$$

$$= \sum_{i=N}^{2N-2} \left(\frac{2+i}{4N^2}\right)\log_2 2 + \sum_{i=N}^{2N-2} \left(\frac{2+i}{4N^2}\right)\log_2\left[\frac{2+i}{5i - 4N + 6}\right]$$

$$\sum_{i=N}^{2N-2} \left(\frac{2+i}{4N^2}\right) = \frac{(N-1)(3N+2)}{8N^2}$$

$$\sum_{i=N}^{2N-2} (2+i) \cdot \log_2\left(\frac{2+i}{5i - 4N + 6}\right)$$

$i = N$ $\quad (N+2)\log_2\left(\frac{2+N}{N+6}\right)$

$i = N+1$ $\quad (N+3)\log_2\left(\frac{3+N}{N+11}\right)$

$i = N+2$ $\quad (N+4)\log_2\left(\frac{4+N}{N+16}\right)$

$$\sum_{i=N}^{2N-2} (2+i) \cdot \log_2\left(\frac{2+i}{5i - 4N + 6}\right) = \sum_{i=1}^{N-1} (N+1+i)\log_2\left(\frac{N+1+i}{5i+N+1}\right)$$

It follows that:

$$f\left(\sum_{i=1}^{N-1} p_i \cdot x_i\right) = \left(\sum_{i=1}^{N-1}(N+1+i)\right) \cdot \log_2\left[\frac{\sum_{i=1}^{N-1}(N+1+i)}{\sum_{i=1}^{N-1}(5i+n+1)}\right]$$

$$\sum_{i=1}^{N-1}(N+1+i) = \frac{(N-1)(3N-2)}{2}$$

$$f\left(\sum_{i=1}^{N-1} p_i \cdot x_i\right) = \frac{(N-1)(3N-2)}{2} \cdot \log_2\left(\frac{3N+2}{7N+2}\right)$$

$$f_Q(a) = f_Q(1) = (N+2) \cdot \log_2\left(\frac{N+2}{N+6}\right)$$

$$f_Q(b) = f_Q(N-1) = (2N) \cdot \log_2\left(\frac{2N}{6N-4}\right)$$

$$f_Q\left(\frac{a+b}{2}\right) = f_Q\left(\frac{N}{2}\right) = \left(\frac{3N+2}{2}\right) \cdot \log_2\left(\frac{3N+2}{7N+2}\right)$$

DI

$$D(Q||M) = \frac{(N-1)(3N-2)}{8N^2} \cdot \left[1 + \log_2\left(\frac{3N+2}{7N+2}\right)\right] + \frac{1}{4N} \cdot \left[f_Q(a) + f_Q(b) - 2f_Q\left(\frac{a+b}{2}\right)\right]$$

Finally, the formula for the computation of the Divergence Index of the distribution $\rho(x)$ is given below:

$$DI(\rho(x)) = \left(\frac{N-1}{4N}\right) + DI(X_{2N-1}) + D(P||M) + D(Q||M)$$

To improve the Bound we multiply the two last terms for ¼.

$$DI(\rho(x)) = \left(\frac{N-1}{4N}\right) + DI(X_{2N-1}) + \frac{1}{4}[D(P||M) + D(Q||M)]$$

It follows that the Upper Bound is given by the following formula:

$$\lim_{N\to\infty} DI(\rho(x)) = \lim_{N\to\infty}\left(\frac{N-1}{4N}\right) + \lim_{N\to\infty} DI(X_{2N-1}) + \lim_{N\to\infty}\frac{1}{4}[D(P||M) + D(Q||M)] \cong 0.34$$

$$\lim_{N\to\infty}\left(\frac{N-1}{4N}\right) = 0.25 \qquad \lim_{N\to\infty} DI(X_{2N-1}) = 0$$

$$\lim_{N\to\infty}\frac{1}{4} \cdot D(P||M) = \frac{1}{4} \cdot \left\{\frac{1}{2}\left[3 + \log_2\left(\frac{1}{7}\right)\right] + \log_2\left(\frac{1}{6}\right) - \log_2\left(\frac{1}{7}\right)\right\} \cong 0.08$$

$$\lim_{N\to\infty}\frac{1}{4} \cdot D(Q||M) = \frac{1}{4} \cdot \left\{\frac{3}{8}\left[1 + \log_2\left(\frac{3}{7}\right)\right] + \frac{1}{2}\log_2\left(\frac{1}{3}\right) - \frac{3}{4}\log_2\left(\frac{3}{7}\right)\right\} \cong 0.01$$

*Q.E.D*

**Lemma A4.4:**

Let $R_U(w)$ be the autocorrelation function of the uniform distribution $U(x) = \left\{x_i = \frac{1}{N} \ \forall\, i = 1, \dots, N\right\}$. Then the compactness of this function is given by the following formula:

$$S_U(N) = \frac{\sum_{i=N}^{3N-2} r_{fu}(w_i)}{N^2} = \frac{11N^3 - 2N^2 + N + 2}{12N^3}$$

*Proof:*

Let $F_U(Y)$ be the BCDF of $U(X)$. Then:

$$F_U(Y) = \begin{cases} \dfrac{Y_i}{N} & i \in \{1, \dots, N\} \\ 1 - \dfrac{Y_i - N}{N} & i \in \{N+1, \dots, 2N-1\} \end{cases}$$

Then the autocorrelation function $R_u(w) = F_U(Y) * F_U(Y)$, is given by the following formula:

$$f_u(t) = \begin{cases} f_1(t) = \displaystyle\sum_{\tau=1}^{t+N} \left(\frac{\tau}{N}\right)\left(1 - \frac{\tau - N}{N} + \frac{t+1-N}{N}\right) : 1 \le t + N \le N \to t \in [1-N \quad 0] \\ \qquad N < t + N \le 2N - 1 \ \to \ t \in [1 \ \ N-1] \\ f_2(t) = \begin{cases} f_{2,1}(t) = \displaystyle\sum_{\tau=1}^{t} \left(\frac{\tau}{N}\right)\left(1 + \frac{\tau - N}{N} - \frac{1+t-N}{N}\right) \\ f_{2,2}(t) = \displaystyle\sum_{\tau=t+1}^{N} \left(\frac{\tau}{N}\right)\left(1 - \frac{\tau - N}{N} + \frac{1+t-N}{N}\right) \\ f_{2,3}(t) = \displaystyle\sum_{\tau=N+1}^{t+N} \left(1 - \frac{\tau - N}{N}\right)\left(1 - \frac{\tau - N}{N} + \frac{t+1-N}{N}\right) \end{cases} \end{cases}$$

Since the autocorrelation function is an even function, we proceed in the computation of the first 2N-1 values.

$$\sum_{\tau=1}^{t+N} \left(\frac{\tau}{N}\right)\left(1 - \frac{\tau - N}{N} + \frac{1+t-N}{N}\right) = \frac{(t+N)(t+N+1)(t+N+2)}{6N^2}$$

It follows that:

$$\sum_{i=1}^{1-N} r_{fu}(w) = \sum_{t=1-N}^{0} \frac{(t+N)(t+N+1)(t+N+2)}{6N^2}$$

$$= \sum_{t=1}^{N-1} \frac{(-t+N)(-t+N+1)(-t+N+2)}{6N^2} + f_1(0) \quad : \quad f_1(0) = \frac{(N+1)(N+2)}{6N}$$

$$= \frac{N^3+6N^2+11N+6}{24N} = f_1(N)$$

Now let's solve the second interval:

$$\sum_{\tau=1}^{t} \left(\frac{\tau}{N}\right)\left(1+\frac{\tau-N}{N}-\frac{1+t-N}{N}\right) = \frac{t(t+1)(3N-t-2)}{6N^2} = f_{2,1}(t)$$

$$f_{2,1}(N) = \sum_{t=1}^{N-1} \frac{t(t+1)(3N-t-2)}{6N^2} = \frac{3N^3-2N^2-3N+2}{24N}$$

$$\sum_{\tau=t+1}^{N} \left(\frac{\tau}{N}\right)\left(1-\frac{\tau-N}{N}+\frac{1+t-N}{N}\right) = \frac{N^2+3N+2}{6N} - \frac{t(t^2+3t+3Nt-3N^2+2)}{6N^2} = f_{2,2}(t)$$

$$f_{2,2}(N) = \sum_{t=1}^{N-1} \frac{N^2+3N+2}{6N} - -\frac{t(t^2+3t+3Nt-3N^2+2)}{6N^2} = \frac{5N^3+6N^2-5N-6}{24N}$$

$$\sum_{\tau=N+1}^{t+N} \left(1-\frac{\tau-N}{N}\right)\left(1-\frac{\tau-N}{N}+\frac{t+1-N}{N}\right) =$$

$$= \frac{7N^2+9N+2}{6N} + \frac{t(4t+5N+5)}{2N} - \frac{(t+N)(t+N+1)(t+7N+2)}{6N^2} = f_{2,3}(t)$$

$$\sum_{t=1}^{\boldsymbol{N-1}} f_{2,3}(t) = \frac{(N-1)(3N^2+N-2)}{24N}$$

Based on the previous results, it follows:

$$\sum_{i=N}^{2N-1} r_{fu}(w) = f_{2,1}(N) + f_{2,2}(N) + f_{2,3}(N) + f_1(0) = \frac{11N^3 + 6N^2 + N + 6}{24N}$$

It follows:

$$\sum_{i=N}^{3N-2} r_{fu}(w) = 2\sum_{i=N}^{2N-2} r_{fu}(w) + r_{fu}(2N-1)$$

Then the maximum of the autocorrelation function of the uniform distribution is:

$$r_{fu}(2N-1) = f_{2,1}(N-1) + f_{2,2}(N-1) + f_{2,3}(N-1) \qquad t = N-1$$

$$f_{2,1}(N-1) = f_{2,3}(N-1) = \frac{(N-1)(2N-1)}{6} \qquad f_{2,2}(N-1) = 1$$

$$r_{fu}(2N-1) = \frac{2N^2+1}{3N}$$

It follows:

$$\sum_{i=N}^{2N-2} r_{fu}(w) = \sum_{i=N}^{2N-1} r_{fu}(w) - r_{fu}(2N-1) = \frac{11N^3 - 10N^2 + N - 2}{24N}$$

Then, we can finally compute:

$$\sum_{i=N}^{3N-2} r_{fu}(w) = \frac{11N^3 - 10N^2 + N - 2}{24N} + \frac{2N^2+1}{3N} = \frac{11N^3 - 2N^2 + N + 2}{12N}$$

$$S_U(N) = \frac{\sum_{i=N}^{3N-2} r_{fu}(w_i)}{N^2} = \frac{11N^3 - 2N^2 + N + 2}{12N^3}$$

As expected

$$\lim_{N \to 1} S_U(N) = 1 \quad \wedge \quad S_U(2) = S_\rho(2) = \frac{7}{8} = 0.875$$

$$\lim_{N \to \infty} S_U(N) = \frac{11}{12} \cong 0.916 > 0.75 = \frac{3}{4} = \lim_{N \to \infty} S_\rho(N)$$

***Q.E.D***

**Lemma A4.5**

Let $DI(X)_U$ be the divergence index of the uniform distribution $X = \{X_1, \dots, X_N\}$. Then

$$\lim_{N\to\infty} DI(X)_U \cong 0.11$$

*Proof:*

Let $\hat{R}_{\delta(x)}$ and $\hat{R}_{U(x)}$ be the autocorrelation function of the impulse and uniform distribution. For sake of simplicity let's denote:

$$\hat{R}_{\delta(x)} = P(w_t) = \begin{cases} 0 & t \in \{1-N, \dots, -1\} \\ \dfrac{t+1}{N^2} & t \in \{0, \dots, N-1\} \end{cases}$$

$$\hat{R}_{U(x)} = Q(w_t) = \begin{cases} \dfrac{(t+N)(t+N+1)(t+N+2)}{6N^4} & t \in \{1-N, \dots, 0\} \\ \dfrac{N^2+3N+2}{N^3} + \dfrac{t(2+t+N)}{2N^3} - \dfrac{3t(t^2+3t+2)}{6N^4} & t \in \{0, \dots, N-1\} \end{cases}$$

Then $M(w_i) = \frac{1}{2}[P(w_i) + Q(w_i)]$

$$M(w_t) = \begin{cases} \dfrac{Q(w_t)}{2} & t \in \{1-N, \dots, -1\} \\ \dfrac{6N^2(t+1) + N(N^2+3N+2) + 3Nt(2+t+N) - 3t(t^2+3t+2)}{12N^4} & t \in \{0, \dots, N-1\} \end{cases}$$

It follows that:

$$DI_U(N) = \sum_{t=1-N}^{N-1} DI(X_t) = 2\left[\sum_{t=1-N}^{-1} DI(X_t) + DI(X_0) + \sum_{t=1}^{N-2} DI(X_t)\right] + DI(X_{N-1})$$

$$\sum_{t=1-N}^{-1} DI(X_t) = \sum_{t=1-N}^{-1} Q(w_t) = \frac{1}{N^2}\left[\frac{N^3+6N^2+11N+6}{24N} - \frac{N^2+3N+2}{6N}\right]$$

Then for $t = 0$

$$DI(X_0) = D_0(P||M) + D_0(Q||M)$$

$$D_0(P||M) = P(w_0) \cdot \log_2\left(\frac{P(w_0)}{M(w_0)}\right) = \frac{1}{N^2}\log_2\left(\frac{12N}{N^2 + 9N + 2}\right)$$

$$D_0(Q||M) = Q(w_0) \cdot \log_2\left(\frac{Q(w_0)}{M(w_0)}\right) = \frac{N^2 + 3N + 2}{6N^3}\log_2\left(\frac{2(N^2 + 3N + 2)}{N^2 + 9N + 2}\right)$$

$t = X_{N-1}$

$$DI(X_{N-1}) = D_{N-1}(P||M) + D_{N-1}(Q||M)$$

$$D_{N-1}(P||M) = P(w_{N-1}) \cdot \log_2\left(\frac{P(w_{N-1})}{M(w_{N-1})}\right) = \frac{1}{N}\log_2\left(\frac{6N^2}{5N^2 + 1}\right)$$

$$D_{N-1}(Q||M) = Q(w_{N-1}) \cdot \log_2\left(\frac{Q(w_{N-1})}{M(w_{N-1})}\right) = \frac{2(2N^2 + 1)}{6N^3}\log_2\left(\frac{2(2N^2 + 1)}{5N^2 + 1}\right)$$

$t \in \{1, \dots, N-2\}$

$$\sum_{t=1}^{N-2} DI(X_t) = \sum_{t=1}^{N-2} D(P||M) + \sum_{t=1}^{N-2} D(Q||M)$$

$$\sum_{t=1}^{N-2} DI(X_t) = \sum_{t=1}^{N-2}\left(\frac{t+1}{N^2}\right)\log_2\left(\frac{12N^2(t+1)}{A+B}\right)$$

$A = N(N^2 + 3N + 2) + 3Nt(2 + t + N) - 3t(t^2 + 3t + 2) \qquad B = 6N^2(t+1)$

Based on the Jensen's inequality, it follows that:

$$\sum_{t=1}^{N-2}\left(\frac{t+1}{N^2}\right)\log_2\left(\frac{12N^2(t+1)}{A+B}\right)$$

$$> \sum_{t=1}^{N-2}\left(\frac{t+1}{N^2}\right)\log_2\frac{\sum(12N^2(t+1))}{\sum A + \sum B}$$

$$= \frac{N^2 - N - 2}{2N^2}\log_2\frac{72N^3 - 72N^2 - 144}{69N^3 - 78N^2 - 105N - 30}$$

$\sum_{t=1}^{N-2}(t+1) = \frac{N^2-N-2}{2} \quad \sum_{t=1}^{N-2} A = \frac{33N^4-42N^3-33N^2-30N}{12} \quad \sum_{t=1}^{N-2} B = 3N^4 - 3N^3 - 6N^2$

$$D(Q||M) = \sum_{t=1}^{N-2} Q(w_t) \log_2 \left[\frac{Q(w_t)}{M(w_t)}\right]$$

$$Q(w_t) = \frac{A}{6N^4} \qquad M(w_t) = \frac{1}{2}\left[\frac{A+B}{6N^4}\right] \qquad \frac{Q(w_t)}{M(w_t)} = \frac{2A}{A+B}$$

$$\sum_{t=1}^{N-2} \frac{A}{6N^4} \log_2 \left[\frac{2A}{A+B}\right] = \frac{1}{6N^4}\left[\sum_{t=1}^{N-2} A + \sum_{t=1}^{N-2} A \log_2 \left(\frac{A}{A+B}\right)\right]$$

Based on the Jensen inequality:

$$\begin{aligned} &\sum_{t=1}^{N-2} A \log_2 \left(\frac{A}{A+B}\right) \\ &\qquad > \sum_{t=1}^{N-2} A \log_2 \left(\frac{\sum A}{\sum A + \sum B}\right) \\ &\qquad = \frac{33N^4 - 42N^3 - 33N^2 - 30N}{12} \log_2 \left(\frac{33N^3 - 42N^2 - 33N - 30}{66N^3 - 78N^2 - 105N - 30}\right) \end{aligned}$$

We finally compute the Upper Bound for the uniform distribution:

$$\lim_{N\to\infty} 2\left[\sum_{t=1-N}^{-1} DI(X_t) + DI(X_0) + \sum_{t=1}^{N-2} DI(X_t)\right] + DI(X_{N-1}) \cong 0.11$$

$$\lim_{N\to\infty} \sum_{t=1-N}^{-1} DI(X_t) = \frac{1}{24} \qquad \lim_{N\to\infty} DI(X_0) = \lim_{N\to\infty} DI(X_{N-1}) = 0$$

$$\lim_{N\to\infty} \sum_{t=1}^{N-2} DI(X_t) = \lim_{N\to\infty} D(P||M) + \lim_{N\to\infty} D(Q||M) = \frac{1}{2}\log_2\left(\frac{72}{69}\right) + \frac{33}{72}\left(1 + \log_2 \frac{33}{69}\right)$$

*Q.E.D*

**Lemma A4.6:**

Let $S_p(N)$ and $S_u(N)$ be the compactness measure of the probability distribution P(X) and U(X). Let's denote $\sigma_p^2$ and $\sigma_u^2$ the variance of $P(X)$ and $U(X)$, then:

$$\frac{S_p(N)}{S_u(N)} = \begin{cases} > 1 \quad \leftrightarrow \ \sigma_p^2 < \sigma_u^2 \qquad \rightarrow DI(p) < DI(u) \ \ Low - \ Variance \\ = 1 \ \ \rightarrow P(X) = U(X) \quad \rightarrow DI(p) = DI(u) \\ < 1 \quad \leftrightarrow \ \sigma_p^2 > \sigma_u^2 \qquad \rightarrow DI(p) > DI(u) \quad High - Variance \end{cases}$$

*Proof*

Let $\sigma_{Rfp}^2$be the variance of the autocorrelation function:

$$\sigma_{Rfp}^2 = E_p[x^2] - (E[x])^2 \ \ : \quad E[x] = 2N - 1 = \mu$$

Since the autocorrelation function is an even function:

$$\sigma_{Rfp}^2 = \frac{1}{N^2}\sum_{i=1}^{4N-3} r_{fp}(w_i)\cdot i^2 - \mu^2 = \sum_{i=-2N+2}^{2N-2} \omega_i \cdot i^2 \quad : \qquad \omega_i = \frac{r_{fp}(\omega_i)}{N^2}$$

In other words, the variance of the distribution is basically the moment of inertia or second moment of a function $f(x)$ about the origin. Therefore, in the analysis of variance, we partition the sum of squares in terms of the compactness measure interval of the autocorrelation function (i.e. the compact support of $\delta(x)$).

$$\begin{cases} \displaystyle\sum_{i=-N+1}^{n-1} \omega_i \cdot i^2 + 2\sum_{i=N}^{2N-2} \omega_i \cdot i^2 = \sigma_{rfp}^2 \\ \displaystyle\sum_{i=-N+1}^{N-1} r_{fp}(\omega_i) + 2\sum_{i=N}^{2N-2} r_{fp}(\omega_i) = N^2 \end{cases}$$

Dividing for $N^2$we express the system of equations in terms of the compactness measure:

$$\begin{cases} P + Q = \dfrac{\sigma_{Rfp}^2}{N^2} \\ S_p(N) + R = 1 \end{cases} \quad P = \frac{1}{N^2}\sum_{i=-N+1}^{N-1} \omega_i \cdot i^2 \quad Q = \frac{2}{N^2}\sum_{i=N}^{2N-2} \omega_i \cdot i^2 \qquad R = \frac{2}{N^2}\sum_{i=N}^{2N-2} r_{fp}(\omega_i)$$

Then

$$S_p(N) + P = \left(1 + \frac{\sigma^2}{N^2}\right) - R - Q$$

$$S_p(\boldsymbol{N}) = 1 \ \rightarrow R = 0$$

$$\rightarrow \sum_{i=N}^{2N-2} r_{fp}(\omega_i) = 0 \ \rightarrow \ r_{fp}(\omega_i) = 0 \ \rightarrow \ r_{fp}(\omega_i) = 0 : i = N, \ldots, 2N - 2$$

Then it follows that:

$$\sum_{i=N}^{2N-2} \omega_i \cdot i^2 = 0 \quad \rightarrow \quad Q = 0$$

Therefore arg $\underset{P(x)}{min}\{\sigma_x^2\} = arg \underset{P(x)}{min}\{Q\} = \delta(x) \rightarrow \sigma_\delta^2 = 0$

Assume $S_p(N) > S_u(N)$, then:

$$1 - R_p > 1 - R_u \ \rightarrow \ R_p < R_u \ \rightarrow \ Q_p < Q_u$$

It follows:

$$Q_p < Q_u \rightarrow \ \frac{\sigma_{Rfp}^2}{N^2} - P_p < \frac{\sigma_{Rfu}^2}{N^2} - P_u$$

$$\frac{Q_p}{\sigma_{Rfp}^2} < \frac{Q_u}{\sigma_{Rfu}^2} \rightarrow \frac{Q_p}{Q_u} < \frac{\sigma_{Rfp}^2}{\sigma_{Rfu}^2}$$

Since $Q_p < Q_u$ then $\sigma_{Rfp}^2 < \sigma_{Rfu}^2$

*Q.E.D*

**LemmaA4.7**

Let $I_k$ be a closed bounded interval over $X=\{X_1,\dots,X_N\}$ such that $|I_k|=k, k\in[2\quad N-1]$. Then the autocorrelation function of the distribution $U^k(x)=\{x_i=\frac{1}{k}:\ x_i\in I_k\}$ is given by the following formula:

$$R_U^k(t)=\begin{cases} 0 & t\in[2-2N\quad 1-k-N] \\ \dfrac{(t+N+k)(t+N+k+1)(t+N+k-1)}{6k^2} & t\in[2-k-N\quad -N] \\ \dfrac{(N+t-k+1)(N+t-k-1)(-t-N+k)}{6k^2}+(t+N) & t\in[1-N\quad k-N-1] \\ t+N & t\in[k-N\quad 1-k] \\ \dfrac{-t^3+t(6k+1)-2k(k^2-3kN-1)}{6k^2} & t\in[2-k\quad 0] \end{cases}$$

*Proof:*

Let $F_U^k(\tau)$ be the BCDF of $U^k(X)$. Then:

$$F_U^k(\tau)=\begin{cases} \dfrac{\tau}{k}+1 & \tau\in[1-k\quad 0] \\ 1 & \tau\in[1\quad N-k] \\ 1-\dfrac{\tau+k-N}{k} & \tau\in[N-k+1\quad N] \\ 0 & Otherwise \end{cases}$$

Then the Autocorrelation function $R_U^k(t)=F_U^k(\tau)*F_U^k(\tau)$ is given by the following formula:

$$1-k\le t+N\le 0 \ \rightarrow\ \boldsymbol{t\in[1-k-N\quad -N]}$$

$$\sum_{\boldsymbol{\tau=1-k}}^{\boldsymbol{t+N}}\left(\frac{\tau}{k}+1\right)\left(1-\frac{\tau+k-N}{k}+\frac{t}{k}\right)=$$

$$=-\frac{1}{k^2}\sum_{\tau=1-k}^{t+N}\tau^2+\left(\frac{N+t}{k^2}-\frac{1}{k}\right)\sum_{\tau=1-k}^{t+N}\tau+\left(\frac{N+t}{k}\right)\sum_{\tau=1-k}^{t+N}1$$

$$\sum_{\tau=1-k}^{t+N} 1 = t + N + k$$

$$\sum_{\tau=1-k}^{t+N} \tau = \sum_{\tau=1}^{t+N+k} \tau - \sum_{\tau=1}^{t+N+k} k = \frac{(t+N+k)(t+N+k+1)}{2} - (t+N+K)k$$

$$\sum_{\tau=1-k}^{t+N} \tau^2 = \sum_{\tau=1}^{t+N+k} \tau^2 - 2k \sum_{\tau=1}^{t+N+k} \tau + \sum_{\tau=1}^{t+N+k} k^2$$

$$\sum_{\tau=1}^{t+N+k} \tau^2 = \frac{(t+N+k)(t+N+k+1)(2t+2N+2k+1)}{6}$$

$$\sum_{\tau=1}^{t+N+k} \tau = \frac{(t+N+k)(t+N+k+1)}{2}$$

$$\sum_{\tau=1}^{t+N+k} k^2 = (t+N+k)k^2$$

$$1 - k \leq t + N - k < 0 \;\; \rightarrow \;\; \boldsymbol{t} \in [\mathbf{1} - \boldsymbol{N} \quad \boldsymbol{k} - \boldsymbol{N} - \mathbf{1}]$$

$$A: \begin{cases} \displaystyle\sum_{\tau=1-k}^{t+N-k} \left(\frac{\tau}{k} + 1\right)(1) & \boldsymbol{A}(\mathbf{1}) \\ \displaystyle\sum_{\tau=t+N-k+1}^{0} \left(\frac{\tau}{k} + 1\right)\left(1 - \frac{\tau + k - N}{k} + \frac{t}{k}\right) & \boldsymbol{A}(\mathbf{2}) \\ \displaystyle\sum_{\tau=1}^{t+N} (1)\left(1 - \frac{\tau + k - N}{k} + \frac{t}{k}\right) & \boldsymbol{A}(\mathbf{3}) \end{cases}$$

$$\boldsymbol{A}(\mathbf{1}) = \frac{\mathbf{1}}{\boldsymbol{k}} \sum_{\tau=1-k}^{t+N-k} \tau + \sum_{\tau=1-k}^{t+N-k} 1$$

$$\sum_{\tau=1-k}^{t+N-k} 1 = \sum_{\tau=1}^{t+N} 1 = t + N$$

$$\sum_{\tau=1-k}^{t+N-k} \tau = \sum_{\tau=1}^{t+N} \tau - \sum_{\tau=1}^{t+N} k = \frac{(t+N)(t+N+1)}{2} - (t+N)k$$

$$\boldsymbol{A}(\mathbf{2}) = -\frac{\mathbf{1}}{\boldsymbol{k}^{\mathbf{2}}} \sum_{\tau=t+N-k+1}^{0} \tau^2 + \left(\frac{N+t}{k^2} - \frac{1}{k}\right) \sum_{\tau=t+N-k+1}^{0} \tau + \left(\frac{N+t}{k}\right) \sum_{\tau=t+N-k+1}^{0} 1$$

$$\sum_{\tau=t+N-k+1}^{0} 1 = \sum_{\tau=0}^{-(t+N-k+1)} 1 = -t - N + k$$

$$\sum_{\tau=t+N-k+1}^{0} \tau = - \sum_{\tau=0}^{-(t+N-k+1)} \tau = -\frac{(-t-N+k-1)(-t-N+k)}{2}$$

$$\sum_{\tau=t+N-k+1}^{0} \tau^2 = \sum_{\tau=1}^{-(t+N-k+1)} \tau^2 = \frac{(-t-N+k-1)(-t-N+k)(2(-t-N+k-1)+1)}{6}$$

$$\boldsymbol{A}(\mathbf{3}) = -\frac{1}{\boldsymbol{k}} \sum_{\tau=1}^{t+N} \tau + \left(\frac{N+t}{k}\right) \sum_{\tau=1}^{t+N} 1$$

$$\sum_{\boldsymbol{\tau}=\mathbf{1}}^{\boldsymbol{t}+\boldsymbol{N}} \boldsymbol{\tau} = \frac{(t+N)(t+N+1)}{2} \qquad \sum_{\boldsymbol{\tau}=\mathbf{1}}^{\boldsymbol{t}+\boldsymbol{N}} \mathbf{1} = t + N$$

$$0 \leq t + N - k \ \wedge \ t < 1 - k \ \rightarrow \ \boldsymbol{t} \in [\boldsymbol{k} - \boldsymbol{N} \quad \mathbf{1} - \boldsymbol{k}]$$

$$\boldsymbol{B}: \begin{cases} \displaystyle\sum_{\boldsymbol{\tau}=\mathbf{1}-\boldsymbol{k}}^{-\mathbf{1}} \left(\frac{\tau}{k} + 1\right)(\mathbf{1}) & \boldsymbol{B}(\mathbf{1}) \\ \displaystyle\sum_{\tau=0}^{t+N-k} (1)(1) & \boldsymbol{B}(\mathbf{2}) \\ \displaystyle\sum_{\boldsymbol{\tau}=\boldsymbol{t}+\boldsymbol{N}-\boldsymbol{k}+\mathbf{1}}^{\boldsymbol{t}+\boldsymbol{N}} (\mathbf{1}) \left(1 - \frac{\tau + k - N}{k} + \frac{t}{k}\right) & \boldsymbol{B}(\mathbf{3}) \end{cases}$$

$$\boldsymbol{B(1)}: \frac{1}{k}\sum_{\tau=1-k}^{-1}\tau + \sum_{\tau=1-k}^{-1}1$$

$$\sum_{\tau=1-k}^{-1}1 = \sum_{\tau=1}^{k-1}1 = k-1 \qquad\qquad \sum_{\tau=1-k}^{-1}\tau = -\sum_{\tau=1}^{k-1}\tau = -\frac{(k-1)k}{2}$$

$$\boldsymbol{B(2)}: \sum_{\tau=0}^{t+N-k}(1) = 1 + \sum_{\tau=1}^{t+N-k}1 = 1 + t + N - k$$

$$\boldsymbol{B(3)}: -\frac{1}{k}\sum_{\tau=t+N-k+1}^{t+N}\tau + \left(\frac{N+t}{k}\right)\sum_{\tau=t+N-k+1}^{t+N}1$$

$$\sum_{\tau=t+N-k+1}^{t+N}1 = \sum_{\tau=0}^{t+N}1 - \sum_{\tau=0}^{t+N-k}1 = k$$

$$\sum_{\tau=t+N-k+1}^{t+N}\tau = \sum_{\tau=1}^{t+N}\tau - \sum_{\tau=1}^{t+N-k}\tau = \frac{(t+N)(t+N+1)}{2} - \frac{(t+N-k)(t+N-k+1)}{2}$$

$$\boldsymbol{t \in [2-k \quad 0]}$$

$$\boldsymbol{C}: \begin{cases} \sum_{\boldsymbol{\tau=1-k}}^{\boldsymbol{t}}\left(\frac{\tau}{k}+1\right)\left(1+\frac{\tau}{k}-\frac{t}{k}\right) & \boldsymbol{C(1)} \\ \sum_{\tau=t+1}^{t+N-k}(1)(1) & \boldsymbol{C(2)} \\ \sum_{\boldsymbol{\tau=t+N-k+1}}^{\boldsymbol{t+N}}\left(1-\frac{\tau+k-N}{k}\right)\left(1-\frac{\tau+k-N}{k}+\frac{t}{k}\right) & \boldsymbol{C(3)} \end{cases}$$

$$\boldsymbol{C(1)}: \frac{1}{k^2}\sum_{\tau=1-k}^{t}\tau^2 + \left(\frac{2}{k}-\frac{t}{k^2}\right)\sum_{\tau=1-k}^{t}\tau + \left(1-\frac{t}{k}\right)\sum_{\tau=1-k}^{t}1$$

$$\sum_{\tau=1-k}^{t}1 = k+t$$

$$\sum_{\tau=1-k}^{t}\tau = \sum_{\tau=1}^{t+k}\tau - \sum_{\tau=1}^{t+k}\tau = \frac{(t+k)(t+k+1)}{2} - (t+k)k$$

$$\sum_{\tau=1-k}^{t} \tau^2 = \sum_{\tau=1}^{t+k} \tau^2 - 2k \sum_{\tau=1}^{t+k} \tau + \sum_{\tau=1}^{t+k} k^2$$

$$\sum_{\tau=1}^{t+k} \tau^2 = \frac{(t+k)(t+k+1)(2t+2k+1)}{6}$$

$$\sum_{\tau=1}^{t+k} \tau = \frac{(t+k)(t+k+1)}{2}$$

$$\sum_{\tau=1}^{t+k} k^2 = (t+k)k^2$$

$$\boldsymbol{C(2)}: \sum_{\tau=t+1}^{t+N-k} 1 = \sum_{\tau=1}^{N} 1 - \sum_{\tau=1}^{k} 1 = N - k$$

$$\boldsymbol{C(3)} = \frac{1}{k^2} \sum_{\tau=t+N-k+1}^{t+N} \tau^2 - \frac{2N}{k^2} \sum_{\tau=t+N-k+1}^{t+N} \tau + \frac{N^2}{k^2} \sum_{\tau=t+N-k+1}^{t+N} 1$$

$$\sum_{\tau=t+N-k+1}^{t+N} 1 = \sum_{\tau=1}^{k} 1 = k$$

$$\sum_{\tau=t+N-k+1}^{t+N} \tau = (t+N) \sum_{\tau=1}^{k} 1 - \sum_{\tau=1}^{k-1} \tau = (t+N)k - \frac{(k-1)k}{2}$$

$$\sum_{\tau=t+N-k+1}^{t+N} \tau^2 = (t^2 + N^2 + 2tN)k + \sum_{\tau=1}^{k-1} \tau^2 - (2t+2N) \sum_{\tau=1}^{k-1} \tau$$

$$\sum_{\tau=1}^{k-1} \tau^2 = \frac{(k-1)k(2k-1)}{6}$$

$$\sum_{\tau=1}^{k-1} \tau = \frac{(k-1)k}{2}$$

***Q.E.D***

## A5: The Concentration Index Specifications CIS(n,k,C)

Let $LF(Y) = \{y_1, \ldots \ , y_N\}$ be the Lorenz curve of the random variable $X = \{x_1, \ldots, x_N\}$ . Then the requirements specification of CI(X) are given by three numbers:

$$S(N,k,C) \ \stackrel{\text{def}}{=} \begin{cases} N \ \stackrel{\text{def}}{=} \ total\ number\ of\ categories \\ k \ \stackrel{\text{def}}{=} number\ of\ categories\ with\ the\ maximum\ Concentration \\ k \ \in \{1, \ldots, N-1\} \\ C = \displaystyle\sum_{i=N-k+1}^{N} \widehat{p_\iota} \ : \ \widehat{p_\iota} \in \ \mathbb{III}(P(X), C, k) \ \stackrel{\text{def}}{=} Distribution\ Concentration \\ C \ \in [{}^{1}/_{N} \ 1) \end{cases}$$

$$\mathbb{III}(P(X), k, C) \in \mathbb{III}(X, k, C) = \begin{cases} X_c \subset X : |X_c| = k \ \wedge \sum p(x_i) = C : \ x_i \in \ X_c \\ p(x_i) \geq p(x_j) \ \ \forall \ x_j \in X \setminus X_c \end{cases}$$

$$\mathbb{III}(X, C) \in \ \mathbb{III}(X) \ \rightarrow \ |\mathbb{III}(X, C)| < |\mathbb{III}(X)| \rightarrow \frac{N!}{(N-C)!} < N!$$

Therefore, $S(N,k,C)$ identifies the set of probability distribution functions with a cumulative percentage (population percentage) equal to the concentration *C* in the *k* most populated groups or categories.

**Theorem A5.1**: Let $S(N,k,C)$ be the requirements specifications of the random variable X. Then the optimal solution to the problem min$\{CI(P(X)\}$ is given by the following formula:

Formula (1)

$$\begin{cases} \widehat{p_i^*} = \dfrac{1-C}{N-k} \quad i \in \{1,\dots,N-k\} \\ \widehat{p_i^*} = \dfrac{C}{k} \quad i \in \{N-k+1,\dots,N\} \\ \qquad \widehat{p_i} \in \mathbb{III}(P(X)) \end{cases} \Leftrightarrow \begin{cases} y_i^* = \dfrac{i(1-C)}{N-k} \quad i \in \{1,\dots,N-k\} \\ y_i^* = (1-C) + \dfrac{(i-N+k)C}{k} \;:\; C \geq \dfrac{k}{N} \\ \qquad i \in \{N-k+1,\dots,N-1\} \end{cases}$$

*Proof*:

Based on Lemma A1.1 and Lemma A1.3 , it follows that:

$$\begin{cases} min \;\; CI(P(X)) \\ \quad \displaystyle\sum_{i=1}^{N} p_i \\ \quad p_i \in [0 \;\; 1] \end{cases} \Leftrightarrow \begin{cases} \quad \max \displaystyle\sum_{i=1}^{N-1} y_i \\ 2y_i \leq y_{i-1} + y_{i+1} \;:\; i = 1,\dots,N-1 \\ y_i \in [0 \;\; 1) \wedge y_0 = 0 \wedge y_N = 1 \end{cases}$$

Since:

$$y_N = \sum_{i=1}^{N} p_i = \sum_{i=1}^{N-k} p_i + \sum_{i=N-k+1}^{N} p_i = 1 \qquad\qquad \sum_{i=N-k+1}^{N} p_i = C$$

$$y_N = y_{N-k} + \sum_{i=N-k+1}^{N} p_i = 1 \xrightarrow{yields} \boldsymbol{y_{N-k} = 1 - C}$$

Then, the objective function can be expressed:

$$\max\sum_{i=1}^{N-1} y_i = \max\sum_{i=1}^{N-k-1} y_i + (1-C) + \ max\sum_{i=N-k+1}^{N-1} y_i$$

It is clear the Lorenz Curve must pass through the fixed point $(x_{N-k}, y_{N-k}) = \left(\frac{N-k}{N}, 1-c\right)$

As we would like to maximize the values of y, such that the straight line with the maximum slope pass through $y_{N-k}$, the optimal values are given by the following system of equations:

$$Y^* = \begin{cases} 2y_i = y_{i-1} + y_{i+1} \\ y_{N-k} = 1 - C \\ i \in \{1, \dots, N-k\} \end{cases} \Rightarrow \ y_i^* = \frac{i(1-C)}{N-k} \Rightarrow \ \hat{p}_i^* = \frac{1-C}{N-k}$$

Given that $\arg\min\{CI(X)\} = CI\{U(X)\} \rightarrow \hat{p}_i^* = \frac{1}{N} \ \wedge C \ \geq \frac{1}{N}$, it follows that:

$$C = \sum_{i=N-k+1}^{N} \widehat{p_i} = \overbrace{\left(\frac{1}{N} + \cdots + \frac{1}{N}\right)}^{k} = \frac{k}{N} \rightarrow \frac{1}{N} = \frac{C}{k} = p_i \ \ : \ \ C \geq \frac{k}{N}$$

It follows that:

$$Y^* = \begin{cases} y_{N-k+1} = (1-C) + \dfrac{C}{k} \\ y_{N-k+2} = (1-C) + 2\dfrac{C}{k} \\ y_{N-k+j} = (1-C) + j\dfrac{C}{k} \end{cases} \rightarrow \begin{cases} y_i^* = (1-C) + \dfrac{(i-N+k)C}{K} \\ i \in \{N-k+1, \dots, N\} \end{cases} \rightarrow \ \hat{p}_i^* = \frac{C}{k}$$

***Q.E.D***

**Theorem A5.2**: Let $S(N,k,C)$ be the requirements specifications of the random variable X. Then the optimal solution to the problem max$\{CI(P(X)\}$ is given by the following formula:

Formula (2):

$$\exists\ j \in \{1,\dots,N-k\}\ \ :\ \ \frac{k}{k+j} \le C < \frac{k}{k+j-1} \qquad C\ \in [\frac{1}{N}\ \ 1)$$

$$\widehat{P^*}\begin{cases} \widehat{p_i^*} = 0 & i\ \in \{1,\dots,N-k-j\} \\ \widehat{p_i^*} = \frac{(1-C)}{j} & i\ \in \{N-k-j+1,\dots,N-1\} \\ \widehat{p_N^*} = 1-(k+j-1)\cdot\frac{(1-C)}{j} & \end{cases}$$

$$Y^*\begin{cases} y_i^* = 0 & i\ \in \{1,\dots,N-k-j\} \\ y_i^* = \dfrac{(i-N+k+j)}{j}\cdot(1-C) & i \in \{N-k-j+1,\dots,N-1\} \\ y_N^* = 1 & \end{cases}$$

*Proof*:

Based on Lemma A1.1 and Lemma A1.3, it follows that:

$$\begin{cases} max\ \ CI(P(X)) \\ \displaystyle\sum_{i=1}^{N} p_i \\ p_i\ \in [0\ \ 1] \end{cases} \Leftrightarrow \begin{cases} \displaystyle\min\sum_{i=1}^{N-1} y_i \\ 2y_i\ \le\ y_{i-1} + y_{i+1}\ :\ i = 1,\dots,N-1 \\ y_i\ \in\ [0\ \ 1)\ \wedge\ y_0 = 0\ \ \wedge\ \ y_N = 1 \end{cases}$$

Since:

$$y_N = \sum_{i=1}^{N} p_i = \sum_{i=1}^{N-k} p_i + \sum_{i=N-k+1}^{N} p_i = 1 \qquad\qquad \sum_{i=N-k+1}^{N} p_i = C$$

$$y_N = y_{N-k} + \sum_{i=N-k+1}^{N} p_i = 1 \xrightarrow{yields} \boldsymbol{y_{N-k} = 1 - C}$$

Then, the objective function can be expressed:

$$\min \sum_{i=1}^{N-1} y_i = \min \sum_{i=1}^{N-k-1} y_i + (1 - C) + \sum_{i=N-k+1}^{N-1} y_i$$

It is clear the objective function has N-2 variables:

$$P = \begin{cases} 2y_i \leq y_{i-1} + y_{i+1} \; : \; i = 1, \dots, N-1 \\ y_{N-k} = 1 - C \\ y_i \in [0 \;\; 1) \wedge y_0 = 0 \;\wedge\; y_N = 1 \end{cases}$$

As we want to find the minimum value of y in the feasible solution space:

$$\sum_{i=1}^{N-k-1} y_i = 0 \;\rightarrow\; y_i = 0 \;:\; i \in \{1, \dots, N-k-1\} \leftrightarrow \frac{k}{k+1} \leq C < 1$$

$$y_{N-k} = y_{N-k-1} + p_{N-k} \;\rightarrow\; p_{N-k} = 1 - C$$

Therefore:

$$Y^* = \begin{cases} y_i = 0 & i \in \{1, \dots, N-k-1\} \\ 2y_i = y_{i-1} + y_{i+1} & i \in \{N-k, \dots, N-2\} \\ y_{N-k} = 1 - C & \\ y_i \in (1 - C \quad 1) & \end{cases}$$

Solving the system of equations:

$$y_i^* = (i - N + k + 1)(1 - C) \quad i \in \{N - k, \dots, N - 1\} \rightarrow \begin{cases} \widehat{p_i^*} = 0 \\ \widehat{p_i^*} = (1 - C) \\ \widehat{p_N^*} = 1 - \sum_{i=N-k+1}^{N-1} \widehat{p_i^*} \rightarrow 1 - k(1 - C) \end{cases}$$

It follows that:

$$\sum_{i=N-k+1}^{N} \widehat{p_i^*} \geq \sum_{i=N-k+1}^{N-1} \widehat{p_i^*} \rightarrow C \geq k(1 - C) \rightarrow \frac{k}{k+1} \leq C < 1$$

In a similar way, solving the following system of equations:

$$\frac{k}{k+2} \leq C < \frac{k}{k+1}$$

$$Y^* = \begin{cases} y_i = 0 & i \in \{1, \dots, N - k - 2\} \\ 2y_i = y_{i-1} + y_{i+1} & i \in \{N - k - 1, \dots, N - 2\} \\ y_{N-k} = 1 - C & \\ y_{N-k-1} \in (0 \quad 1 - C) & \\ y_i \in (1 - C \quad 1) & \end{cases}$$

$$y_i^* = \frac{(i-N+k+1)(1-C)}{2} \quad i \in \{N-k-1, \dots, N-1\} \rightarrow \begin{cases} \widehat{p_i^*} = 0 \\ \widehat{p_i^*} = \frac{(1-C)}{2} \\ \widehat{p_N^*} = 1 - \sum_{i=N-k+1}^{N-1} \widehat{p_i^*} \rightarrow 1 - \frac{k(1-C)}{2} \end{cases}$$

This implies that:

$$\frac{k}{N} \leq C < \frac{k}{N-1}$$

$$y_i^* = \frac{i(1-C)}{N-k} \quad i \in \{1, \dots, N-1\} \rightarrow \begin{cases} \widehat{p_i^*} = \frac{(1-C)}{N-k} \\ \widehat{p_N^*} = 1 - \frac{k(1-C)}{N-k} \end{cases}$$

This result can be easily computed in the Formula (1) for j=N-k. $\boldsymbol{Q.E.D}$

**Corollary A5.1**: Let $S(N,k,C)$ be the requirements specifications of the random variable X. Then, the Lower Bound of the Concentration Index is:

$$LB\{CI(X)\} = \left(\frac{N}{N-1}\right)\cdot\left(C-\frac{k}{N}\right) : \begin{cases} C \geq \dfrac{k}{N} \\ k = 1,\dots,N-1 \end{cases}$$

*Proof*:

$$LB\{CI(X)\} = \widehat{LZ}^{-1}\cdot R \qquad R = \left[1-\frac{1}{N}\left(1+2\sum_{i=1}^{N-1} y_i^*\right)\right] \qquad \widehat{LZ}^{-1} = \frac{N}{N-1}$$

$$\sum_{i=1}^{N-1} y_i^* = \sum_{i=1}^{N-k} y_i^* + \sum_{i=N-k+1}^{N-1} y_i^*$$

Based on theorem A5.1:

$$P = \sum_{i=1}^{N-k} y_i^* = \sum_{i=1}^{N-k}\frac{i(1-C)}{N-k} = \frac{(1-C)}{N-k}\cdot\sum_{i=1}^{N-K} i = \frac{(1-C)(N-k+1)}{2}$$

$$Q = \sum_{i=N-k+1}^{N-1} y_i^* = \sum_{i=N-k+1}^{N-1}\left[(1-C)+\frac{(i-N+k)\cdot C}{k}\right]$$

It is worth noting $(N-1) \geq N-k+1 \ \rightarrow k \geq 2$

$$Q = \sum_{i=N-k+1}^{N-1}(1-C)+\frac{C}{k}\cdot\sum_{i=N-k+1}^{N-1}(k-N)+\frac{C}{k}\cdot\sum_{i=N-k+1}^{N-1} i$$

$$\boldsymbol{Q} = \ (k-1)(1-C)+\frac{(k-1)C(k-N)}{k}+\frac{C}{k}\cdot\sum_{i=N-k+1}^{N-1} i$$

$$\sum_{i=N-k+1}^{N-1} i = \sum_{i=1}^{N-1} i - \sum_{i=1}^{N-k} i = \frac{(N-1)N}{2} - \frac{(N-k)(N-k+1)}{2}$$

It follows that:

$k = 1$

$$R = 1 - \frac{1}{N}[1 + 2 \cdot P] = 1 - \frac{1}{N}[1 + N(1 - C)] = C - \frac{1}{N} \ : \ C \geq \frac{1}{N}$$

$k = 2$

$$R = 1 - \frac{1}{N}[1 + 2 \cdot (P + Q)] = C - \frac{2}{N} \ : \ C \geq \frac{2}{N}$$

Therefore:

$$R = 1 - \frac{1}{N}[1 + 2 \cdot (P + Q)] = C - \frac{k}{N} \ : \ C \geq \frac{k}{N}$$

It is worth noting:

$$\lim_{C \to 1}\left(C - \frac{1}{N}\right) \approx \left(\frac{N-1}{N}\right) = \widehat{LZ} \to \lim_{C \to 1} LB\{CI(X)\} \approx \widehat{LZ}^{-1} \cdot \widehat{LZ} = 1 \quad \wedge \quad LB\{CI(X)\} = 0 \quad C = \frac{k}{N}$$

$\boldsymbol{Q.E.D}$

**Corollary A5.2**: Let $S(N, k, C)$ be the requirements specifications of the random variable X. Then, the Upper Bound of the Concentration Index is given by the following formula:

$$\exists \;\; j \in \{1, \dots, N-k\} \;:\; \frac{k}{k+j} \leq C < \frac{k}{k+j-1} \qquad C \in [{}^{1}/_{N} \;\; 1)$$

Then:

$$UB\{CI(X)\} = 1 - \frac{(1-C)(k+j-1)(k+j)}{j(N-1)}$$

*Proof*:

Based on Theorem A5.2:

$$\sum_{i=1}^{N-1} y_i^* = \sum_{i=1}^{N-k-j} y_i^* + \sum_{i=N-k-j+1}^{N-1} y_i^* \qquad\qquad \sum_{i=1}^{N-k-j} y_i^* = 0$$

$$= \sum_{i=N-k-j+1}^{N-1} y_i^* \qquad : \; \frac{k}{k+j} \leq C < \frac{k}{k+j-1} \quad \wedge \quad j \in \{1, \dots \; N-k\}$$

Assuming $j = 1 \;\; \rightarrow \;\; \frac{k}{k+1} \leq C < 1$

$$Y^*: \begin{cases} y_i^* = 0 & i \in \{1, \dots, N-k-1\} \\ y_i^* = (i - N + k + 1)(1-C) & i \in \{N-k, \dots, N-1\} \end{cases}$$

Since $k = \{1, \dots, N-1\}$, it follows that:

$k = 1$

$$y_{N-1} = (1-C) \;\; \wedge \;\; y_i = 0 \quad \forall\, i < N-1 \qquad\qquad \rightarrow \sum y_i^* = (1-C)$$

$k = 2$

$$y_{N-2} = \;\; (1-C) \;\; \wedge \;\; y_i = 0 \quad \forall\, i < N-2$$

$$y_{N-1} = \; 2(1-C) \qquad\qquad \rightarrow \sum y_i^* = 3(1-C)$$

$k = 3$

$$y_{N-3} = \;\; (1-C) \;\; \wedge \;\; y_i = 0 \quad \forall\, i < N-3$$

$$y_{N-2} = \; 2(1-C)$$

$$y_{N-1} = 3(1-C) \qquad \rightarrow \sum y_i^* = 6(1-C)$$

Therefore, for $j = 1$:

$$\sum y_i^* = \sum_{i=1}^{k} (1-C) \cdot i \qquad : \quad \frac{k}{k+1} \leq C < 1$$

In a similar way, solving by inspection for $j = 2$:

$$\sum y_i^* = \sum_{i=1}^{k+1} \frac{(1-C) \cdot i}{2} \qquad : \frac{k}{k+2} \leq C < \frac{k}{k+1}$$

It then follows:

$$\sum_{i=1}^{N-1} y_i^* = \sum_{i=1}^{k+j-1} \frac{(1-C) \cdot i}{j} = \frac{(1-C)(k+j-1)(k+j)}{2j}$$

Based on the definition of $CI(X)$:

$$UB\{CI(X)\} = \widehat{LZ}^{-1} \cdot R \qquad R = \left[1 - \frac{1}{N}\left(1 + 2\sum_{i=1}^{N-1} y_i^*\right)\right] \qquad \widehat{LZ}^{-1} = \frac{N}{N-1}$$

$$UB\{CI(X)\} = \left(\frac{N}{N-1}\right) \cdot \left[\left(\frac{N-1}{N}\right) - \frac{(1-C)(k+j-1)(k+j)}{Nj}\right]$$

$$= 1 - \frac{(1-C)(k+j-1)(k+j)}{(N-1)j}$$

Solving by inspection for

$$j = 1 \xrightarrow{yields} UB\{CI(X)\} = 1 - \frac{(1-C)(k)(k+1)}{(N+1)} \qquad : \quad \frac{k}{k+1} \leq C < 1$$

$$j = 2 \xrightarrow{yields} UB\{CI(X)\} = 1 - \frac{(1-C)(k+1)(k+2)}{2(N+1)} \qquad : \quad \frac{k}{k+1} \leq C < \frac{k}{k+1}$$

$$j = 3 \xrightarrow{yields} UB\{CI(X)\} = 1 - \frac{(1-C)(k+1)(k+2)}{3(N+1)} \qquad : \quad \frac{k}{k+1} \leq C < \frac{k}{k+1}$$

It follows that:

$$UB\{CI(X)\} = 1 - \frac{(1-C)(k+j-1)(k+j)}{j(N-1)} \quad : \qquad \frac{k}{k+j} \leq C < \frac{k}{k+j-1}$$

It worth noting:

$$\lim_{C\to 1} UB\{CI(X)\} = 1$$

$$\lim_{C\to\frac{k}{N}} UB\{CI(X)\} = 0 \quad \to \quad j = N-k \qquad \frac{k}{N} \leq C < \frac{k}{N-1}$$

$$\lim_{C\to\frac{k}{N}} UB\{CI(X)\} = 1 - \frac{\left(1-\frac{k}{N}\right)(N-1)N}{(N-k)(N-1)} = 1 - \frac{(N-k)}{(N-k)} = 0$$

$\boldsymbol{Q.E.D}$

# Appendix B - Location Index (LI)

Let $I_{x_0} = [x_m \quad x_M]$ be a closed bounded intervalof the random variable $X = \{x_1, \dots, x_N\} \rightarrow |X| = N$. Then $\forall x_0 \in X \; \exists \; I_{x_0}$: $x_m \leq x_0 \leq x_M \; \wedge \; x_m, x_0, x_M \; \in \{1, \dots, N\}$. It follows that $\left|I_{x_0}\right| = M - m + 1 = b$. In the case of a symmetric closed bounded interval $b = 2k + 1$.

$$I_{x_0,k} = [x_0 - k \qquad x_0 + k] \;\; \rightarrow \;\; \left|I_{x_0}\right| = 2k + 1$$

Let's denote the set of nested compact subset of $x_0$ with the following notation:

$$I_{x_0} = \bigcup_{k=0}^{\infty} I_{x_0,k} \rightarrow \bigcap_{k=0}^{\infty} I_{x_0,k} \neq \{\emptyset\} \rightarrow X \subset I_{x_0}$$

Then, let's define the functions $\varphi$ (***Likelihood Central Region function)*** and $\gamma$ (***Level Set Density function***).

$$\varphi(x_0, k) = \sum_{i=x_0-k}^{x_0+k} P(X = x_i) \;\; : \;\; k \; \in \{0, \dots, N-1\}$$

$$\gamma(x_0, s) = \sum_{k=0}^{s} \sum_{i=x_0-k}^{x_{0+k}} P(X = x_i)$$

Then the ***Bin Concentration function*** (BCF), is given by the following formula:

$$BCF = \gamma(x_0) = \gamma(x_0, N-1) = \sum_{k=0}^{N-1} \sum_{i=x_0-k}^{x_0+k} P(X = x_i)$$

Let $\Gamma(X) = \{\gamma_1, \dots, \gamma_i, \dots, \gamma_N\}$ be the *BCF* vector of the random varaible X.

$$\gamma_i = \gamma(x_i) = \gamma(x_i, N-1) = \sum_{k=0}^{N-1} \sum_{i=x_0-k}^{x_0+k} P(X = x_i)$$

Then, the **Location Index** of this random variable is given by the formula:

$\lambda$: $\mathbb{R}^N \rightarrow$ N x N $\;\; \lambda = (\lambda_1, \lambda_2)$. $\;\;$ N $= \{1, \dots, N\}$

$$\lambda_1 = arg \min_{i \in \mathrm{N}} \{ \, Max \, \gamma_i \; \in \; \Gamma(X)\}$$
$$\lambda_2 = arg \max_{i \in \mathrm{N}} \{ \, Max \, \gamma_i \; \in \; \Gamma(X)\}$$

**Theorem B1.1:** Let $\Phi_{x_0} = \{\varphi_0, \dots , \varphi_k, \dots , \varphi_{N-1}\}$ be a sequence in $X = \{x_1, \dots , x_N\}$, such that:

$$\varphi_k = \varphi(x_0, k) = \sum_{i=x_0-k}^{x_0+k} P(X = x_i)$$

Then $\forall x_0 \in X \; \exists \; k_m: \varphi(x, k) = 1 \;\; \forall \; k \geq k_m$.

*Proof:*

Let $I_x$ and $I_{x0,X}$ be two collections of non empty closed bounded intervals in $\mathbb{Z}$ such that

$$I_x = [x_1 \;\; x_N] \quad and \quad I_{x0,X} = \bigcup_{k=0}^{N-1} I_{x_0,k}$$

Since $|I_x| = N$ and $|I_{x_0,X}| = 2N - 1$, it follows that:

$$\forall x_0 \in X \; \exists \; k_m: I_x \subseteq \bigcup_{j=0}^{k} I_{x_0,j} \;\; \forall \;\; k \geq \; k_m \quad \rightarrow \quad E = I_x \bigcap \bigcup_{j=0}^{k_m} I_{x_0,j} \;\; \neq \{\emptyset\}$$

Then

$$\bigcup_{j=0}^{k} I_{x_0,j} = E \bigcup \bar{E}$$

It follows that

$$E = X = \{x_1, \dots, x_N\} \rightarrow \sum_{i=1}^{N} P(X = x_i) = 1 \quad \wedge \;\; \bar{E} = \bar{X} \rightarrow \sum_{i \in \bar{X}} P(X = x_i) = 0$$

Therefore,

$$\varphi_{k_m} = \varphi(x_0, k_m) = \varphi(x_0, k) = \sum_{i=1}^{N} P(X = x_i) + \sum_{i \in \bar{X}} P(X = x_i) = 1$$

$$\varphi(x_0, k_m) = \sum_{i=x_0-k_m}^{x_0+k_m} P(X = x_i) = \sum_{i=x_0-k}^{x_0+k} P(X = x_i) \qquad \forall \;\; k \; \geq k_m$$

***Q.E.D***

**Lemma B1.1** Let $\varphi(j,k)$ be the *Likelihood Central Region Function* and $\gamma(x_j)$ be the BCF of the random variable $X = \{x_1, \dots, x_N\}$. Then the BCF is a linear convex combination such that:

$$\gamma(x_j) = \sum_{i=1}^{N} w_i \cdot p_i \;:\quad w_i \in \{1,2,\dots,N\}$$

$$\gamma(x_j) = \begin{cases} \displaystyle\sum_{i=0}^{N-1} (N-i) \cdot p_{i+1} & j = 1 \\ \displaystyle\sum_{i=0}^{j-1} (i+N-j+1) \cdot p_{i+1} + \sum_{i=j}^{N-1} (N-i+j-1) \cdot p_{i+1} & 1 < j < N \\ \displaystyle\sum_{i=0}^{N-1} (i+1) \cdot p_{i+1} & j = N \end{cases}$$

*Proof:*
Let $j = 1$, then the *Likelihood Central Region Function* is given by the following formula:

$$\varphi(1,k) = \sum_{i=1-k}^{1+k} P(X = x_i) : \quad k \in \{0, \dots, N-1\}$$

Since $P(X = x_i) = 0 \;\; \forall\, i < 1$, it follows that:

$$\varphi(1,k) = \sum_{i=1}^{k} P(X = x_i) = \sum_{i=0}^{k-1} P(X = x_{i+1}) = \sum_{i=0}^{k-1} p_{i+1} \quad : \; k \in \{1, \dots, N-1\}$$

Then the *Level Set Density Function* is given by the following equation:

$$\gamma(1,s) = \sum_{k=1}^{s} \sum_{i=0}^{k-1} p_{i+1} \quad : \; s \in \{1, \dots, N\}$$

$\gamma(1,1) = p_1$
$\gamma(1,2) = 2 \cdot p_1 + p_2$
$\vdots$
$\gamma(1,s) = s \cdot p_1 + (s-1) \cdot p_2 + \cdots + (N-i) \cdot p_{i+1} + \cdots + p_N$

Then for $s = N$

$\gamma(1,N) = N \cdot p_1 + (N-1) \cdot p_2 + \cdots + (N-i) \cdot p_{i+1} + \cdots + p_N$

It follows that:

$$\gamma(1,N) = \sum_{i=0}^{N-1}(N-i)\cdot p_{i+1}$$

*Q.E.D*

Let's prove for $j = N$:

$$\varphi(N,k) = \sum_{i=N-k}^{N+k} P(X = x_i) : \quad k \in \{0, \dots, N-1\}$$

Since $P(X = x_i) = 0 \ \forall\, i > N$, it follows that:

$$\varphi(N,k) = \sum_{i=N-k}^{N} P(X = x_i) = \sum_{i=N-k}^{N-1} P(X = x_{i+1}) = \sum_{i=N-k}^{N-1} p_{i+1} \quad : \ k \in \{1, \dots, N\}$$

Then the *Level Density Function* is given by the following equation:

$$\gamma(N,s) = \sum_{k=1}^{s}\sum_{i=N-k}^{N-1} p_{i+1} \ : \ s \in \{1, \dots, N\}$$

$\gamma(N,1) = p_N$
$\gamma(N,2) = 2\cdot p_N + p_{N-1}$
$\vdots$
$\gamma(N,s) = s\cdot p_N + (s-1)\cdot p_{N-1} + \cdots + p_{N-s}$

Then for $s = N$

$$\gamma(N,N) = N\cdot p_N + (N-1)\cdot p_{N-1} + \cdots + (N-i)\cdot p_{N-i} + \cdots + p_1$$

It follows that:

$$\gamma(N,N) = \sum_{i=0}^{N-1}(i+1)\cdot p_{i+1}$$

*Q.E.D*

Let $2 \le j \le N-1$

The *Likelihood Central Region Function* is given by the following formula:

$$\varphi(j,k) = \sum_{i=k-1}^{k+1} P(X = x_i) : \quad k \in \{1, \dots, N\}$$

Since $P(X = x_i) = 0 \ \forall\, i < 1 \ \wedge \ i > N$. It follows from the previous results that:

$$\varphi(j,k) = \sum_{i=j-k}^{j-1} P(X = x_{i+1}) + \sum_{i=j}^{j+k-2} P(X = x_{i+1}) : \quad k \in \{1, \dots, N\}$$
$$= \sum_{i=j-k}^{j-1} p_{i+1} + \sum_{i=j}^{j+k-2} p_{i+1}$$

Then the *Level Set Density Function* is given by the following equation:

$$\gamma(j,s) = \sum_{k=1}^{s} \sum_{i=j-k}^{j-1} p_{i+1} + \sum_{k=2}^{s} \sum_{i=j}^{j+k-2} p_{i+1} \quad : \quad s \in \{1, \dots, N\}$$

Let's solve the first term:

$$\sum_{k=1}^{s} \sum_{i=j-k}^{j-1} p_{i+1}$$

$\gamma(j,1) = p_j$
$\gamma(j,2) = 2 \cdot p_j + p_{j-1}$
$\vdots$
$\gamma(j,s) = s \cdot p_j + (s-1) \cdot p_{j-1} + \cdots + p_{j-s}$

Then for $s = N$

$\gamma(j,N) = N \cdot p_j + (N-1) \cdot p_{j-1} + \cdots + p_{j-N}$

Since $P(X = x_i) = 0 \quad \forall \ i < 1$

$\gamma(j,N) = N \cdot p_j + (N-1) \cdot p_{j-1} + \cdots + (N-j+1) \cdot p_1$

It follows that:

$$\gamma(j,N) = \sum_{i=0}^{j-1} (i + N - j + 1) \cdot p_{i+1}$$

The second term:

$$\sum_{k=2}^{s} \sum_{i=j}^{j+k-2} p_{i+1}$$

$\gamma(j,2) = p_{j+1}$
$\gamma(j,3) = 2 \cdot p_{j+1} + p_{j+2}$
$\vdots$
$\gamma(j,s) = (s-1) \cdot p_j + (s-2) \cdot p_{j-1} + \cdots + p_{j+s-1}$

Then for $s = N$

$\gamma(j, N) = (N-1) \cdot p_j + (N-2) \cdot p_{j-1} + \cdots + p_{j+s-1}$

Since $P(X = x_i) = 0 \quad \forall \ i > N$

$\gamma(j, N) = (N-1) \cdot p_{j+1} + (N-2) \cdot p_{j+2} + \cdots + j \cdot p_1$

It follows that:

$$\gamma(j, N) = \sum_{i=j}^{N-1} (N - i + j - 1) \cdot p_{i+1}$$

***Q.E.D***

**Corollary B1.1**: Let $\mu_U$ be the expected value of the uniform distribution and $m = \lfloor \mu_u \rfloor$ be the rounded floor value. Then: $W = \{1, \dots, N\}$

$$arg \min_{w_i}\{x_j\} = \begin{cases} j & j \in [1 \ \ m] \\ N - j + 1 & j \in [m+1 \quad N] \end{cases} \quad \wedge \quad arg \max_{w_i}\{x_j\} = N$$

*Proof:*

$arg \max_{w_i}\{x_j\} = N$

For $j = 1$

$$\begin{cases} \max f(i) \\ i \in \{0, \dots, N-1\} \end{cases} \rightarrow \max(N - i) = N \qquad i^* = 0$$

For $1 < j < N$

$$max \begin{Bmatrix} \max(i + N - j + 1), \max(N - i + j - 1) \\ i \in \{0, \dots, j-1\}, \quad i \in \{j, \dots, N-1\} \end{Bmatrix} = N \quad i^* = j - 1$$

For $j = N$

$$\begin{cases} \max f(i) \\ i \in \{0, \dots, N-1\} \end{cases} \rightarrow \max(i + 1) = N \qquad i^* = N - 1$$

$arg \min_{w_i}\{x_j\}$

For $j = 1$

$$\begin{cases} \min f(i) \\ i \in \{0, \dots, N-1\} \end{cases} \rightarrow \min(N-i) = 1 = j \quad i^* = N-1$$

For $1 < j < N$

$$min \begin{cases} \min(i+N-j+1), \min(N-i+j-1) \\ i \in \{0,\dots,j-1\}, \quad i \in \{j,\dots,N-1\} \end{cases} = \begin{cases} j & j \in \{2,\dots,m\} \quad i^* = N-1 \\ N-j+1 & j \in \{m+1 \;\; N\} \quad i^* = 0 \end{cases}$$

For $j = N$

$$\begin{cases} \min f(i) \\ i \in \{0, \dots, N-1\} \end{cases} \rightarrow \min(i+1) = 1 \quad i^* = 0$$

$\min(i+1) = \min(i+N-j+1) \xrightarrow{j=N} 1$

***Q.E.D***

**Lemma B1.2:** Let $\Gamma(X) = [\gamma_1, \dots, \gamma_i, \dots, \gamma_N]$ be the BCF vector of the random variable X and $m = \lfloor \mu_u \rfloor$. Then

$$arg \max_{\gamma_i}\{\Gamma(X)\} = \gamma_j \leftrightarrow \gamma_j : \begin{cases} \sum_{i=1}^{j-1} p_i \le 1/2 \\ \sum_{i=j+1}^{N} p_i \le 1/2 \end{cases} \wedge \begin{cases} \sum_{i=1}^{j} p_i \ge 1/2 & j \in [1 \;\; m] \\ \sum_{i=j}^{N} p_i \ge 1/2 & j \in [m+1 \;\; N] \end{cases}$$

*Proof:*

According to the *Corollary A6.1* :

$$\gamma(x_j) = \sum_{i=1}^{N} w_i \cdot p_i \quad \begin{cases} w_i \in [1 \;\; N] : & j \in [1 \;\; m] \\ w_i \in [N-j+1 \;\; N] : & j \in [m+1 \;\; N] \end{cases}$$

Let’s prove

$$arg \max_{\gamma_i}\{\Gamma(X)\} = \gamma_1 = x_1 \leftrightarrow p_1 \ge 1/2$$

It follows that:

$\gamma(x_1) = N \cdot p_1 + (N-1) \cdot p_2 + \cdots + (N-i) \cdot p_{i+1} + \cdots + p_N$
$\gamma(x_1) = (N-1) \cdot p_1 + N \cdot p_2 + \cdots + (N-i+1) \cdot p_{i+1} + \cdots + 2 \cdot p_N$

So $\gamma(x_1) \geq \gamma(x_2) \leftrightarrow \gamma(x_1) - \gamma(x_2) \geq 0$

For sake of simplicity let's denote $P = [p_1, \dots, p_N]$, then:

$arg \max_P\{\gamma(x_1)\} = \{p_1 = 1 \ \wedge \ p_i = 0 \ \ \forall\, i \ \neq 1\}$

$arg \min_P\{\gamma(x_1)\} : \quad \gamma(x_1) \geq \gamma(x_2)$

$$Let \ \ p_1 = {}^1\!/_2 \rightarrow \gamma(x_1) = \begin{cases} \widehat{\gamma_1} = \dfrac{N}{2} + \displaystyle\sum_{i=1}^{N-1} (N - i) \cdot p_{i+1} \\ \displaystyle\sum_{i=2}^{N} p_i = {}^1\!/_2 \end{cases}$$

It follows:

$$max\ \widehat{\gamma_1} = \frac{N}{2} + \frac{N-1}{2} \quad : \ p_2 = {}^1\!/_2$$

$$\gamma(x_2) = \begin{cases} \widehat{\gamma_2} = \dfrac{N-1}{2} + N \cdot p_2 + \displaystyle\sum_{i=2}^{N-1} (N - i + 1) \cdot p_{i+1} \\ \displaystyle\sum_{i=2}^{N} p_i = {}^1\!/_2 \end{cases}$$

$$max\ \widehat{\gamma_2} = \frac{N-1}{2} + \frac{N}{2} \quad : \ p_2 = {}^1\!/_2$$

It follows that $\gamma(x_1) \geq \gamma(x_2) \leftrightarrow p_1 \geq {}^1\!/_2 \rightarrow \gamma(x_1) \geq \gamma(x_j) \leftrightarrow p_1 \geq {}^1\!/_2 \ \forall \ j \in \{2, .., N\}$

Assume

$$arg \ \max_{\gamma_i}\{\Gamma(X)\} = \gamma_2 = x_2$$

Then

$\gamma(x_2) \geq \gamma(x_1) \ \leftrightarrow p_2 \geq p_1$

$\gamma(x_2) \geq \gamma(x_3) \ \leftrightarrow p_1 + p_2 \geq {}^1\!/_2$

It follows that:

$\gamma(x_2) \geq \gamma(x_j) \ \leftrightarrow p_1 + p_2 \geq {}^1\!/_2 \quad \forall \quad j \in \{3, \dots, N\}$

Therefore, for $j \ \in [1, \dots, m]$

$$\gamma(x_j) \geq \gamma(x_i) \quad \leftrightarrow \sum_{i=1}^{j-1} p_i \leq {}^1\!/_2 \ \wedge \ \sum_{i=1}^{j} p_i \geq {}^1\!/_2 \ \wedge \ \sum_{i=j+1}^{N} p_i \leq {}^1\!/_2$$

Similarly, for $j \in [m+1, \dots, N]$

$$\gamma(x_j) \geq \gamma(x_i) \quad \leftrightarrow \sum_{i=1}^{j-1} p_i \leq 1/2 \;\wedge\; \sum_{i=j}^{N} p_i \geq 1/2 \;\wedge\; \sum_{i=j+1}^{N} p_i \leq 1/2$$

***Q.E.D***

**Lemma B1.3** Let $\lambda = (\lambda_1, \lambda_2)$ be the LI of the random variable $X = \{x_1, \dots, x_N\}$ and $P(X) = \{x_1, \dots, x_N\}$ its probability mass function. Then:

$$arg \max_{P(X)} \{\lambda(P(X)\} = \delta(X) \;\rightarrow\; Max\{\lambda(P(X))\} = N$$
$$arg \min_{P(X)} \{\lambda(P(X)\} = \rho(X) \;\rightarrow\; min\{\lambda(P(X))\} = \frac{N+1}{2}$$

*Proof:*

Since the BCF is a linear convex combination:

$$\gamma(x_j) = \sum_{i=1}^{N} w_i \cdot p_i : \qquad arg \max_{w_i} \{\gamma(x_j)\} = N = w_j$$

It follows that the maximum value is attained when the coefficients $p_i$ of the polynomial are all set to zero execpt for $p_j = 1$. In other words:

$$Max\{\gamma(x_j)\} = N \quad \leftrightarrow \quad p_j = 1 \;\wedge\; p_i = 0 \quad \forall \; i \neq j \quad \rightarrow \lambda = (j, j)$$

On the other hand, the minimum value is attained when all the bins have exactly the same concentration value. This situation occurs when $p_1 = p_N = 0.5$.

$\gamma(1, N) = N \cdot p_1 + (N-1) \cdot p_2 + \cdots + 2 \cdot p_{N-1} + p_N$
$\vdots$
$\gamma(j, N) = (N - j + 1) \cdot p_1 + \cdots + (N-1) \cdot p_{j-1} + N \cdot p_j + \cdots . + (j-1) \cdot p_{N-1} + j \cdot p_N$
$\vdots$
$\gamma(N, N) = p_1 + 2 \cdot p_2 + \cdots + (N-1) \cdot p_{N-1} + N \cdot p_N$

Substituiting $p_1 = p_N = 0.5$.

$\gamma(1, N) = N/2 + 1/2 = \frac{N+1}{2}$
$\gamma(2, N) = (N-1)/2 + 1 = \frac{N+1}{2}$
$\vdots$
$\gamma(j, N) = (N-j+1)/2 + j/2 = \frac{N+1}{2}$
$\vdots$

$$\gamma(N,N) = {(N-j)}/{2} + {(j+1)}/{2} = \frac{(N+1)}{2} \quad \rightarrow \lambda = (1,N)$$

***Q.E.D***

**Lemma B1.4:** Let $\lambda = (\lambda_1, \lambda_2)$ be the LI of the random variable $X = \{x_1, \dots, x_N\}$ and $C = \gamma(\lambda)$ be the Bin concentration value. Then the normalized value is given by the following formula:

$$Bin\ Compactness\colon\ C_\lambda = \frac{2C-N-1}{N-1} \qquad Bin\ Compactness\ Deviation\colon \overline{C_\lambda} = \frac{2(N-C)}{N-1}$$

*Proof:*
According to Lemma A6.3:

$$Max\{\lambda(P(X)\} = N \quad \min\{\lambda(P(X)\} = \frac{(N+1)}{2}$$

It follows that:

$$C_\lambda = \frac{C - \min\{\lambda(P(X)\}}{Max\{\lambda(P(X)\} - \min\{\lambda(P(X)\}} = \frac{2C-N-1}{N-1}$$

It follows that

$$C_\lambda = 1 \quad \leftrightarrow \quad P(X) = \delta(X) \ \rightarrow C = N$$
$$C_\lambda = 0 \quad \leftrightarrow \quad P(X) = \rho(X) \ \rightarrow C = \frac{N+1}{2}$$

As for the Bin compactness deviation:

$$\overline{C_\lambda} = 1 - C_\lambda = \frac{2(N-C)}{N-1}$$

***Q.E.D***

**Lemma B1.5**: Let $\gamma(x_j)$ be the BCF of the random variable $X = \{x_1, \dots, x_j, .., x_N\}$ and $\delta(x_j)$ be the average absolute deviation. Then:

$$arg\max_j\{\gamma(x_j)\} = arg\min_j\{\delta(x_j)\} \quad j \in \{1, \dots, N\}$$

*Proof:*

For sake of simplicity let’s dentoe:

$$\hat{\delta}(x_j) = \sum_{i=1}^{N} P(X = x_i) \cdot |i-j|$$

It follows that:

$$\delta(x_j) = \frac{1}{N}\hat{\delta}(x_j)$$

$$\hat{\delta}(x_j) = \begin{cases} \sum_{i=0}^{N-1} i \cdot p_{i+1} & j = 1 \\ \sum_{i=0}^{j-1}(j-1-i) \cdot p_{i+1} + \sum_{i=j}^{N-1}(i-j+1) \cdot p_{i+1} & 1 < j < N \\ \sum_{i=0}^{N-1}(N-1-i) \cdot p_{i+1} & j = N \end{cases}$$

Let $j = 1 \rightarrow \hat{\delta}(x_1) = \sum_{i=0}^{N-1} i \cdot p_{i+1}$

$$\gamma(x_1) = N \cdot p_1 + (N-1) \cdot p_2 + \cdots + (N-1+1) \cdot p_i + \cdots + p_N$$
$$= N \cdot \sum_{i=1}^{N} p_i - \sum_{i=0}^{N-1} i \cdot p_{i+1} = N - \hat{\delta}(x_1)$$

Let $p_i = P(X = x_i)$ and $P = \{p_1, \dots, p_i, \dots, p_N\}$ be the mass probability of the random variable $X = \{x_1, \dots, x_i, \dots, x_N\}$. Then, since $\max f(x) = \min f(-x)$, it follows that:

$\arg\max_P\{\gamma(x_1)\} = arg\,\min_P\{\gamma(-x_1)\} = arg\,\min_P\{\hat{\delta}(x_1)\}$ $\qquad \gamma(-x_1) = N + \hat{\delta}(x_1)$

Let $j = N \rightarrow \hat{\delta}(x_N) = \sum_{i=0}^{N-1}(N-1-i) \cdot p_{i+1}$

$$\gamma(x_N) = p_1 + 2p_2 + \cdots + i \cdot p_i + \cdots + (N-1)p_{N-1} + \cdots + Np_N$$
$$= N \cdot \sum_{i=1}^{N} p_i - \sum_{i=0}^{N-1}(N-1-i) \cdot p_{i+1} = N - \hat{\delta}(x_N)$$

It follows that:

$$arg\,\max_P\{\gamma(x_N)\} = arg\,\min_P\{\gamma(-x_N)\} = arg\,\min_P\{\hat{\delta}(x_N)\}$$

Let $1 < j < N$

$$\hat{\delta}(x_j) = \sum_{i=0}^{j-1}(j-1-i) \cdot p_{i+1} + \sum_{i=j}^{N-1}(i-j+1) \cdot p_{i+1}$$

$$\gamma(x_j) = \sum_{i=0}^{j-1}(i+N-j+1) \cdot p_{i+1} + \sum_{i=j}^{N-1}(N-1+j-1) \cdot p_{i+1}$$

$$\sum_{i=0}^{j-1}(i+N-j+1)\cdot p_{i+1} = (N-j+1)\cdot p_1 + (N-j+2)\cdot p_2 + \cdots + N\cdot p_j$$

$$= N\sum_{i=1}^{j} p_i - \sum_{i=0}^{j-1}(j-1-i)\cdot p_{i+1}$$

$$\sum_{i=j}^{N-1}(N-i+j-1)\cdot p_{i+1} = (N-1)\cdot p_{j+1} + (N-2)\cdot p_{j+2} + \cdots + (j-1)\cdot p_{N-1} + j\cdot p_N$$

$$= N\cdot\sum_{i=j+1}^{N} p_i - \sum_{i=j}^{N-1}(i-j+1)\cdot p_{i+1}$$

$$\begin{aligned}\gamma(x_j) &= N\cdot\sum_{i=0}^{N} p_i - \sum_{i=0}^{j-1}(j-1-i)\cdot p_{i+1} - \sum_{i=j}^{N-1}(i-j+1)\cdot p_{i+1} \\ &= N - \sum_{i=0}^{j-1}(j-1-i)\cdot p_{i+1} - \sum_{i=j}^{N-1}(i-j+1)\cdot p_{i+1} \\ &= N - \hat{\delta}(x_j)\end{aligned}$$

It follows that:

$$arg\max_P\{\gamma(x_j)\} = arg\min_P\{\gamma(-x_j)\} = arg\min_P\{\hat{\delta}(x_j)\}$$

Where:

$\gamma(-x_j) = N + \hat{\delta}(x_j)$

***Q.E.D***

# Appendix C - PWAVGS

We constructed the IRSD for the SA3 geography using population weigthed averages of the constituent SA1s. We used the following formula:

$$AVGS_{SA3_j} = \frac{\sum_{i=1}^{N}(INDEX_{SA1_i} \times Pop_{SA1_i})}{Pop_{SA3_j}}$$

Where:

$INDEX_{SA1_i} = index\ for\ each\ SA1\ \in SA3_j$
$Pop_{SA1_i} \quad = population\ for\ each\ SA1\ \in SA3_j\ that\ recieved\ an\ index\ score$
$N \qquad = total\ number\ of\ SA1s\ (with\ index\ score)\ in\ the\ SA3_j$
$Pop_{SA1_j} \quad = the\ sum\ of\ the\ populations\ from\ the\ constituent\ SA1s\ that\ recieved\ an\ index\ score.$

Caution: Although the SA3 level index were constructed from standardised SA1 level index, they were not standardise themselves. Therefore, the SA3 level area index does not have a mean of 1000 and standard deviation of 100 just like the SA1s distribution. For this reason we standardized the new score and ranked the SA3s following the ABS score table.
More precisely, we computed the Z-score first and then standardized with a mean of 1000 and standard deviation of 100:

$$Z = \frac{X - \mu}{\sigma} \quad \rightarrow \quad \hat{X} = 1000 + 100 \cdot Z$$

Where

$X$: $AVG_{SA3_j}$ $is\ the\ raw\ score\ of\ the\ SA3$
$\mu$: $is\ the\ mean\ of\ the\ SA3\ level\ area\ index$ (995.14)
$\sigma$: $is\ the\ std\ of\ the\ SA3\ level\ area\ index$ (65.03)

| | $\hat{X}$ |
|---|---|
| Decile | Score |
| 1 | 121 to < 874 |
| 2 | 874 to < 931 |
| 3 | 931 to < 968 |
| 4 | 968 to < 997 |
| 5 | 997 to < 1020 |
| 6 | 1020 to < 1041 |
| 7 | 1041 to < 1061 |
| 8 | 1061 to < 1081 |
| 9 | 1081 to < 1104 |
| 10 | 1104 to < 1194 |

# Appendix D - Skewness Classification

The Pearson's moment coefficient of skewness of a data is given by the following formula:

$$\gamma_1 = \frac{E[(X-\mu)^3]}{\left(E[(X-\mu)^2]\right)^{3/2}} = \frac{\mu_3}{\sigma^3}$$

Where:

$$\mu_3 = \sum_{j=1}^{N} p_j \cdot (j-\mu)^3 \qquad \mu = \sum_{j=1}^{N} p_j \cdot j \qquad \sigma = \sqrt{\sum_{j=1}^{N} p_j \cdot (j-\mu)^2}$$

If $\gamma_1 = 0$, the data are perfectly symmetrical: But a skeweness of exactly zero is quite unlikely for real-world data, so Bulmer (1979) (Bulmer, Principles of Statistics p.63, 1979) suggests this rule of thumb:

- $\gamma_1 < -1 \quad \vee \quad \gamma_1 > 1$ : *Highly Skewed Distribution*
- $-1 \leq \gamma_1 < -0.5 \quad \vee \quad 0.5 < \gamma_1 \leq 1$ : *Moderatly Skewed Distribution*
- $-0.5 \leq \gamma_1 \leq 0.5$ : *Approximately Symmetric*

**Approximately Symmetric classification**

- $-0.2 \leq \gamma_1 \leq 0.2$ : *Highly Symmetric*
- $-0.3 \leq \gamma_1 < -0.2 \quad \vee \quad 0.2 < \gamma_1 \leq 0.3$ : *Moderately Symmetric*
- $-0.4 \leq \gamma_1 < -0.3 \quad \vee \quad 0.3 < \gamma_1 \leq 0.4$ : *Marginal Symmetric*
- $-0.5 \leq \gamma_1 < -0.4 \quad \vee \quad 0.4 < \gamma_1 \leq 0.5$ : *Approximately Skewed*

# Appendix E - Entropy Index

Let $P(X) = \{p_1, \dots, p_i, \dots, p_N\}$ be the probability mass function of the random variable $X = \{x_1, \dots, x_i, \dots., x_N\}$. Then the Entropy index of the random variable $X$ is given by the following formula:

$$E(X) = 1 - H(X) \qquad H(X) = -\sum_{i=1}^{N} p_i \cdot \log_N p_i$$

# Appendix F - IRSD variable specifications

This appendix gives description of each variable considered for inclusion in the IRSD. The description of the variable proportion is followed by two bullet points; the first is a description of the numerator, the second is a description of the denominator. The square brackets contain specifications for creating the numerator/denominator from Census data items, according to the mnemonics used in the *Census Dictionary, 2011, cat. No. 2901, ABS, Canberra.* The variables are arranged by socio-economic dimension.

Income variables:

| | |
|---|---|
| INC_LOW | % PEOPLE WITH STATED ANNUAL HOUSEHOLD EQUIVALISED INCOME BETWEEN $1 AND $20,799 (approx.. 1st and 2nd deciles)<br>• Number of people living in classifiable occupied private dwellings with stated annual household equivalized income between $1 and $20,799 [HEID = 03-05].<br>• Number of people living in classifiable occupied private dwellings with stated household equivalized income [HEID = 01-12]. |

Education variables:

| | |
|---|---|
| NOEDU | % PEOPLE AGRED 15 YEARS AND OVER WHO HAVE NO EDUCATIONAL ATTAINMENT. |
| | • Number of people aged 15 years and over whose highest level of education is no educational attainment [AGEP > 14 and HEAP = 998].<br>• Number of people aged 15 years and over (excluding highest level of education not stated) [AGEP > 14 and HEAP ne 001, @@@,VVV, &&&] |
| NOYR12ORHIGHER | % PEOPLE AGED 15 YEARS AND OVER WHOSE HIGHEST LEVEL OF EDUCATION IS YEAR II OR LOWER.<br>• Number of people aged 15 years and over whose highest level of education is year 11 or lower (includes certificate I and II qualifications; excludes those still at secondary school) [AGEP>14 and HEAP=50,52,613,621,622,067,998 and TYPP ne 31,32,33].<br>• Number of people aged 15 years and over (excluding highest level of education not stated) [AGEP>14 and HEAP ne 001,@@@,VVV,&&&] |

Employment variables:

| | |
|---|---|
| UNEMPLOYED | % PEOPLE (IN THE LABOUR FORCE) WHO ARE UNEMPLOYED. |
| | • Number of people aged 15 years and over who are unemployed and looking for work [LSF=4-5]<br>• Number of people aged 15 years and over in the labour force [LSF=1-5] |

Occupation variables:

| | |
|---|---|
| OCC_DRIVERS | % EMPLOYED PEOPLE CLASSIFIED AS MACHINERY OPERATORS AND DRIVERS.<br>• Number of employed people classified as Machinery operators and Drivers [OCCP=7]<br>• Number of employed people with a stated occupation [OCCP=1-8] |
| OCC_LABOUR | % EMPLOYED PEOPLE CLASSIFIED AS LABOURERS.<br>• Number of employed people classified as Labourers [OCCP=8]<br>• Number of employed people with a stated occupation [OCCP=1-8] |
| OCC_SERVICE_L | % EMPLOYED PEOPLE CLASSIFIED AS LOW-SKILL COMMUNITY AND PERSONAL SERVICE WORKERS.<br>• Number of employed people classified as Low-skill Community and Personal Service Workers [OCCP=4 and Skill Level=4-5]. |
| | • Number of employed people with a stated occupation [OCCP=1-8] |

Housing variables:

| | |
|---|---|
| LOWRENT | % OCCUPIED PRIVATE DWELLINGS PAYING LESS THAN $166 PER WEEK IN RENT (EXCLUDING $0 PER WEEK)<br>• Number of rented classifiable occupied private dwellings with rent payments less than $166 per week (excluding rent-free and renting from employer) [RNTD=1-165 and HHCD=11-32 and LLDD ne 51,52].<br>• Number of classifiable occupied private dwellings (excluding those with tenure not stated, mortgage not stated and rent not stated]{TEND ne &,@,MRED ne &&&&, RNTD ne &&&& and HHCD=11-32] |
| OVERCROWD | % OCCUPIED PRIVATE DWELLINGS REQUIRING ONE OR MORE EXTRA BEDROOMS (BASED ON CANADIAN NATIONAL OCCUPANCY STANDARD)<br>• Number of classifiable occupied private dwellings needing one or more extra bedrooms (based on Canadian National Occupancy Standard) [Housing utilization ='one or more extra bedrooms needed' and HHCD=11-32]. |
| | • Number of classifiable occupied private dwellings (excluding dwellings where housing utilization cannot be determined or is not stated)[Housing utilization ne 'Not applicable' 'Unable to be determined', 'Not stated' and HHCD=11-32] |

Other

| | |
|---|---|
| CHILDJOBLESS | % FAMILIES WITH CHILDREN UNDER 15 YEARS OF AGE AND JOBLESS PARENTS.<br>• Number of families with children aged under 15 years and jobless parents [ FMCF=21,31andLFSF=16,17,19,25,26]<br>• Number of families (excluding not applicable and not stated) [FMCF ne @@@@ and LFSF ne 06,11,15,18,20,21,27,@@] |
| DISABILITYU70 | % PEOPLE AGED UNDER 70 WHO NEED ASSISTANCE WITH CORE ACTIVITIES.<br>• Number of people aged under 70 years needing assistance in one or more of the three core activity areas of self-care, mobility and communication, because of a disability, long term health condition (lasting six months or more) or old age [AGEP < 70 and ASSNP = 1].<br>• Number of people aged under 70 years (excluding need for assistance not stated) [AGEP < 70 and ASSNP=1-2]. |
| ENGLISHPOOR | % PEOPLE WHO DO NOT SPEAK ENGLISH WELL.<br>• Number of people aged 5 years and over who speak English either not well or not at all [AGEP > 4 and ENGLP=4,5].<br>• Number of people aged 5 years and over (excluding those who did not state their English proficiency or main language) [AGEP > 4 and ENGLP=1-5] |
| ONEPARENT | % FAMILIES THAT ARE ONE PARENT FAMILIES WITH DEPENDENT OFFSPRING ONLY.<br>• Number of families that are one parent families with dependent offspring only [FMCF = 3112,3122,3212].<br>• Number of families [FMCF ne @@@@] |
| NOCAR | % OCCUPIED PRIVATE DWELLINGS WITH NO CARS.<br>• Number of classifiable occupied private dwellings which did not have a registered motor vehicle at or near the dwelling [VEHD=0 and HHCD=11-32].<br>• Number of classifiable occupied private dwellings (excluding number of vehicles not stated) [VEHD ne &&, @@ and HHCD=11-32]. |
| SEP_DIVORCED | % PEOPLE AGED 15 AND OVER WHO ARE SEPARATED OR DIVORCED. |
| | • Number of people aged 15 years and over who are separated or divorced [MSTP=3,4].<br>• Number of people aged 15 years and over [MSTP=1-5]. |

# Glossary

- **Unwarranted variation:** unwarranted variation (geographic variation) in health care service delivery refers to medical practice pattern variation that cannot be explained by illness, medical need, or the dictates of evidence-based medicine.
- **Population distribution of area-based measures**: proportion of people (probability) of the contextual variable (age groups, socioeconomic status, remoteness category, etc) in the area of residence.
- **Census unit**: census geographic units are the administrative division defined and used by federal government statistics bureau.
- **Place effects**: place effects are contextual or environmental factors that influence health outcomes.
- **AS**: Approximately-Symmetric distribution
- **BCDF**: Bilateral Cumulative Distribution Function
- **BCDFA**: Bilateral Cumulative Distribution Autocorrelation Function
- **CDF**: Cumulative Distribution Function
- **CI:** Concentration Index
- **DI:** Divergence Index
- **ERP:** Estimated Residential Population is an estimate of all people who usually live in area at a given date.
- **HS:** Highly-Skewed distribution
- **MAD:** Mean Absolute Deviation
- **MS:** Moderately-Skewed distribution
- **PWAVGS:** Population Weighted Average Score